\documentclass[superscriptaddress,groupedaddress,nofootnoteinbib,11pt,notoc]{article}
\pdfoutput=1
\usepackage{jcappub}

\usepackage{graphicx,color}
\usepackage{appendix}
\usepackage{latexsym,amsmath,amssymb,graphicx,booktabs}
\usepackage{epsfig,latexsym}
\usepackage{hyperref}
\numberwithin{equation}{section}
\usepackage{url}

\definecolor{MyBlue}{rgb}{0.15,0.15,0.70}

\hypersetup{
colorlinks=true,
citecolor=MyBlue,
linkcolor=MyBlue,
urlcolor=MyBlue
}

\definecolor{lightgray}{gray}{0.9}

\newcommand{\dgw}{d_L^{\,\rm gw}}
\newcommand{\dem}{d_L^{\,\rm em}}

\newcommand{\DP}{\Delta_4}

\newcommand{\nn}{\nonumber}

\newcommand{\iBox}{\Box^{-1}}
\newcommand{\Stu}{St\"uckelberg }

\newcommand{\Fmn}{F_{\mu\nu}}
\newcommand{\FMN}{F^{\mu\nu}}
\newcommand{\Am}{A_{\mu}}
\newcommand{\An}{A_{\nu}}

\newcommand{\AMU}{A^{\mu}}

\renewcommand\({\left(}
\renewcommand\){\right)}
\renewcommand\[{\left[}
\renewcommand\]{\right]}

\newcommand\n{{\mbox {\boldmath $\nabla$}}}
\newcommand{\ra}{\rightarrow}

\def\lsim{\raise 0.4ex\hbox{$<$}\kern -0.8em\lower 0.62
ex\hbox{$\sim$}}

\def\gsim{\raise 0.4ex\hbox{$>$}\kern -0.7em\lower 0.62
ex\hbox{$\sim$}}

\def\lbar{{\hbox{$\lambda$}\kern -0.7em\raise 0.6ex
\hbox{$-$}}}

\newcommand\eq[1]{eq.~(\ref{#1})}
\newcommand\eqs[2]{eqs.~(\ref{#1}) and (\ref{#2})}
\newcommand\Eq[1]{Equation~(\ref{#1})}
\newcommand\Eqs[2]{Equations~(\ref{#1}) and (\ref{#2})}

\newcommand\eqst[2]{eqs.~(\ref{#1})--(\ref{#2})}
\newcommand\Eqst[2]{Eqs.~(\ref{#1})--(\ref{#2})}
\newcommand\pa{\partial}
\newcommand\p{\partial}

\newcommand\ee{\end{equation}}
\newcommand\be{\begin{equation}}
\def\bea{\begin{array}}
\def\eea{\end{array}}\def\ea{\end{array}}
\newcommand\ees{\end{eqnarray}}
\newcommand\bees{\begin{eqnarray}}
\def\nn{\nonumber}

\def\a{\alpha}

\def\s{\sigma}
\def\g{\gamma}

\def\d{\delta}

\def\eps{\epsilon}

\def\dslash{\hspace{-1mm}\not{\hbox{\kern-2pt $\partial$}}}
\def\Dslash{\not{\hbox{\kern-2pt $D$}}}
\def\pslash{\not{\hbox{\kern-2.1pt $p$}}}
\def\kslash{\not{\hbox{\kern-2.3pt $k$}}}
\def\qslash{\not{\hbox{\kern-2.3pt $q$}}}

\newcommand{\vac}{|0\rangle}
\newcommand{\cav}{\langle 0|}

\newcommand{\vk}{{\bf k}}
\newcommand{\vx}{{\bf x}}

\def\p1{{\bf p}_1}
\def\p2{{\bf p}_2}
\def\k1{{\bf k}_1}
\def\k2{{\bf k}_2}

\newcommand{\emn}{\eta_{\mu\nu}}

\newcommand{\eMN}{\eta^{\mu\nu}}
\newcommand{\eRS}{\eta^{\rho\sigma}}
\newcommand{\eMR}{\eta^{\mu\rho}}
\newcommand{\eNS}{\eta^{\nu\sigma}}
\newcommand{\eMS}{\eta^{\mu\sigma}}
\newcommand{\eNR}{\eta^{\nu\rho}}

\newcommand{\gmn}{g_{\mu\nu}}

\newcommand{\gMN}{g^{\mu\nu}}
\newcommand{\gRS}{g^{\rho\sigma}}

\newcommand{\gbMN}{\bar{g}^{\mu\nu}}

\newcommand{\hmn}{h_{\mu\nu}}
\newcommand{\hrs}{h_{\rho\sigma}}
\newcommand{\hmr}{h_{\mu\rho}}

\newcommand{\hnr}{h_{\nu\rho}}

\newcommand{\hMN}{h^{\mu\nu}}
\newcommand{\hRS}{h^{\rho\sigma}}

\newcommand{\hTTij}{h_{ij}^{\rm TT}}

\newcommand{\hatx}{\hat{\bf x}}
\newcommand{\haty}{\hat{\bf y}}
\newcommand{\hatz}{\hat{\bf z}}

\newcommand{\xim}{\xi_{\mu}}
\newcommand{\xin}{\xi_{\nu}}

\newcommand{\pam}{\pa_{\mu}}

\newcommand{\pan}{\pa_{\nu}}
\newcommand{\parho}{\pa_{\rho}}
\newcommand{\pas}{\pa_{\sigma}}
\newcommand{\paM}{\pa^{\mu}}
\newcommand{\paN}{\pa^{\nu}}
\newcommand{\paR}{\pa^{\rho}}
\newcommand{\paS}{\pa^{\sigma}}

\newcommand{\Rmn}{R_{\mu\nu}}
\newcommand{\Gmn}{G_{\mu\nu}}
\newcommand{\RMN}{R^{\mu\nu}}

\newcommand{\Tmn}{T_{\mu\nu}}
\newcommand{\Smn}{S_{\mu\nu}}

\newcommand{\TMN}{T^{\mu\nu}}

\newcommand{\tmn}{t_{\mu\nu}}

\newcommand{\dddM}{\kern 0.2em \raise 1.9ex\hbox{$...$}\kern -1.0em \hbox{$M$}}
\newcommand{\dddQ}{\kern 0.2em \raise 1.9ex\hbox{$...$}\kern -1.0em \hbox{$Q$}}
\newcommand{\dddI}{\kern 0.2em \raise 1.9ex\hbox{$...$}\kern -1.0em\hbox{$I$}}
\newcommand{\dddJ}{\kern 0.2em \raise 1.9ex\hbox{$...$}\kern-1.0em
\hbox{$J$}}
\newcommand{\dddcalJ}{\kern 0.2em \raise 1.9ex\hbox{$...$}\kern-1.0em
\hbox{${\cal J}$}}

\newcommand{\dddO}{\kern 0.2em \raise 1.9ex\hbox{$...$}\kern -1.0em
\hbox{${\cal O}$}}
\def\dddz{\raise 1.5ex\hbox{$...$}\kern -0.8em \hbox{$z$}}
\def\dddd{\raise 1.8ex\hbox{$...$}\kern -0.8em \hbox{$d$}}
\def\dddbd{\raise 1.8ex\hbox{$...$}\kern -0.8em \hbox{${\bf d}$}}
\def\ddbd{\raise 1.8ex\hbox{$..$}\kern -0.8em \hbox{${\bf d}$}}
\def\dddx{\raise 1.6ex\hbox{$...$}\kern -0.8em \hbox{$x$}}

\newcommand{\msun}{M_{\odot}}

\newcommand{\Sch}{Schwarzschild }

\newcommand{\mpl}{M_{\rm Pl}}
\newcommand{\mplr}{m_{\rm Pl}}

\newcommand{\ode}{\Omega_{\rm DE}}
\newcommand{\oma}{\Omega_{M}}
\newcommand{\ora}{\Omega_{R}}

\newcommand{\ola}{\Omega_{\Lambda}}

\newcommand{\rde}{\rho_{\rm DE}}
\newcommand{\wde}{w_{\rm DE}}

\graphicspath{ {./Graphs/} }

\title{Gravity in the infrared and effective nonlocal models}
\author[a]{Enis Belgacem,}
\author[b]{Yves Dirian,}
\author[a]{Andreas Finke,}
\author[a]{Stefano Foffa}
\author[a]{and Michele Maggiore}

\affiliation[a]{D\'epartement de Physique Th\'eorique and Center for Astroparticle Physics,\\
Universit\'e de Gen\`eve, 24 quai Ansermet, CH--1211 Gen\`eve 4, Switzerland}
\affiliation[b]{Center for Theoretical Astrophysics and Cosmology, Institute for Computational Science,
University of Z\"urich, CH-8057 Z\"urich, Switzerland}

\abstract{We provide a systematic and  updated discussion of a research line carried out by our group over the last few years, in which  gravity is modified at cosmological distances by the introduction of nonlocal terms, assumed to emerge 
at an effective level from the infrared behavior of the quantum theory. The requirement of producing a viable cosmology turns out to be very stringent and basically selects a unique  model,  in which the nonlocal term describes an effective mass for the conformal mode. We discuss how such a specific structure could emerge from a fundamental local theory of gravity, and we perform a detailed comparison of this model with the most recent cosmological datasets, confirming that it  fits current data at the same level as $\Lambda$CDM. 

Most notably, the model  has striking predictions in the sector of tensor perturbations, leading to a very large effect in the propagation of gravitational wave (GWs) over cosmological distances. At the redshifts relevant for  the next generation of GW detectors such as  Einstein Telescope, Cosmic Explorer and LISA, this leads to deviations from 
GR that could be as large as $80\%$, and could be verified with the detection of  just a single coalescing binary with electromagnetic counterpart. This would also have potentially important consequences for the search of the counterpart since, for a given luminosity distance to the source, as inferred through the GW signal,  the actual source redshift  could be  significantly different from that predicted by $\Lambda$CDM. At the redshifts
relevant for  advanced LIGO/Virgo/Kagra the effect is smaller, but still potentially observable over a few years of runs at target sensitivity.}

\begin{document}
\maketitle
\flushbottom

\section{Introduction}

The infrared (IR) dynamics of quantum field theories with massless particles can in general be highly non-trivial. In General Relativity (GR), where the coupling constant $G_N=1/\mpl^2$ has the dimension of inverse mass squared, perturbation theory is organized in powers of $G_NE^2$, with $E$ a typical energy scale, and one might think that the IR limit $E\ra 0$ is fully perturbative, contrary to the large energy limit, where an 
UV completion is needed at the latest at $E\simeq \mpl$. In fact, several lines of investigation indicate that this conclusion might be too naive. Already for pure gravity in flat space a non-trivial vacuum structure at infinity emerges, relating the asymptotic symmetries of flat space-time (the BMS group) to soft theorems and memory effects, leading to the conclusion that the vacuum in GR is not unique~\cite{Strominger:2017zoo}. In de~Sitter space the Feynman propagator grows without bounds for large separations, leading to IR divergences in gauge-invariant scattering processes~\cite{Antoniadis:1986sb}; the strongest divergence comes from the propagator of the conformal mode, a point that will be relevant in the following. Infrared divergences also  appear when performing computation of physical quantities during inflation~\cite{Tsamis:1994ca}.  
When gravity is coupled to 
massless fields,  further non-trivial dynamics for the conformal mode of the metric, $\sigma$, arises through the conformal anomaly. The latter gives a contribution to the quantum effective action that, in flat four-dimensional space-time, is proportional to $(\Box\sigma)^2$, i.e. is fourth-order in the derivatives. In turn, this leads to a propagator for the conformal mode  $G_{\sigma}(x,x')\propto \log (x-x')^2$,  which again grows without bounds in the IR, leading to the possibility of  non-trivial large-distance dynamics, including the possibility of non-trivial IR fixed points~\cite{Antoniadis:1991fa,Antoniadis:2006wq}. IR divergences are the sign that something is missing in our understanding of the physics, and that  the long-distance behavior of the theory could be different from that suggested by a perturbative analysis.  A nontrivial IR dynamics could of course be relevant for understanding the origin of dark energy, and  
indeed the idea that quantum effects in gravity could have cosmological relevance
already appeared in older works~\cite{Taylor:1989ua}.

In the presence of strong IR effects a system  often reacts by generating  a mass scale dynamically. At first, it might seem that this is precluded for the gravitational field, since a mass term would be forbidden by diffeomorphism invariance. However, once quantum effects enter into play, the relevant quantity is no longer the fundamental action of the theory, but rather the quantum effective action. Whenever the theory has massless particles, such as the graviton in GR,  the quantum effective action unavoidably develops nonlocal terms. As we will review in section~\ref{sect:NLmass}, with nonlocal terms it is  possible to construct gauge-invariant mass terms for gauge fields, and diffeomorphism-invariant mass terms for different modes of the gravitational field. 
The question then arises whether, in the IR limit of GR, a nonlocal  mass term of this form could emerge. This question is very difficult to answer, since it basically involve non-perturbative physics,\footnote{Although a possibly simpler alternative is provided by theories with extra dimensions, that, when projected onto a four-dimensional brane, can indeed induce nonlocal terms relevant in the IR, as we will illustrate below with the example of the DGP model.} and non-perturbative techniques for gravity are still far from being fully established. After recalling  in
Section~\ref{sect:aspects} some generic features that are important for understanding the proper way of dealing with nonlocal terms, in section~\ref{sect:mech} we will discuss tentative evidence, both from lattice gravity and from functional renormalization group equations, that suggests that the generation of such a mass scale is in principle possible. We will also stress that one can generate very specific nonlocal structures, rather than the most general nonlocal theory. If one assumes the generation of nonlocal terms relevant in the IR as a sensible working hypothesis, the next question is what terms can give a viable cosmology, consistent with the wealth of current data. Of course, ideally one would like  to derive the form of the nonlocal terms from first principle, and then study their consequences. Given the difficulty of such a top-down approach, it make sense to start by a study of the cosmological consequences of possible nonlocal terms. This turns out to be  already a rather interesting study; summarizing previous work our group, we will indeed see that the condition of obtaining a viable cosmology are very stringent, and basically select a single model (the so-called RT model), among a large class of alternatives that have been studied. In this model, the nonlocal term corresponds to a diff-invariant mass for the conformal mode.  We will then  perform in Section~\ref{sect:pheno} an updated and detailed study of the phenomenological consequences of the RT model, including comparison with the most recent cosmological datasets, as well as predictions for gravitational-wave (GW) detectors that could turn out to be the smoking gun of this model.

\section{Nonlocal mass terms for gauge and gravitational fields}\label{sect:NLmass}

In this section we review how one can construct nonlocal mass terms for gauge fields and for different components of the gravitational field, while preserving gauge or diffeomorphism invariance, respectively. Even if the formal manipulations might seem to hold already at the level of the classical action,  we will recall in section~\ref{sect:aspects} that these nonlocal terms only make sense at the level of quantum effective action, i.e. must be considered as generated by quantum effects (we will recall in section~\ref{sect:causality} some basic properties of the quantum effective action). Indeed, writing such nonlocal terms directly at the level of the fundamental action, one would  run into fatal problems with ghost-like degrees of freedom and with causality. In contrast, nonlocal terms routinely appears at the level of  quantum effective actions, and,  in this context,  create no problems with causality or with ghosts.

\subsection{Nonlocal and gauge-invariant mass term for gauge fields}\label{sect:massAmu}

The simplest example is given by massive electrodynamics~\cite{Dvali:2006su}. 
Consider the  Proca action for a massive photon coupled to an external conserved current $j^{\mu}$
\be\label{1Lemmass}
S=\int d^4x\( -\frac{1}{4}F_{\mu\nu}F^{\mu\nu}-\frac{1}{2}m_{\g}^2\, A_{\mu}A^{\mu}
-j_{\mu}A^{\mu}\) \, .
\ee
The equation of motion derived from this action is $\pam F^{\mu\nu}-m_{\g}^2A^{\nu}=j^{\nu}$.
Acting with $\pan$ on both sides and using  $\pan j^{\nu}=0$ one finds
$m_{\g}^2\, \pan A^{\nu}=0$, so, for  $m_{\g}\neq 0$, we get 
$\pan A^{\nu}=0$. Note that this condition, that in massless electrodynamics can be obtained as a choice of gauge, is here obtained dynamically from the equations of motion; indeed, the action (\ref{1Lemmass}) is not gauge invariant, and there is no gauge freedom to fix.
Then, using this condition, the equation of motion becomes $(\Box -m_{\g}^2) A^{\mu}=j^{\mu}$. In summary, the equations of motion derived from \eq{1Lemmass} can be written as
\be\label{BoxmA0}
(\Box -m_{\g}^2) A^{\mu}=j^{\mu}\, , \qquad \pan A^{\nu}=0\, ,
\ee
and describe the three degrees of freedom of a massive spin-1 particle. 
Now, let us compare this with the nonlocal action
\be\label{Lnonloc}
S=\int d^4x\,\[ -\frac{1}{4}\Fmn \(1-\frac{m_{\g}^2}{\Box}\)\FMN
-j_{\mu}A^{\mu}\]\, .
\ee
The corresponding equation of motion is 
\be\label{eqnonlocFMN}
\(1-\frac{m_{\g}^2}{\Box}\)\pam\FMN=j^{\nu}\, .
\ee
The action (\ref{Lnonloc}) is nonlocal but  gauge invariant. We can therefore impose $\pam A^{\mu}=0$ as a choice of gauge, and then \eq{eqnonlocFMN} reduces to $(\Box -m_{\g}^2) A^{\mu}=j^{\mu}$. We therefore get back the two equations in  (\ref{BoxmA0}), showing that \eqs{1Lemmass}{Lnonloc} provide two equivalent formulations of the same classical theory.
The equivalence of the two theories can also be shown by using  the ``\Stu trick", as originally done in \cite{Dvali:2006su}. One
 introduces the \Stu  field $\varphi$ and  
replaces 
\be\label{eqStu}
\Am\ra\Am+(1/m_{\g})\pam\varphi\, ,
\ee 
in the Proca action (\ref{1Lemmass}), which becomes
\be\label{LagStu}
S[\Am,\varphi]=\int d^4x\, \[-\frac{1}{4}F_{\mu\nu}F^{\mu\nu}-\frac{1}{2}m_{\g}^2A_{\mu}A^{\mu}
-\frac{1}{2}\pam\varphi\paM\varphi-m_{\g}\AMU\pam\varphi
-j_{\mu}A^{\mu}\]\, .
\ee
We have  added a new degrees of freedom $\varphi$, but we have gained a gauge symmetry, defined by the transformation
\be
\Am\ra \Am -\pam\theta\, ,\qquad \varphi\ra\varphi+m_{\g}\theta\, ,
\ee
since the combined transformation leaves invariant the right-hand side of \eq{eqStu}.
The equations of motion obtained from the action (\ref{LagStu})  are
\bees
\pam F^{\mu\nu}&=&m_{\g}^2A^{\nu}+m_{\g}\paN\varphi+j^{\nu}\, ,\label{QEDeqmotion1}\\
\Box\varphi&=&-m_{\g}\pam\AMU\, .\label{QEDeqmotion2}
\ees
\Eq{QEDeqmotion2} can be formally solved by
$\varphi(x)=-m_{\g}\Box^{-1}(\pam\AMU)$. Inserting this  into 
\eq{QEDeqmotion1} we get \eq{eqnonlocFMN}
[or, equivalently, inserting it into $S[\Am,\varphi]$, \eq{LagStu}, we get \eq{Lnonloc}]. Thus, \eq{Lnonloc} provides an alternative description of a massive photon which is explicitly gauge invariant, but nonlocal. 
An equivalent way of expressing the same result is to observe that the Ward identities of QED do not forbid a photon mass term~\cite{Porrati:2001db}. Indeed, they only imply that the photon self-energy $\Sigma_{\mu\nu}$ is transverse, so that, in momentum space, it can be written as 
\be
\Sigma_{\mu\nu}(p)=\(\gmn-\frac{p_{\mu}p_{\nu}}{p^2} \) F(p^2)\, .
\ee
If, in the limit $p^2\ra 0$, $F(p^2)\neq 0$, then the photon acquires a nonzero mass, which is precisely the one described by the nonlocal term in \eq{Lnonloc}.

It is interesting to rewrite the nonlocal mass term in a way that will be useful to make contact with the gravitational case~\cite{Jaccard:2013gla}. We separate the gauge field into its transverse and longitudinal parts,
\be\label{separAMU}
\Am=\Am^{\rm T}+\pam\a\, ,
\ee
where $\paM \Am^{\rm T} = 0$. Under a gauge transformations $\Am\ra\Am-\pam\theta$, we have 
$\a\ra\a-\theta$ and  $\Am^{\rm T}\ra \Am^{\rm T}$, so $\Am^{\rm T}$ is gauge invariant. To invert \eq{separAMU} we take the divergence, which gives $\paM \Am=\Box\a$. This can be formally inverted 
as $\a=\iBox \paM \Am$.
Substituting this into $\Am^{\rm T}=\Am-\pam\a$ we get
\be\label{defAhat}
\Am^{\rm T}=\Am-\frac{1}{\Box}\pam\paN\An\equiv
P_{\mu}^{\nu}A_{\nu}\, ,
\ee 
where 
\begin{equation}  \label{defPMunu}
P_{\mu}^{\nu} \equiv \delta_{\mu}^{\nu} - \frac{\partial_{\mu} \partial^{\nu}}{\Box}\, 
\end{equation}
is a nonlocal operator. The transverse part $\Am^{\rm T}$ is therefore a gauge-invariant and nonlocal functional of the  gauge field $A_{\mu}$. In terms of $\Am^{\rm T}$, it is straightforward to check that the action (\ref{Lnonloc}) can be rewritten as
\be\label{eq:massMaxw}
S = \int d^4 x \left( -\frac{1}{4} F_{\mu\nu} F^{\mu\nu} - \frac{1}{2} m_{\g}^2 \Am^{\rm T} A^{T\mu} \right) -j^{\mu}A_{\mu} \, .
\ee
We can further replace $A_{\mu}$ with $\Am^{\rm T}$ in the kinetic term, since
$F_{\mu\nu}=\pam A_{\nu}-\pan A_{\mu}=
\pam A_{\nu}^T-\pan A_{\mu}^T$, and in $j^{\mu}A_{\mu}$, since $j^{\mu}\pam\alpha$ vanishes upon integrating by parts and using current conservation. This is a consequence  of the fact that  $\alpha$ is a pure gauge degree of freedom and can be set to zero with a gauge transformation.

In the case of massive electrodynamics the nonlocality is only apparent, since we have seen that the nonlocal term in the equation of motion (\ref{eqnonlocFMN}) can be made local with the gauge choice $\pam\AMU=0$. In this sense, the above manipulations can be performed even at the level of fundamental action, since anyhow the nonlocality can be gauged away. 
Less trivial examples can be constructed where the nonlocality is genuine (and therefore, as we will review in Section~\ref{sect:aspects}, only makes sense at the level of quantum effective action). In particular, the non-abelian generalization of the nonlocal mass term in \eq{Lnonloc} is
\be\label{Fmn2}
\frac{m_g^2}{2} {\rm Tr}\, \int d^4x\,   F_{\mu\nu} \frac{1}{\Box}F^{\mu\nu}\, ,
\ee
where  $F_{\mu\nu}=F_{\mu\nu}^aT^a$, $\Box^{ab}=D_{\mu}^{ac}D^{\mu,cb}$ and $D_{\mu}^{ab}=\d^{ab}\pam-gf^{abc}A_{\mu}^c$ is the covariant derivative. This nonlocal term corresponds to giving a mass $m_g$ to the non-abelian gauge bosons, plus extra  nonlocal interaction terms that, altogether, reconstruct a gauge-invariant quantity. 
This nonlocal mass term  cannot be reduced to a local term with a gauge choice, and has been postulated to appear in the quantum effective action of QCD, in order to reproduce non-perturbative  results on the  running of the strong coupling constant and on the gluon propagator in the IR, obtained from operator product expansions and from lattice QCD~\cite{Boucaud:2001st,Capri:2005dy,Dudal:2008sp}. It is therefore an example of a nonlocal term that can appear in a quantum effective action because of strong IR effects.

\subsection{Nonlocal and diff-invariant mass term for the conformal mode}

\subsubsection{Linearized GR in nonlocal variables}

We now discuss possible generalizations of the above construction to the gravitational field. A possible route is to begin with gravity linearized over Minkowski space. First of all, it is useful to see how linearized gravity can be rewritten in terms of nonlocal variables, analogous to  $A_{\mu}^T$ of the previous section (we follow the discussion in~\cite{Jaccard:2013gla,Foffa:2013sma}).
We begin by  writing $\gmn=\emn+\kappa \hmn$, where $\kappa=(32\pi G)^{1/2}$. To quadratic level, the Einstein-Hilbert action becomes 
\be\label{Squadr}
S_{\rm EH}^{(2)}=\frac{1}{2}\int d^{4}x \,
\hmn{\cal E}^{\mu\nu,\rho\sigma}\hrs\, ,
\ee
where 
${\cal E}^{\mu\nu,\rho\sigma}$ is the Lichnerowicz operator,\footnote{The Lichnerowicz operator is defined by ${\cal E}^{\mu\nu,\rho\sigma}\equiv
\frac{1}{2}(\eMR\eNS+\eMS\eNR-2\eMN\eRS) \Box
+ (\eRS\paM\paN+\eMN\paR\paS)
-\frac{1}{2}\( \eMR\paS\paN+\eNR\paS\paM+\eMS\paR\paN+\eNS\paR\paM\)$, where $\Box=\eMN\pam\pan$ is the flat-space d'Alembertian.
We use the signature $\emn=(-,+,+,+)$ and MTW~\cite{MTW} sign conventions.} while the interaction with matter with energy-momentum tensor $\TMN$, to linear order in $\hmn$, is given by
\be
S_{\rm int}^{(1)}=\frac{\kappa}{2}\int d^4x\, \hmn\TMN\, .
\ee
The linearized equations of motion derived from $S_{\rm EH}^{(2)}+S_{\rm int}^{(1)}$ are therefore
\be\label{lineqmotion}
{\cal E}^{\mu\nu,\rho\sigma}\hrs=-\frac{\kappa}{2}\TMN\, .
\ee
We next decompose  the metric as
\be\label{decomphmn}
\hmn =\hmn^{\rm TT}+\frac{1}{2}(\pam \eps_{\nu}+\pan \eps_{\mu}) +\frac{1}{3}\emn s\, ,
\ee
where $\hmn^{\rm TT}$ is transverse ($\paM \hmn^{\rm TT}=0$) and traceless  ($\eMN \hmn^{\rm TT}=0$), and therefore has five independent components.\footnote{We assume here $3+1$ spacetime dimensions. See \cite{Foffa:2013sma} for the corresponding equations in $d+1$ spacetime dimensions, with $d$ generic.}
 We have therefore decomposed the 10 independent components of the symmetric tensor $\hmn$ into the  five  components of $\hmn^{\rm TT}$, the
four components of $\eps_{\mu}$, and the scalar $s$. 
Under a linearized diffeomorphism
$\hmn\ra\hmn -(\pam\xin+\pan\xim)$ we have $\eps_{\mu}\ra\eps_{\mu}-\xi_{\mu}$ while  
$\hmn^{\rm TT}$ and $s$ are gauge invariant. Thus $\eps_{\mu}$ describes the four pure gauge degrees of freedom, while  $s$ plus the five components of the TT tensor $\hmn^{\rm TT}$ describe the six gauge-invariant degrees of freedom of the gravitational field. Notice that, at this linearized level, $s$ is equivalent to the conformal mode of the metric. Indeed, restricting  to the scalar sector (i.e. setting $\eps_{\mu}=0$ and $\hmn^{\rm TT}=0$) and writing $\gmn =e^{2\sigma}\emn$, comparison with \eq{decomphmn} shows that, at the linear level, $2\sigma=s/3$.

Similarly to the electromagnetic case of section~\ref{sect:massAmu}, the quantities that appear in the right-hand side of \eq{decomphmn} are nonlocal functionals of the original metric perturbation $\hmn$. 
The inversion of  \eq{decomphmn} is straightforward~\cite{Jaccard:2013gla}. It is convenient to further separate $\eps_{\mu}$ into its transverse and longitudinal parts,
$\eps_{\mu}=\eps_{\mu}^{\rm T}+\pam\a$, where $\paM \eps_{\mu}^{\rm T}=0$. Then, 
taking the trace of 
\eq{decomphmn} we get
 $h=(4/3)s+\Box\a$, while contracting \eq{decomphmn} with $\paM\paN$, gives
 $\paM\paN\hmn=\Box [s/3+\Box\a]$. Combining these equations we get 
 \be\label{defshmn}
 s=\(\eMN-\frac{1}{\Box}\paM\paN\)\hmn\, ,\qquad\qquad
 \a=-\frac{1}{3}\, \frac{1}{\Box}\(\eMN-\frac{4}{\Box}\paM\paN\)\hmn\, .
 \ee
We can now extract $\eps_{\mu}^{\rm T}$ by applying $\paM$ to \eq{decomphmn} and using the above expressions for $\a$ and $s$. This gives $\eps_{\mu}^T=2\iBox P_{\mu}^{\rho}\paS\hrs$.
Finally, substituting these expressions into \eq{decomphmn} we get
\bees
\hmn^{\rm TT}&=&\hmn -\frac{1}{3}\(\emn -\frac{\pam\pan}{\Box}\)h
-\frac{1}{\Box}(\pam\paR\hnr+\pan\paR\hmr)+
\frac{1}{3}\,\emn \frac{1}{\Box}\paR\paS\hrs \nn\\
&&+\frac{2}{3}\frac{1}{\Box^2}\pam\pan\paR\paS\hrs\, .\label{defhath}
\ees
These results can be written more compactly using the projector $P^{\mu\nu}=\eMN-(\paM\paN/\Box)$. In particular,
\bees
s&=&P^{\mu\nu}\hmn\, ,\label{invert:s}\\
\hmn^{\rm TT}&=&\(  P_{\mu}^{\rho}P_{\nu}^{\sigma}  -\frac{1}{3}P_{\mu\nu}P^{\rho\sigma}\)\hrs\, .\label{invert:h}
\ees
The fact that this expression for $\hmn^{\rm TT}$ is indeed transverse and traceless is easily checked by using the properties of $P_{\mu\nu}$,
$\paM P_{\mu\nu}=0$, $\eMN P_{\mu}^{\rho}P_{\nu}^{\sigma}=P^{\rho\sigma}$ and $\eMN P_{\mu\nu}=3$.

Plugging the decomposition (\ref{decomphmn}) into the action (\ref{Squadr}) one finds that $\eps_{\mu}$ cancels (an obvious consequence of  the fact that it  is a pure gauge mode), and~\cite{Foffa:2013sma}
\be\label{SEH2nl}
S_{\rm EH}^{(2)}=\frac{1}{2}\int d^{4}x \,\[\hmn^{\rm TT}\Box (h^{\mu\nu })^{\rm TT}
-\frac{2}{3}\, s\Box s\]\, .
\ee
Performing the same decomposition as in (\ref{decomphmn}) for  the energy-momentum tensor, the linearization of the interaction term becomes
\be\label{Sint2}
S_{\rm int}^{(1)} 
=\frac{\kappa}{2}\int d^4x\,\[ \hmn^{\rm TT}(\TMN)^{\rm TT}+\frac{1}{3}sT\]
\, ,
\ee
where $T=\eMN\Tmn$.
The equations of motion (\ref{lineqmotion}) derived from $S_{\rm EH}^{(2)}+S_{\rm int}^{(1)}$ can then be rewritten as 
\be\label{BoxhTT}
\Box\hmn^{\rm TT}=-\frac{\kappa}{2}\Tmn^{\rm TT}\, ,\qquad\qquad
\Box s=\frac{\kappa}{4}T\, .
\ee
At first, \eq{BoxhTT} can be surprising, because it seems to imply that  $\hmn^{\rm TT}$ and  $s$ describe six radiative gauge-invariant degrees of freedom. Of course,  we know that in GR only the two degrees of freedom associated to the helicities $\pm 2$ are radiative, while the remaining four gauge-invariant degrees of freedom are non-radiative and satisfy Poisson equations.
Furthermore, the sign of the kinetic term of $s$ in \eq{SEH2nl} is such that
the scalar $s$ seems to be a ghost! As discussed in \cite{Foffa:2013sma}, the resolution of this apparent paradox is related to the nonlocal relation between the original metric perturbation $\hmn$ and the variables
$\{\hmn^{\rm TT},s\}$. The fact that this relation is nonlocal in time, and not only in space, implies that  the initial data assigned on $\hmn$ on a given time slice are not sufficient to provide initial data on $\{\hmn^{\rm TT},s\}$, so a naive counting of degrees of freedom in terms of $\{\hmn^{\rm TT},s\}$ goes wrong.\footnote{A simple  example to understand  what exactly goes wrong, again discussed in  \cite{Foffa:2013sma}, is provided by a scalar field $\phi$ that satisfies a Poisson equation  $\n^2\phi=\rho$. If one introduces a  field $\tilde{\phi}$ related to $\phi$ by a nonlocal relation such as  $\tilde{\phi}=\iBox\phi$,  the original  Poisson  equation  can be rewritten as $\Box\tilde{\phi}=\tilde{\rho}$, where $\tilde{\rho}\equiv\n^{-2}\rho$,
so now $\tilde{\phi}$ looks like a propagating degree of freedom. However, for $\rho=0$ the original equation $\n^2\phi=\rho$ (with vanishing boundary conditions at infinity) only has the solution $\phi=0$. If we want to rewrite this equation in terms of $\tilde{\phi}$ without introducing spurious degrees of freedom we must therefore supplement the equation $\Box\tilde{\phi}=\tilde{\rho}$ with the condition that, when $\tilde{\rho}=0$, the only acceptable solution is $\tilde{\phi}=0$, which precisely kills the radiative solution.
\label{foot:spurious}}
Notice that this is different from what happens in the standard $3+1$ decomposition of the metric perturbations over flat space,
\bees
h_{00}&=&2\psi\, ,\qquad\qquad
h_{0i}=\beta_i+\pa_i\gamma\nn\\
h_{ij}&=&-2\phi\d_{ij}+\(\pa_i\pa_j-\frac{1}{3}\d_{ij}\n^2\)\lambda 
  +\frac{1}{2}(\pa_iv_j+\pa_jv_i)
 +H_{ij}^{\rm TT}\, ,\label{hijphilambdaeps}
\ees
where $v^i$ and $\beta^i$ are transverse spatial vectors, $\pa_i\beta^i=0$ and $\pa_iv^i=0$, and $H_{ij}^{\rm TT}$ is transverse and traceless with respect to the spatial indices,
$\pa^jH_{ij}^{\rm TT}=0$ and
$\d^{ij}H_{ij}^{\rm TT}=0$. Indeed, this decomposition  only involves spatial derivatives and therefore its inversion is nonlocal in space but local in time. From these variables, one can form six variables that are invariant under linearized gauge transformations: 
the  two Bardeen variables, 
$\Phi=-\phi-(1/6)\n^2\lambda$ and
$\Psi = \psi -\dot{\gamma}+(1/2)\ddot{\lambda}$, that are scalars under spatial rotations; 
the spatial vector 
$\Xi_i=\beta_i-(1/2)\dot{v}_i$, which, being transverse, has only two independent  components; and the spatial tensor $H_{ij}^{\rm TT}$, which is already gauge-invariant (again, at the linearized level) and, being transverse and traceless (and carrying only spatial indices, contrary to $\hmn^{\rm TT}$), also has only two independent components. Standard analysis (see e.g. \cite{Flanagan:2005yc,Jaccard:2012ut} or
chapter~18 of \cite{Maggiore:2018sht}) then shows that, after performing the same decomposition for the energy-momentum tensor, namely
\bees
T_{00}&=&\rho\, ,\qquad\qquad
T_{0i}=S_i+\pa_iS\, ,\nn\\
T_{ij}&=&P\d_{ij}+\(\pa_i\pa_j-\frac{1}{3}\d_{ij}\n^2\)\Sigma   +\frac{1}{2}(\pa_i\Sigma_j+\pa_j\Sigma_i)+\Sigma_{ij} \label{Tij3plus1}\, ,
\ees
where $\pa_i\Sigma^i=0$,  $\pa_i S^i=0$, 
$\pa^i\Sigma_{ij}=0$ and $\d^{ij}\Sigma_{ij}=0$, the linearized equations of motion can be rewritten as
\bees
\n^2\Phi &=&-4\pi G\rho\, ,\qquad\qquad
\n^2\Psi =-4\pi G (\rho-2\n^2\Sigma)\, ,\label{n2Phin2Psi}\\
\n^2\Xi_i&=&-16\pi G S_i\, ,\qquad\qquad
\Box H_{ij}^{\rm TT}=-16\pi G \Sigma_{ij}\label{hijsij}\, .
\ees
We then get the standard result that only the  two degrees of freedom of the tensor perturbations obey a wave equation, while the remaining  gauge-invariant degrees of freedom described by $\Phi$, $\Psi$ and $\Xi_i$
obey  Poisson equations, and therefore are non-radiative. 

Comparing the decompositions (\ref{decomphmn}) and (\ref{hijphilambdaeps}) one finds that 
the field $s$ can be written explicitly as a nonlocal function of the Bardeen variables as~\cite{Foffa:2013sma}
\be\label{siBoxPhi}
s=6\Phi- 2\iBox\n^2(\Phi+\Psi)\, .
\ee
Just as in the example discussed in footnote~\ref{foot:spurious}, the apparent radiative nature of $s$ in \eq{BoxhTT} 
is an artifact due to this nonlocal relation, that introduces a spurious degree of freedom associated to the homogeneous equation $\Box s=0$. Indeed, from \eq{n2Phin2Psi}, $\n^2(\Phi+\Psi)$ is fully determined by the source terms, and vanishes if the latter vanish.
Thus, in order to eliminate this spurious degree of freedom   we must supplement
\eq{BoxhTT} with the condition that $s=0$ when $T=0$, i.e. we must discard again the homogeneous solution of \eq{BoxhTT} (and similarly for the helicities  $0,\pm 1$ of $\hmn^{\rm TT}$)
At the quantum level, this implies that there are no creation and annihilation operators associated to $s$, and  $s$ cannot appear on the external legs of a Feynman diagram. Therefore, the apparent ghost-like nature of $s$ in \eq{SEH2nl} is fictitious and, of course, in General Relativity there is no actual ghost.

\subsubsection{Nonlocal mass terms at the linearized level}

As we have seen,  the use of the variables $\{\hmn^{\rm TT},s\}$ is not convenient if we want to count the independent degrees of freedom of the theory and  determine their radiative/non-radiative nature; for those purposes it is better to work directly with the original metric perturbation $\hmn$, or with the variables of the 3+1 decomposition (\ref{hijphilambdaeps}), or with the  ADM decomposition. However, the variables $\{\hmn^{\rm TT},s\}$ have the advantage that one can very easily see   how a  diff-invariant nonlocal mass term can be naturally written for  different modes of the gravitational field, at the linearized level. Quite trivially, we can just modify \eq{SEH2nl} into
\be\label{Gamma2hhssm}
\Gamma^{(2)}=\frac{1}{2}\int d^{4}x \,\[\hmn^{\rm TT}(\Box-m_1^2) (h^{\mu\nu })^{\rm TT}
-\frac{2}{3}\, s(\Box+m_2^2) s\]\, ,
\ee
for some masses $m_1$ and $m_2$.\footnote{We have chosen the signs in front of $m_1^2$ and $m_2^2$ so that $m_1^2>0$ and $m_2^2>0$ corresponds to `non-tachyonic' masses, independently of the signs in front of the $\Box$ operators.}  This is  analogous to \eq{eq:massMaxw} in the case of a massive gauge field.
These mass terms are clearly diff-invariant, since $\hmn^{\rm TT}$ and $s$ are diff-invariant (again, at the linearized level). On the other hand, because of the relations (\ref{invert:s},\ref{invert:h}), once rewritten in terms of $\hmn$ they will be nonlocal.  We have indeed used the notation $\Gamma$, rather than $S$, to stress that, because of the nonlocality, this modification makes sense at the level of  the quantum effective action $\Gamma$, rather than for the fundamental action $S$.

To go beyond the linearized approximation, we can  search for covariantizations of these expressions, as we will do in Section~\ref{sect:cov}. The second term gives a mass to $s$ or, equivalently, to the conformal mode. 
The models that we will study in the following will be covariantizations of the above expression, with
$m^2_1=0$ and $m_2^2\equiv m^2 >0$. We are therefore assuming that there exists a mechanism that, in the quantum effective action, generates a mass  for the conformal mode, while leaving  $\hmn^{\rm TT}$ massless. At the phenomenological level this is required by the fact that, as we will review in app.~\ref{sect:difficulties}, among a large class of models explored, only those of this form appear to have a viable cosmological evolution. At the theoretical level, this is also suggested by various arguments, that will be discussed in section~\ref{sect:mech}, that identify the conformal mode as the main candidate for producing strong IR quantum effects. Thus, we will look for a covariantization of a quantum effective action that, at quadratic level, has the form
\be\label{SEH2nlmass}
\Gamma^{(2)}=\frac{1}{2}\int d^4x \,\[\hmn^{\rm TT}\Box (h^{\mu\nu })^{\rm TT}
-\frac{2}{3}\, s(\Box+m^2) s\]\, ,
\ee
so that the  linearized equations of motion (\ref{BoxhTT}) are modified into\footnote{As we will recall in section~\ref{sect:causality}, the quantum effective action actually generates the equations of motion for the vacuum expectation values of the corresponding operators.}
\be\label{BoxhTTmassive}
\Box\hmn^{\rm TT}=-\frac{\kappa}{2}\Tmn^{\rm TT}\, ,\qquad\qquad
(\Box +m^2)s=\frac{\kappa}{4}T\, .
\ee
To perform the covariantization, it is now convenient to go back to the original metric perturbation $\hmn$. 
Using \eq{invert:s}, we immediately see that \eq{SEH2nlmass} can be rewritten   as
\be
\Gamma^{(2)}=\frac{1}{2}\int d^{4}x \,
\[ \hmn{\cal E}^{\mu\nu,\rho\sigma}\hrs 
-\frac{2}{3}\, m^2(P^{\mu\nu}\hmn)^2  \]\, ,
\ee
while \eq{BoxhTTmassive} is equivalent to
\be\label{eqmotnl}
{\cal E}^{\mu\nu,\rho\sigma}\hrs 
-\frac{2}{3}\, m^2 P^{\mu\nu}P^{\rho\sigma}\hrs =-\frac{\kappa}{2}\TMN\, .
\ee
In view of the covariantization, it is also convenient to rescale $\hmn\ra \hmn/\kappa$, so that now $\gmn=\emn+\hmn$, 
\be\label{SquadrG}
\Gamma^{(2)}=\frac{1}{64\pi G}\int d^{4}x \,
\[ \hmn{\cal E}^{\mu\nu,\rho\sigma}\hrs 
-\frac{2}{3}\, m^2 (P^{\mu\nu}\hmn)^2  \]\, ,
\ee
and 
\be\label{eqmotnlG}
{\cal E}^{\mu\nu,\rho\sigma}\hrs 
-\frac{2}{3}\, m^2 P^{\mu\nu}P^{\rho\sigma}\hrs =-16\pi G\, \TMN\, .
\ee
\subsubsection{Covariantizations: RT and RR models}\label{sect:cov}

We now look for possible covariantizations of the above expressions. Covariantizations, when they exists, are in general not unique. However, some choices can be more natural than others. We will see that, starting from the equation of motion (\ref{eqmotnlG}) or from the quantum effective action (\ref{SquadrG}), one ends up quite naturally with two different covariantizations, that  define two possible models.

Let us start from the covariantization of \eq{eqmotnlG}. The linearization of the Einstein tensor $\Gmn$ is 
$\Gmn^{(1)} =-(1/2){\cal E}_{\mu\nu,\rho\sigma}\hRS$, so the term
${\cal E}_{\mu\nu,\rho\sigma}\hRS=-2\Gmn^{(1)}$ in \eq{eqmotnlG} is uniquely promoted to $-2\Gmn$ in the full covariant theory, by the requirement that we recover GR for $m=0$. The nontrivial part is the covariantization of the mass term. At linear level the Ricci scalar becomes 
$R^{(1)} =-(\eRS\Box-\paR\paS)\hrs$, that can be rewritten as $R^{(1)} =-\Box (P^{\rho\sigma}\hrs)$, so
\be\label{PiBoxR1}
P^{\rho\sigma}\hrs=-\iBox R^{(1)}\, . 
\ee
Therefore \eq{eqmotnl} is equivalent to
\be\label{Gmn1iBoxR}
-2\Gmn^{(1)}+\frac{2}{3}\, m^2 P_{\mu\nu} \iBox_{\eta} R^{(1)}=-16\pi G\, \Tmn\, ,
\ee
where the notation $\Box_{\eta}$ stresses that, until now, the $\Box$ operator was the one with respect to the flat metric $\emn$.
After promoting $\Gmn^{(1)}$ to $\Gmn$, if we want to preserve energy-momentum conservation $\n^{\mu}\Tmn=0$, we must promote $P_{\mu\nu} \iBox_{\eta} R^{(1)}$ to a transverse tensor, whose covariant derivative vanishes. To this purpose it is useful to observe that, in a generic Riemannian manifold, any symmetric tensor $\Smn$ can be decomposed as
\be\label{transv}
S_{\mu\nu}= S_{\mu\nu}^{\rm T}+\frac{1}{2}(\n_{\mu}S_{\nu}+\n_{\nu}S_{\mu})\, ,
\ee
where  
$\n^{\mu}S_{\mu\nu}^{\rm T}=0$ \cite{Deser:1967zzb,York:1974}. The extraction of the transverse part of a tensor is itself a nonlocal  operation. 
In flat space, where $\n_{\mu}\ra\pam$, proceeding as we have done in the derivation of \eqst{defshmn}{defhath}, one finds that
\be\label{Tflat}
S_{\mu\nu}^{\rm T}=\Smn
-\frac{1}{\Box_{\eta}}(\pam\paR S_{\rho\nu}+\pan\paR S_{\rho\mu})
+\frac{1}{\Box_{\eta}^2}\pam\pan\paR\paS S_{\rho\sigma}\, .
\ee
Using this expression we can easily check that, in flat space, for a tensor $S_{\mu\nu}$ of the form $S_{\mu\nu}(x)=\emn A(x)$, we have $S_{\mu\nu}^{\rm T}=P_{\mu\nu}A(x)$.\footnote{This could be derived even more simply by observing that, in flat space, where $\pam$ commutes with $\Box_{\eta}$ and therefore with $\iBox_{\eta}$, we can write 
\be
\emn A=(\emn -\iBox_{\eta} \pam\pan)A +(1/2)\pam (\iBox_{\eta}\pan A)+(1/2)\pan (\iBox_{\eta}\pam A)
=P_{\mu\nu}A +(1/2)(\pa_{\mu}S_{\nu}+\pa_{\nu}S_{\mu})\, ,
\ee
where $S_{\mu}=\iBox_{\eta}\pam A$. Since $\paM (\emn -\iBox_{\eta} \pam\pan)A=(\pan-\pan) A=0$,  $P_{\mu\nu}A$ is transverse, so
$P_{\mu\nu}A=S_{\mu\nu}^{\rm T}$.}
 Thus, to linear order in an expansion over flat space, the term $P_{\mu\nu} \iBox_{\eta} R^{(1)}$ in 
\eq{Gmn1iBoxR} is the same as the transverse part of the tensor $(\emn \iBox_{\eta} R^{(1)})$, that we denote as $(\emn \iBox_{\eta} R^{(1)})^T$, and \eq{Gmn1iBoxR} is the same as
\be
\Gmn^{(1)}-\frac{1}{3}\, m^2 (\emn \iBox_{\eta} R^{(1)})^{\rm T}=8\pi G\, \Tmn\, .
\ee
In this form, there is a natural covariantization given by
\be\fbox{\parbox{8cm}
{$$\label{RT}
\Gmn -\frac{1}{3}\, m^2\(\gmn \iBox R\)^{\rm T}=8\pi G\,\Tmn\, ,
$$}}\ee
where now $\iBox$ is the inverse of the covariant $\Box$  operator with respect to the generic metric $\gmn$, and the operation of taking the transverse part is  the fully covariant operation defined by \eq{transv}. \Eq{RT} defines the so-called RT model, where R stands for the occurrence of the Ricci scalar and T for the extraction of the transverse part. This is the model that was first proposed in \cite{Maggiore:2013mea} (through a rather different route that we will review in app.~\ref{sect:difficulties}). It was the first model of this class of nonlocal theories that looked cosmologically viable and even today, after the study of many alternative possibilities, it turns out to be the only viable one; the reasons that gradually eliminated all other alternatives will be discussed in app.~\ref{sect:difficulties}. This model will therefore be the main focus of this paper. Notice that it is defined at the level of a nonlocal equation of motion rather than by a (quantum effective) action. Indeed, there is no known nonlocal  action from which \eq{RT} can be derived.

A different covariantization emerges naturally if we rather start from the quantum effective action (\ref{Squadr}). As usual,
$d^4x\, (1/4)\hmn{\cal E}^{\mu\nu,\rho\sigma}\hrs$ becomes $d^4x\sqrt{-g}R$ while, 
using \eq{PiBoxR1}, $(P^{\mu\nu}\hmn)^2$ is the same as $(\iBox_{\eta} R^{(1)})^2$, which is naturally covariantized into $(\iBox R)^2$. Thus, a natural covariantization of \eq{SquadrG} is
\bees\label{RR}
\Gamma_{\rm RR}&=&\frac{1}{16\pi G}\int d^{4}x \sqrt{-g}\, 
\[R-\frac{m^2}{6}  ({\iBox} R)^2\]\nn\\
&=&\frac{1}{16\pi G}\int d^{4}x \sqrt{-g}\, 
\[R-\frac{m^2}{6} R\frac{1}{\Box^2} R\]\, ,\,
\ees
where in the last line we have integrated $1/\Box$ by parts.\footnote{Note that the inversions of the $\Box$ operator (and therefore also the integration by parts above) until now have been somewhat formal operations. We will justify them in more detail in section~\ref{sect:causality}.}
This gives the model that was first proposed in \cite{Maggiore:2014sia}; we will refer to it as the RR model, after the two occurrences of the Ricci scalar in the nonlocal term.
 
The RT and RR model by construction coincide to linear order in an expansion over flat space. However, they are otherwise different, and have different cosmological predictions. As we will see in app.~\ref{sect:difficulties}, the RR model shared most of the phenomenologically attractive properties of the RT model, such as viable cosmological background evolution, stable cosmological perturbations, good fit to Cosmic Microwave Background (CMB), Baryon Acoustic Oscillations (BAO), type Ia Supernovae (SNe) and structure formation data. However, we will also see that it does not pass the constraints from Lunar Laser Ranging, contrarily to the RT model, which is completely immune to it and, to date,  passes all the observational tests. Thus, in this paper we will mostly focus on the RT model. Still, the RR model can be useful to illustrate some concepts in a somewhat simpler setting, also because of its relatively simple quantum effective action, and we will therefore also occasionally use it for pedagogical purposes.

\subsection{Aspects of effective nonlocal theories}\label{sect:aspects}

A correct treatment of nonlocal terms involves a few subtle points. It is particularly important to make clear that the nonlocality that we have introduced is not fundamental, i.e. it does not appear at the level of the fundamental action of the theory, that in our case could even simply be Einstein-Hilbert gravity. Indeed, fundamental actions with nonlocal terms have problems with causality, and extra (typically ghost-like) degrees of freedom. However, even when the fundamental theory is local, if it contains massless particles (such as the graviton in GR) the corresponding quantum effective action will be unavoidably nonlocal, and this sort of nonlocality is not associated to any pathology. The same happens when nonlocal terms appear from a fundamental higher-dimensional theory, as in the DGP example, see \eq{eqgravNL2noT} below. These issues have been reviewed at length in~\cite{Maggiore:2016gpx,Belgacem:2017cqo} (see also \cite{Tsamis:1997rk,Deser:2007jk,Barvinsky:2011rk,Deser:2013uya}). Here, for completeness, we summarize them briefly.

\subsubsection{Localization and degrees of freedom}\label{sect:dof}
A nonlocal quantum effective action can be rewritten in local form by introducing auxiliary fields (see also \cite{Nojiri:2007uq,Jhingan:2008ym,Koshelev:2008ie,Koivisto:2008dh,Koivisto:2009jn,Barvinsky:2011rk,Deser:2013uya}). This is quite convenient for working  out the predictions of the theory (e.g. for studying the equations of motions of the theory, the cosmological perturbations, etc.), but requires some care at the level of interpretation, in order not to confuse the auxiliary fields with actual  degrees of freedom of the theory. As a simple example, consider the theory of a massive photon discussed in section~\ref{sect:massAmu}. We have seen that it can be formulated as a local  but non gauge-invariant theory, as in \eq{1Lemmass}, or as a  gauge-invariant theory at the price of nonlocality, as in \eq{Lnonloc}. One might  also  get   a theory that is at the same time local and gauge-invariant,  by introducing an auxiliary anti-symmetric tensor field $U^{\mu\nu}$ defined by $U^{\mu\nu}=\iBox \FMN$. In this way, one gets a local and gauge-invariant action written in terms of the two fields $A^{\mu}$ and $U^{\mu\nu}$. The equations of motion of the theory can then be rewritten as\footnote{This can be easily seen by implementing the definition $U^{\mu\nu}=\iBox \FMN$  by adding to the action a term $\lambda_{\mu\nu} (\Box U^{\mu\nu} -\FMN)$, where  $\lambda_{\mu\nu}$ is a Lagrange multiplier, and taking the variations with respect to $A_{\mu}, U^{\mu\nu}$ and $\lambda_{\mu\nu}$. A combination  of the two latter equations gives $\lambda^{\mu\nu}=-(m^2/4)U^{\mu\nu}$, and the remaining two equations give \eq{ProcawithU}. Of course, \eq{ProcawithU} can also be verified  more simply by comparison with \eq{eqnonlocFMN}.}
\be\label{ProcawithU}
\pam \FMN =j^{\nu}+m^2 \pam U^{\mu\nu}\, ,\qquad \qquad \Box U^{\mu\nu} =\FMN\, .
\ee
While the steps leading to \eq{ProcawithU} are formally correct, this local and gauge-invariant formulation seems to suggest that the theory has many more degrees of freedom than the Proca theory of a massive photon that was our starting point: we apparently have  a massless gauge-invariant vector field $A^{\mu}$, which carries two  degrees of freedom, interacting with  an antisymmetric tensor field $U^{\mu\nu}$, which apparently carries  six degrees of freedom. This seems very different from   the three degrees of freedom of a massive vector field from which we started. Of course, new degrees of freedom cannot pop out from nowhere, and the delicate point here is the passage from an equation such as $U^{\mu\nu}=\iBox \FMN$ to the equation $\Box U^{\mu\nu} =\FMN$, i.e. the inversion of the $\Box$ operator. By itself, the most general solution of an equation such as $\Box U^{\mu\nu} =\FMN$ is given by a solution of the inhomogeneous equation plus the most general solution of the associated homogeneous equation $\Box U^{\mu\nu} =0$. The latter carries with itself the six degrees of freedom associated to $U^{\mu\nu}$. Clearly, if we want this local and gauge-invariant formulation to be equivalent to the original Proca theory, we cannot accept the most general solution of $\Box U^{\mu\nu} =\FMN$. In other words, the initial condition of the auxiliary  field $U^{\mu\nu}$ cannot be taken as independent, but must be fixed in terms of the initial condition of the two transverse and the longitudinal components of $A^{\mu}$, so that the theory indeed still has three independent degrees of freedom. In this sense,  $U^{\mu\nu}$ is just an  auxiliary field, and does not carry independent degrees of freedom.  In particular, at the quantum level there are no creation/annihilation operators associated to it.
 
A similar example, worked out in detail in \cite{Maggiore:2016gpx}, is given by the Polyakov quantum effective action in two dimensions. For two-dimensional gravity coupled to conformal matter  it is possible to compute exactly the quantum effective action by integrating the conformal anomaly. This  leads to the famous Polyakov quantum effective action, which can be written, in terms of the conformal mode, in a local form which is not explicitly invariant under  diffeomorphism;  equivalently, one can write it  in a form which 
is nonlocal  but diff-invariant. In the latter form the Polyakov quantum effective action is
proportional to  $R\iBox R$. One could further rewrite  the theory in a form which is both local and diff-invariant 
by introducing an auxiliary field $U=-\iBox R$. However, in this case where the computation of the quantum effective action can be performed explicitly, it is easy to check that $U$ is not an independent degree of freedom  that popped out from nowhere; rather, its initial conditions are fixed in terms of the initial conditions of the conformal factor $\sigma$, the precise relation being simply $U_{\rm in}=2\sigma_{\rm in}$, $\dot{U}_{\rm in}=2\dot{\sigma}_{\rm in}$~ \cite{Maggiore:2016gpx}.

In the following we will use a similar localization procedure for the RR and RT models. As in the examples above, the auxiliary fields that will be introduced are not new independent degrees of freedom; rather, their  initial conditions  should be understood as  fixed in terms of the initial conditions on the metric, and there are no creation/annihilation operators associated to them (and, therefore, no issues of ghosts at the quantum level). 
If one had an explicit derivation of the nonlocal  term from a fundamental theory, one would in principle be able to determine explicitly their initial conditions in terms of those on the metric. In practice, lacking such a derivation, these initial conditions must be taken as free phenomenological parameters. One might fear that this significantly reduces  the predictive power of the theory. However,  we will see  in section~\ref{sect:pheno} that, in the cosmological context in which we are interested, this introduces only very limited freedom, both at the level of background evolution and of cosmological perturbations, since these initial conditions turn out to be associated mostly to irrelevant directions in parameter space.

\subsubsection{Localization of the RR and RT models}\label{sect:locRRRT}

We next show how to write nonlocal gravity in a local form. We write the equations both for the RR model, and for the RT model that will eventually be our main focus, since the comparison between the two models can be instructive, and also the manipulations of the equations of the RR model are somewhat simpler.
To write the RR model in a local form we introduce two auxiliary fields $U$ and $S$, 
defined by $U=-\iBox R$ and $S=-\iBox U$~\cite{Maggiore:2014sia}.
This can be implemented at the Lagrangian level by introducing two Lagrange multipliers $\xi_1,\xi_2$ into \eq{RR},
\bees\label{S2}
\Gamma_{\rm RR}=\frac{1}{16\pi G}\int d^4x \sqrt{-g}\, \[ R\( 1-\frac{m^2}{6} S\)-\xi_1(\Box U+R)-\xi_2 (\Box S+U)\]\nn
\, .
\ees
The  variation  with respect to $\hmn$ gives
$\Gmn=(m^2/6) K_{\mu\nu}+8\pi G\Tmn$,
where  
\be\label{defKmn}
K^\mu_\nu \equiv 2 S G^\mu_\nu - 2 \nabla^\mu \partial_\nu S + 2 \delta^\mu_\nu \Box_g S + \delta^\mu_\nu \partial_\rho S \partial^\rho U - \frac{1}{2} \delta^\mu_\nu U^2 - \big( \partial^\mu S \partial_\nu U + \partial_\nu S \partial^\mu U \big)\, ,
\ee
while the variation with respect to the  Lagrange multipliers $\xi_1,\xi_2$  gives
$\Box U=-R$ and
$\Box S =-U$.
Thus, the RR model is formally written as a scalar-tensor theory, with two scalar fields $U$ and $S$, although, as we have discussed in section~\ref{sect:dof}, $U$ and $S$ are not independent degrees of freedoms, and their initial conditions are in principle fixed in terms of the initial conditions of the metric. In particular, there are no independent solutions associated to the homogeneous equations $\Box U=0$ and $\Box S=0$, and no corresponding quanta at the quantum level.

For the RT model the localization proceeds by defining again $U=-\iBox R$.
We also introduce
$S_{\mu\nu}=-U\gmn=\gmn \iBox R$ and we extract its transverse part   $S^T_{\mu\nu}$ by using \eq{transv}. Thus, \eq{RT} is localized in terms of an auxiliary scalar field $U$ and the auxiliary four-vector field $S_{\mu}$ that enters through \eq{transv}. The equations of motion then read~\cite{Maggiore:2013mea,Kehagias:2014sda}
\bees
\Gmn +\frac{m^2}{6}\, \(2U\gmn + \n_{\mu}S_{\nu}+\n_{\nu}S_{\mu}\)&=&8\pi G\,\Tmn\, ,
\label{v2loc1}\\
\Box U&=&-R\label{BoxUR}\, ,\\
(\d^{\mu}_{\nu}\Box +\n^{\mu}\n_{\nu})S_{\mu}&=&-2\pan U\, , \label{panU}
\ees
where \eq{panU} is obtained by taking the divergence of \eq{transv} with $S_{\mu\nu}=-U\gmn$.
The equations of motion of the RT model have a suggestive property in connection with the cosmological constant problem.  Let us  perform  a shift $U(x)\ra U(x)+u_0$, with $u_0$  a constant. 
\Eqs{BoxUR}{panU} are invariant while \eq{v2loc1} becomes
\be\label{u0cosmconst}
\Gmn +\frac{m^2}{6}\, \(2U\gmn + \n_{\mu}S_{\nu}+\n_{\nu}S_{\mu}\)
=8\pi G\,\(\Tmn-\lambda\gmn\)\, .
\ee
where $\lambda=m^2u_0/(24\pi G)$.
Thus, $u_0$ (or, equivalently, the initial condition on $U$) generates a cosmological constant, and one could chose $u_0$  to cancel any vacuum energy term in $\Tmn$. 

It is also instructive to  consider  the  equations of motion of the RR and RT models linearized over flat space,
\eq{eqmotnlG}, that were our starting point, and write them in terms of the auxiliary fields and of  the metric variables of the $3+1$ decomposition (\ref{hijphilambdaeps}).  Since, by construction, the RR and RT model coincide when linearized over flat space, we use the   RR model, whose localization is slightly simpler, since it involves two scalar fields $U$ and $S$, rather than $U$ and $S_{\mu}$ for the RT model.
One then finds that \eqst{n2Phin2Psi}{hijsij} are modified into~\cite{Maggiore:2014sia}
\bees
\n^2\[\Phi-(m^2/6) S\]&=&-4\pi G\rho\, ,\qquad\qquad
\Phi-\Psi-(m^2/3) S= -8\pi G\Sigma\, ,\label{dof2}\\
\n^2\Xi_i&=&-16\pi G S_i\, ,\qquad\qquad
\Box H_{ij}^{\rm TT}=-16\pi G \Sigma_{ij}\, ,,\label{dof3}\\
(\Box+m^2)U&=&-8\pi G (\rho-3P)\, ,\qquad\qquad
\Box S=-U\, ,\label{dof4}
\ees
\Eq{dof4} is needed to close the system, since $S$ appears in \eq{dof2}.
\Eqs{dof2}{dof3} shows that the original metric perturbations $\Phi$, $\Psi$ and $\Xi_i$ remain non-radiative variables that satisfy Poisson equations, just as in GR.\footnote{This should be contrasted with what happens when one linearizes massive gravity with a Fierz-Pauli mass term, in which case $\Phi$ becomes a radiative field that satisfies 
$(\Box-m^2)\Phi=$~source term \cite{Deser:1966zzb,Alberte:2010it,Jaccard:2012ut}. The fact that, for $m=0$, $(\Box-m^2)\Phi$   does not reduces to its GR counterpart $\n^2\Phi$ is a reflection of the van~Dam-Veltman-Zakharov (vDVZ) discontinuity of linearized massive gravity.} The auxiliary fields $U$ and $S$ satisfy Klein-Gordon equations, but, as we have seen, their initial conditions are fixed in terms of the initial conditions on the metric, and therefore are not free radiative degrees of freedom either. 
From these equations it is also clear that the conformal mode $s$ remains a non-propagating degree of freedom also in the RT or RR models. Indeed, combining the two equations in  (\ref{dof2}) we get
\bees
\n^2(\Phi+\Psi)&=&2\n^2\( \Phi-\frac{m^2}{6}S \)+8\pi G\n^2\Sigma \nn\\
&=& -8\pi G\n^2 (\rho-\Sigma)\, .
\ees
Then, from \eq{siBoxPhi} we get (again at the linearized level over flat space)
\bees\label{siBoxPhiRT}
s&=&6\Phi- 2\iBox\n^2(\Phi+\Psi)\nn\\
&=&6\Phi+16\pi G\iBox\n^2(\rho-\Sigma)
\, .
\ees
We see that the nonlocal term in $s$  is fully determined by the energy-momentum tensor,  in particular by the density $\rho$ and by the anisotropic stress $\Sigma$ that enters in $T_{ij}$ through \eq{Tij3plus1}. Thus, $s$ remains a non-radiative degree of freedom, exactly as in GR, and vanishes if $\rho=0$ and $\Sigma=0$.

\subsubsection{Causality and the quantum effective action}\label{sect:causality}

We next discuss why nonlocal terms would induce problems with causality if added at the level of a fundamental action, while they do not in a quantum effective action.

To illustrate the problem with causality of a nonlocal fundamental action, consider for instance an action with  a nonlocal term proportional to $(1/2)\int d^4x \,\varphi\iBox\varphi$ where $\varphi$ is a scalar field~\cite{Foffa:2013sma}. To complete the definition of this term we must specify the Green's function $G(x,x')$ used to define  $\iBox$, 
and then 
\be
\frac{1}{2}\int d^4x \,\varphi(x) (\iBox\varphi)(x) \equiv\frac{1}{2}\int d^4x d^4x'  \,\varphi(x) G(x,x')\varphi(x')\, .
\ee
Consider now the contribution of this term to the equation of motion. Taking the variation with respect to $\varphi$, we get
\be\label{symGreen}
\frac{1}{2}\frac{\d}{\d\varphi(x)} \int dx' dx'' \varphi(x') G(x';x'') \varphi(x'')
=\frac{1}{2}\int dx' [G(x;x')+G(x';x)] \varphi(x')\equiv \iBox_{\rm sym}\varphi\, , 
\ee
where $\iBox_{\rm sym}$ is the inverse d'Alembertian with respect to the symmetrized Green's function
$[G(x;x')+G(x';x)]/2$.
Thus, independently of the choice of $G(x,x')$, in the equations of motion we end up with a symmetric 
Green's function. Since the retarded Green's function is not symmetric, it  cannot be obtained from such a variation. The equations of motion  obtained from a nonlocal classical action are therefore  in general acausal. This is one of the reasons why a fundamental action must be local.

The situation is however completely different for the quantum effective action.  Let us recall,
following  standard textbook material, that, for a scalar field  $\varphi(x)$ with fundamental action $S[\varphi]$, the quantum effective action is obtained by introducing an auxiliary source $J(x)$ and defining the generating functional of the connected Green's function $W[J]$ from
\be\label{defWJ}
e^{iW[J]}\equiv\int D\varphi\,\,  e^{iS[\varphi]+i\int J\varphi}\, ,
\ee
where $\int J\varphi$ is a notation for $\int d^4x \, J(x)\varphi(x)$. Then 
$\d W[J]/\d J(x)=\cav \varphi(x)\vac_J$. We will use the notation 
$\cav \varphi(x)\vac_J\equiv \phi[J]$ for 
 the vacuum expectation value of the field $\varphi(x)$ in the presence of the source $J(x)$. The quantum effective action $\Gamma[  \phi]$ is defined as a functional of  $\phi$ (rather than of the original field $\varphi$), obtained by performing the Legendre transform,
$\Gamma[  \phi]\equiv W[J] -\int  \phi J$,
where $J=J[ \phi]$ is obtained by inverting $\phi=\phi[J]$. As a consequence, one immediately finds that 
\bees\label{dGJ}
\d\Gamma[  \phi]/\d  \phi(x)=-J(x)\, .
\ees
From the path integral representation (\ref{defWJ}) it is also easy to show that
\be\label{Gammapathint}
e^{i\Gamma[ \phi]}=\int D\varphi\,\,  e^{iS[ \phi+\varphi]-i\int \frac{\d\Gamma[  \phi]}{\d  \phi}\varphi }
\, .
\ee
Thus,  the physical meaning of the quantum effective action $\Gamma[\phi]$ is that it is a functional of $\phi(x)=\cav \varphi(x)\vac$, obtained by integrating out the quantum fluctuations around it. From  \eq{dGJ} we also see that $\Gamma[\phi]$ is the quantity whose variation gives the exact equations of motion for the expectation values of the field, which by construction include (in principle, exactly, if one were able to compute $\Gamma$ exactly)  the contribution of the quantum fluctuations.

It is clear  a priori that  the quantum effective action  obtained from a  local and causal fundamental action cannot have problems with causality. To see explicitly how this comes out, one must take into account that, as we have seen, $\Gamma$  does not give the equations of motion of the original field, but rather of its vacuum expectation value. We  must however  distinguish between the in-out and the in-in expectation values.
The effective action computed using the  standard Feynman path integral gives the equations of motion of the in-out vacuum expectation value, which are indeed acausal, because they involve the $\iBox$ operator constructed with the Feynman propagator. However, there is nothing wrong with this, since in-out matrix elements are not directly observable. Rather, they  just appear in  intermediate steps of the computation of observables, such as  scattering cross section, and indeed the Feynman propagator appears everywhere in quantum field theory computations. 
In contrast, the 
in-in matrix elements of the field are observables; for instance,  $\langle 0_{\rm in}|\varphi (t,\vx)|0_{\rm in}\rangle$ is the vacuum expectation value of the quantum field $\varphi$ at a given time $t$.
To obtain the equations of motion of the  in-in matrix elements one must evaluate the path integral in $\Gamma$ by using 
the Schwinger-Keldish prescription. As a result, the in-in matrix elements  automatically  obey  causal equations of motions in which the retarded propagator appears~\cite{Jordan:1986ug,Calzetta:1986ey,Mukhanov:2007zz}. In practice, the result of the  computation with the 
Schwinger-Keldish path integral turns out to be equivalent to that obtained by just performing a formal variation of the quantum effective action, without specifying the Green's function used to define $\iBox$, and then replacing the resulting occurrences of $\iBox$ in the equations of motion with the $\iBox$ operator defined with respect to the retarded Green's function (see  Section~12.1.6 of \cite{Mukhanov:2007zz}  for a pedagogical discussion in the quantum mechanical case, and \cite{Barvinsky:1987uw} for a proof valid for the one-loop quantum effective action in curved space).\footnote{This also justifies the integration by parts of $\iBox$ that we have performed when constructing the RR model in \eq{RR}. At the level of the quantum effective action we can simply define $\iBox$ with a symmetric Green's function,
$G(x,x')=G(x',x)$, which ensure a formal hermiticity of the action and for which the validity of the integration by parts of $\iBox$ is easily established (see app.~A of \cite{Jaccard:2013gla}). In any case, the equations of motion for the in-in expectation values will come automatically with a retarded Green's function.}

\subsection{Possible mechanisms for the generation of an IR  mass scale}\label{sect:mech}

\subsubsection{Perturbative loop corrections}
We next discuss possible mechanisms for the generation of these nonlocal terms. We begin by observing that perturbative  loop corrections due to massive matter fields cannot be responsible for them~\cite{Maggiore:2016fbn}. In gravity the one-loop corrections induced by   matter fields can  produce  nonlocal form factors in the quantum effective action, associated to terms quadratic in the curvature~\cite{'tHooft:1974bx,Barvinsky:1985an, Barvinsky:1987uw,Barvinsky:1990up,Gorbar:2002pw,Gorbar:2003yt} (see   \cite{Birrell:1982ix,Buchbinder:1992rb,Shapiro:2008sf} for reviews). The resulting quantum effective action has the form
\be\label{formfact}
\Gamma_{\rm one-loop}=\int d^{4}x \sqrt{-g}\, \bigg[\frac{\mplr^2}{2}R - R \,k_R(\Box) R 
-C_{\mu\nu\rho\sigma}k_W(\Box)C^{\mu\nu\rho\sigma} +GB\bigg]\, ,
\ee
where $\mplr^2=1/(8\pi G)$, $C_{\mu\nu\rho\sigma}$ is the Weyl tensor and `GB' denotes a similar nonlocal term that reduces to the topological Gauss-Bonnet term when its  form factor is set to one. Consider the contribution to the form factor from a particle of mass $M$. When the particle is very massive compared to the 
energies or curvatures involved (so $M$ much heavier than the center of mass energy $E$  in a scattering experiment, or $M$ much larger than the Hubble parameter $H(t)$ in a cosmological setting), according to the usual decoupling theorem, the particle decouples and its contribution to the 
form factor is local and suppressed by a factor ${\cal O}(\Box/M^2)\ll 1$. A nonlocal contribution instead emerges when the particle is light compared to the energy scale involved.
 In that case, the result has  the form~\cite{Donoghue:1994dn,Gorbar:2002pw,Gorbar:2003yt,Codello:2015mba} 
\be\label{expan}
k_R\(\frac{-\Box}{M^2}\)=\alpha\log\(\frac{-\Box}{M^2}\)+\beta \(\frac{M^2}{-\Box}\)
+\gamma \(\frac{M^2}{-\Box}\)\log\(\frac{-\Box}{M^2}\)+\delta \(\frac{M^2}{-\Box}\)^2+\ldots \, ,
\ee
and similarly for $k_W$. In \cite{Codello:2015pga} it was  observed that the logarithmic terms and the term $(M^2/\Box)$ have little effect on the cosmological evolution in the present epoch, so one might hope that the leading term is actually given by the term  $M^4/\Box^2$, which is the operator that appears in the RR model. Comparison with \eq{RR} then shows that we must have $M^4={\cal O}(m^2\mplr^2)$ and therefore $m={\cal O}(M^2/\mplr )$. Since the expansion
(\ref{expan}) is  valid, today,  only if $M\ll H_0$, such loop corrections could only generate a nonlocal term 
$m^2R\Box^{-2}R$ with $m={\cal O}(M^2/\mplr )\ll H_0 (H_0/\mplr)$.
In contrast, we will see that the requirement of obtaining a dynamical dark energy density today  of the order of the observed value fixes $m$ to be of order $H_0$. Thus, loop corrections from light particles, i.e. (hypothetical) massive particles with masses $M\ll H_0$, fall short from providing the required value of $m$ by a factor ${\cal O}(H_0/\mplr)\sim 10^{-52}$.
 On the other hand,  in the present cosmological epoch particles with a mass $M\gg H_0$ are heavy compared to the relevant curvature scale fixed by $H_0$  and only give local 
contributions to the form factor, furthermore suppressed by  ${\cal O}(H_0^2/M^2)\ll 1$. Thus, perturbative loop corrections are  totally irrelevant to the IR dynamics of gravity.\footnote{Note that this cannot be cured by including a large number $N$ of particles with $M\ll H_0$ in the loops, as has been suggested. Since, for large $N$, the form factor $k_{R}(\Box)$ is proportional to $N$, in that case we rather get $m^2\sim N M^4/\mplr^2$ and therefore, given that $M\ll H_0$,  we now get $m \ll \sqrt{N} H_0 (H_0/\mplr)$. To obtain  $m\sim H_0$ we would then need $N\sim (\mplr/H_0)^2\sim 10^{104}$. Apart from the  fact that  existence of such a huge number of hypothetical particles with $m< H_0$ is very implausible, this would result in enhancing all loop corrections of gravity by this factor $\sqrt{N}$, and therefore the scale of quantum gravity would become $\mplr/\sqrt{N}\sim H_0$ in all observables.}
Furthermore, they produce a generic nonlocal structure such as that given in \eq{formfact},
while we have already anticipated that, for phenomenological reasons, we need a very specific nonlocal structure such as that in (\ref{RT}).

\subsubsection{Nonlocal terms from extra dimensions}\label{sect:extradim}

The above discussion shows that we must look for a different mechanism for the generation of nonlocal terms relevant in a cosmological setting. The Dvali-Gabadadze-Porrati (DGP) model~\cite{Dvali:2000hr}, even if by now ruled out phenomenologically, still  provides an instructive example of how a theory with a four-dimensional brane in a space with infinite extra dimensions can be rewritten as a four-dimensional covariant theory with nonlocal terms. The DGP action is 
\be\label{SDGP}
S_{\rm DGP}=\frac{M_5^3}{2}\int d^5X\sqrt{-G}\,R(G)+
\frac{M_4^2}{2}\int d^4x\sqrt{-g}\,R(g)+S_M\, ,
\ee
where $X^A=\{x^{\mu},y\}$  are the five-dimensional bulk coordinates, $G_{AB}(X)$ is the 5d metric, and
$M_5$ is the 5d Planck mass; the 4d coordinates, metric and Planck mass are denoted as $x^{\mu}$, $\gmn(x)$ and $M_4$, respectively. The 4d metric $\gmn(x)$ is defined as the pullback of the 5d metric, $\gmn(x)=G_{AB}[X(x)]\pam X^A\pan X^B$. The matter action, $S_M$, is localized on the 4d brane.

One can expand the action to quadratic order over flat space, writing $G_{AB}(x,y)=\eta_{AB}+H_{AB}(x,y)$.  Away from the brane, the corresponding equations of motions are just the 5d linearized Einstein equations in vacuum. One then finds that
it is possible to  write explicitly the solutions of the 5d equations of motion for $H_{AB}(x,y)$ in terms of the 4d metric perturbation on the brane, $\hmn(x)$, which  plays the role of a boundary value in the equation of motion for $H_{AB}(x,y)$
(the computation is described in detail in Section~IX.A of ref.~\cite{Hinterbichler:2011tt}).
For instance, for the $(A=\mu,B=\nu)$ components of $H_{AB}$, the result (for a flat brane located at $y=0$) is of the form 
\be\label{HmnsqrtBox}
H_{\mu\nu}(x,y)=e^{-|y|\sqrt{-\Box}}\,\hmn(x)\, ,
\ee
where $\sqrt{-\Box}$ is the formal square root of the d'Alembertian operator. 
Expanding the action (\ref{SDGP}) to quadratic order in $H_{AB}$, substituting the  solution for 
$H_{AB}(x,y)$ in terms of the boundary value $\hmn(x)$,  and integrating the explicit $y$ dependence, one obtains an equivalent nonlocal four-dimensional linearized action,
\be\label{S2dgp}
S^{(2)}_{\rm DGP}=\frac{1}{64\pi G}\int d^{4}x \,
\[ \hmn{\cal E}^{\mu\nu,\rho\sigma}\hrs -m(\hmn\sqrt{-\Box}\,\hMN-h\sqrt{-\Box}\,h) \]
+\frac{1}{2}\int d^{4}x \,\hmn\TMN\, ,
\ee
where $m=2M_5^3/M_4^2$ and, as usual,   $M_4^2\equiv \mplr^2=1/(8\pi G)$.
This has the form of Fierz-Pauli massive gravity, with the mass term $m^2$  replaced by $m\sqrt{-\Box}$.\footnote{The meaning of the formal expression $\sqrt{-\Box}$ can be better understood by looking at the momentum dependence of the corresponding propagator in momentum space, $-i/(p^2+m\sqrt{p^2})$. This propagator has a branch cut that corresponds to a continuum of resonances, the so-called `resonance graviton'~\cite{Dvali:2000rv,Gabadadze:2003ck,Dvali:2007kt}.}
We can now rewrite this expression in a form that involves only the linearized Einstein tensor.\footnote{We put in a more precise form a result discussed in~\cite{Dvali:2002fz,Dvali:2002pe}.} The computation can be nicely performed following the steps  in sect.~3 of 
\cite{Jaccard:2013gla}. That computation was  done for  Fierz-Pauli massive gravity, but goes through without any changes in our case, with the replacement  $m^2\ra m\sqrt{-\Box}$.  The strategy, which is analogous to that used in Section~\ref{sect:massAmu} to rewrite the Proca action in nonlocal form, is to introduce a \Stu field $A_{\mu}$ by the replacement
$\hmn\ra \hmn  +(1/m)(\pam\An+\pan\Am)$, to obtain a theory that is explicitly invariant under linearized diffeomorphisms $\hmn\ra\hmn  -(\pam\xin+\pan\xim)$, $\Am\ra \Am+m\xim$, and then integrate out $\Am$  using its own equations of motion, as in  \cite{Dvali:2006su,Dvali:2007kt,Porrati:2002cp,Hinterbichler:2011tt}. The result can then be read from eq.~(3.19) of \cite{Jaccard:2013gla}, by replacing  $m^2\ra m\sqrt{-\Box}$. Using furthermore ${\cal E}_{\mu\nu,\rho\sigma}\hRS=-2\Gmn^{(1)}$ (and rescaling $\hmn\ra\hmn/\kappa$) we get 
\be\label{eqgravNL2noT}
\(1+\frac{m}{\sqrt{-\Box}}\)\Gmn^{(1)}=
8\pi G\, \(\Tmn-\frac{1}{3}P_{\mu\nu} T\) \, .
\ee
At the linearized level, this nonlocal four-dimensional equation of motion is completely equivalent to the local five-dimensional DGP model. Notice that the extra term on the right-hand side, that survives in the limit $m=0$, is a reflection of the vDVZ discontinuity of the linearized theory.

One could then in principle look for the correct covariantization that would give back  the DGP model at the  full nonlinear level.\footnote{A tempting guess for the correct covariantization is obtained by observing that the term $P_{\mu\nu} T$ can be eliminated in favor of $P_{\mu\nu}R^{(1)}$, where $R^{(1)}$ is the linearized Ricci tensor, by taking the trace of \eq{eqgravNL2noT}, so that \eq{eqgravNL2noT} can be rewritten as
$
\(1+m/\sqrt{-\Box}\)[ \Gmn^{(1)}-\frac{1}{6}P_{\mu\nu} R^{(1)}]=
8\pi G\, \Tmn\, .
$
As we already observed below \eq{Tflat}, at linear level the transverse part
of a tensor $\emn A(x)$ is $P_{\mu\nu}A(x)$, so $P_{\mu\nu}A=(\emn A)^T$. Then, one is naturally led to
$
\[ \(1+\frac{m}{\sqrt{-\Box}}\)\( \Gmn-\frac{1}{6}g_{\mu\nu} R\)\] ^T=
8\pi G\, \Tmn\, .
$
}
Independently of the correct covariantization (that should also reproduce the absence of the vDVZ discontinuity of the theory in the nonlinear theory), for our purposes the above analysis is instructive because it shows how a nonlocal term, relevant in the IR, can in principle emerge from a theory with infinite extra dimensions. It also shows that, with a mechanism of this kind, one will generate a very specific and peculiar nonlocal structure, rather than the most general expression quadratic in the curvatures that is obtained  from perturbative corrections, as in \eq{formfact}.

Another interesting example of this type is given by the Karch-Randall model~\cite{Karch:2000ct}, which is a five-dimensional  theory of gravity with a negative  cosmological constant, giving rise to an AdS$_5$ spacetime, 
in which is embedded a 4d brane such that the induced metric on the brane is AdS$_4$.
The peculiar feature of this compactification is that, despite the fact that  four-dimensional general covariance is preserved, still it does not have a massless spin-2 state, but rather a tower of massive spin-2 states. As discussed in \cite{Karch:2001jb}, from the point of view of an effective four-dimensional action the corresponding mass term cannot be obtained from terms quadratic in the curvature, and it was suggested that the proper description is in terms of a nonlocal effective action.

Observe also that, in the example of DGP, the mass scale $m$ is given by $m=2M_5^3/M_4^2$, i.e. is a combination of the five-dimensional and four-dimensional Planck masses, that were already explicitly present in the fundamental action (\ref{SDGP}). We will next explore a different possibility, namely that such a mass is generated dynamically by non-perturbative IR effects in gravity.

\subsubsection{Dynamical mass generation}\label{sect:dynmass}

In this subsection we will discuss indications, from various non-perturbative techniques, in favor of  the possibility of a dynamical mass generation in the IR limit of four-dimensional quantum gravity, in particular in relation to the conformal mode.

\vspace{2mm}\noindent
{\bf Lattice gravity}.
A possible non-perturbative tool  is provided  by   lattice gravity, based  either on a simplicial decomposition of the space-time manifold in Euclidean space (see  \cite{Hamber:2009zz} for review), or on causal dynamical triangulations~(see \cite{Ambjorn:2012jv,Loll:2019rdj} for reviews).

In Euclidean quantum gravity one starts from a lattice discretization of the  path integral over all Euclidean metrics, weighted with the Euclidean version of   the Einstein-Hilbert action with  bare cosmological constant $\Lambda_0$ and bare Newton constant $G_0$, and  a suitable choice of the lattice measure (and possibly  terms quadratic in the curvature). Euclidean lattice gravity is not at the same level of development as, say, lattice QCD, due to the difficulty of finding  clear evidence for UV fixed points where one could take a nontrivial continuum limit,
so the results should be taken with some qualifications. Still, numerical simulations indicate the existence of a critical coupling $G_c$ such that, for $G_0<G_c$, the lattice collapses into a degenerate   collection of long, elongated simplices, and the
four-dimensional geometry 
collapses into an effective two-dimensional manifold \cite{Hamber:2009zz}. This phase, that takes place for $G_0<G_c$ and therefore also in the perturbative regime $G_0\ra 0$, is interpreted as a result of the conformal mode instability in the Euclidean path integral for gravity. As the manifold collapses it reaches an effective dimension equal to two, where the Einstein-Hilbert action becomes a topological invariant, so the instability shuts off and the geometry does not collapse further. In contrast, for $G_0>G_c$ the system is in a smooth phase. This is interpreted as an effect of the integration measure, that at sufficiently strong coupling suppresses singular  spike-like curvature singularities that, in the phase $G_0<G_c$, trigger the conformal mode instability.
At the critical point the correlation length diverges, so a continuum limit can be taken, and, in the vicinity of the critical point  the renormalized Newton's constant runs as~\cite{Hamber:1999nu,Hamber:2004ew,Hamber:2005dw,Hamber:2013rb,Hamber:2015jja} (see  also \cite{Hamber:2009zz,Hamber:2017pli} for reviews)
\be\label{Gdik2}
G(k^2)=G_N\[ 1+  \(\Lambda_{\rm grav}^2/k^2\)^{\frac{1}{2\nu}}+
{\cal O} \(\Lambda_{\rm grav}^2/k^2\)^{\frac{1}{\nu}} \, \]\, ,
\ee
where  $\nu$ is a critical index which, within the numerical accuracy, turns out to be  consistent with $\nu=1/3$, and $\Lambda_{\rm grav}$ is a renormalization-group invariant  mass scale which is dynamically generated, analogous to $\Lambda_{\rm\scriptscriptstyle QCD}$ in QCD.  The expression (\ref{Gdik2}) is only valid  in the far UV regime $|k^2|\gg \Lambda_{\rm grav}^2$, and is not directly applicable to the IR regime relevant for cosmology. 

The above results point toward the possibility of dynamical mass generation in the IR regime of quantum gravity, but do not yet give hints on what would be the precise role of this mass scale. Recent work using causal dynamical triangulation (CDT), however, indicates precisely the dynamical generation of a mass for the conformal mode~\cite{Knorr:2018kog}. In CDT one defines the path integral from a sum over Lorentzian geometries, weighted with the factor  $e^{iS_L}$, where $S_L$ is the Einstein-Hilbert action in Lorentzian signature,  discretized through triangulations of space-time in terms of simplices with time-like and space-like edges. 
The four-dimensional lengths of the space-like and time-like edges are defined as $\ell_s^2=a^2$ and $\ell_t^2=-\alpha a^2$, respectively, where $a$ is the lattice spacing and $\alpha >0$. The analytic continuation to $\alpha <0$ transforms the factor $e^{iS_L}$ in the Lorentzian path integral into $e^{-S_E}$, 
where $S_E$ is the Euclidean Einstein-Hilbert action,
allowing the use of  tools from statistical physics and Monte Carlo techniques for the numerical evaluation of the path integral. Note however that the sum now is not over all (discretized) Euclidean geometries, but only over those that have a causal, Lorentzian, origin, i.e. those that, in the above sense of analytic continuation in $\alpha$, can be obtained from  a Wick rotation of discretized Lorentzian geometries. Thus, the approach to quantum gravity of causal dynamical triangulation is a priori different from  that of Euclidean quantum gravity, meant as a sum over all Euclidean geometries.\footnote{The unboundedness of the  Euclidean Einstein-Hilbert action, due to the  fact that the conformal mode has a kinetic term with the `wrong' sign, is now regularized by the lattice spacing $a$. In the limit $a\ra 0$ it would again reappear; however,  similarly to what we have seen in the case of Euclidean gravity, near a nontrivial fixed point it can happen that  configurations with unbounded action are suppressed by the integration measure and play no role in the continuum limit, and this is indeed what happens near the fixed points obtained from causal dynamical triangulations~\cite{Ambjorn:2012jv}. This competition between configuration with unbounded action and entropy is precisely what  gives rise to the Kosterlitz-Thouless transition in the two-dimensional XY-model.}    Using numerical simulation of CDT it is possible to measure  non-perturbatively the   two-point correlation function of the fluctuations of the spatial three-volumes. The latter is related to the two-point function of the conformal mode, and  ref.~\cite{Knorr:2018kog} showed that the numerical results provide evidence for a massive conformal mode, i.e. for a linearized nonlocal quantum effective action of the form (\ref{SEH2nlmass}) [or, equivalently (\ref{SquadrG})], whose  covariantizations can be provided by the RR or RT models. A caveat of the result is that the simulation was performed at a single value of the coupling $(\kappa_0,\Delta)$ of the theory (which are related to the bare Newton constant and the parameter $\alpha$), but the approach to the continuum was not studied. Still, 
this is a first  indication that a mass for the conformal mode could indeed  be generated dynamically in quantum gravity.

\vspace{2mm}\noindent
{\bf Functional renormalization group equations}. In quantum field theory, exact renormalization group (RG) equations, such as the Polchinski equation~\cite{Polchinski:1983gv} and the Wetterich equation~\cite{Wetterich:1992yh},  provide, in principle,  an equivalent way of computing exactly a path integral, by transforming the functional integration into a functional differential equation. As such, both the path integral formulation and the functional renormalization group equations can be taken as equivalent  non-perturbative definitions of a quantum field theory.
In practice, just as the evaluation of the functional integral for an interacting theory requires approximations methods (perturbation theory, semiclassical methods such as instantons, etc.) or numerical evaluation through a lattice formulation, the functional RG equation, to be  reduced to a manageable form, requires a truncation of the space of action functionals, projecting the intrinsically infinite-dimensional RG flow onto a manifold of finite (and manageable) dimension.  In the end, the reliability of the non-perturbative results obtained depends on  whether the truncation catches the most important features, and is the main uncertainty of the method. Still, 
functional RG method can provide important insight into the non-perturbative behavior of a theory.
For gravity,  functional renormalization group  techniques have been developed particularly in connection with the asymptotic safety program, i.e. the search for a non-trivial UV fixed point (see \cite{Reuter:2012id} for review). More recently, these tools are being applied to the study of the IR behavior of gravity. It should be pointed out that the study of the IR behavior of Einstein-Hilbert gravity at the quantum level is completely independent from the issue of its UV completion. Independently of whether the latter is given by a non-trivial UV fixed point, string theory, or other options, we know that gravity, at, say, the laboratory or solar system scales, is very well described by the Einstein-Hilbert action, and we ask how this theory evolves with the RG flow as we run toward even lower energy scales. A number of recent functional RG studies, with different approximations,  have  found indication of strong quantum gravity effect in the IR~\cite{Wetterich:2017ixo,Knorr:2017mhu,Morris:2018mhd,Wetterich:2018poo}.
 The possibility of  dynamical mass generation, in the functional RG language, is signaled by the fact that, running toward the IR,  the RG flow encounters a singularity at some momentum scale $k$. An instance of this phenomenon was already found for some RG trajectories in \cite{Reuter:2004nx}, where, in a truncation of the theory including only the Einstein-Hilbert term  $\int d^4x\sqrt{-g}\, R$ and the cosmological constant term $\int d^4x\sqrt{-g}$, it was found that, evolving the RG flow toward the IR,  for some trajectories the running of Newton's constant hits a singularity and terminates at a finite scale $k_{\rm term}$. Of course, in general the singularity can be an artifact of the truncation. The same happens using functional RG equations in QCD; in that case, a simple truncation of the space of possible terms in the action is not sufficient, and a reliable description of the IR limit involves also nonlocal terms in the truncation ansatz~\cite{Reuter:2004nx}, such as the one in \eq{Fmn2}. Thus, by itself a RG flow that, within some truncation, becomes singular in the IR, can be a hint that a mass scale is generated and that at this scale nonlocal terms, that have not been  included in the truncation, become important. For our purposes, an interesting observation is that  a dynamical scale  also  appears using functional renormalization group equations because of the dynamics of the conformal mode. Indeed,  in the functional RG approach, the would-be `wrong' sign of the kinetic term of the conformal mode  leads to functional differential equations that are perfectly well defined, contrary to the Euclidean path integral formulation, but `backward-parabolic', i.e. the resulting flow toward the  IR is not well defined, and reaches a singularity at a finite energy scale~\cite{Dietz:2016gzg,Morris:2018mhd}. It is quite natural to expect that, at this scale, nonlocal terms associated to the conformal mode, such as those defining the RR or RT models, become important to resolve the singularity and allow for a smooth flows that extends in the IR down to $k\ra 0$.

A related interesting result is the one discussed in
ref.~\cite{Wetterich:2017ixo}, where it is found that, truncating the theory so to include only fluctuations of the transverse-traceless modes, there are strong non-perturbative infrared renormalization effects, that screen the cosmological constant. Together, these results can suggest a scenario where the cosmological constant is screened by strong IR effects due to the TT modes fluctuations; at the same time, the conformal mode fluctuations are responsible for generating a new IR mass scale and the nonlocal term that defines the RT model (\ref{RT}), that, as we have seen, at the linearized level is simply a mass term for the conformal mode. This nonlocal term, in turn, generates a viable dynamical dark energy model, as we will see in section~\ref{sect:pheno}.

It is also important to stress that, contrary to the perturbative corrections as in \eq{expan}, that unavoidably induce the most general structures consistent with the symmetries of the theory, it is perfectly conceivable, and indeed quite natural, that a non-perturbative phenomenon such as a dynamical mass generation could produce a mass for the conformal mode while still leaving massless the $\hmn^{\rm TT}$ mode, as in
\eq{SEH2nlmass}. Indeed, even in the usual Higgs mechanism of the Standard Model, the photon remains massless while  the $W^{\pm}$ and $Z^0$ get  a mass. In our context, this is particularly natural because the conformal mode appears to be the most `problematic' one, both because it is the mode of the gravitational field with  the strongest IR divergences in de~Sitter space~\cite{Antoniadis:1986sb} and because of the conformal mode instability in Euclidean quantum gravity.\footnote{Indications for dynamical generation of a mass scale may also come from the running of the coupling constant associated to terms quadratic in the curvature~\cite{Maggiore:2015rma}, and in particular the Gauss-Bonnet term, whose coupling is asymptotically free and generates an IR scale through dimensional transmutation, exponentially suppressed, with respect to the Planck mass, by instanton effects~\cite{Einhorn:2014gfa,Tong:2014era}.}

Finally, we comment on the naturalness and the numerical value of the mass scale that would be generated dynamically. For simplicity, we illustrate the argument using the RR model, so that we can explain the argument in the more immediate language  of the quantum effective action. \Eq{RR} can be rewritten as
\bees\label{RR2}
\Gamma_{\rm RR}&=&
\frac{\mplr^2}{2}\int d^{4}x \sqrt{-g}\, 
\[R-\frac{1}{6} m^2R\frac{1}{\Box^2} R\]\nn\\
&=&\int d^{4}x \sqrt{-g}\, 
\[\frac{\mplr^2}{2} R-R\frac{\Lambda_{\rm\scriptscriptstyle RR}^4}{\Box^2} R\]\, ,
\ees
where $\Lambda_{\rm\scriptscriptstyle RR}=(1/12)m^2\mplr^2$. In this form, it is clear that 
$\Lambda_{\rm\scriptscriptstyle RR}$ should be taken as the fundamental scale generated dynamically, 
corresponding to a dimensionless form factor $k_R(\Box)=\Lambda_{\rm\scriptscriptstyle RR}^4/\Box^2$ in $Rk_R(\Box)R$, while the parameter $m$ is just a derived quantity introduced for convenience. The value of a scale generated 
generated dynamically in this way cannot be predicted, just as we cannot predict the value of 
$\Lambda_{\rm\scriptscriptstyle QCD}$, and can only be obtained by comparison with the observation. 
In our case, as we will see below, we need $m=O(H_0)$ in order to have a dark energy that becomes important near the present epoch.\footnote{More precisely, we will see in Section~\ref{sect:DeltaN} that
 the model has a significantly different evolution if initial conditions of order one are set during radiation dominance or during an earlier inflationary phase. In the former case $m\sim H_0$ with a numerical coefficient of order one, while in the latter case $m$ can be numerically much smaller than 
 $H_0$. \Eq{LRTmeV} therefore only holds in the former case.
 \label{foot:msmaller}}
Therefore,
\be\label{LRTmeV}
\Lambda_{\rm\scriptscriptstyle RR}=O(H_0\mplr)^{1/2}=O({\rm meV})\, .
\ee
The same holds for the RT model,  as we see by rewriting
\eq{RT} as
\be
\mplr^2\Gmn -\Lambda_{\rm\scriptscriptstyle RT}^4 \(\gmn \iBox R\)^{\rm T}=\Tmn\, ,
\ee
where we have now defined $\Lambda_{\rm\scriptscriptstyle RT}^4=(1/3) \mplr^2 m^2$. 
Thus, in the RR or RT model dark energy can be explained   by the dynamical generation of an energy scale whose value, of the order of the milli-eV (or some orders of magnitude smaller, see footnote~\ref{foot:msmaller}), even if cannot be predicted, is not particularly surprising from the point of view of quantum field theory.
This is different from attempts at explaining dark energy through the introduction of some particle of mass $m$,  in which case $m$ is the fundamental scale and 
should be fixed to the extremely small value $m\sim H_0\sim 10^{-33}$~eV. Notice also that there is no  problem of technical naturalness associated to  a scale such as $\Lambda_{\rm\scriptscriptstyle RT}$ since, just as $\Lambda_{\rm\scriptscriptstyle QCD}$, a mass scale which is generated dynamically in this way is  a renormalization group invariant.

\section{Phenomenology of the RT model}\label{sect:pheno}

We now have all the elements for working out the predictions of nonlocal gravity. We focus on the RT model, that eventually turns out to be the most interesting phenomenologically.  In order to make the paper self-contained, we begin by reviewing 
material on the background evolution already discussed and reviewed in~\cite{Maggiore:2013mea,Foffa:2013vma,Maggiore:2016gpx,Belgacem:2017cqo}. We will then move to a detailed discussion of the perturbations and an updated comparison with the cosmological data. We will finally discuss GW propagation in the RT model and show that this leads to very interesting effects that could be detected in the near future with GW detectors.


\subsection{Background evolution}\label{sect:bkg}

\subsubsection{Equations in FRW}
We consider a spatially flat Friedman-Robertson-Walker (FRW) background, $ds^2=-dt^2+a^2(t)d\vx^2$.
For symmetry reasons the spatial component $S_i$ of the auxiliary field $S_{\mu}$ must vanish, since there is no preferred spatial direction,\footnote{A recent paper~\cite{Tian:2019jcq} has studied the evolution of the RT model in FRW by setting $S_{\mu}(t) =(S_0(t),v(t),v(t),v(t) )$ and claimed that this is the most general  ansatz consistent with the rotational invariance of FRW. This is clearly wrong, since this ansatz  selects a privileged spatial direction ${\bf S}(t)=v(t) (\hatx+\haty+\hatz)$, and therefore breaks the rotational invariance  of FRW. The authors of ~\cite{Tian:2019jcq} appear to have made  confusion with the fact that, for a perfect fluid in FRW, $T^{\mu}_{\nu}$ has  the form ${\rm diag}(-\rho,p,p,p)$. Obviously, the trace $T^i_i$ of a tensor is invariant under spatial rotations, while a spatial vector $S_i$ is not! With a rotation we can bring the unit vector $ (\hatx+\haty+\hatz)/\sqrt{3}$ onto the $\hatz$ axis, and in this frame the choice of \cite{Tian:2019jcq} becomes  ${\bf S}(t)=\sqrt{3} v(t) \hatz$. This ansatz therefore is not consistent with the isotropy of FRW at the background level. Perturbations over FRW do not have to respect the isotropy,  so $S_i$ will be non-vanishing at the perturbative level.
As we will discuss in section~\ref{sect:formalism},  the vector $S_i$ contributes to scalar perturbations through fluctuations of the form $S_i=\pa_i (\delta S)$.
Since the $i$-th component of the vector  ${\bf S}=(v(t),v(t),v(t) )$ can be written as $S_i=\pa_i [v(t)r]$, where $r=|x|$, what ref.~\cite{Tian:2019jcq} is actually doing is to add to the background solution of the model an unphysical scalar perturbation $\delta S(t,r)=v(t)r$ that grows in space without bounds, radially from an arbitrarily chosen origin. Treating $S_i$ correctly as a perturbation, the  full equation for its evolution is not the one given in eq.~(2.4) of ref.~\cite{Tian:2019jcq}. Rather, it involves all other first-order quantities [see eqs.~(A.6)-(A.10) of \cite{Dirian:2014ara} for the full set of equations] and, as we will review in section~\ref{sect:formalism}, the corresponding perturbations are stable.}
and the only variables are $U(t)$ and $S_0(t)$, together with the FRW scale factor $a(t)$.
\Eqst{v2loc1}{panU} then become~\cite{Maggiore:2013mea}
\bees
H^2-\frac{m^2}{9}(U-\dot{S}_0)&=&\frac{8\pi G}{3}\rho\,
\label{loc1} \\
\ddot{U}+3H\dot{U}&=&6\dot{H}+12H^2\, ,\label{loc2}\\
\ddot{S}_0+3H\dot{S}_0-3H^2S_0&=&\dot{U}\, ,\label{loc3}
\ees
where we have written $T^{\mu}_{\nu}={\rm diag}(-\rho,p,p,p)$, and the dot denotes the derivative with respect to cosmic time $t$. It is convenient to define 
$Y=U-\dot{S}_0$, $h=H/H_0$, and 
$\Omega_i (t)=\rho_i(t)/\rho_c(t)$, where $i={\rm M}, {\rm R}, {\rm DE}$ labels radiation, matter and dark energy, respectively,  and $\rho_c(t)=3H^2(t)/(8\pi G)$. We will also use the standard notation  $\Omega_M\equiv \Omega_M (t_0)$, 
$\Omega_R\equiv \Omega_R (t_0)$ and  $\ode \equiv \ode(t_0)$ 
(where $t_0$ is the present value of cosmic time) for the present values of $\Omega_i(t)$. We henceforth  use
the dimensionless variables
\be
x\equiv \ln a(t)
\ee
instead of cosmic time $t$, and we denote $df/dx=f'$. 
Then the Friedmann equation (\ref{loc1})  reads
\be\label{hLCDM}
h^2(x)=\Omega_M e^{-3x}+\Omega_R e^{-4x}+\g Y(x)
\, ,\\
\ee
where 
\be\label{defgamma}
\g\equiv  m^2/(9H_0^2)\, .
\ee 
This shows that  there is an effective DE density
\be
\rde(t)=\rho_0\g Y(x)\, , 
\ee
where $\rho_0=3H_0^2/(8\pi G)$. Using  $U(x)$ and $Y(x)$, \eqs{loc2}{loc3} take the form
\bees
&&\hspace*{-5mm}Y''+(3-\zeta)Y'-3(1+\zeta)Y=3U'-3(1+\zeta)U\, ,\label{sy1}\\
&&\hspace*{-5mm}U''+(3+\zeta)U'=6(2+\zeta)\label{sy3}\, ,
\ees
where, using \eq{hLCDM},
\be
\zeta(x)\equiv\frac{h'}{h}=-\, \,
\frac{3\Omega_M e^{-3x}+4\Omega_R e^{-4x}
-\g Y' }{2(\Omega_M e^{-3x}+\Omega_R e^{-4x}+\g Y)}\label{syz}\, .
\ee

\subsubsection{Initial conditions; the parameter $\Delta N$}\label{sect:DeltaN}

As a next step, we discuss the initial conditions on the auxiliary fields (we follow refs.~\cite{Maggiore:2016gpx,Cusin:2016mrr}). To get a first analytic understanding we observe that, in any given cosmological epoch, such as radiation dominance (RD), matter dominance (MD), or an earlier inflationary de~Sitter (dS) phase,  $\zeta(x)$ has an approximately constant value $\zeta_0$, with  $\zeta_0=0$ in dS, $\zeta_0=-2$ in RD and $\zeta_0=-3/2$ in MD. In the approximation of constant $\zeta$ \eq{sy3} can be integrated analytically, and has the solution~\cite{Maggiore:2013mea} 
 \be \label{pertU}
U(x)=\frac{6(2+\zeta_0)}{3+\zeta_0}x+u_0
+u_1 e^{-(3+\zeta_0)x}\, .
\ee
The first term on the right-hand side is a particular solution of the inhomogeneous equation, while $u_0$ and $u_1$ parametrize the most general solution of the homogeneous equation $\Box U=U''+(3+\zeta_0)U=0$.
The initial conditions on $U$, i.e. $U(x_{\rm in})$ and  $U'(x_{\rm in})$, are in one-to-one correspondence with the choice of the solutions of homogeneous equation, i.e. with $u_0$ and $u_1$. 
The constant $u_0$ corresponds to the reintroduction of a cosmological constant, as we have seen in \eq{u0cosmconst}. Our  aim is to see if we can obtain a self-accelerated evolution from the nonlocal term, without introducing by hand a cosmological constant, and we will therefore set $u_0=0$. A non-vanishing $u_0$  could always be reintroduced later, and, not surprisingly,  produces an evolution that is intermediate between that of the RT model with $u_0=0$ and  that $\Lambda$CDM, see section~7.4 of \cite{Maggiore:2016gpx}.
The other solution of the homogeneous equation, proportional to $e^{-(3+\zeta_0)x}$, is instead a decaying mode, in all cosmological phases. Thus, 
the solution with initial conditions $U(x_{\rm in})=U'(x_{\rm in})=0$ has a marginally stable direction, corresponding to the possibility of reintroducing a cosmological constant, and  a stable direction, i.e. is an attractor in the $u_1$ direction.
Consider next \eq{sy1}. Using \eq{pertU} and solving for $Y(x)$ 
we get \cite{Maggiore:2013mea}
\bees
Y(x)&=&-\frac{2(2+\zeta_0)\zeta_0}{(3+\zeta_0)(1+\zeta_0)}
+\frac{6(2+\zeta_0)}{3+\zeta_0} x+u_0
-\frac{6(2+\zeta_0) u_1 }{2\zeta_0^2+3\zeta_0-3}e^{-(3+\zeta_0)x} 
\nn\\
&& +a_1 e^{\a_{+}x}+ a_2 e^{\a_{-}x}\, ,\label{pertY}
\ees
where 
$\a_{\pm}=(1/2)[-3+\zeta_0\pm \sqrt{21+6\zeta_0+\zeta_0^2}]$.
In both RD and MD we have $\alpha_+<0$ and $\alpha_-<0$, so both modes are decaying.
This means that, if we start the evolution deep in the RD phase, with $u_0=0$ in order not to have a cosmological constant, and $u_1\sim a_1\sim a_2\sim{\cal O}(1)$, the solution will quickly approach the one obtained with initial conditions $U(x_{\rm in})=U'(x_{\rm in})=Y(x_{\rm in})=Y'(x_{\rm in})=0$. We will refer to this solution as the `minimal' RT model.

The situation becomes more interesting if we start the evolution during a primordial phase of de~Sitter-like inflation, before RD. In dS there is a growing mode with  $\alpha_+=(-3+\sqrt{21})/2\simeq 0.79$. Then $Y$ will grow during dS (exponentially in $x$, so as a power of the scale factor), and will then decrease again during RD and MD. In general, a growing mode during MD or the late RD phase would be fatal to the viability of the model, because any perturbation of the initial conditions would result in an activation of the unstable mode, and would bring the solution very far from a FRW solution  driven by $\Tmn$, as in standard cosmology (this is indeed a criterium that ruled out several other nonlocal  models, as we will recall  in app.~\ref{sect:difficulties}). For the evolution during an early dS phase the situation is, however, different~\cite{Maggiore:2016gpx,Cusin:2016mrr}. Indeed, let us
denote by $x_{\rm in}$ the value of  $x=\ln a$ at a time, during inflation, when we set  initial conditions  $u_1\sim a_1\sim a_2\sim{\cal O}(1)$, and by $x_{\rm end}$ the value  when inflation ends and RD begins (we neglect for simplicity an intermediate reheating phase). We use the notation  
\be
x_{\rm end}-x_{\rm in}=\log \(a_{\rm end}/a_{\rm in}\)\equiv \Delta N\, ,
\ee 
so $\Delta N$ is the number of e-folds from the  time where we set initial condition of order one,
to the end of a de~Sitter phase of inflation. Thus, if $Y(x_{\rm in})$ has a generic value of order one (i.e., is not fine-tuned to zero), by the end of inflation 
\be\label{YexpDeltaN}
Y(x_{\rm end})\simeq \exp\{\alpha_+^{\rm dS} \Delta N\}\simeq \exp\{0.79\Delta N\}\, . 
\ee
The evolution of $U$ can be computed similarly, using \eq{pertU}. During a quasi-de~Sitter phase of inflation, starting from a value of order one, we get 
\be\label{Uxend}
U(x_{\rm end})\simeq 4\Delta N\, . 
\ee
The important point is that, despite the exponential growth in \eq{YexpDeltaN}, even for very large values of  $\Delta N$  the corresponding  DE density $\rde(x)=\rho_0\g Y(x)$  has no effect on the inflationary dynamics. This is due to the fact that $\rho_0=3H_0^2/(8\pi G)\sim (10^{-3} {\rm eV})^4$ is extremely small compared to the energy density during inflation. For instance, if  $Y(x_{\rm in})={\cal O}(1)$ and we take $\Delta N=60$, at the end of inflation we get $Y(x_{\rm end})={\cal O}(10^{20})$.
Even with such a large value of $Y$, we have
\be
[\rho_0 Y(x_{\rm end})]^{1/4}\sim 10^{-3} {\rm eV}\times Y^{1/4}(x_{\rm end})\sim 10^2
\, {\rm eV}\, .
\ee   
This is totally negligible  compared to the inflationary scale  $M$, that has  typical values, say, of order $10^{13}$~GeV. Thus, during the inflationary phase the evolution of the scale factor is the same  as in standard GR without the nonlocal term. So, the important conclusion is that, at the level of background evolution, there is no  
evident pathology associated  with the exponential growth of $Y(x)$. Rather, one will have to study in detail the  evolution through dS, RD and MD to see if it gives a viable and interesting background cosmology. As we will recall below, following~\cite{Maggiore:2016gpx,Cusin:2016mrr}, indeed the corresponding background evolution is viable, and also quite interesting. As discussed in \cite{Cusin:2016mrr}, 
even at the level of cosmological perturbations this growth during de~Sitter is innocuous, again because of the smallness of the scale associated to the nonlocal  term with respect to the inflationary scale.

\Eqs{YexpDeltaN}{Uxend} give the values of $Y(x)$ and $U(x)$ when they enter the subsequent RD phase
(apart from some minor modification due to reheating).  As we will see explicitly in  section~\ref{sect:results}, even if in the RD and MD phases the solution obtained with vanishing initial conditions is an attractor, the fact that $Y(x)$ enters the RD phase with an exponentially large value gives an evolution that is sensibly different from that of the minimal model, simply because there is not enough time to relax to zero this exponentially large value by the end of the MD phase and the beginning of the current DE-dominated phase, when (having chosen   $m$ of order $H_0$) the energy scale associated to $Y$ eventually becomes comparable to the total energy density. Thus, there is a residual dependence of the dark energy evolution near the present cosmological epoch, from the value $\Delta N$ that determines, through \eqs{YexpDeltaN}{Uxend}, the values of the auxiliary field when the enter  the RD phase.

The conclusion is that, at the level of the background cosmological evolution, our ignorance on the initial conditions of the auxiliary fields can be reabsorbed into a single parameter $\Delta N$, that gives the number of e-folds from the moment when these fields have initial conditions ${\cal O}(1)$ during inflation, until the end of inflation (plus the parameter $u_0$, that corresponds to reintroducing a cosmological constant, and that we will set to zero). 

As discussed in \cite{Belgacem:2017cqo}, no further freedom emerges at the level of cosmological perturbations. Indeed, at the perturbation level we must consider all Fourier modes of the perturbations, so in principle we should assign the initial conditions on 
$\d U_{\vk}(x)$, $\d Y_{\vk}(x)$ and on their first time derivatives, at an initial time $x_{\rm in}$. 
The fact that the auxiliary fields do not represent arbitrary degrees of freedom but are fixed in terms of the metric means that the initial conditions for the perturbations of the auxiliary fields will be of order of the metric perturbations. One can therefore ask what happens if we start with initial conditions of this order of magnitude. The explicit numerical study in \cite{Belgacem:2017cqo} shows that the effect of such a change in the initial conditions of the perturbations is completely negligible. 

In the rest of this paper we will study the predictions of the RT model for a  few values of $\Delta N$. For this purpose, it is useful to recall that, for inflation taking place at a scale 
$M_{\rm infl}=(\rho_{\rm infl})^{1/4}$, assuming instantaneous reheating,  the minimum number of e-folds required to solve the flatness and horizon problems is 
(see e.g. sect.~21.1 of \cite{Maggiore:2018sht})
\be\label{DeltaNMinfl}
(\Delta N)_{\rm min}\simeq 64-\log\frac{10^{16}\, {\rm GeV}}{M_{\rm infl}}\, .
\ee
In the following, beside the `minimal' model defined by initial conditions of order one during RD, which is equivalent  to setting $\Delta N=0$ (or, equivalently,  $\Delta N$ of order one), we will study also the cases 
$\Delta N=34,50,64$ that, according to \eq{DeltaNMinfl}, approximately correspond to  the minimum value of $\Delta N$ for $M_{\rm infl}=\{10^3,10^{10},10^{16}\}$~GeV, respectively. These values are chosen because
$M_{\rm infl}=10^3$~GeV corresponds to inflation at the electroweak scale, which is on the lower range of possible inflationary scales,
while $M_{\rm infl}=10^{16}$~GeV is the highest value consistent with the non-detection of tensor perturbations in the CMB anisotropies, and $M_{\rm infl}=10^{10}$~GeV is an intermediate value which is quite often considered as a typical inflationary scale. 

Of course, the number of e-folds during inflation at a given scale does not need to be the minimum required to solve the flatness and horizon problems and, for a given  value of $M_{\rm infl}$, we could chose a higher value of $\Delta N$. We have therefore studied also how the results change increasing $\Delta N$ for a fixed value of $M_{\rm infl}$. As already pointed out in \cite{Cusin:2016mrr}, increasing $\Delta N$ the results eventually saturate to a limiting curve (as a function of redshift). In particular, setting $M_{\rm infl}=10^{16}$~GeV, we find that this limiting curve is reached, within  sub-percent level accuracy, already for $\Delta N\simeq 70$. In the following, beside the cases $(M_{\rm infl}=10^3\, {\rm GeV}, \Delta N=34)$,
$(M_{\rm infl}=10^{10}\, {\rm GeV}, \Delta N=50)$ and $(M_{\rm infl}=10^{16}\, {\rm GeV}, \Delta N=64)$, we will also show the results for $(M_{\rm infl}=10^{16}\, {\rm GeV}, \Delta N=100)$, that represents the limiting curve for the various background quantities as a function of redshift. 
For brevity, we will refer to these cases as the RT model with $\Delta N=34,50,64$ and 100, respectively.
We have checked that the same limiting curve is obtained starting from a different value of $M_{\rm infl}$ and raising again sufficiently $\Delta N$. This behavior is due to a scaling property of the equations when $Y$ starts from a very large value at the beginning of RD~\cite{Cusin:2016mrr}.

A  quite interesting aspect of the cosmological evolution of the RT model with initial conditions set during inflation, that will emerge clearly from the discussion below,  is that the behavior of dark energy at the present epoch depends on the existence and duration (as quantified by $\Delta N$) of a phase of primordial inflation, providing an unexpected connection between early- and late-time cosmology.

\subsubsection{Results: $\rde(z)$, $\wde(z)$, $H(z)$}\label{sect:results}

\begin{figure}[t]
 \centering
\includegraphics[width=0.42\textwidth]{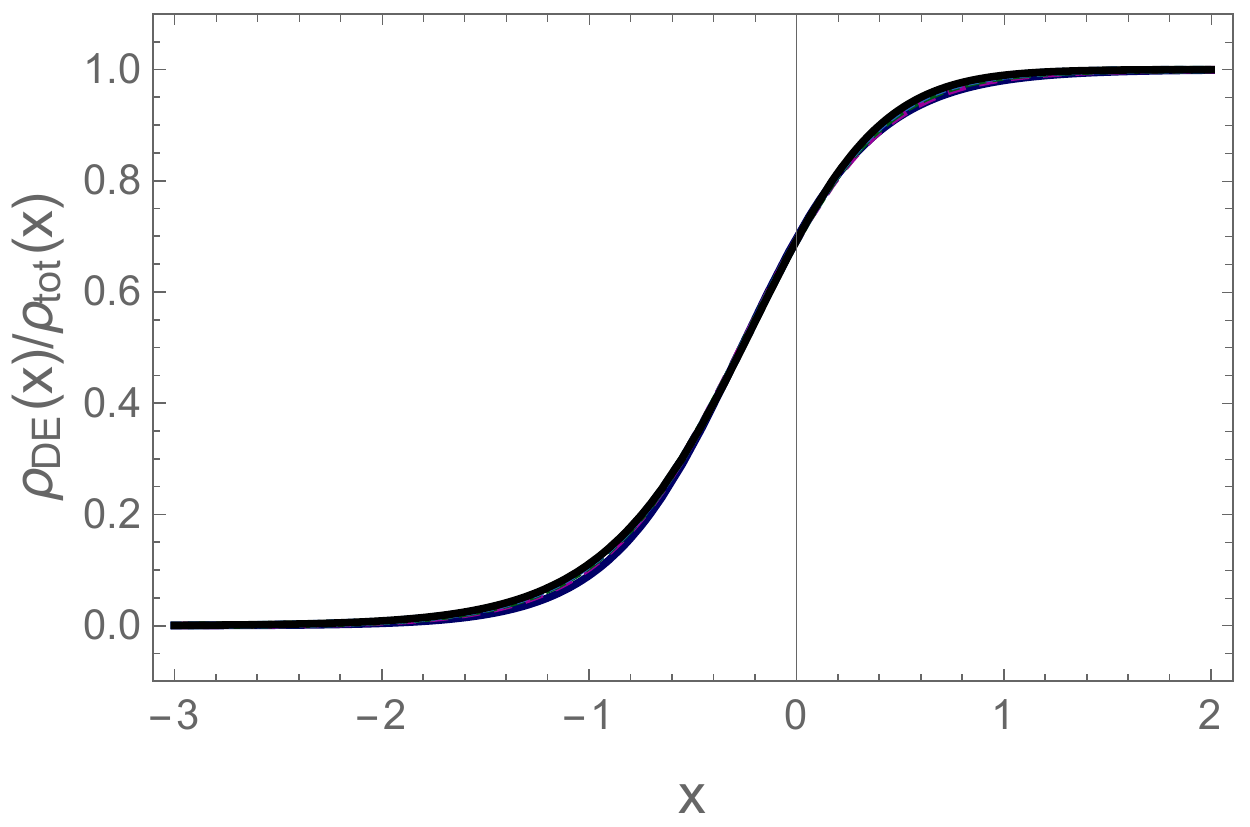} 
\includegraphics[width=0.42\textwidth]{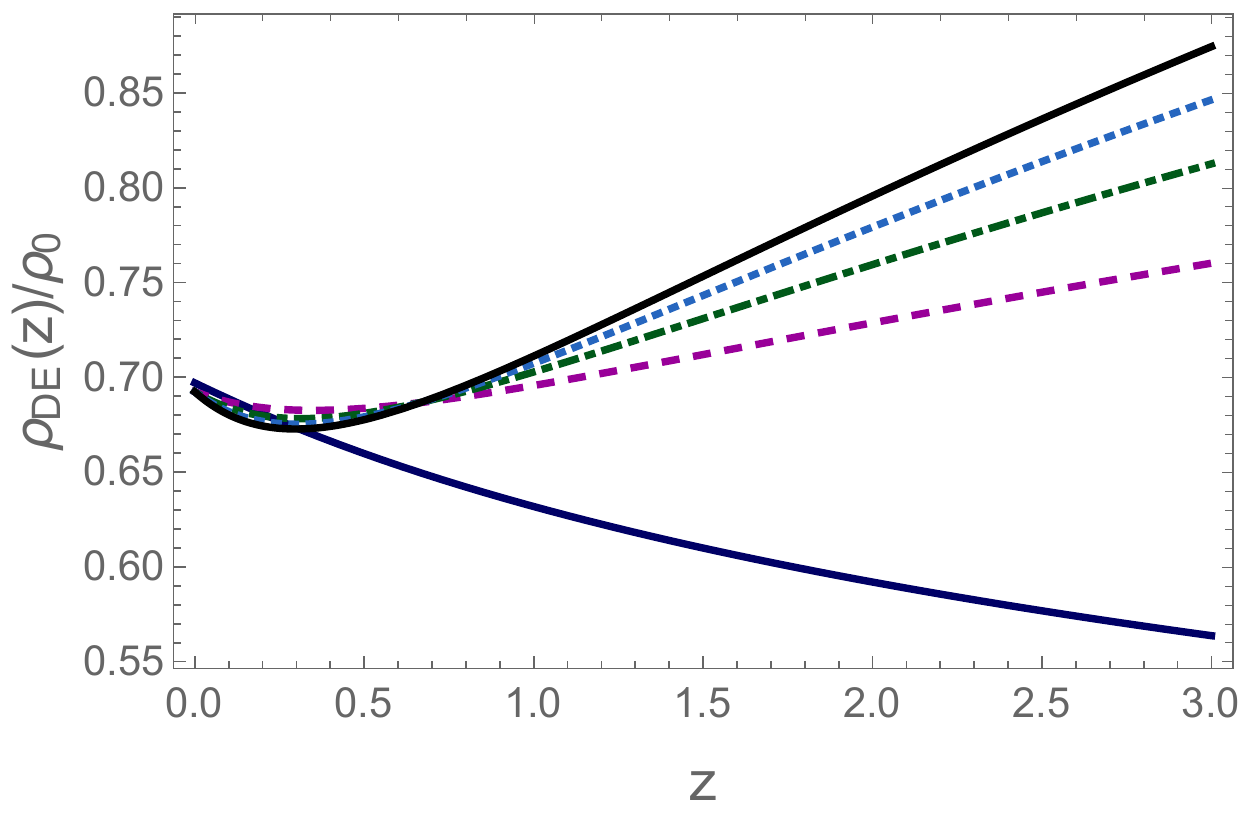} 
\includegraphics[width=0.42\textwidth]{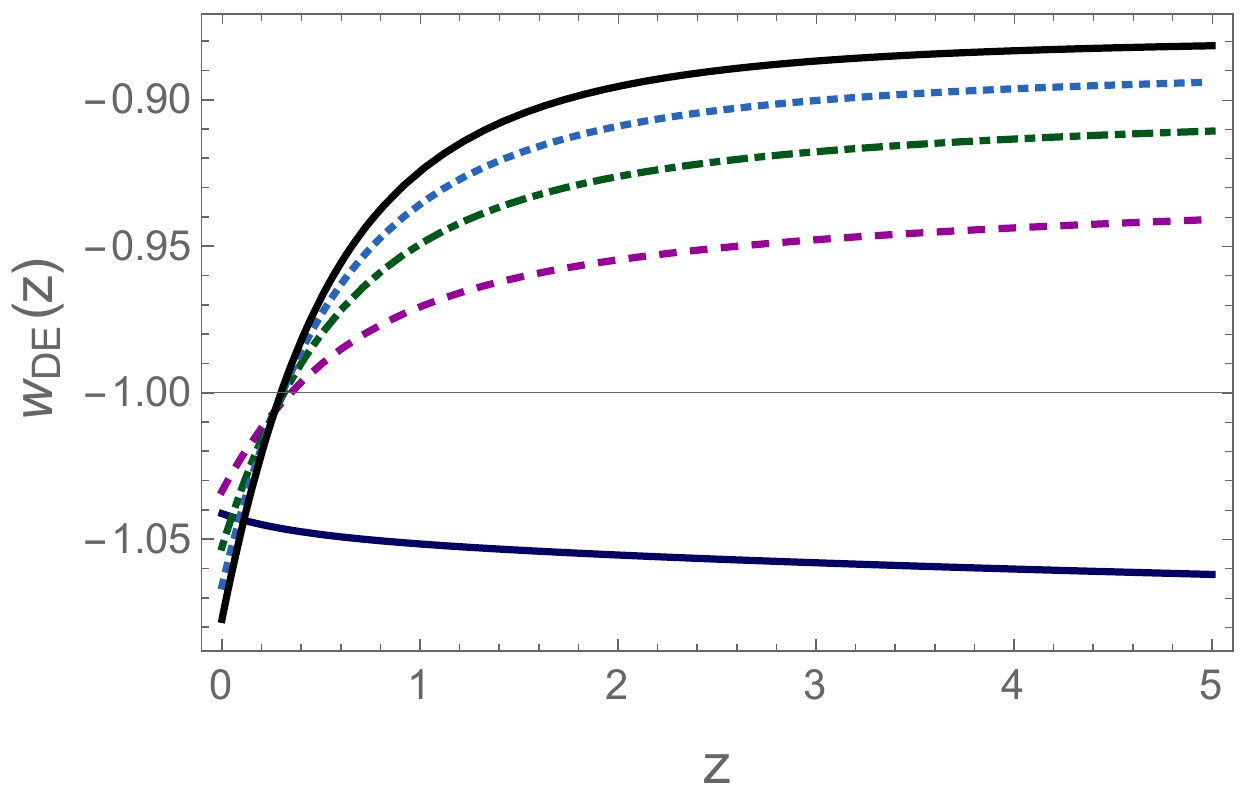}
 \caption{Upper left panel:  $\rde(x)$ normalized to the total energy density $\rho_{\rm tot}(x)$ as a function of $x$. Upper right panel:  $\rde(z)$ normalized to the critical energy density today, $\rho_0$, as a function of  redshift $z$.  Lower panel:
the DE equation of state $\wde(z)$ as a function of redshift. The curves correspond to 
 the minimal RT model (blue solid line) and  the RT model with  $\Delta N=34$ (magenta, dashed), $\Delta N=50$ (green, dot-dashed), $\Delta N=64$ (cyan, dotted) and  $\Delta N=100$ (black solid line).}
  \label{fig:wde}
\end{figure}

Given the initial conditions and a choice of values for the cosmological parameters $\oma$ and $h_0$ (defined as usual from $H_0 = 100 h_0 \, \rm{km} \, \rm{s}^{-1} \rm{Mpc}^{-1}$), the numerical integration of the equations for the background evolution, \eqst{hLCDM}{sy3}, is straightforward.\footnote{In practice, in the numerical implementation of our integration routine, we consider that the transition between inflation and RD takes place when, extrapolating backward in time 
the present energy density in radiation $\rho_{R,0}$,  the energy density in radiation $\rho_{R,0}/a^4$ becomes equal to $M^4_{\rm infl}$, i.e. when the scale factor has the value $a_{\star}$ given by
$a_{\star}=\rho^{1/4}_{R,0}/M_{\rm infl}$ (notice that the quantity $M_{\rm infl}$ defined in this way  corresponds to the actual inflationary scale only in the approximation of instantaneous reheating). Using $\rho^{1/4}_{R,0}\simeq 2.41\times 10^{-4}$~eV,
the corresponding value of $x=\log a$ is 
$x_{\star}\simeq -65.9+\log (10^{16}\, {\rm GeV}/M_{\rm infl})$. Assuming that initial conditions of order one have been set $\Delta N$ e-folds earlier, at the inflation-RD transition we take $Y=\exp\{0.79\Delta N\}$  and $U=4\Delta N$. The numerical integration through the full RD phase would be numerically difficult, and not necessary, since we know that, until we are deep in RD, the solution for $Y$ evolves according to the slowest-decaying mode, which decays as $\exp\{-0.70 x\}$ and the solution for $U$ stays constant. Thus, at a value $x_0$ still deep into RD (we take $x_0=-15$;   RD-MD equilibrium is at $x\simeq -8.1$) we have $U(x_0)=4\Delta N$, $U'(x_0)=0$,
$Y(x_0)=\exp\{0.79\Delta N-0.70 (x_0-x_{\star}) \}$ and  $Y'(x_0)=-0.70 Y(x_0)$. At this point we start the numerical evolution with these initial conditions. To produce Fig.~\ref{fig:wde}, for the minimal model and for
 $\Delta N=34,50,64$ we have used the respective mean values for $\oma$ and $h_0$ from Table~\ref{tab:results}, obtained from our MCMC chains.  For the limiting curves $\Delta N=100$  we have not rerun our MCMC and we have used the same values as for $\Delta N=64$, which is an excellent approximation since we see from 
Table~\ref{tab:results} that, for large $\Delta N$, the variation in the parameters are very small (and would give effects totally unappreciable on the scale of the figures). A final detail is that,
in $\Lambda$CDM, assuming flatness and fixing $\oma$ and $\ora$, directly fixes $\ola$ from $\oma+\ora+\ola=1$, and one can immediately integrate the evolution equations. In contrast, in the nonlocal model, once fixed $\oma$ and $\ora$ (and assuming flatness), the remaining parameter in the equations is $\gamma$, which is fixed by trials and errors  until the value of the dark energy energy fraction today, $\ode$, obtained from the solution of the equations,  satisfies the condition $\oma+\ora+\ode=1$, i.e. $\ode\simeq 0.7$. The corresponding values of  $\gamma$ turn out to be $\gamma\simeq 5.13555\times 10^{-2}$ for the minimal model, and
$\gamma\simeq \{2.69512\times 10^{-3},1.0321\times 10^{-3}, 3.73915\times 10^{-4}, 1.94944\times 10^{-11}\}$ for
$\Delta N=34,50,64,100$, respectively. For the mass $m$ this means $m/H_0\simeq 0.68$ for the minimal model, and $m/H_0\simeq \{0.16,0.10,0.06,4.2\times 10^{-8}\}$ for  $\Delta N=34,50,64,100$. We perform the numerical integration of the differential equations both with Mathematica and with CLASS, and we check the consistency of the results.
\label{foot:valuesgamma}} 
In the following figures we show the results  for the minimal RT model and for the RT model with $\Delta N=34,50,64,100$.

The upper left panel of Fig.~\ref{fig:wde} shows the evolution of the dark energy density $\rde(x)$, normalized to the total energy density $\rho_{\rm tot}(x)=\rho_M(x)+\rho_R(x)+\rde(x)$, as a function of $x$ [recall that here $x=\ln a$, and we normalize the scale factor so that $a(t_0)=1$]. For orientation, matter-radiation equilibrium is at $x\simeq -8.1$, at the present epoch  $x=0$, and $x>0$ corresponds to the cosmological future. We see from the plot that the DE density due to the nonlocal term is negligible until the relatively recent cosmological epoch, when  eventually dominates.  

When $\rde(x)$ is normalized to $\rho_{\rm tot}(x)$, which includes the contribution of $\rde(x)$ itself, the result for  the minimal model and for the RT models with large $\Delta N$ look all very similar, and the various curves are basically indistinguishable. However, the individual behaviors of $\rde(x)$ are quite different. This is shown in the upper right panel of Fig.~\ref{fig:wde}, where $\rde$ is shown as a function of the redshift $z$ [related to $x$ by $x=-\log (1+z)$], and normalized to the constant critical energy density  today $\rho_0$. We see that, as we approach the present epoch from large $z$, in the minimal model $\rde$ increases, until it reaches the present value
$\rde/\rho_0\simeq 0.7$, which is fixed by our choice of $\oma\simeq 0.3$.
In contrast, for large $\Delta N$, $\rde$ starts from a very large value deep in RD (a consequence of the large value of the auxiliary field $Y$ at the end of inflation), and then decreases for most of its evolution, until the present epoch. This behavior can be understood observing that, for $\Delta N=0$, the evolution of $Y$ is determined by the particular solution of the inhomogeneous equation (\ref{sy1}), which stays close to zero during RD and then starts to increases with time during MD, until we enter in a regime dominated by DE; in contrast,  for large $\Delta N$ the solution starts from a very large initial value at the beginning of RD and then decays according to the decaying modes of the associated homogeneous equation, until, close to the recent epoch, the decaying modes have become smaller than the solution of the inhomogeneous equation, that takes over, so the solution for $Y$ starts to rise again. 

As mentioned before, for sufficiently large $\Delta N$, the results saturate toward a limiting curve, independent of the chosen value of $M_{\rm infl}$.  As explained in \cite{Cusin:2016mrr}, this is due to the fact that,
for sufficiently large $\Delta N$, an increase in the initial values of $Y$ at the beginning of RD is exactly compensated by a decrease in $\gamma$, and we end up on the same solution. This limiting curve is shown as the black solid line in Fig.~\ref{fig:wde}, obtained for definiteness setting  $(M_{\rm infl}=10^{16}\, {\rm GeV}, \Delta N=100)$. For instance, in this and in all similar plots below, on the scale of the figure all the curves with $M_{\rm infl}=10^{16}$~GeV and $\Delta N\,\gsim\, 70$ are indistinguishable, and fall on this asymptotic curve.

\begin{figure}[t]
\centering
\includegraphics[width=0.5\textwidth]{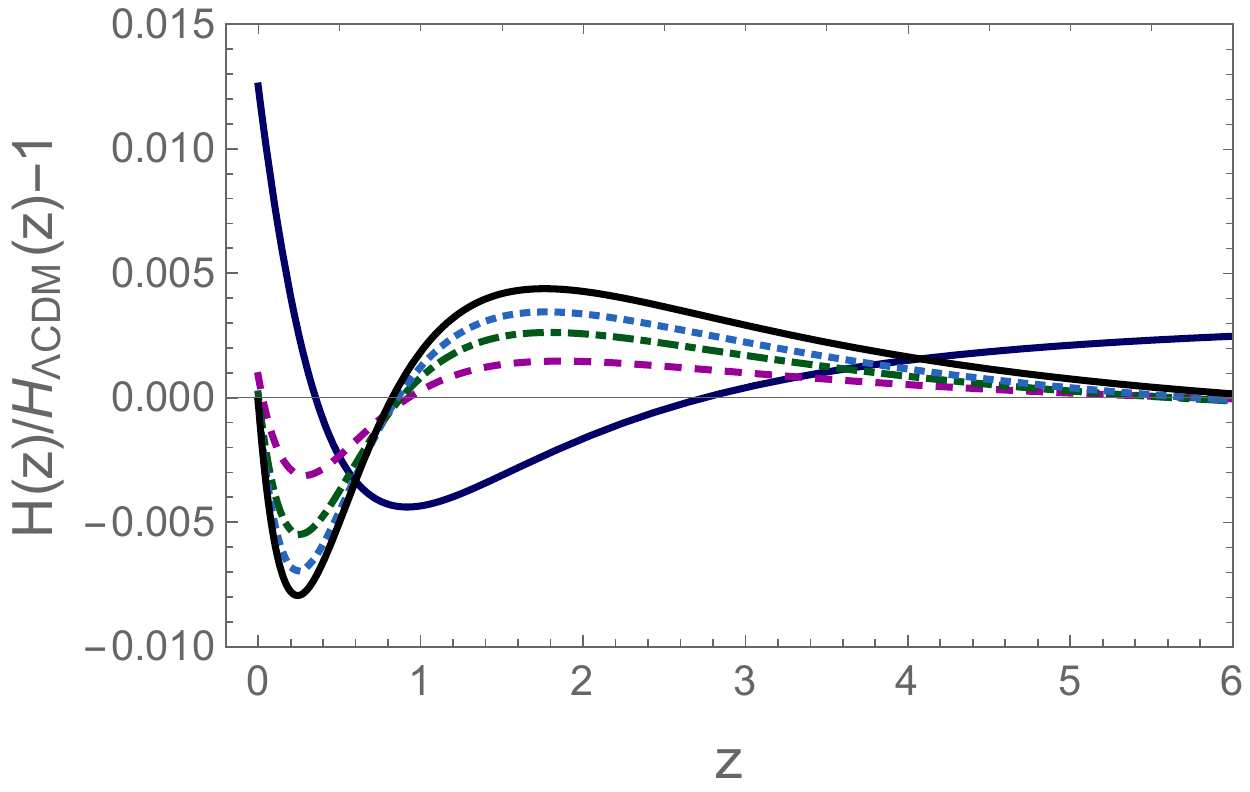}
\caption{Relative difference of Hubble rate with respect  to $\Lambda$CDM
 for the minimal RT model (blue solid line) and  for RT with $\Delta N=34$ (magenta, dashed), $\Delta N=50$ (green, dot-dashed), $\Delta N=64$ (cyan, dotted) and  $\Delta N=100$  (black solid line).
\label{fig:Hubble}}
\end{figure}

The lower panel in Fig.~\ref{fig:wde} shows the DE equation of state, defined as usual from the conservation equation
\be\label{consrho}
\dot{\rho}_{\rm DE}+3H(1+w_{\rm DE})\rho_{\rm DE}=0\, .
\ee
The  different evolutions of $\rde$ for the minimal model and for  large $\Delta N$ result in different, and quite distinctive behaviors of $\wde$ as a function of redshift. For the minimal model $\wde(z)$ is always on the `phantom' side, $\wde(z)<-1$, while, for large $\Delta N$, the evolution exhibits `phantom crossing' at
$z\simeq 0.30-0.35$. 
In all cases, we see that the DE density starts to dominate near the present cosmological epoch, and its equation of state corresponds to accelerated expansion. Thus,  the nonlocal term generates a dynamical  DE density that drives an accelerated expansion of the Universe at the current cosmological epoch. This is already a very non-trivial result: it means that giving a mass to the conformal mode, and covariantizing it as discussed in section~\ref{sect:cov}, provides an explanation for the  observed accelerated expansion of the Universe.

\begin{figure}[t]
 \centering
\includegraphics[width=0.42\textwidth]{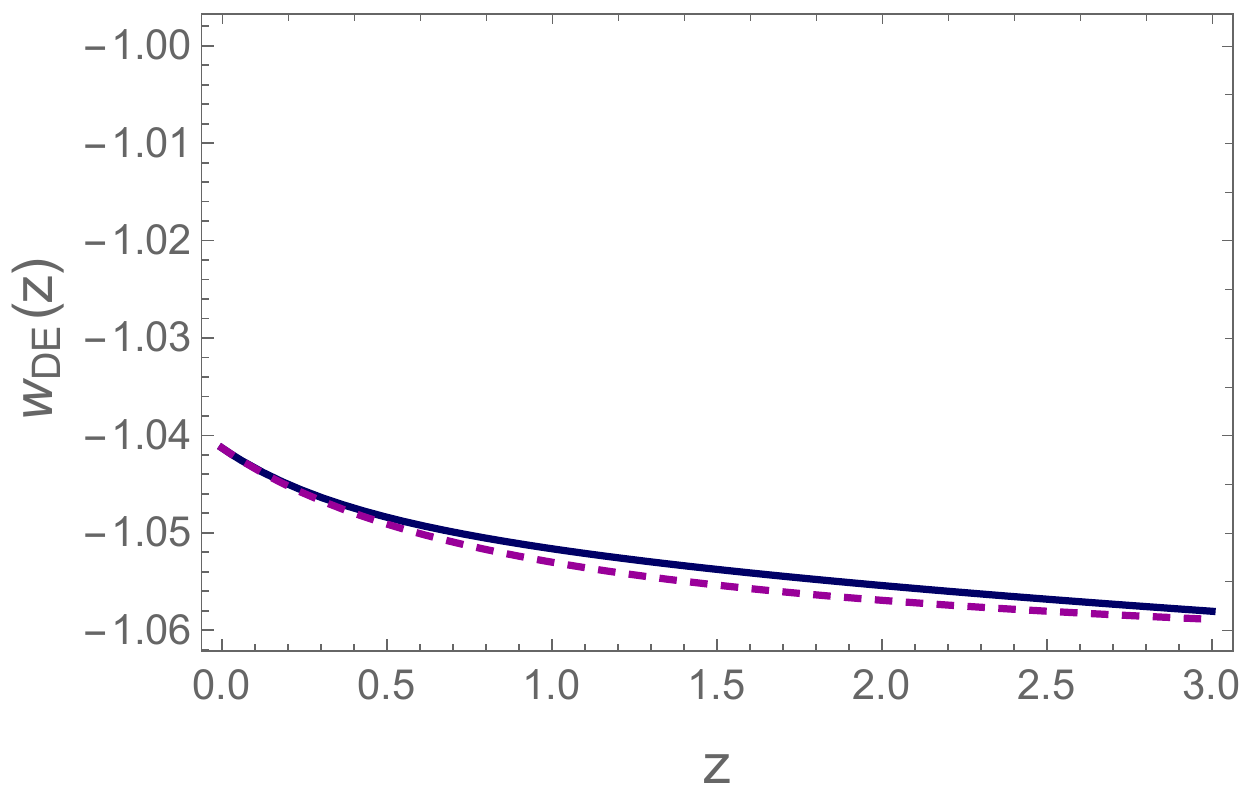} 
\includegraphics[width=0.42\textwidth]{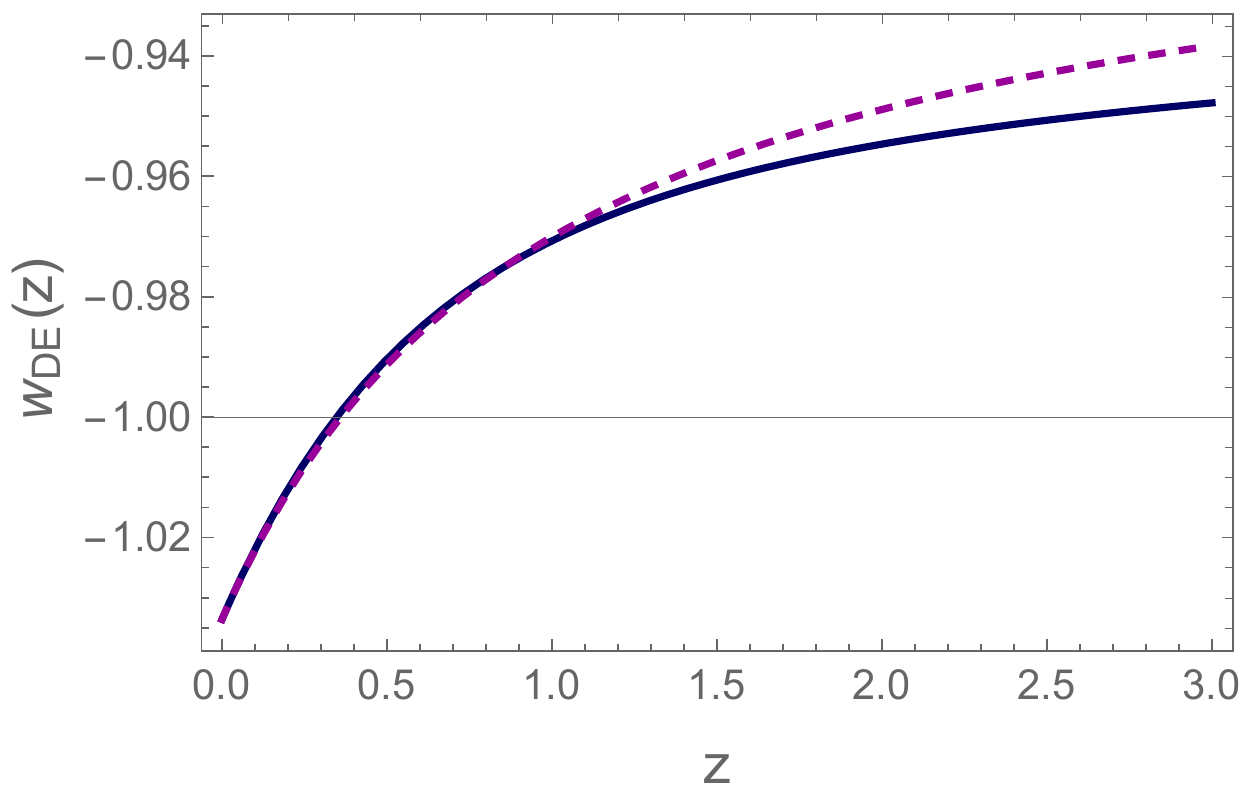} 
\includegraphics[width=0.42\textwidth]{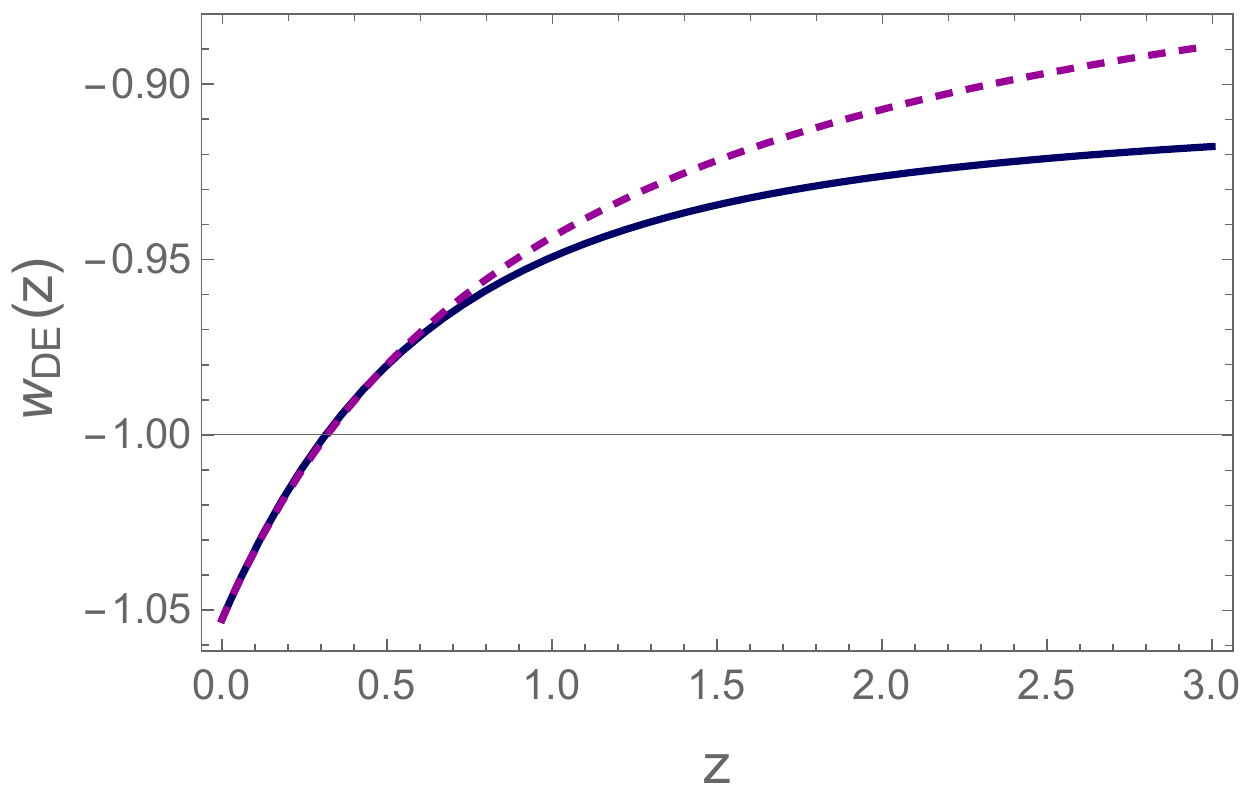}
\includegraphics[width=0.42\textwidth]{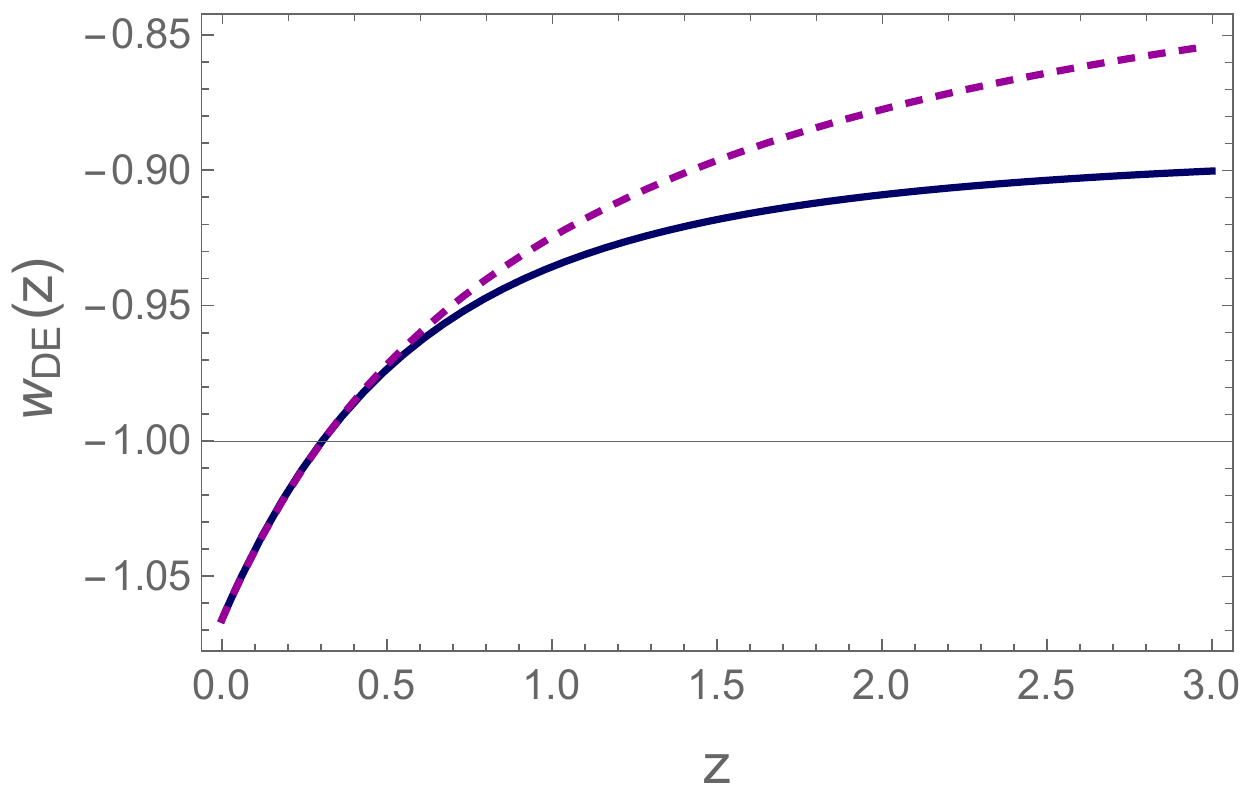}
\caption{The DE equation of state $\wde(z)$ from the numerical integration of the equations (blue solid lines), compared with the parametrization (\ref{ChevLind}) (magenta dashed lines) for RT minimal (upper left panel) and RT with $\Delta N=34$ (upper right), $\Delta N=50$ (lower left) and $\Delta N=64$ (lower right).}
\label{fig:fitwde}
\end{figure}

\begin{table*}[t]
\centering
\begin{tabular}{|c|c|c|c|c|c|c|}
\hline  
&  RT, minimal & $\Delta N=34$ & $\Delta N=50$ &  $\Delta N=64$ &  $\Delta N=100$\\ \hline
$w_0$   & $-1.041$ & $-1.034 $     & $-1.053$    & $-1.066$    & $-1.077$      \\ 
$w_a$   & $-0.023$ & $+0.127$     & $+0.218$   & $+0.283$    & $+0.335$       \\         
\hline
\end{tabular}
\caption{Values of $w_0$ and $w_a$ for the RT model, minimal and with various values of $\Delta N$.\label{tab:w0wa} }
\end{table*}

Fig.~\ref{fig:Hubble} shows the relative difference  $[H_{\rm RT}(z)-H_{\Lambda{\rm CDM}}(z)]/H_{\Lambda{\rm CDM}}(z)$ between each RT model (minimal and with $\Delta N=34,50,64,100$) and $\Lambda$CDM. Once again, the predictions of each model are computed using the respective mean  values of the cosmological parameters in Table~\ref{tab:results}.
At $z=0$ the difference between the various curves is due to the different mean values for $H_0$, and at large $z$ (but still within MD) it is determined by the different mean values for $\oma$. We see that, at $z=0$, the minimal RT model differs from $\Lambda$CDM by about $1\%$,  while the RT models with large $\Delta N$ give a prediction for $H_0$ basically indistinguishable from that of $\Lambda$CDM. Away from $z=0$, $|\Delta H(z)|/H(z)$ is of order $0.5\%$ or less.
The evolution with redshift is, however, quite distinctive, with   $\Delta H(z)/H(z)$ oscillating and changing sign  as $z$ increases. 
These differences with respect to $\Lambda$CDM can be compared to a compilation of measurements of $H(z)$ at different redshifts. We will perform this test in section~\ref{sect:Comp}, after having performed the Bayesian parameter estimation for the models.

It is interesting to compare the actual predictions of the model to the results obtained with  the standard $(w_0,w_a)$ parametrization $w_{\rm DE}(a)= w_0+(1-a) w_a$ \cite{Chevallier:2000qy,Linder:2002et},
or, in terms of redshift,
\be\label{ChevLind}
\wde(z)=w_0+\frac{z}{1+z} w_a\, .
\ee
Setting $w_0\equiv w(a=1)$ and $w_a \equiv -(dw/da)_{|a=1}$ we get the values of $w_0$ and $w_a$ given in Table~\ref{tab:w0wa}. In Fig.~\ref{fig:fitwde} we
compare the actual  numerical result for $w(z)$ to the fit provided by this parametrization. We see
that, for  large $\Delta N$, the parametrization (\ref{ChevLind}) is not  very accurate beyond some value of $z$, with the range in $z$  shrinking as $\Delta N$ increases.

\subsection{Scalar perturbations}\label{sect:scalarpert}

\subsubsection{Formalism}\label{sect:formalism}
Cosmological scalar perturbations for the RR and RT model (in the minimal case) have been studied in  detail in \cite{Nesseris:2014mea,Dirian:2014ara} (see also \cite{Maggiore:2016gpx} for review). Here, after recalling the basic formalism,  we will  extend the results to the RT model with large $\Delta N$ and we will present updated results on various indicators of cosmological perturbations, using the values of the cosmological parameters that will be determined in section~\ref{sect:Comp} by the comparison with observations.
We work  in the  Newtonian gauge, where, in the scalar perturbation sector, the perturbed FRW metric  has the form
\be\label{defPhiPsi}
ds^2 =  -(1+2 \Psi) dt^2 + a^2(t) (1 + 2 \Phi) \delta_{ij} dx^i dx^j\, ,
\ee
where $\Phi$ and $\Psi$ are the Bardeen variables. We similarly perturb the auxiliary fields, writing
\be
U(t,\vx)=\bar{U}(t)+\d U(t,\vx)\, ,\qquad S_{\mu}(t,\vx)=\bar{S}_{\mu}(t) +\d S_{\mu}(t,\vx)\, ,
\ee
where, in this section, background quantities are denoted with an overbar.
In FRW, 
$\bar{S}_{i}$ vanishes because at the background level  there is no preferred spatial direction, but  its perturbation $\d S_i$ is non-vanishing. As with any vector, we can decompose it into a transverse and longitudinal part, $\d S_i=\d S_i^{\rm T}+\pa_i (\d S)$, where $\pa_i(\d S_i^{\rm T})=0$. Since we  are considering scalar perturbations, we only retain $\d S$. Thus, in the RT model the metric perturbations in the scalar sector  are described  by $\Psi,\Phi,\d U,\d S_0$ and $\d S$. It is convenient to trade $S_0$ and $S$ for  
\be\label{defVZ}
V=H_0S_0\, ,\qquad
Z=H_0^2 S\, , 
\ee
so we eventually  work with the variables $\{\Psi,\Phi,\d U,\d V,\d Z\}$.\footnote{Note that here we are using coordinates $(t,\vx)$, where $t$ is cosmic time, and $S_0\equiv S_t$ is the $\mu=0$ component of $S_{\mu}$ with respect to these coordinates. If one rather uses  conformal time $\eta$, defined as usual by $dt=a(\eta)d\eta$, then the corresponding $\mu=0$ component  $S_{\eta}$ is related to $S_t$ by 
$S_{\eta}=aS_t$ and then $V=H_0a^{-1}S_{\eta}$. In app.~A of \cite{Dirian:2014ara}, where the perturbation equations for the RT model where first computed, the equations are written in conformal time and  the notation $S_0$ is used for $S_{\eta}$.
\label{foot:notationV}}
We similarly perform  the usual expansion of the energy-momentum tensor, writing

\begin{figure}[t]
\begin{center}
\includegraphics[width=0.42\columnwidth]{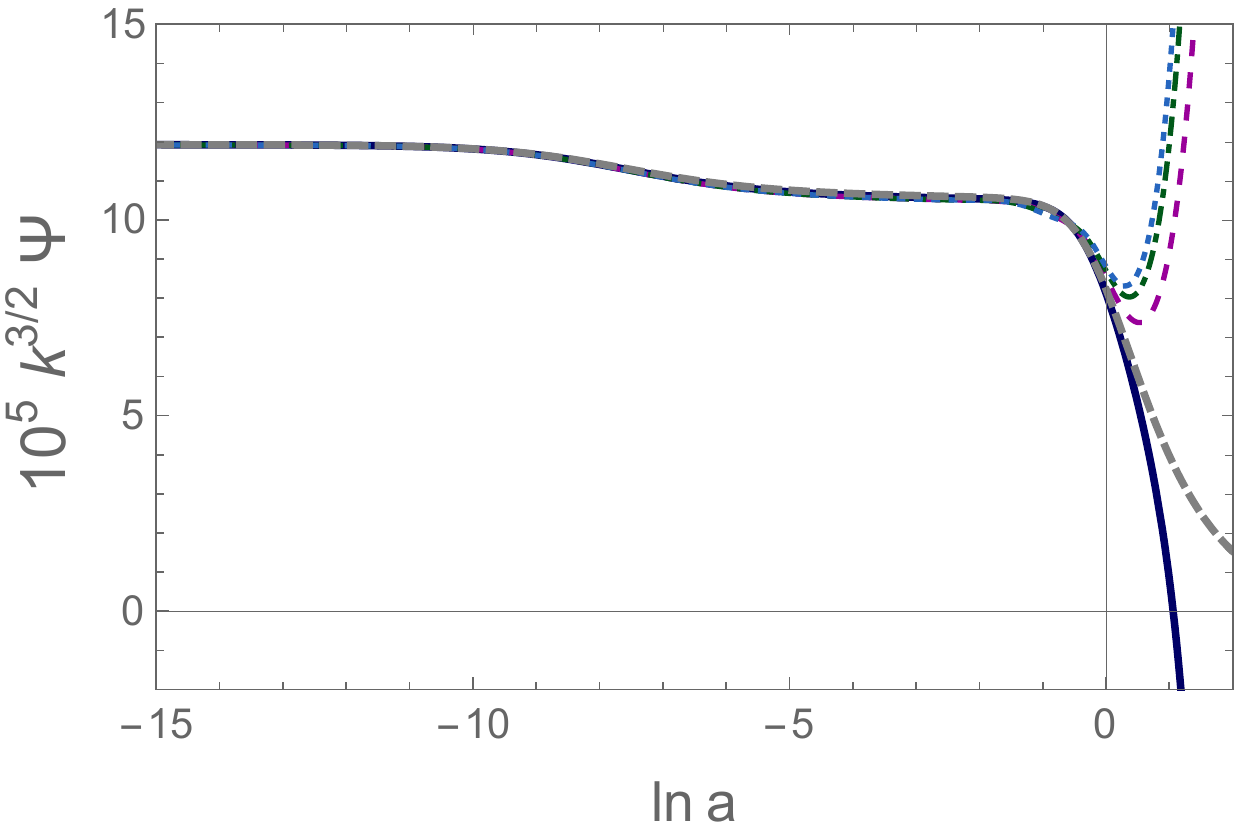}
\includegraphics[width=0.42\columnwidth]{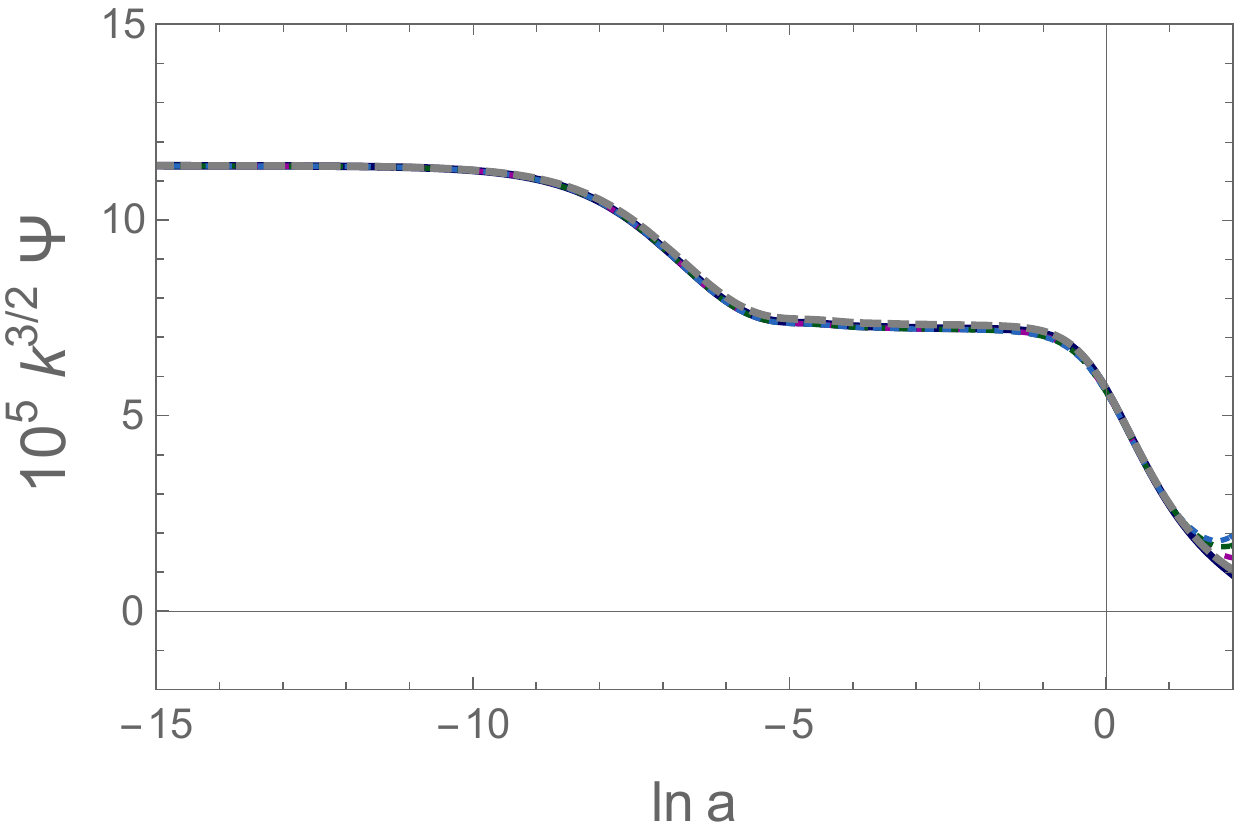}
\includegraphics[width=0.42\columnwidth]{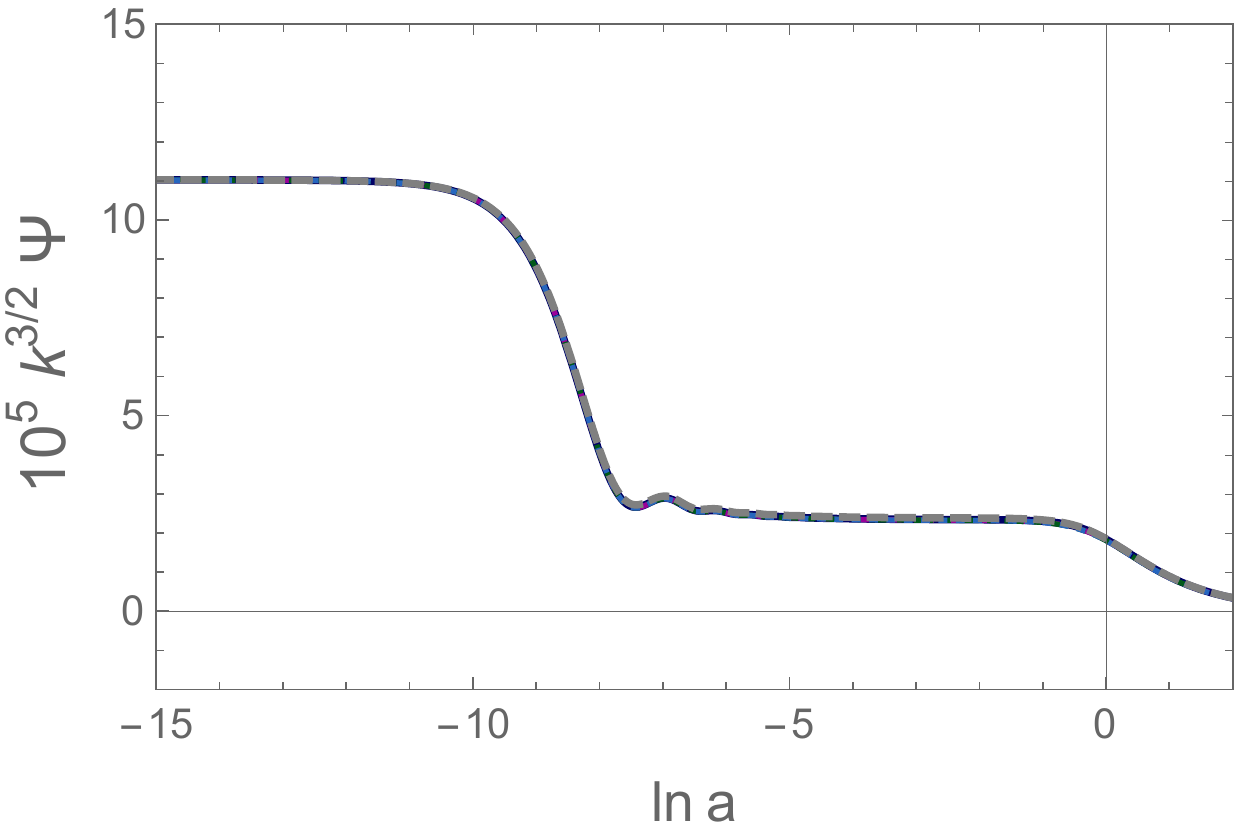}
\caption{\label{fig:PsiRT}  $k^{3/2}\Psi(a;k)$  in 
$\Lambda$CDM (gray dashed line), 
 the minimal RT model (blue solid line) and   RT with $\Delta N=34$ (magenta, dashed), $\Delta N=50$ (green, dot-dashed) and $\Delta N=64$ (cyan, dotted) for $\kappa=0.1$ (upper left panel), $\kappa=1$ (upper right) and $\kappa=5$ (lower panel). 
On the scale of these figures, the results for $\kappa=5$ are indistinguishable among the models, while for $\kappa=1$ one can barely distinguish some small differences in the cosmological future, $x>0$. 
Observe that the quantity that we plot is $k^{3/2}\Psi(a;k)$ multiplied by a factor $10^5$. Matter-radiation equilibrium is at $x\simeq -8.1$, and in this region one sees the usual transition between two different plateaux in $\Psi$.
}
\end{center}
\end{figure}

\be\label{Tij}
T^0_0 = -(\bar{\rho} +\delta \rho)\, ,\qquad
T^0_i =  (\bar{\rho} + \bar{p})v_i\, ,\qquad
T^i_j = (\bar{p} + \delta p) \delta^i_j + \Sigma^i_j\, ,
\ee
where $\bar{\rho}$ and $\bar{p}$ are the unperturbed density and pressure.
The matter perturbation variables are therefore $\delta \rho, \delta p$, $v_i$, and 
the anisotropic stress tensor $\Sigma^i_j$, which is symmetric  and  traceless, 
$\Sigma^i_i = 0$. The pressure perturbations can be written as $\d p =c_s^2\d \rho$, where $c_s^2$ is the speed of sound of the fluid, and we define as usual $\delta \equiv \delta \rho / \bar{\rho}$ and $\theta \equiv \delta^{ij} \partial_i v_j$, with $\delta_R,\theta_R$ referring to radiation and  $\delta_M,\theta_M$ to matter.
We only consider the contribution to $\Tmn$ from radiation and non-relativistic matter, so $\Sigma^i_j=0$. We transform the perturbation equations to Fourier space and we denote  comoving momenta by $k$. We further define
\be
\hat{k}=k/(aH)\, ,\qquad \hat{\theta}=\theta/(aH)\, .
\ee 
We also use
\be
\kappa \equiv k/k_{\rm eq}\, , 
\ee
where
$k_{\rm eq}=a_{\rm eq} H_{\rm eq}$ is the wavenumber of the mode that enters the horizon at matter-radiation equilibrium.  Numerically, $k_{\rm eq}\simeq 0.014 \, h_0\, {\rm Mpc}^{-1}\simeq 0.010 \, {\rm Mpc}^{-1} $.
To illustrate our numerical results, we  
use as reference values $\kappa = 0.1,1$ and $5$ (or just $\kappa = 0.1$ and 1, when the results for $\kappa=5$ turn out to be graphically indistinguishable from $\kappa=1$).  The mode with $\kappa=5$ entered  inside the horizon already during RD, while 
the mode  $\kappa=1$ reentered at matter-radiation equality. In contrast,  the mode with $\kappa=0.1$  was  outside the horizon during RD and most of MD, and re-entered at $z\simeq 1.5$.

The full set of  equations for the perturbations are given by eqs.~(A.6)-(A.10) of \cite{Dirian:2014ara}.
In  Fig.~\ref{fig:PsiRT}  we show the time evolution of the Fourier modes of the Bardeen variable $\Psi_{\vk}$   for the RT model (minimal and with $\Delta N=34,50,64$), obtained from the numerical integration  of these perturbation equations, and we compare with the result  in 
$\Lambda$CDM, for  $\kappa=0.1$, $\kappa=1$ and $\kappa=5$. 
We actually plot $k^{3/2}\Psi_k$, whose square gives the variance of the field per unit logarithmic interval of momentum. We see  that, up to the present time $x=0$, the evolution of the scalar perturbations is well-behaved, and very close to that of $\Lambda$CDM, and become closer and closer as $k$ increases.  
This can be understood from the fact that any  instability  induced by the nonlocal term on the cosmological evolution can only develop on a timescale $t$ such that $mt$ is (much) larger than one. However, we have seen that $m$ is of order $H_0$, and in fact numerically smaller, with $m\simeq 0.68 H_0$ for the minimal RT model and even smaller for large $\Delta N$, see footnote~\ref{foot:valuesgamma}. Thus,  any instability induced by the nonlocal term can  only develop on a timescale larger or equal than to a few times  $H_0$, and therefore in the cosmological future, where these modes could eventually enter a non-linear regime.

\subsubsection{Indicators of deviations from GR: $(G_{\rm eff}, \eta)$ and $(\mu,\Sigma)$}\label{sect:indicators}

The full set of   perturbation equations is needed for implementing the model into a Boltzmann code and comparing its predictions to CMB, BAO and SNe observations, as we will do in section~\ref{sect:Comp}. For a first qualitative understanding, however, it is convenient to introduce some simpler indicators of deviations from $\Lambda$CDM. One such quantity is the effective Newton's constant, which is defined so that the modified Poisson equation for the Fourier modes $\Phi_{\vk}$   can be rewritten as in GR, with $G$ replaced by $G_{\rm eff}(x,k)$ [recall that here $x\equiv \ln a(t)$ is used to parametrize the time evolution, and should not be confused with a spatial variable],
\bees\label{PhiGeff}
k^2\Phi_{\vk}(x) &=&4\pi G_{\rm eff}(x;k) a^2 \rho_0\\
&&\times \[ \Omega_R e^{-4x} \bigg( \delta_{R,\vk}(x) + \frac{4}{\hat{k}^2} \hat{\theta}_{R,\vk}(x) \bigg) + \Omega_M e^{-3x} \bigg( \delta_{M,\vk}(x) + \frac{3}{\hat{k}^2} \hat{\theta}_{M,\vk}(x) \bigg) \]\, .\nn
\ees
Its explicit expression in terms of the perturbed fields can be read from eq.~(A.6) of \cite{Dirian:2014ara},
\bees
\frac{G_{\rm eff}(x;k)}{G}=1+\gamma\frac{\delta U_{\vk} +h\left(2\Psi_{\vk} \bar{V}'+\Psi_{\vk}' \bar{V}-\delta V_{\vk}'\right)+3 h^2\left(\delta Z_{\vk} -\frac12\delta Z_{\vk}' \right)+3 h\left(\Psi_{\vk} \bar V-\frac12 \delta V_{\vk}\right)}{\Omega_R e^{-4x} \bigg( \delta_{R,\vk}(x) + \frac{4}{\hat{k}^2} \hat{\theta}_{R,\vk}(x) \bigg) + \Omega_M e^{-3x} \bigg( \delta_{M,\vk}(x) + \frac{3}{\hat{k}^2} \hat{\theta}_{M,\vk}(x)\bigg)}.\nonumber\\
\ees
From this  expression one finds  that, for sub-horizon modes, i.e. 
in the limit $\hat{k}\gg 1$, we have~\cite{Nesseris:2014mea,Dirian:2014ara}
\be\label{GeffGRTlargek}
\frac{G_{\rm eff}(x;k)}{G}=1+{\cal O}\(\frac{1}{\hat{k}^2}\)\, .
\ee
As we will see in section~\ref{sect:LLR}, this property, which is not shared by other modified gravity models and in particular by the RR nonlocal model, is crucial, since it allows the RT model to evade limits on the time variation of the (effective) Newton's constant obtained from Lunar Laser Ranging.

Together with $G_{\rm eff}$,  a second useful 
indicator is \cite{Amendola:2007rr}
\be\label{defeta}
\eta(x;k) =\frac{\Phi_{\vk}(x)+\Psi_{\vk}(x)}{\Phi_{\vk}(x)}\, ,
\ee
which, in GR, vanishes in the absence of anisotropic stress.
Alternatively, two useful quantities are the functions $\mu(x;k)$ \cite{Daniel:2010ky} and $\Sigma(x;k)$
\cite{Amendola:2007rr} which are defined through\footnote{In the literature the quantity that we call $1+\mu$ is sometimes denoted by $\mu$, and similarly our $1+\Sigma$ is sometimes denoted by $\Sigma$. Our definitions are such that, in GR, $\mu=\Sigma=0$.}
\be\label{defmu}
\Psi=[1+\mu(x;k)]\Psi_{\rm GR}\, \qquad
\Psi-\Phi=[1+\Sigma(x;k)] (\Psi-\Phi)_{\rm GR}\, ,
\ee
where the subscript denotes the same quantities computed in GR, assuming a $\Lambda$CDM model with the same value of $\oma$ as the modified gravity model. The advantage of this  parametrization is that it separates  the modifications to the motion of non-relativistic particles, which is described by $\mu$, from the modification to light propagation, which is encoded in  $\Sigma$. Therefore $\mu$ is sensitive to structure formation and $\Sigma$ is sensitive to lensing.

\begin{figure}[t]
\centering
\includegraphics[width=0.42\textwidth]{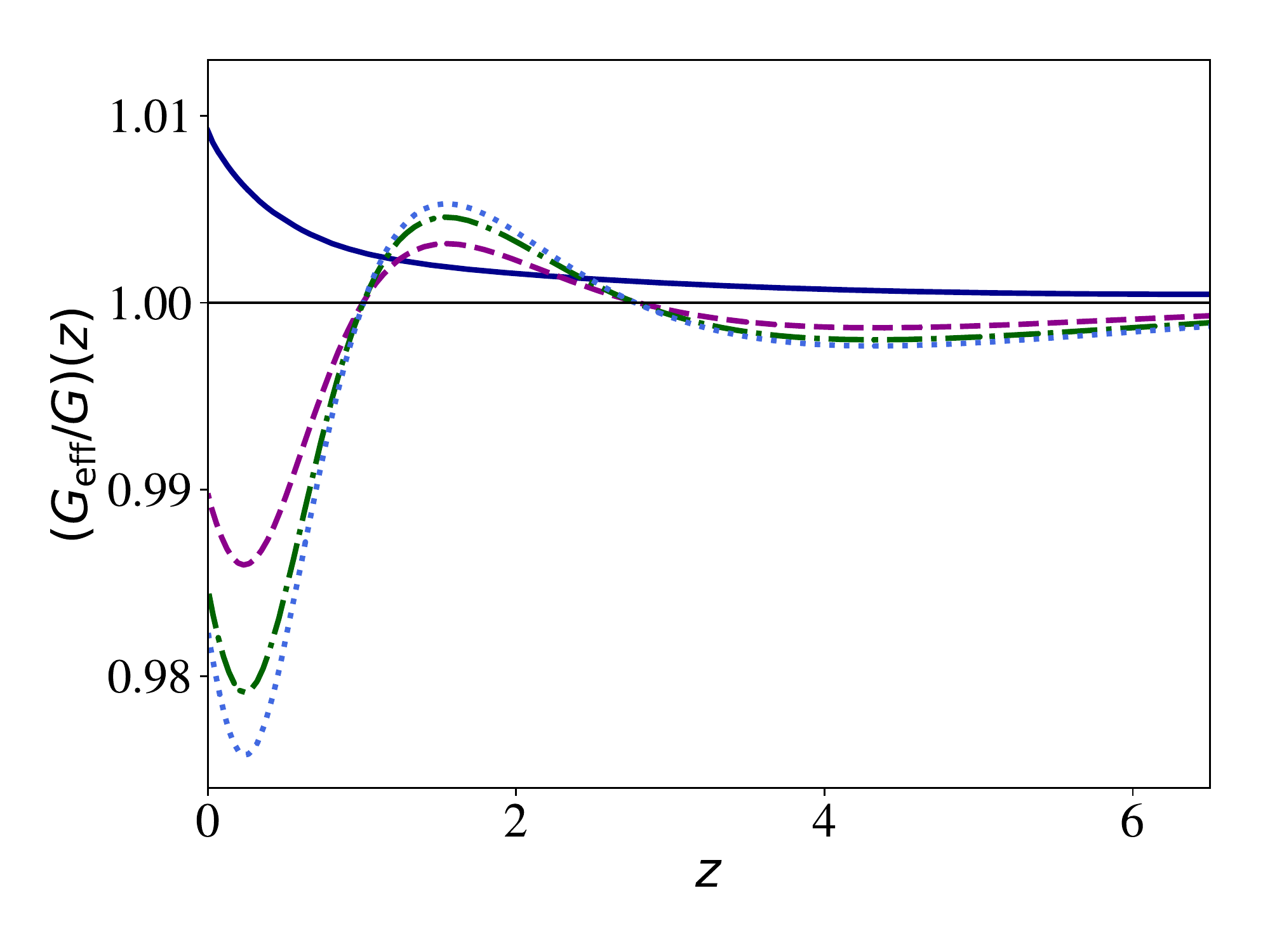}
\includegraphics[width=0.42\textwidth]{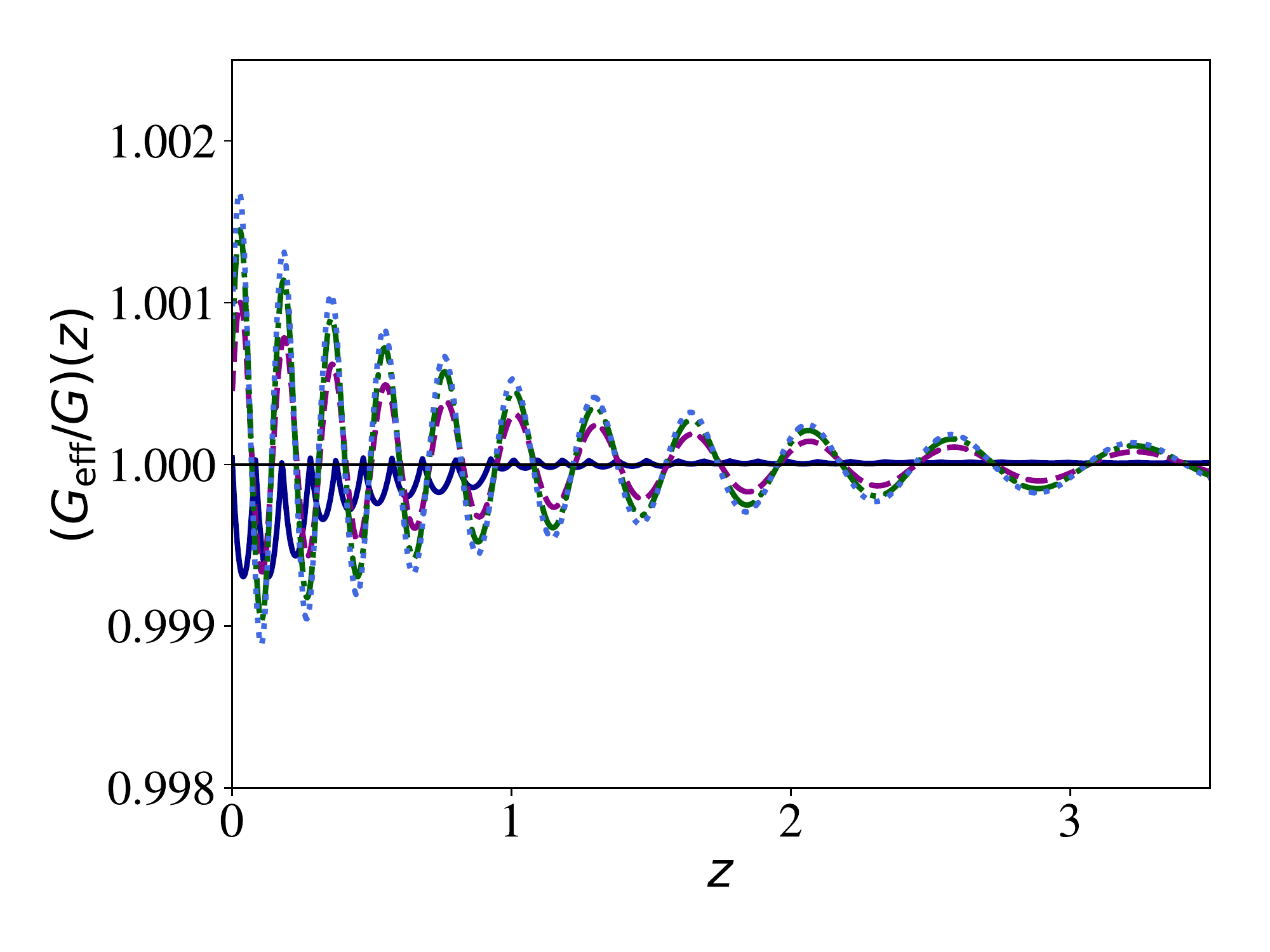}
\caption{ $G_{\rm eff}/G$ as a function of $z$ for fixed $\kappa$, 
for the minimal RT model (blue solid line) and  for RT with $\Delta N=34$ (magenta, dashed), $\Delta N=50$ (green, dot-dashed) and $\Delta N=64$ (cyan, dotted), for  $\kappa=0.1$  (left panel) and $\kappa =1$ (right panel).
}
\label{fig:Geff}
\end{figure}

\begin{figure}[t]
\centering
\includegraphics[width=0.42\textwidth]{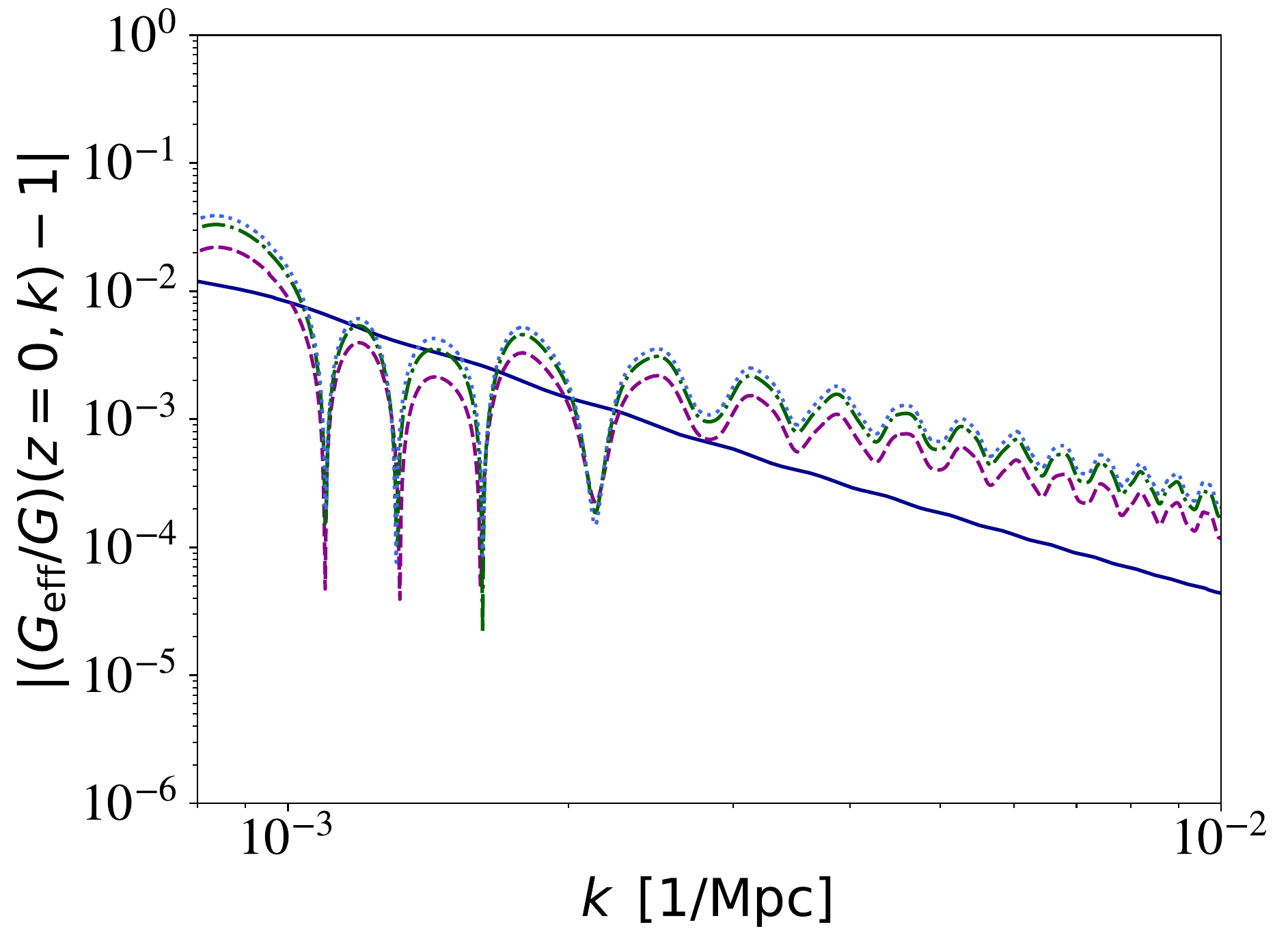}
\includegraphics[width=0.42\textwidth]{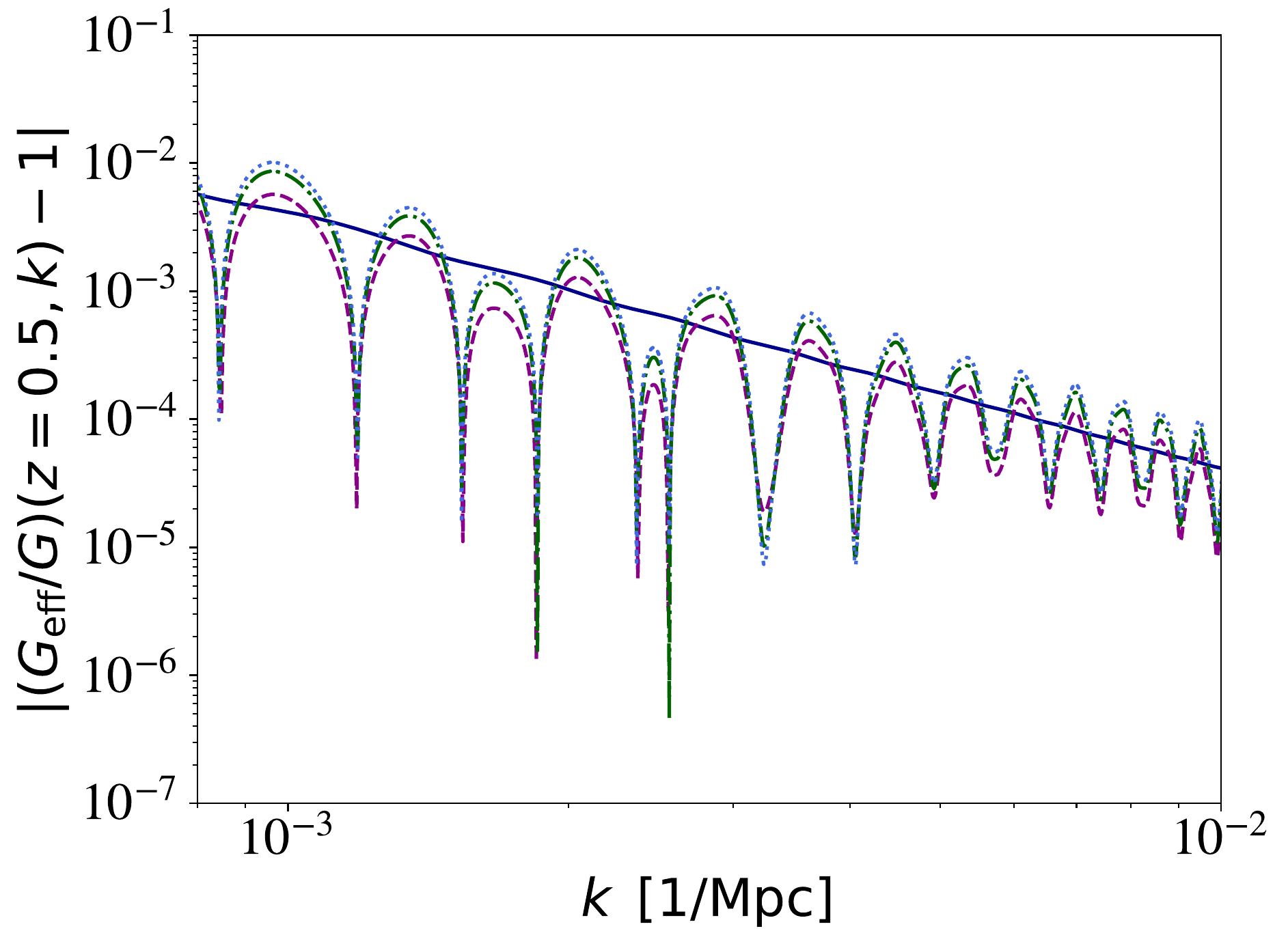}
\includegraphics[width=0.42\textwidth]{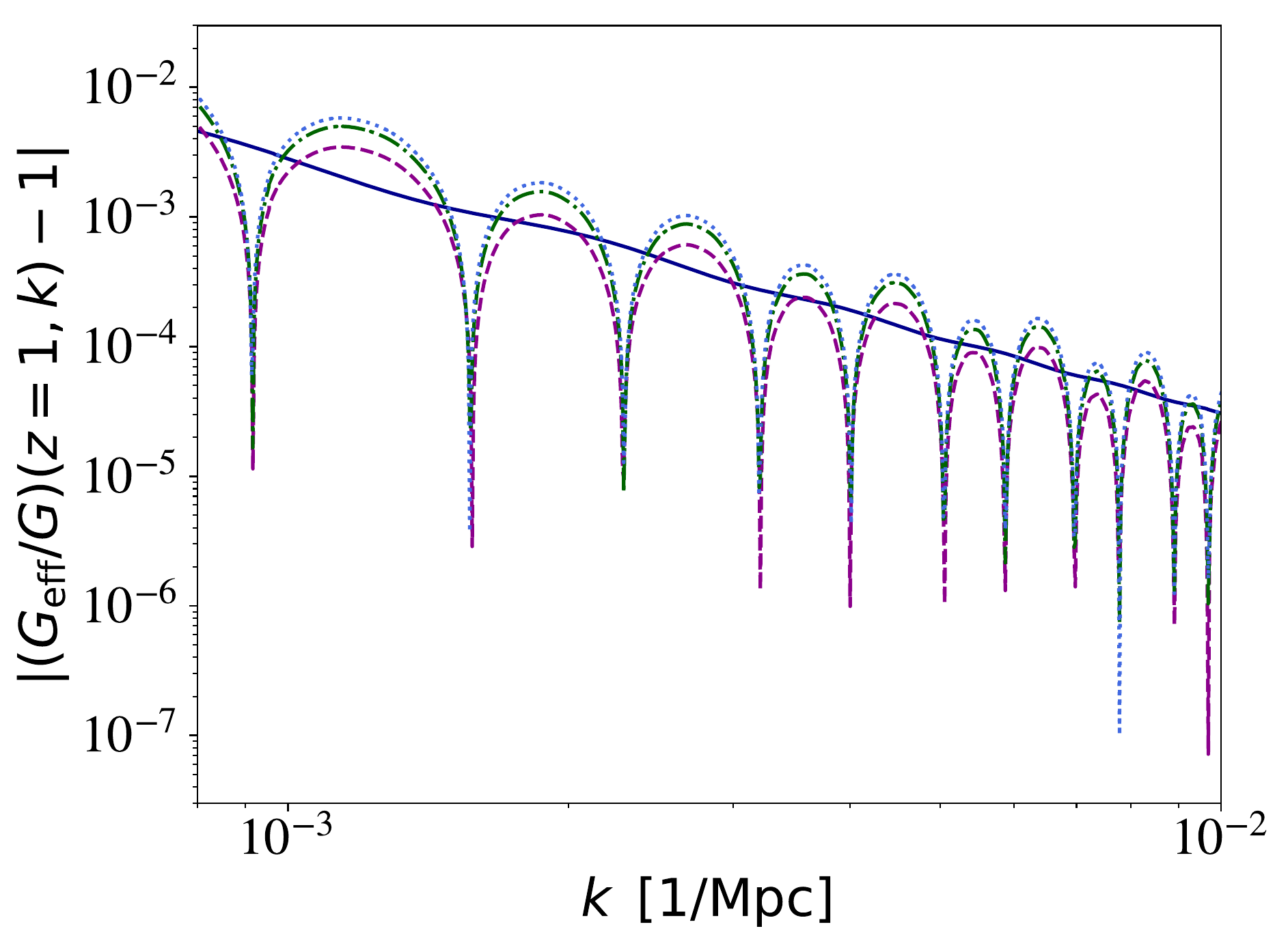}
\caption{ $|(G_{\rm eff}/G) -1|$ as a function of $k$ for fixed $z$, 
for the minimal RT model (blue solid line) and  for RT with $\Delta N=34$ (magenta, dashed), $\Delta N=50$ (green, dot-dashed) and $\Delta N=64$ (cyan, dotted), on a logarithmic scale. The three panels refers to $z=0$ (upper left panel), $z =0.5$ (upper right) and $z=1$ (lower panel). The sign of $(G_{\rm eff}/G) -1$ is such that, close to the vertical axis, $G_{\rm eff}/G>1$ for the minimal model and $G_{\rm eff}/G<1$ for the other cases, and the sign changes each time the logarithmic plot has a downward spike. 
}
\label{fig:Geffvsk}
\end{figure}

\begin{figure}[t]
\centering
\includegraphics[width=0.42\textwidth]{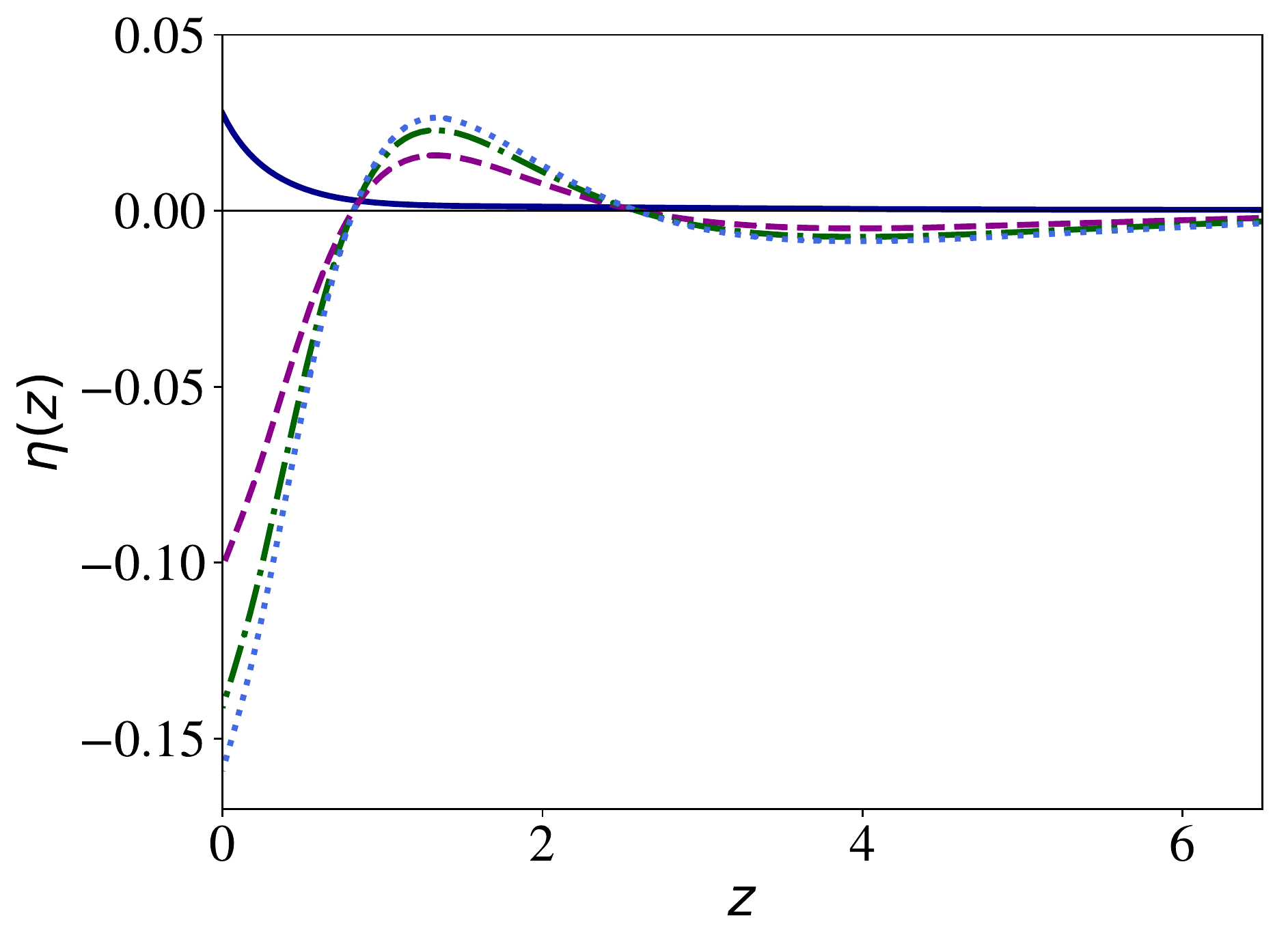}
\includegraphics[width=0.42\textwidth]{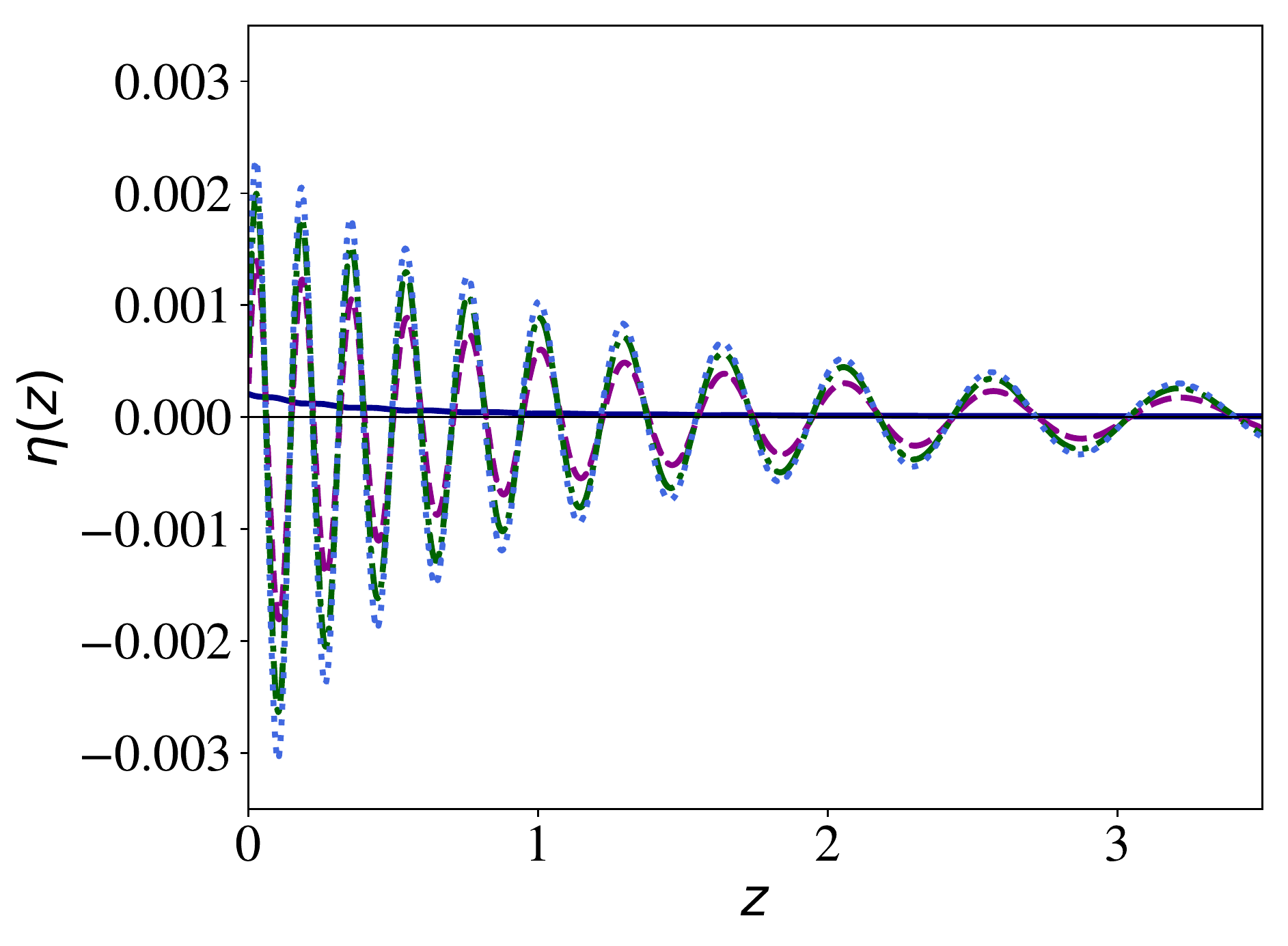}
\caption{ $\eta$ as a function of $z$, 
for the minimal RT model (blue solid line) and  for RT with $\Delta N=34$ (magenta, dashed), $\Delta N=50$ (green, dot-dashed) and $\Delta N=64$ (cyan, dotted),  for  $\kappa=0.1$  (left panel) and $\kappa =1$ (right panel).
}
\label{fig:eta}
\end{figure}

\begin{figure}[t]
\centering
\includegraphics[width=0.42\textwidth]{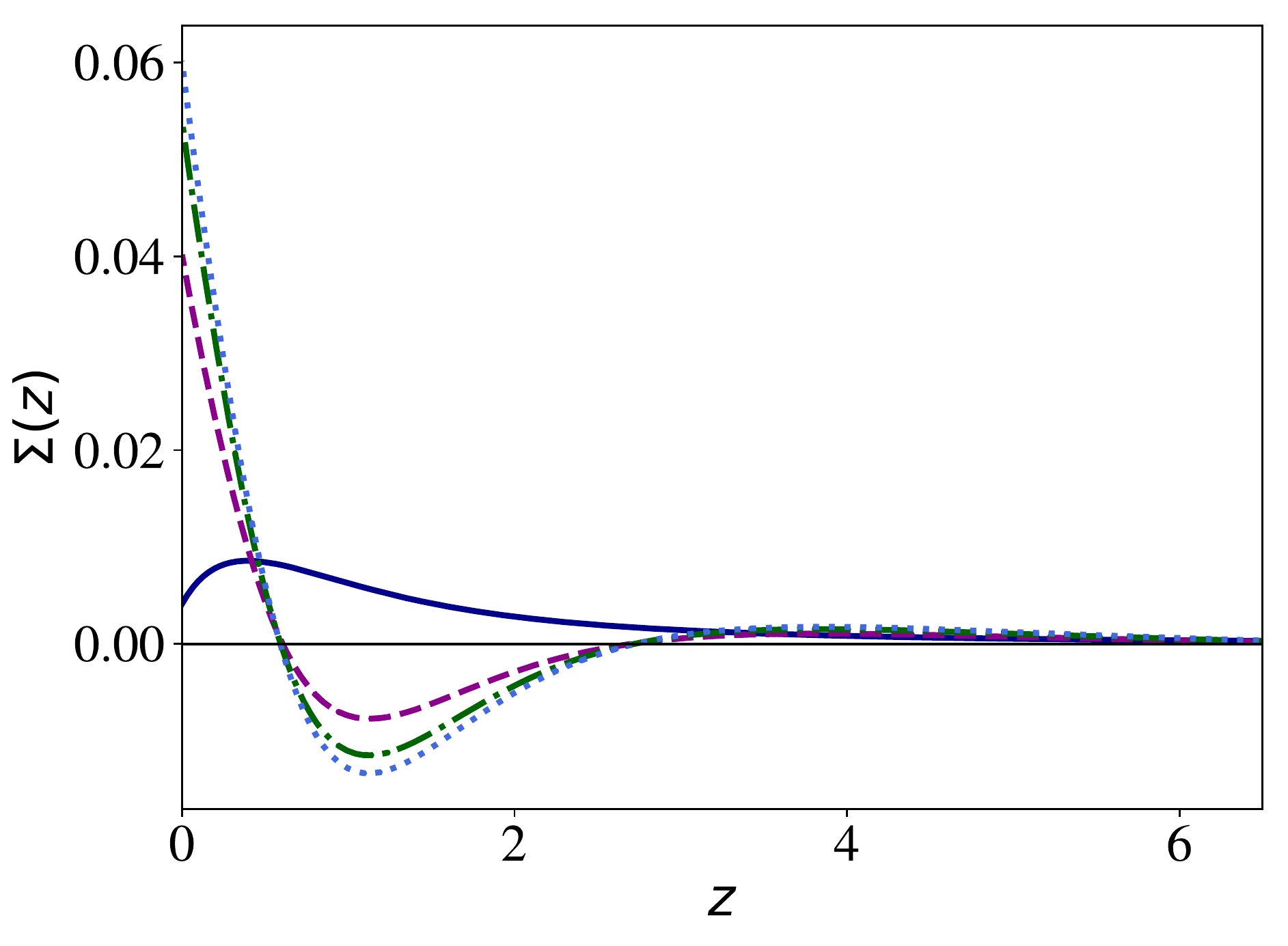}
\includegraphics[width=0.42\textwidth]{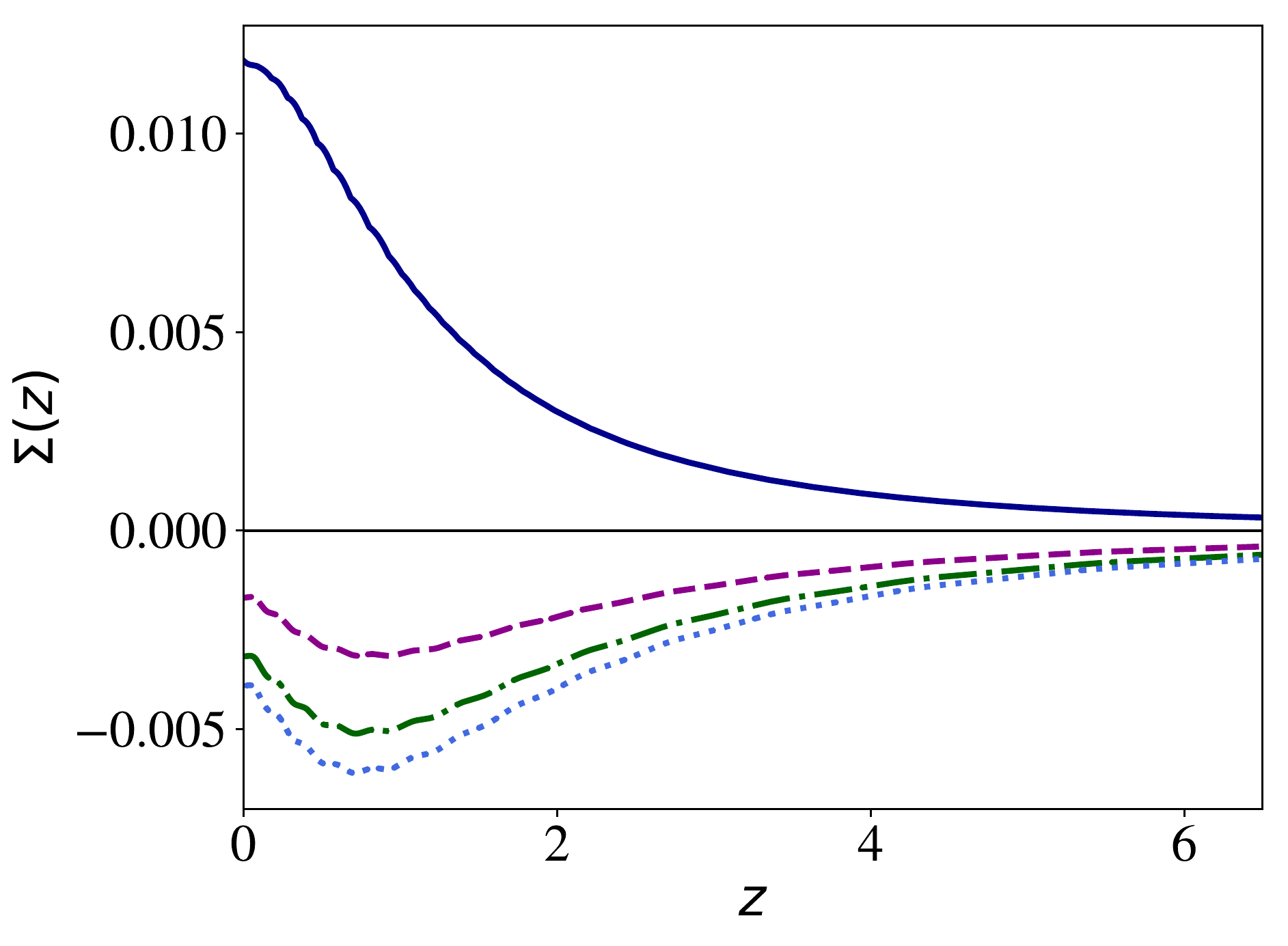}
\caption{ $\Sigma$ as a function of $z$, 
for the minimal RT model (blue solid line) and  for RT with $\Delta N=34$ (magenta, dashed), $\Delta N=50$ (green, dot-dashed) and $\Delta N=64$ (cyan, dotted),  for  $\kappa=0.1$  (left panel) and $\kappa =1$ (right panel).
}
\label{fig:Sigma}
\end{figure}

\begin{figure}[t]
\centering
\includegraphics[width=0.42\textwidth]{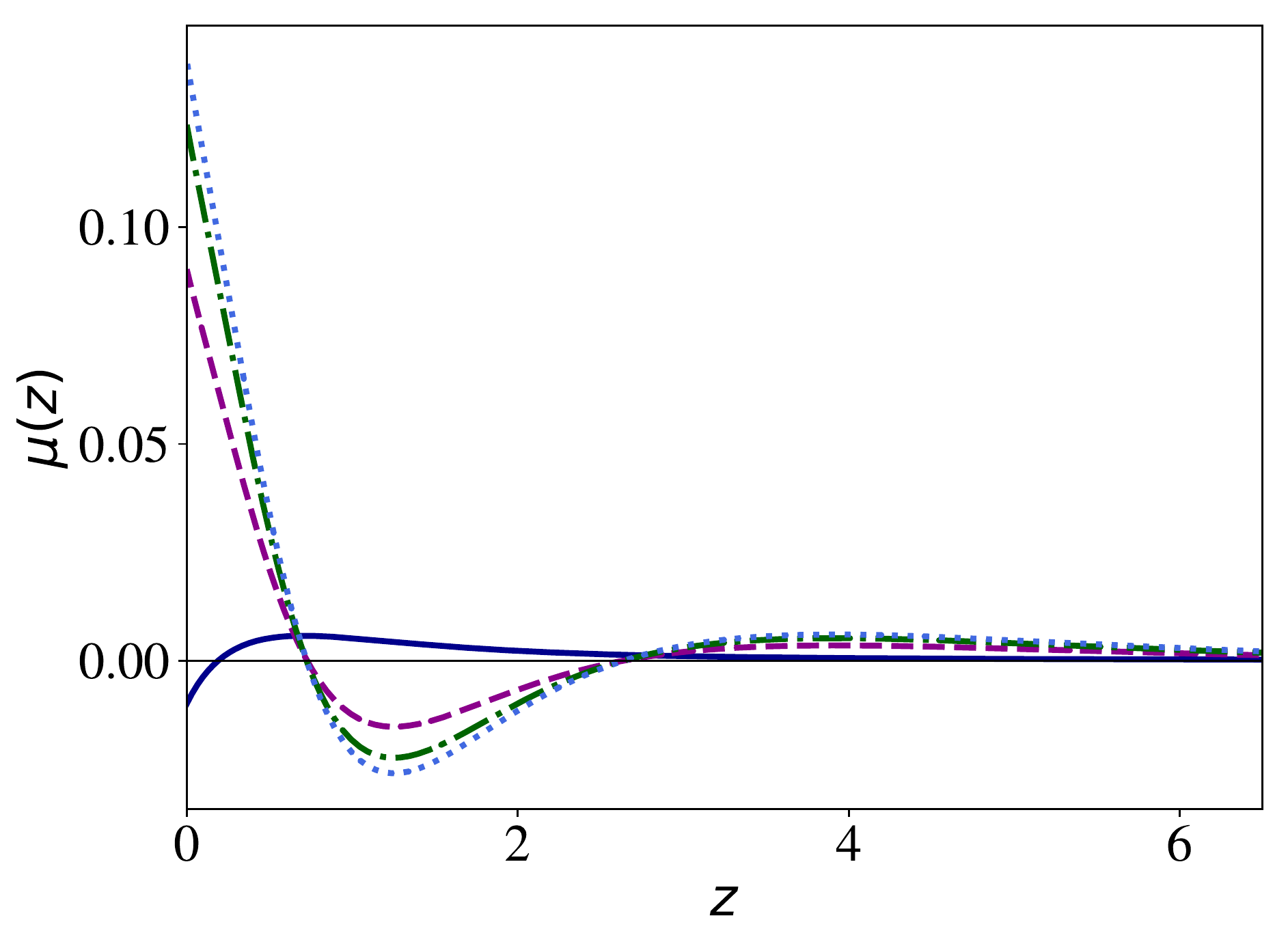}
\includegraphics[width=0.42\textwidth]{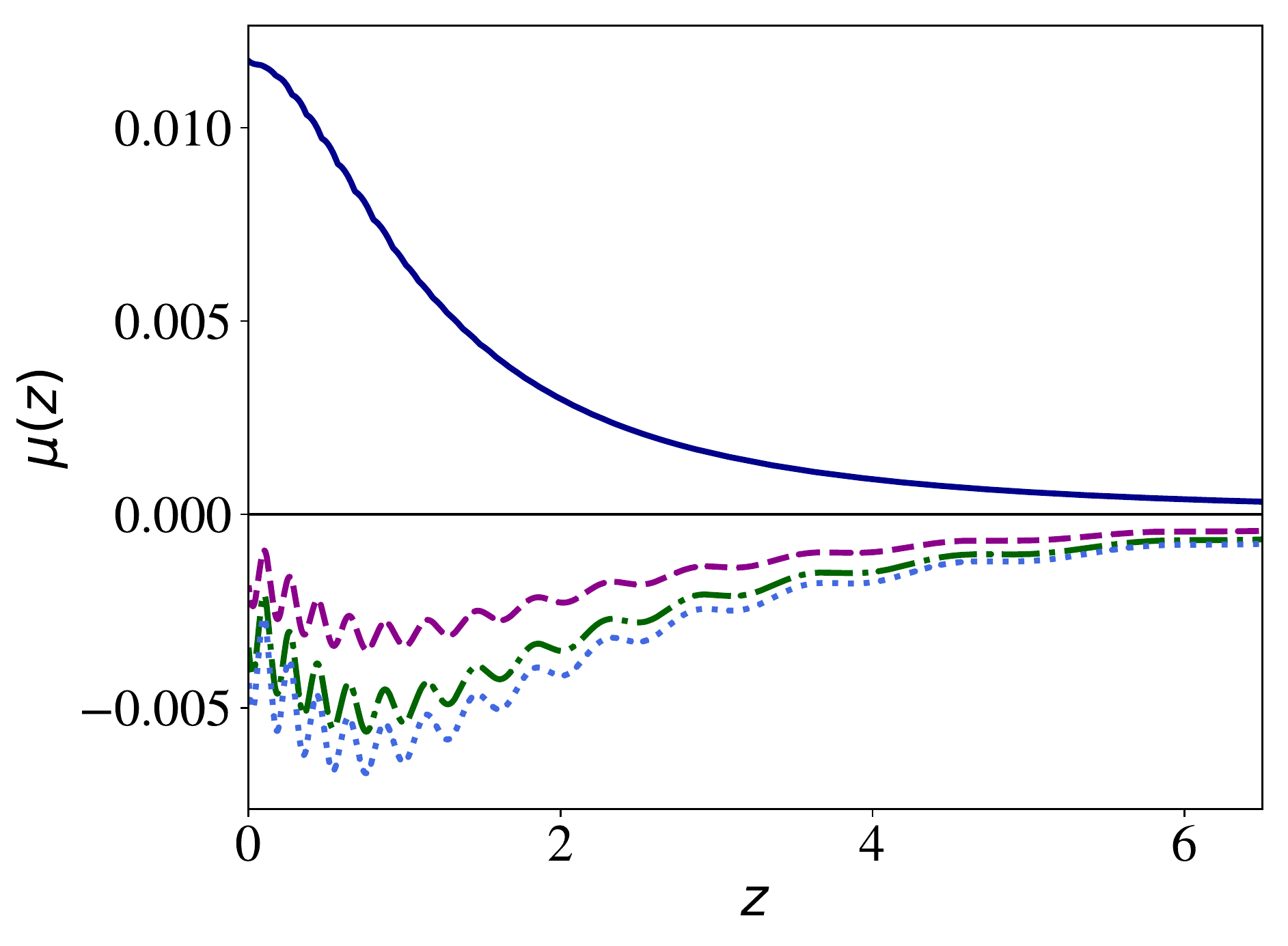}
\caption{ $\mu$ as a function of $z$, 
for the minimal RT model (blue solid line) and  for RT with $\Delta N=34$ (magenta, dashed), $\Delta N=50$ (green, dot-dashed) and $\Delta N=64$ (cyan, dotted), for  $\kappa=0.1$  (left panel) and $\kappa =1$ (right panel).}
\label{fig:mu}
\end{figure}

In  Fig.~\ref{fig:Geff} we show the numerical results for the effective Newton constant as a function of redshift,   for  the minimal RT model and for the RT model with $\Delta N=34,50,64$, for $\kappa=0.1$ and 1. 
We see that, already for $\kappa=0.1$ (i.e. $k=0.1 k_{\rm eq}\simeq 0.001 \, {\rm Mpc}^{-1}$), $G_{\rm eff}$ differs by $G$  by less than $1\%$, and, for higher values of $k$,  $G_{\rm eff}$ goes quickly to $G$, in agreements with \eq{GeffGRTlargek} (for instance, in the plot for $\kappa =5$, $|G_{\rm eff}/G|$ would always be below
1.001). For these values of $k$,  there are also  some oscillations as a function of $z$ and, for given $z$, the envelop of the oscillations reproduces  the $1/k^2$ behavior found analytically in \eq{GeffGRTlargek}. 
Notice that, because of \eq{GeffGRTlargek}, on small scales  $G_{\rm eff}$ reduces to the standard Newton's constant $G$ probed by solar system or by laboratory experiments. However, at typical cosmological scales such as $k\sim k_{\rm eq}$, its value is different, even at $z=0$. In particular, in the RT models with large $\Delta N$, on these scales $G_{\rm eff}<G$, i.e. gravity is weakened on cosmological scales, while for the minimal RT model it is strengthened.
Fig.~\ref{fig:Geffvsk} shows, on a logarithmic scale,  the dependence of $|(G_{\rm eff}/G)-1|$ on the wavenumber $k$, for three different values of the redshift, $z=0,0.5$ and 1. 

Fig~\ref{fig:eta} shows  $\eta$ as a function of $z$, again for $\kappa=0.1$ and 1,
while in Figs.~\ref{fig:Sigma} and \ref{fig:mu} we show the same results for the indicators $\Sigma$ and $\mu$. Notice in particular that both $\Sigma$ and $\mu$ have a rather non-trivial dependence on $k$ for cosmological scales $k\sim k_{\rm eq}$. We see from the plots that, at small redshifts, in the RT model with large $\Delta N$,   for $k=0.1k_{\rm eq}$ both $\Sigma$ and $\mu$ are positive  (with $1+\Sigma$ higher by about  $5\%$ than the  $\Lambda$CDM  value of unity,   and $\mu$ by about $10\%$ in $z=0$), while for $k=k_{\rm eq}$ or larger the situation is reversed and $\Sigma$ and $\mu$ become negative at small $z$.

\begin{figure}[t]
\centering
\includegraphics[width=0.42\textwidth]{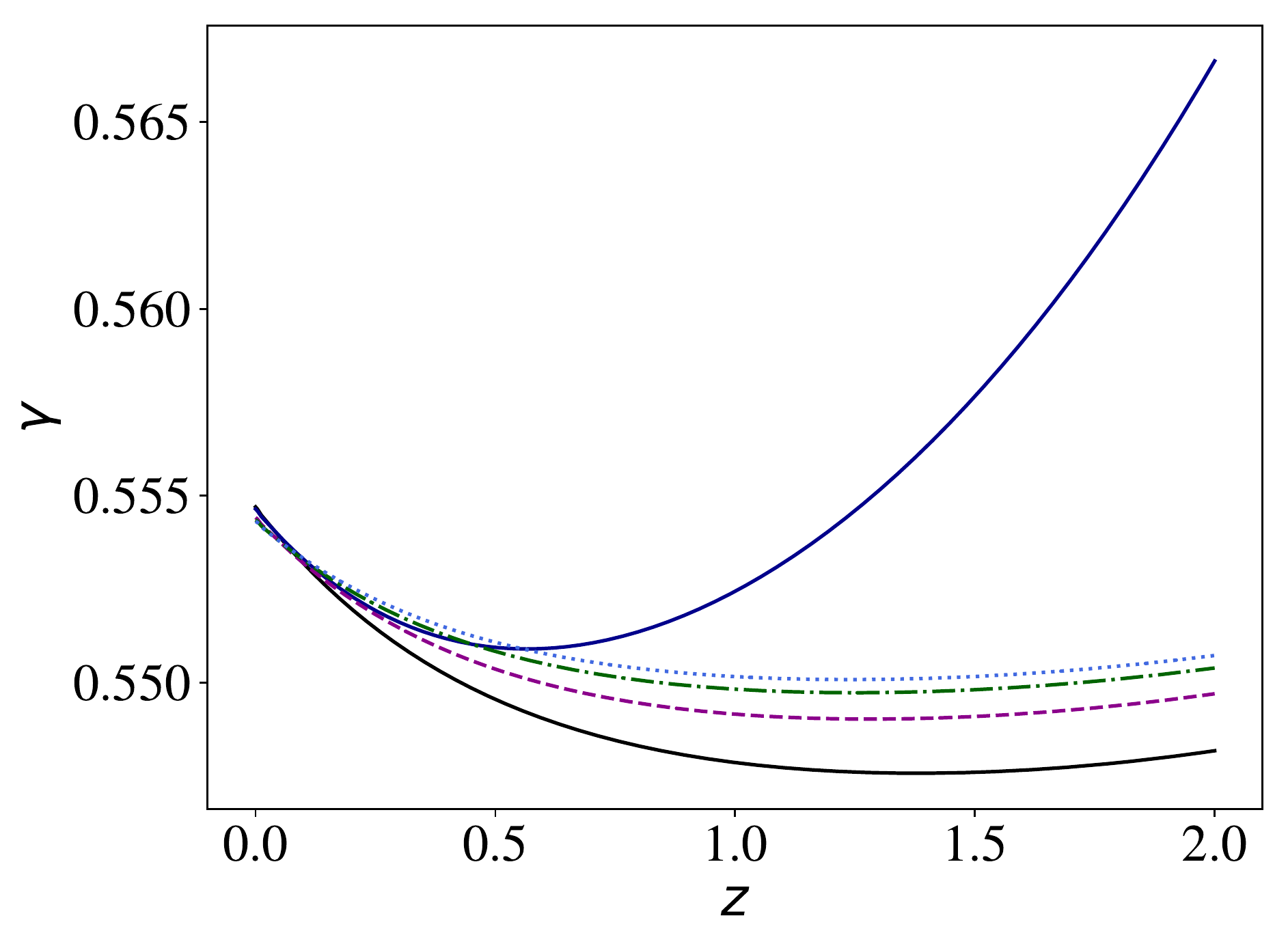}
\caption{The function $\gamma(z)$ related to the growth rate $f(z)$ by $f(z)=[\oma(z)]^{\gamma(z)}$,
for  $\Lambda$CDM (gray solid line), for the minimal RT model (blue solid line) and  for RT with $\Delta N=34$ (magenta, dashed), $\Delta N=50$ (green, dot-dashed) and $\Delta N=64$ (cyan, dotted).}
\label{fig:growthindex}
\end{figure}

Another useful derived quantity is the growth rate $f(z,k)\equiv d \log \delta_M/d\ln a$.
As is well known, in $\Lambda$CDM, for the typical wavenumbers relevant for structure formation, $f(z,k)$ is basically independent of wavenumber $k$ and very well fitted by $f(z)=[\oma(z)]^{\gamma}$ with $\gamma$ a constant, numerically close to 0.55. More precisely, writing $f(z)=[\oma(z)]^{\gamma(z)}$, the function $\gamma(z)$ for 
$\Lambda$CDM is shown as the gray solid line in Fig.~\ref{fig:growthindex}, so it is indeed approximately constant and 
given numerically by $\gamma\simeq 0.55$, within percent level accuracy.
We find that the fit $f(z)=[\oma(z)]^{\gamma(z)}$
also  holds for the RT model, again with a function $\gamma(z)$ independent of the wavenumber $k$. The corresponding functions  $\gamma(z)$ are shown in Fig.~\ref{fig:growthindex} for the RT model, minimal and with large $\Delta N$. We see that, for large values of $\Delta N$, $\gamma(z)$ is indeed independent of $z$ within  percent level accuracy, just as in $\Lambda$CDM, and again is given numerically by $\gamma\simeq 0.55$.\footnote{Of course the growth index, traditionally denoted by $\gamma$, should not be confused with the parameter $\gamma$ of the RT model, defined in \eq{defgamma}!}
For the minimal RT model the variation of $\gamma(z)$ with redshift is somewhat larger, but still it stays  between
$0.55-0.56$ up to $z=2$. Notice that the growth index $\gamma$ is a useful quantity only as long as we are in the epoch where DE is still important. When we are deep into MD, $\oma(z)\ra 1$, and $[\oma(z)]^{\gamma}\ra 1$ independently of $\gamma$.

Two main conclusions emerge from this study of the cosmological perturbations of the RT model in the scalar sector. First,   they are  well-behaved.
This is already a rather non-trivial result. Several modified gravity models have indeed been ruled out by the presence of instabilities in their perturbations. This was for instance   the case for the DPG model~\cite{Dvali:2000hr}, which opened the way to the study of IR modifications of GR and  has a  self-accelerated solution \cite{Deffayet:2000uy,Deffayet:2001pu} but  had a ghost-like instability on the self-accelerated branch~\cite{Luty:2003vm,Nicolis:2004qq,Gorbunov:2005zk,Charmousis:2006pn,Izumi:2006ca}. 
Massive gravity~\cite{deRham:2010ik,deRham:2010kj,Hassan:2011hr} has difficulties already in obtaining a viable background FRW evolution~\cite{DAmico:2011eto}, while  
in bigravity~\cite{Hassan:2011zd} a  background FRW solutions exist, but, in a  branch of solutions that has a dynamical dark energy, the cosmological perturbations have instabilities in both the scalar and  tensor sectors~\cite{Konnig:2014xva,Lagos:2014lca,Cusin:2014psa,Akrami:2015qga,Cusin:2015pya,Schmidt-May:2015vnx,Cusin:2015tmf} (see \cite{Ishak:2018his} for a recent comprehensive review of modifications of GR at the cosmological scale). Thus, already the fact of producing quite naturally a viable cosmological background evolution with self-acceleration, and stable scalar perturbations, is a non-trivial results.

The second conclusion that  emerges from this study is that, both in the background evolution and in the  scalar perturbations, the RT model is very close to $\Lambda$CDM, with deviations of at most  a few percent, for all $\Delta N$. This already indicates that the model is a good candidate for fitting well  the current cosmological observations.
In the next section we will confirm this conclusion by comparing the RT model with $\Lambda$CDM from the point of view of the quality of the fit to the cosmological observations, and  we will perform Bayesian parameter estimation for the  values of the cosmological parameters.

\subsection{Comparison with cosmological observations}\label{sect:Comp}

We now perform a detailed comparison with cosmological observations, using the most recent cosmological datasets in order to update the results presented in~\cite{Dirian:2014bma,Dirian:2016puz} for the minimal RT model, and in \cite{Belgacem:2019lwx} for the RT model with large $\Delta N$.
As in these previous works, we implement  the perturbations of the RT model computed in~\cite{Dirian:2014ara} into  the CLASS cosmological Boltzmann code \cite{Blas:2011rf} (v2.7), that we have modified so to describe the background evolution and scalar perturbations of the RT model. Our  code
has been tested against other Einstein-Boltzmann solvers in \cite{Bellini:2017avd}, and the most recent version is
publicly available on GitHub \cite{git_nonlocal} (evolved from \cite{git_nonlocalYD}).

\subsubsection{Datasets and methodology}

For $\Lambda$CDM, the {\em Planck} baseline analysis  uses six independent
cosmological parameters:  the Hubble parameter today
$H_0 = 100 h_0 \, \rm{km} \, \rm{s}^{-1} \rm{Mpc}^{-1}$, the physical baryon and cold dark matter density fractions today $\omega_b = \Omega_b h_0^2$ and $\omega_c = \Omega_c h_0^2$, respectively, the amplitude  $A_s$ and tilt $n_s$ of the primordial scalar perturbations,    and  the reionization optical depth  $\tau_{\rm re}$. Note that, assuming flatness, the energy fraction $\ola$ associated to a cosmological constant is a derived parameter, fixed by the flatness condition. In the RT model we have a  mass scale $m$ [or, equivalently, the dimensionless parameter $\gamma$, \eq{defgamma}] which replaces the cosmological constant, and again can be taken as  
a derived parameter, fixed by the flatness condition. Thus, for the  RT model, we can take the same six independent cosmological parameters, as in $\Lambda$CDM.

An important extension, however, is provided by  the sum of neutrino masses, $\sum_{\nu}m_{\nu}$. 
As discussed in \cite{Dirian:2017pwp}, their inclusion can  a priori be important when comparing a modified gravity model to $\Lambda$CDM. 
Oscillation experiments give a lower limit   $\sum_{\nu}m_{\nu}\, \gsim \, 0.06$ eV \cite{GonzalezGarcia:2012sz}
(assuming a normal mass hierarchy dominated by the heaviest neutrino mass eigenstate).  In the {\em Planck} baseline analysis the sum of neutrino masses is kept fixed to this minimum allowed value.
As discussed in the {\em Planck} papers~\cite{Planck_2015_CP,Aghanim:2018eyx}, there is actually no compelling theoretical reason for this choice, and there are other possibilities, including a degenerate hierarchy with 
$\sum_{\nu}m_{\nu}\,\gsim\, 0.1$~eV. The choice of fixing the sum of neutrino masses to the minimum allowed values
is justified by the fact that, in $\Lambda$CDM, letting the sum of neutrino masses as a free parameter, one finds that its marginalized posterior is peaked in zero, and if we  let it vary 
with the prior $\sum_{\nu}m_{\nu}\, \geq \, 0.06$~eV the data drive $\sum_{\nu}m_{\nu}$ back to the prior (see Fig.~34 of \cite{Aghanim:2018eyx}). In contrast, in a modified gravity model, the posterior for $\sum_{\nu}m_{\nu}$ could be peaked at a value higher than the lower bound 0.06~eV.\footnote{This is indeed what happens in the RR nonlocal model~\cite{Dirian:2017pwp,Belgacem:2017cqo}, and we  will find that this also happens for the minimal RT model.}
A uniform comparison of a modified gravity model with $\Lambda$CDM therefore requires to let $\sum_{\nu}m_{\nu}$ as a free parameter in both models, as we have done in \cite{Belgacem:2017cqo} and as  we will do below. We will denote by $\nu\Lambda$CDM the $\Lambda$CDM model in which $\sum_{\nu}m_{\nu}$ is added to the list of free parameters.

In summary,  we will perform Bayesian parameter estimation for both $\nu\Lambda$CDM and the RT model (minimal, and with $\Delta N=34,50,64$), and we will compare the quality of their fits to the  datasets  discussed below, using, as free parameters, 
\be \label{base_param}
\theta = \left\{H_0, \omega_b, \omega_c, A_s, n_s, \tau_{\rm re}, \mbox{$\sum_{\nu}$}m_{\nu}\right\} \, .
\ee
For CMB, SNe and BAO we use the following likelihoods:

\begin{itemize}

\item For CMB we use the {\em Planck} 2018 data release, using the
low-$\ell$ temperature-only likelihood, the low-$\ell$ EE likelihood, and the
high-$\ell$ temperature and polarization \textsc{plik} likelihood described in ref.~\cite{Aghanim:2019ame}, as well as the lensing likelihood  based on temperature+polarization map-based lensing reconstruction~\cite{Aghanim:2018oex}. This provides a significant update of our previous studies based on the {\em Planck} 2015 data release.

\item For type Ia supernovae we use the likelihood of the Pantheon type Ia supernova sample \cite{Scolnic:2017caz}, which includes data from the Pan-STARRS1 (PS1) Medium Deep Survey. This updates our previous study based on the JLA~\cite{Betoule:2014frx}  dataset.

\item For BAO we still use the likelihoods of the BAO detection of the 6dF Galaxy Survey \cite{Beutler:2011hx} and  the BAO scale measurement of SDSS DR7 Main Galaxy Sample \cite{Ross:2014qpa}, and we update the SDSS data using  the power spectrum of BAO  from the Data Release 12
\cite{Alam:2016hwk}.

\end{itemize}

For the RT model  the initial conditions of the perturbations of the auxiliary fields  $\d U$ and $\d S_{\mu}$
are set to zero. As we have already  shown in \cite{Belgacem:2017cqo}, taking different initial conditions, of the order of the metric perturbations (which is their natural scale, since, as discussed in section~\ref{sect:dof}, the initial conditions on the auxiliary fields are in principle fixed by the initial conditions on the metric perturbations) has a totally negligible  effect.

After having determined in this way the mean values of the  parameters of the models ($\Lambda$CDM and RT, minimal and with various $\Delta N$), we will use these values to compare the models with further datasets, namely measurements of $H(z)$ (``cosmic chronometers'') and $f\sigma_8$ data.\footnote{Some  technical details on our MCMC.
We use the statistical framework Cobaya 2.0.2 (\url{https://github.com/CobayaSampler/cobaya}, developed by Jesus Torrado and Antony Lewis) to let Markov chains sample the posterior distribution for the cosmological parameters. Cobaya uses the sampler developed for CosmoMC \cite{Lewis:2002ah,Lewis:2013hha} tailored for parameter spaces with a speed hierarchy
(it also implements the ``fast dragging" procedure described in \cite{Neal:2005}). 
We determine the best-fit cosmological parameters as follows. For each model, we select from its Markov Chain samples the $N$ samples that are closest to the highest-posterior sample and fit a generic quadratic function using least squares. In practice, we choose $N=5 d^2$ where $d = 28$ is the total number of parameters including the fiducial ones and the factor $5$ is chosen as a compromise between locality and numerical stability. The norm used to determine the closest samples is Euclidean after normalization of the sample coordinates by their standard deviations as estimated from all samples of the chain.
 We then identify the convex subspace of the quadratic fit using an eigen-decomposition of the Hessian. Finally, we minimize the quadratic fit within the convex subspace under the constraint that some of the parameters must be positive (for example the neutrino mass). Given the Markov Chain samples, this procedure gives a best-fit candidate within seconds of runtime.
The posterior is then evaluted at the candidate point predicted by this procedure. Typically, the prediction and the actual evaluation are close. The values for $\chi^2$ given here are always corresponding to the true evaluation at the predicted minimum.  We consistently get better results following this method than using Cobaya's BOBYQA~\cite{cartis2019improving, cartis2018escaping, powell2009bobyqa} minimizer, which takes into account the previous samples only via their covariance matrix. A python notebook is available at 
\url{https://github.com/AndreasFinke/quadfit}.}

\subsubsection{Comparison with CMB, BAO, SNe, cosmic chronometers  and $f\sigma_8$.}

\vspace{2mm}\noindent
{\bf Fit to CMB+BAO+SNe and Bayesian parameter estimation}.
Table~\ref{tab:results} shows the results for  the Bayesian parameter estimation and the resulting $\chi^2$ for  
$\nu\Lambda$CDM and the RT  model (minimal, and with $\Delta N=34, 50,64$), using the combined CMB+SNe+BAO data.  Beside the values of the seven fundamental independent parameters  given in (\ref{base_param}), we also give some useful derived parameters, namely  $\oma$,  the reionization redshift $z_{\rm re}$, and  the amplitude of matter density fluctuations in spheres of radius $8h_0^{-1}$~Mpc, $\sigma_8$. In the last line we show the  differences in $\chi^2$, with respect to the value for $\nu\Lambda$CDM. 
We recall that, for models with the same number of free parameters, as $\nu\Lambda$CDM and the RT models, the conventional interpretation is that a  difference $|\Delta \chi^2| \leq  2$ implies  statistical equivalence between the two models, while $2\,\lsim\, |\Delta \chi^2|\,\lsim\, 6$ suggests ``weak evidence'' in favor of the model with lower $\chi^2$, and $|\Delta \chi^2|\gtrsim 6$ indicates ``strong evidence'' in favor of the model with lower $\chi^2$. Thus, all models considered fit the data at a statistically equivalent level.\footnote{Note that $\Delta N$ is not a free parameter varied so to minimize the $\chi^2$. Rather, we have used a very limited sample of values of $\Delta N$, chosen a priori on the basis of the fact that, according to the relation (\ref{DeltaNMinfl}), they correspond to significant choices for the inflationary scale $M_{\rm infl}$, see the discussion below \eq{DeltaNMinfl}. }

\begin{table*}[t]
\centering
\resizebox{\columnwidth}{!}{
\begin{tabular}{|l||c|c|c|c|c|}
\hline
		Parameter & $\nu\Lambda$CDM & RT, minimal & RT, $\Delta N=34$ & RT, $\Delta N=50$ & RT, $\Delta N=64$ \\ \hline
		$H_0$ \phantom{\Big|}& $67.89\pm 0.47             $ & $68.74^{+0.59}_{-0.51}     $ & $67.95\pm 0.48             $ & $67.90\pm 0.47             $ & $67.88\pm 0.48             $\\
		$\sum_{\nu}m_{\nu}\ [{\mathrm eV}]$ & $< 0.057$ (at $1\sigma$) & $0.071^{+0.024}_{-0.066}  $ & $< 0.048$ (at $1\sigma$) & $< 0.044$ (at $1\sigma$) &$< 0.041$ (at $1\sigma$)\\
		$\omega_c$ \phantom{\Big|} & $0.1193\pm 0.0009 $ & $0.1120\pm 0.0009$ & $0.1191 \pm 0.0009$ & $0.1190 \pm 0.0009$ & $0.1189\pm 0.0009$\\
		100$\omega_b$ \phantom{\Big|} & $2.242\pm 0.013 $  & $2.237 \pm 0.013$ & $2.243 \pm 0.013 $ & $2.244 \pm 0.013 $ & $2.244  \pm 0.013$\\
		$\ln (10^{10} A_s)$\phantom{\Big|} & $3.045 \pm 0.014 $ & $3.043\pm 0.014$ & $3.047\pm 0.014$ & $3.048^{+0.013}_{-0.015}$ & $3.049\pm 0.014$\\
		$n_s$ \phantom{\Big|} & $0.9665 \pm 0.0036$ & $0.9649\pm 0.0036$ & $0.9670\pm 0.0036$ & $0.9673\pm 0.0035$ & $0.9672\pm 0.0035$\\
		$\tau_{\mathrm re}$ \phantom{\Big|} & $0.0555\pm0.0072$ & $0.0537 \pm 0.0072$ & $0.0565\pm 0.0073$ & $0.0572^{+0.0065}_{-0.0075}$ & $0.0575\pm 0.0071$\\
		\hline
		$\Omega_M$\phantom{\Big|} & $0.3085\pm 0.0060$ & $0.3029_{-0.0070}^{+0.0061}$ & $0.3075\pm 0.0061$ & $0.3076\pm 0.0060$ & $0.3076 \pm 0.0060$\\
		$z_{\mathrm re}$ & $7.76 \pm 0.72$ & $7.60 \pm 0.73$ & $7.86\pm 0.72$ & $7.93 \pm 0.70$ & $7.96\pm0.70$\\
		$\sigma_8$ \phantom{\Big|} & $0.8164^{+0.0097}_{-0.0068}$ & $0.823^{+0.0130}_{-0.0087}$ & $0.8141^{+0.0089}_{-0.0067}$ & $0.8134^{+0.0088}_{-0.0064}$ & $0.8129^{+0.0084}_{-0.0066}$\\
		\hline
		$\Delta\chi^2$\phantom{\big|} & 0 & 1.30 & -0.48 & -0.20 & -0.00\\
		\hline
\end{tabular}
}
\caption{\label{tab:results} Mean values (with $1\sigma$ errors) of the parameters  for 
$\nu\Lambda$CDM  and the RT model (minimal, and with $\Delta N$= 34, 50, 64), using CMB, BAO and SNe. $H_0$ is in units of ${\rm km}\, {\rm s}^{-1}\, {\rm Mpc}^{-1}$. The last line gives the difference in the $\chi^2$ of each given model with respect to $\nu\Lambda$CDM. The RT model with $\Delta N=34$ or with $\Delta N=50$ fits the data slightly better than $\nu\Lambda$CDM, but the difference is not statistically significant.}
\end{table*}

The result of Bayesian parameter estimation shows that all models with large $\Delta N$ give predictions extremely close to those of $\nu\Lambda$CDM, consistently with the analysis of the previous sections, that showed that these models are very close to $\Lambda$CDM both in the background evolution and in the cosmological perturbations.
The minimal RT model differs a bit more, and in particular predicts a slightly higher value of $H_0$, which in any case is not enough to significantly relieve the tension with the local $H_0$ measurement~\cite{Riess:2019cxk,Wong:2019kwg}. Indeed, as discussed in~\cite{Poulin:2018zxs,Aylor:2018drw},   it might not be possible to solve  the $H_0$ tension, together with other potential tensions within $\Lambda$CDM,  with a modification
of only the late-Universe dynamics (as in our nonlocal model). The other difference of the minimal RT model is that it predicts a non-zero value for the sum of the neutrino masses, while all other models considered only give an upper bound.

\begin{figure}[t]
 \centering
 \includegraphics[width=0.42\textwidth]{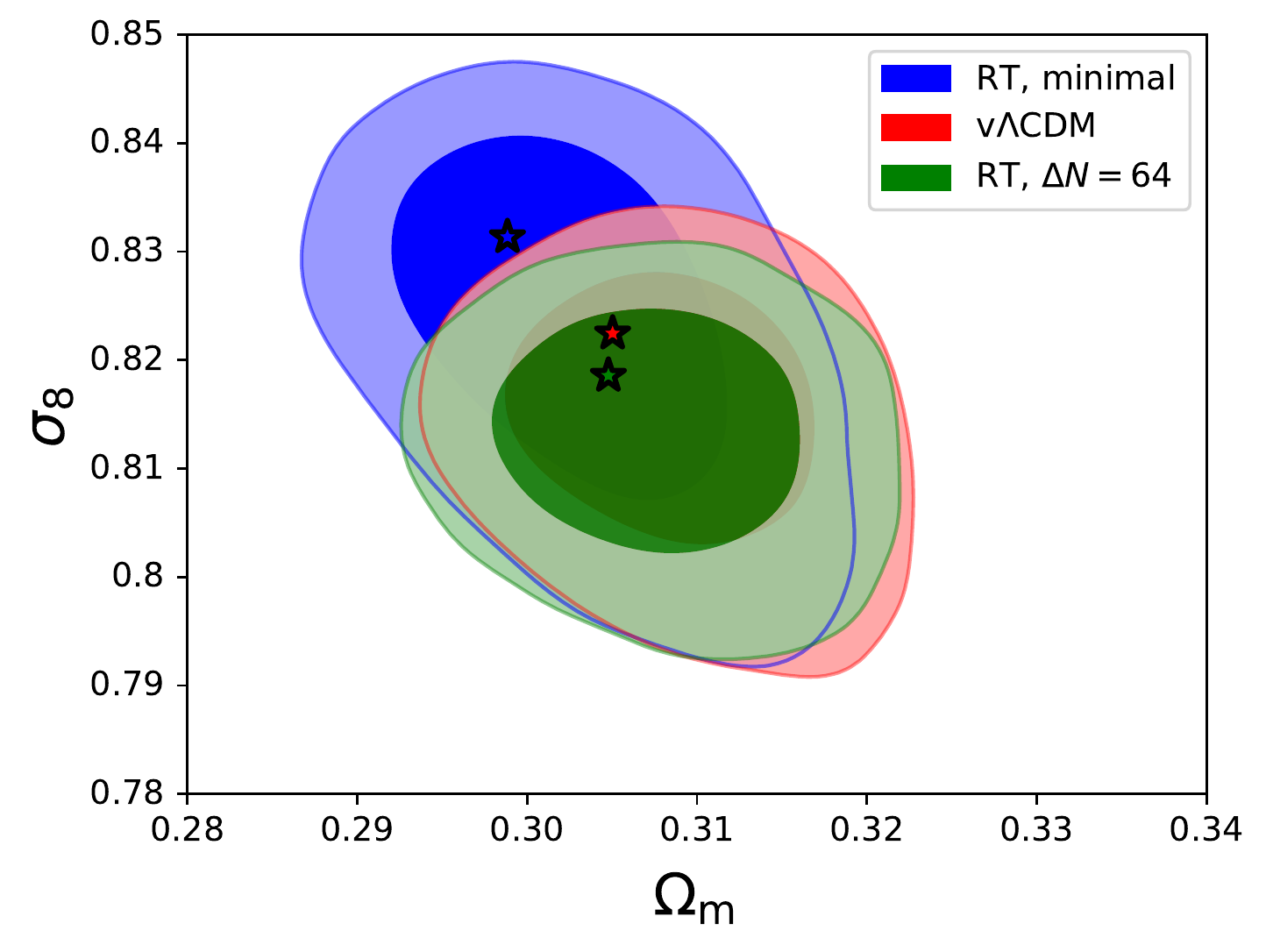}
 \includegraphics[width=0.42\textwidth]{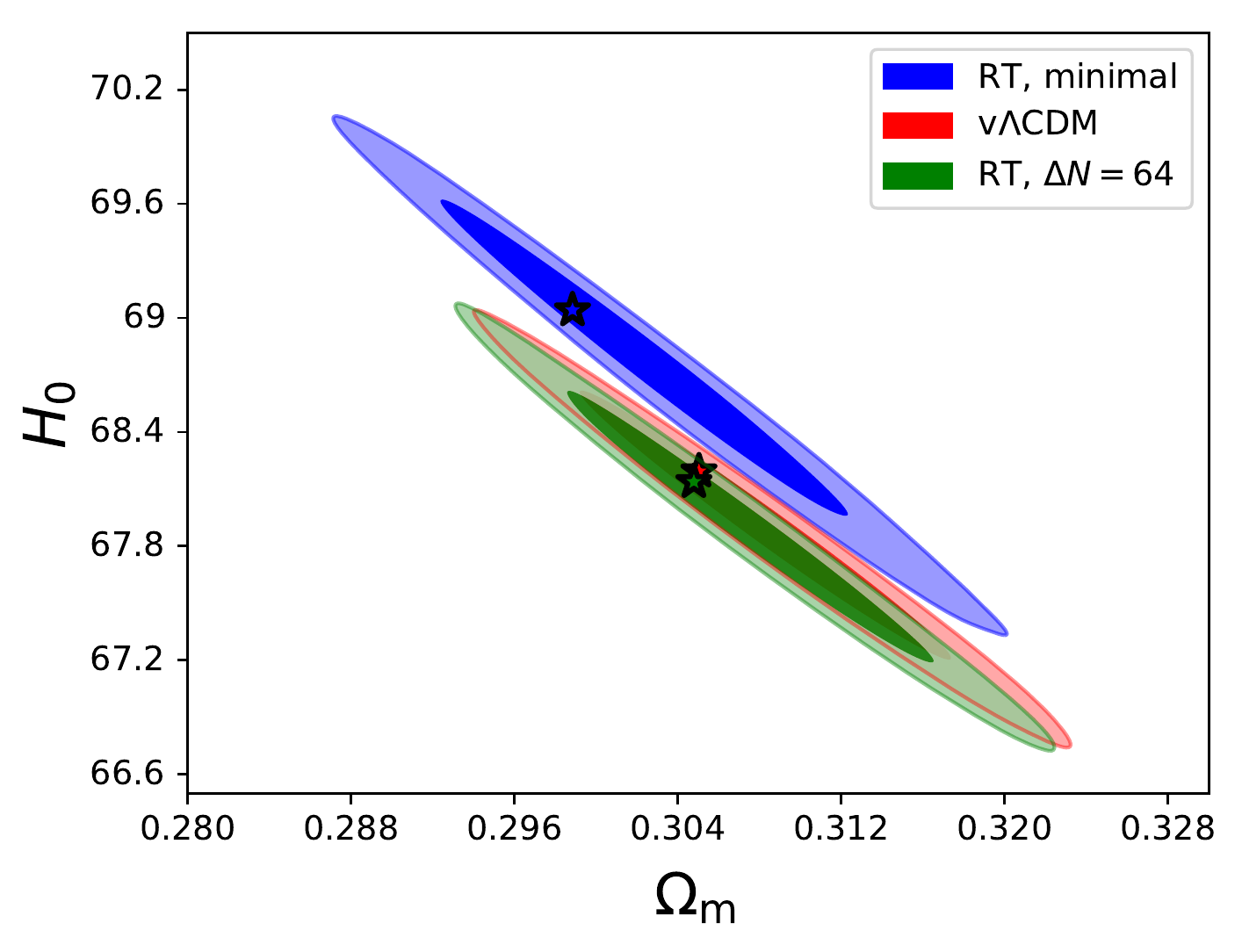}
 \caption{Left panel: the two-dimensional likelihood in the $(\oma,\sigma_8)$ plane for $\nu\Lambda$CDM (red), the minimal RT model (blue) and the RT model with $\Delta N=64$ (green). The stars are the best-fit values of the parameters (note that the values reported in Table~\ref{tab:results} are rather the mean values).
Right panel: the same for $(\oma,H_0)$.}
  \label{fig:omsig8}
\end{figure}

\begin{figure}[t]
\centering
\includegraphics[width=0.42\textwidth]{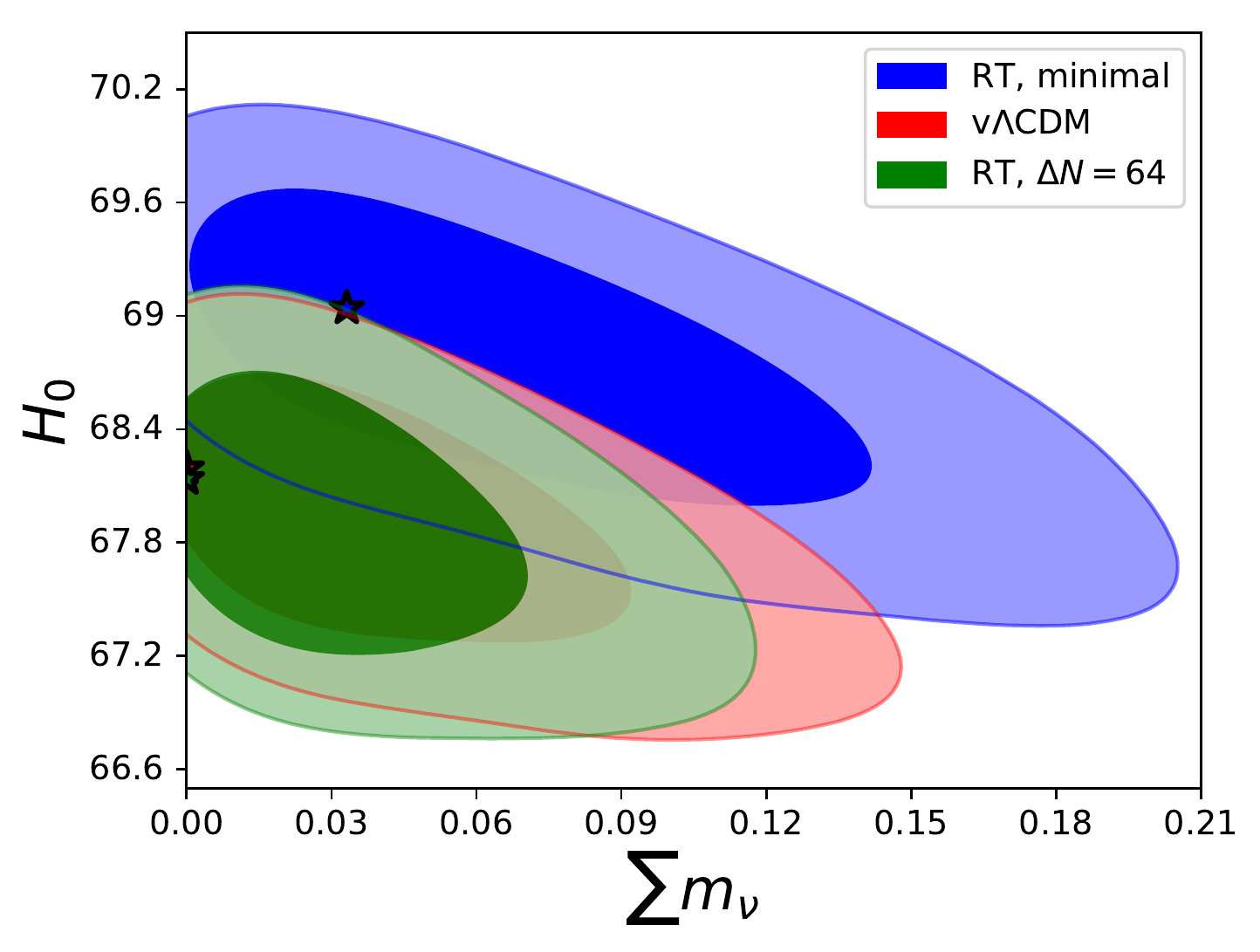}
\caption{As in Fig.~\ref{fig:omsig8}, for  $(\sum_{\nu}m_{\nu},H_0)$.}
\label{fig:mnuH0}
\end{figure}

Figures \ref{fig:omsig8} and \ref{fig:mnuH0} show the two-dimensional likelihoods for
$(\oma,\sigma_8)$, $(\oma,H_0)$ and $(\sum_{\nu}m_{\nu},H_0)$. The pattern that emerges, from this and similar plots, is that the RT model with large values of $\Delta N$ is extremely close to $\Lambda$CDM, as we already saw from Table~\ref{tab:results}, while the minimal RT model has some more significant differences, such as a slightly higher value of $H_0$ (although, as mentioned above, not enough to significantly decrease the tension with local measurements), of $\sigma_8$, and of the sum of neutrino masses.

\vspace{2mm}\noindent
{\bf Cosmic chronometers}. Another useful observational test is provided by
measurements of $H(z)$ at different redshifts (``cosmic chronometers"~\cite{Moresco:2016nqq}). We use a compilation of 36 measurements of $H(z)$ between $z=0.07$ and $z=2.34$, given in Table~I of \cite{Nesseris:2019fwr}.
Using the respective  prediction for $H(z)$ in  $\Lambda$CDM and in the RT models  (with the respective mean values of $\oma$ and $H_0$ from Table~\ref{tab:results}, obtained from the MCMC comparison to CMB+BAO+SNe) to fit these $H(z)$ measurements,
we find the difference in $\chi^2$, with respect to $\nu\Lambda$CDM, shown in Table~\ref{tab:chrono}.
The corresponding reduced $\chi^2$, all of order $0.63-0.64$, show that, by themselves, all the models fits these data well.

\begin{table*}[t]
\centering
\begin{tabular}{|c|c|c|c|c|c|}
\hline  
                        & $\nu\Lambda$CDM  &  RT, minimal & $\Delta N=34$ & $\Delta N=50$ &  $\Delta N=64$ \\ \hline
$\Delta\chi^2$   & 0& $-1.13$ & $0.22 $     & $0.42$    & $0.57$        \\     
\hline
\end{tabular}
\caption{Values of $\Delta\chi^2$, with respect to $\nu\Lambda$CDM, for the RT model, minimal and with various values of $\Delta N$, from the fit to a compilation of measurements of $H(z)$.\label{tab:chrono} }
\end{table*}

\begin{table*}[t]
\centering
\begin{tabular}{|c|c|c|c|c|c|}
\hline  
                        & $\nu\Lambda$CDM  &  RT, minimal & $\Delta N=34$ & $\Delta N=50$ &  $\Delta N=64$ \\ \hline
$\Delta\chi^2$   & 0& $1.41$ & $-0.05 $     & $-0.18$    & $-0.28$        \\   
\hline
\end{tabular}
\caption{Values of $\Delta\chi^2$, with respect to $\nu\Lambda$CDM, for the RT model, minimal and with various values of $\Delta N$, from the fit to a compilation of measurements of $f\sigma_8$.
\label{tab:fsigma8} }
\end{table*}
 
\vspace{2mm}\noindent
{\bf Structure formation and  $f\sigma_8$ data}. The properties of the models with respect to structure formation are already partly tested by the inclusion of BAO in our MCMC analysis. We further 
compare the models to a set of measurements of $f\sigma_8$, using the  
datapoints that we already used in \cite{Belgacem:2017cqo}.\footnote{Actually, many more measurement exists:  ref.~\cite{Kazantzidis:2018rnb} provides a compilation of 63 measurement of $f\sigma_8$ from 2006 to 2018. However, as stressed in \cite{Kazantzidis:2018rnb}, many of these datapoints are correlated, due to overlap in the galaxy samples used, and no covariance matrix is available for the full dataset, nor for most of its subsets. Furthermore,  one must also take care of the fact that different datapoints have been obtained with different fiducial cosmologies, and that survey systematic may vary with time of publication and lead to inhomogeneities in the data. Therefore the  use of the full dataset, without the appropriate covariance matrix and corrections, would lead to results of dubious interpretation.}

\begin{figure}[t]
\centering
\includegraphics[width=0.6\columnwidth]{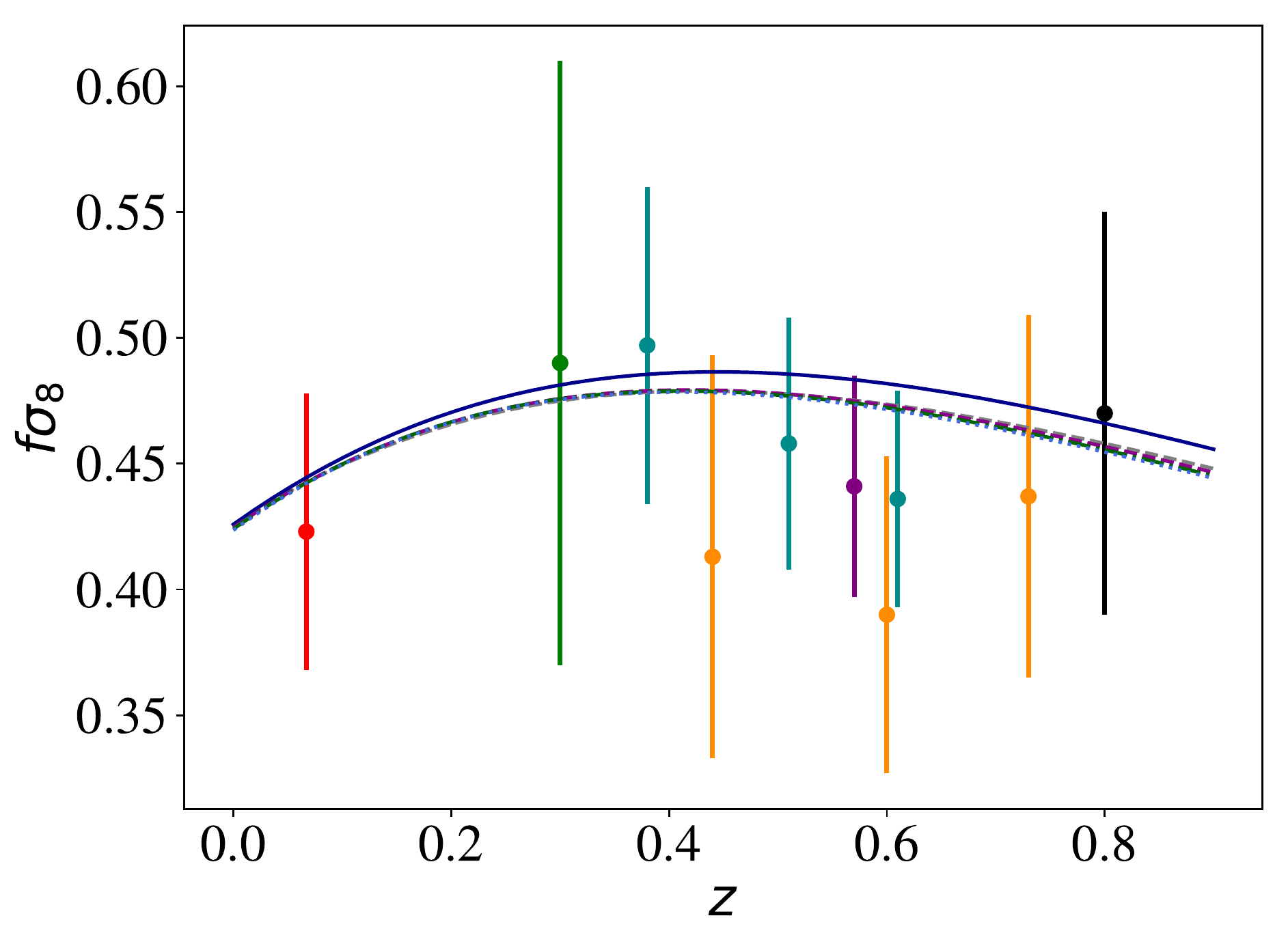}
\caption{A collection of  measurements of $f\sigma_8$  and the corresponding predictions of 
$\nu\Lambda$CDM and  
of the RT model, minimal and with  $\Delta N=34, 50$ and $64$. The curve for the minimal RT model is the upper one, while all others are almost indistinguishable on this scale. The data points are from
 6dF GRS \cite{Beutler:2012px} (red),  SDSS LRG \cite{Oka:2013cba} (green), BOSS CMASS \cite{Samushia:2013yga} (purple), WiggleZ \cite{Blake:2012pj} (orange), VIPERS \cite{delaTorre:2013rpa} (black) and  BOSS DR12 \cite{Alam:2016hwk} (cyan).
 \label{fig:fsig8}}
\end{figure}

Fig.~\ref{fig:fsig8} shows the data and the predictions of $\Lambda$CDM and  
of the RT model, minimal and with  $\Delta N=34,\Delta N=50,\Delta N=64$, obtained using for each model the respective mean values of $\oma$ and $H_0$ from Table~\ref{tab:results}. The corresponding differences of $\chi^2$, with respect to the value in $\nu\Lambda$CDM,
are given in  Table~\ref{tab:fsigma8}. We see that, once again, the differences between $\Lambda$CDM and the RT model with various $\Delta N$ are not statistically significant.  From  the
plots of $G_{\rm eff}$ in Fig.~\ref{fig:Geff} we see that at low $k$ (upper left panel) the minimal RT  model predicts $G_{\rm eff}/G >1$, while the RT models with large $\Delta N$  predict $G_{\rm eff}/G <1$. The data  favor a weakening of gravity at these scales, so the RT models with large $\Delta N$ are slightly preferred with respect to $\Lambda$CDM, and the minimal RT model is slightly disfavored, but in all cases   at a statistically insignificant level.

\begin{figure}[t]
\centering
\includegraphics[width=0.42\textwidth]{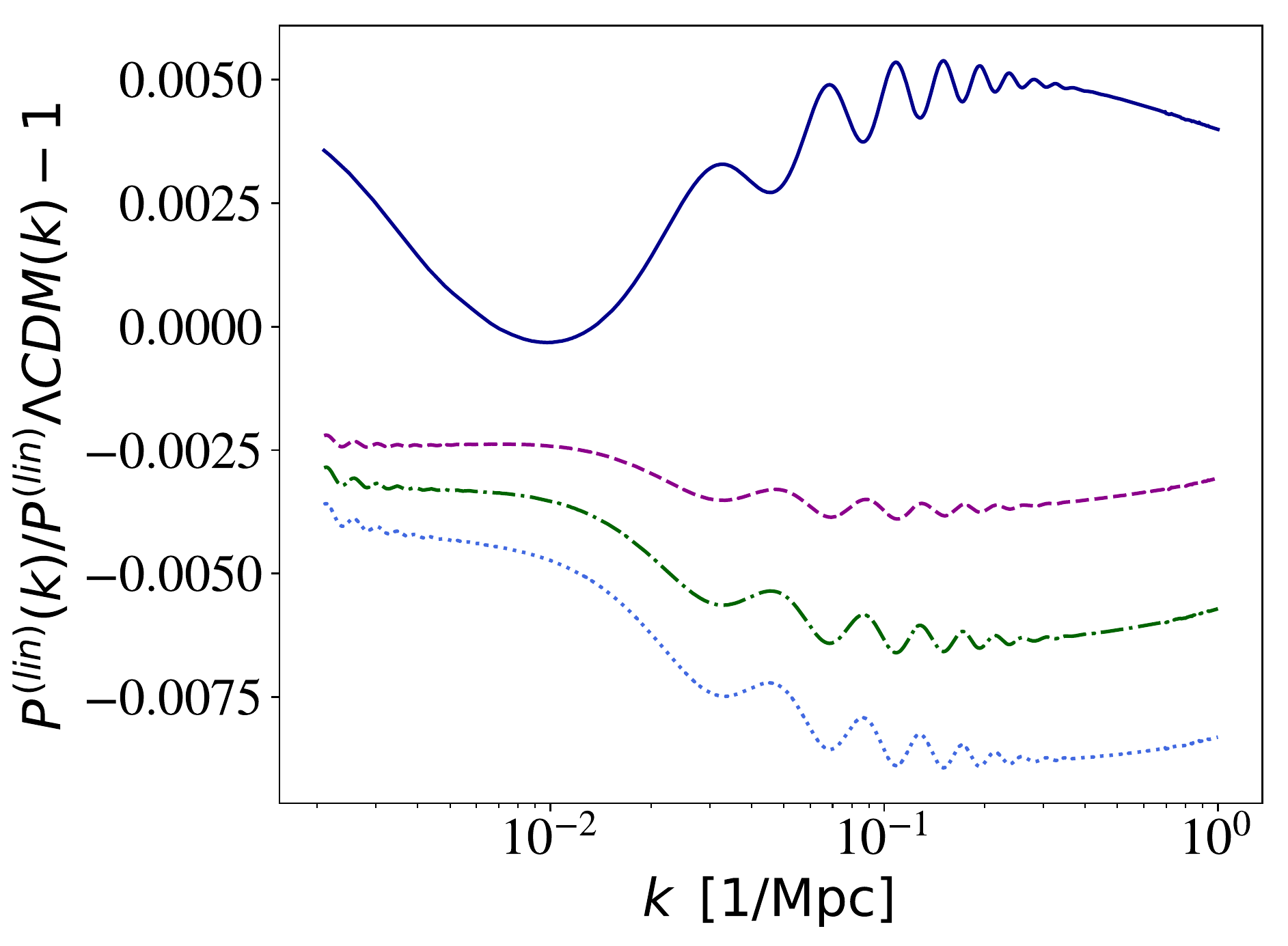}
\includegraphics[width=0.42\textwidth]{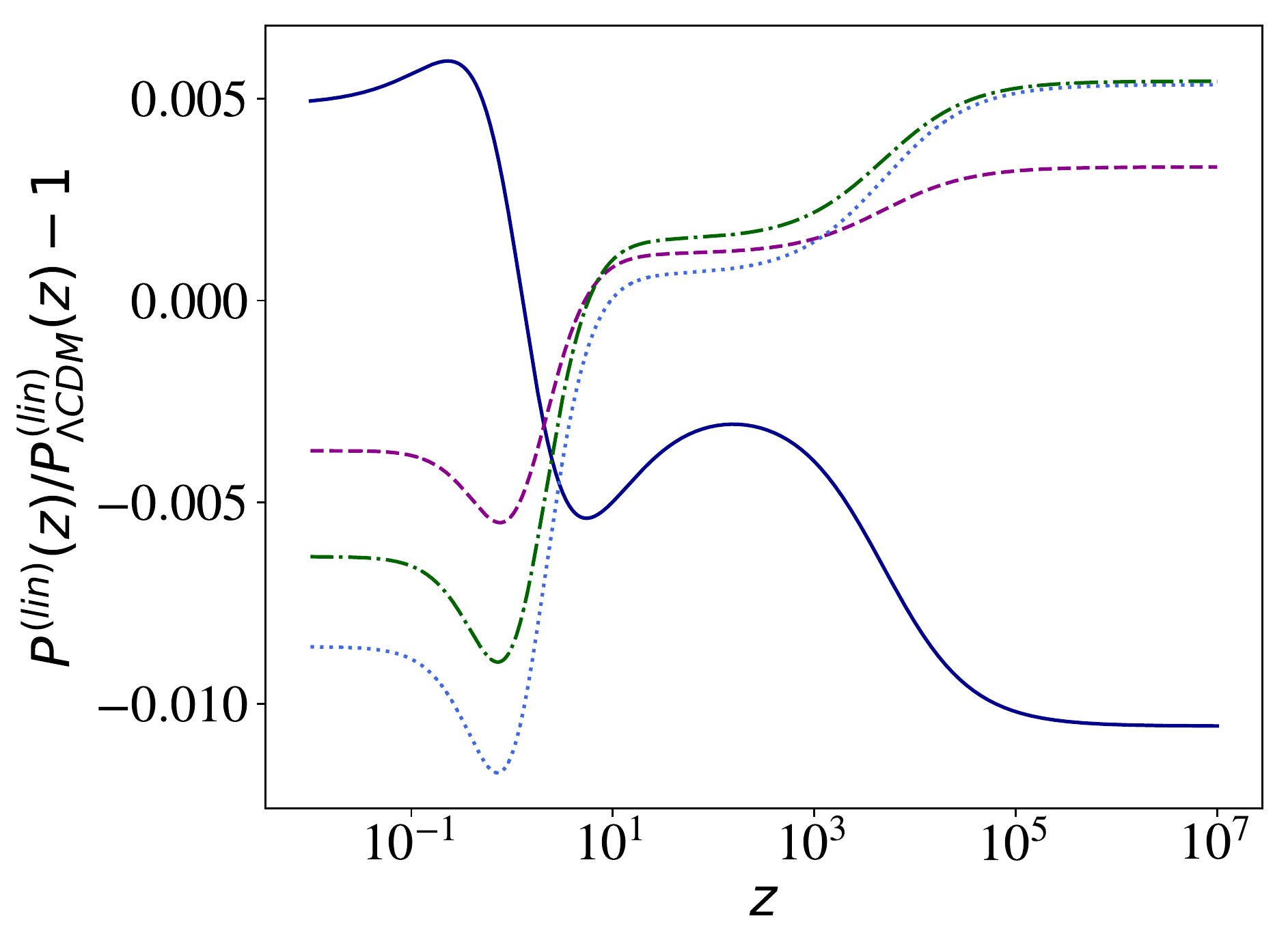}
\caption{Relative difference of total linear matter power spectrum, with respect to best-fit $\Lambda$CDM, for the 
minimal RT model (blue solid line) and  for RT with $\Delta N=34$ (magenta, dashed), $\Delta N=50$ (green, dot-dashed) and $\Delta N=64$ (cyan, dotted).
Left panel: as a function of $k$, at $z=0$. Right panel: as a function of redshift, for the mode with $k=0.1 /{\rm Mpc}$.
}
\label{fig:growth}
\end{figure}

Finally, Fig.~\ref{fig:growth} shows the relative difference in the linear power spectrum of the RT models with respect to $\Lambda$CDM (each one computed using their respective mean values of the cosmological parameters) as a function of $k$, for $z=0$ (left panel), and as  as a function of $z$, for the mode with $k=0.1 /{\rm Mpc}$ (right panel).

The conclusion of this analysis is that, on the one hand, the RT model, for all values of $\Delta N$, is very close to $\Lambda$CDM at the level of background evolution and scalar perturbations, and fits the observations at the same level as $\Lambda$CDM. On the other hand, the deviations, which for both the background and scalar perturbations are typically 
at the percent or sub-percent level, could in principle be within reach for future missions. For instance,  assuming that the function  $\mu(a,k)$ that characterizes deviations from the Poisson law is scale independent and parametrizing its dependence on the scale factor as
$ \mu(a)=\mu_s a^s$, a future survey such as {\sc Euclid}~\cite{Amendola:2012ys}, for fixed cosmological parameters, is  expected to measure $\mu_s$ with an  error $\Delta\mu_s=0.0046$ for $s=1$ and 
$\Delta\mu_s=0.014$ for $s=3$~\cite{Song:2010fg}. 
The RT model has indeed been selected by the Dark Energy Science Collaboration (DESC) of the Large Synoptic
Survey Telescope (LSST), among a few modified gravity models, for further studies and development of 
dedicated pipelines~\cite{Ishak:2019aay}.

\subsection{Recovery of GR at short scales}

\subsubsection{Solar system constraints and absence of vDVZ discontinuity}

Any cosmological model that modifies GR on cosmological scale must also be able to reproduce the successes of GR at  much smaller scales, such as the solar system and laboratory scales. In theories that introduces extra fields, as in scalar-tensor theories, or extra polarization of the gravitons, as in massive gravity, this is highly non-trivial. The linearized theory does not reduce to GR, and screening mechanisms involving the non-linearities of the theory are needed. In the RT model, in contrast, the situation is much simpler. Already at the linear level the theory reduces smoothly to GR, and there is no discontinuity such as the vDVZ discontinuity of massive gravity.  These issues have been discussed at length in~\cite{Maggiore:2013mea,Kehagias:2014sda} and here we summarize these results, for completeness.

\vspace{1mm}\noindent
{\bf GR limit in the linearization over Minkowski space}. Let us consider first the GR limit for the RT model linearized over flat space.
In this case \eq{RT}  reduces to \eq{eqmotnlG}.
In order to compute the matter-matter interaction induced by this coupling of $\Tmn$ with $\hmn$ we proceed as follows~\cite{Maggiore:2013mea}. We  use the gauge invariance of the linearized theory to fix the De~Donder gauge $\paM[\hmn -(1/2)h\emn]=0$. Going in momentum space, the resulting equation can be solved for  the Fourier transform 
$\tilde{h}_{\mu\nu}(k)$, obtaining
\be
\tilde{h}_{\mu\nu}(k)=\frac{16\pi G}{k^2}\[
\tilde{T}_{\mu\nu}(k)-
\frac{\emn k^2}{2(k^2-m^2)}\tilde{T}(k)+\frac{m^2}{3 (k^2-m^2)}\(\emn-\frac{k_{\mu}k_{\nu}}{k^2}\)
\tilde{T}(k)\]\, ,
\ee
where $T=\eMN\Tmn$.
Plugging this result  into the linearized interaction term\footnote{In \cite{Maggiore:2013mea} the  overall factor $1/2$  was missed in $S_{\rm int}$.}

\be
S_{\rm int}=\frac{1}{2}\int d^4x\,\hmn\TMN\, ,
\ee 
and using $k^{\mu}\tilde{T}_{\mu\nu}(k)=0$ we get
\be\label{SintTDT}
S_{\rm int} =8\pi G\int \frac{d^4k}{(2\pi)^{4}}\, \tilde{T}_{\mu\nu}(-k)\Delta^{\mu\nu\rho\sigma}(k)
\tilde{T}_{\rho\sigma}(k)\, , 
\ee
where 
\be\label{Delta}
\Delta^{\mu\nu \rho\s}(k)=\frac{1}{2k^2}\, 
\( \eMR\eNS +\eMS\eNR-\eMN\eRS \) 
+ \frac{1}{6}\, \[ \frac{1}{k^2}-\frac{1}{k^2-m^2}\]\eMN\eRS\, .
\ee
The term in the first line is the usual GR result  due to the exchange of the helicities $\pm 2$ of a massless graviton. The term in brackets vanishes for $m\ra 0$. Therefore the RT model has no vDVZ discontinuity, and
reduces smoothly to GR as $m\ra 0$. In the regime  where a linearization over flat space is adequate, for modes with  $|k^2|\gg m^2$   the predictions of the RT model  differ from the predictions of GR by a factor $1+{\cal O}(m^2/k^2)$. We have seen that the comparison with cosmological observations fixes
$m\sim H_0$ (or even smaller for large $\Delta N$, see footnote~\ref{foot:valuesgamma}). For  $|k|=(1\, {\rm a.u.})^{-1}$ (as appropriate to solar system experiments),
$m^2/k^2\sim (1\, {\rm a.u.}/H_0^{-1})^2\sim 10^{-30}$, and the predictions of the RT model are indistinguishable from that of  GR.

The absence of vDVZ discontinuity can also be understood observing that the  term in bracket in \eq{Delta} induces a matter-matter interaction
\be\label{TT}
8\pi G\int \frac{d^4k}{(2\pi)^{4}}\,
\frac{1}{6}\tilde{T}(-k)\[ \frac{1}{k^2}-\frac{1}{k^2-m^2}\]
\tilde{T}(k)\, .
\ee
Comparing with \eq{SEH2nlmass} one  realizes that the two terms in brackets corresponds to the exchange of the helicity zero component of $\hmn^{\rm TT}$ and of the trace mode $s$. In GR, where $s$ is massless and both 
$\hmn^{\rm TT}$ and  $s$ appear with a $\Box$ factor in the quadratic lagrangian
[see \eq{SEH2nl}] these two terms cancel exactly, while here the cancellation is only partial but is recovered for $m\ra 0$. Notice that both the helicity zero component of $\hmn^{\rm TT}$ and  $s$ are non-propagating degrees of freedom  in GR and remain non-propagating in the RT model. We have indeed seen in \eqst{dof2}{dof4} that, in the RR or RT models linearized over flat space, the only radiative degree of freedom of the metric are still given by the helicity $\pm 2$ modes described by $H_{ij}^{\rm TT}$ and $s$ remains non-radiative, see \eq{siBoxPhiRT}. Exactly as in GR,
the negative sign in front of the $1/(k^2-m^2)$ term in \eq{TT}, which would correspond to a ghost if it were due to a propagating particle, is therefore  innocuous from the point of view of quantum vacuum stability. The  helicity zero component of $\hmn^{\rm TT}$ and  $s$ are not associated to creation/annihilation operators and cannot appear on the external lines of a Feynman diagram.\footnote{Notice that,  in GR,  vacuum stability it is not related to the fact that the contribution of $s$ to the interaction (\ref{SintTDT}) is canceled by the contribution of the helicity zero component of $\hmn^{\rm TT}$. This is a cancelation that, in the language of Feynman graphs, takes place only in the internal lines. However, if $s$ were a propagating degree of freedom it would also appear in external lines, where it would induce vacuum decay into negative-energy ghost states plus positive-energy particles (and this, of course, cannot be canceled by graphs with the helicity zero mode of 
$\hmn^{\rm TT}$ on external lines, since these contribute to different $S$-matrix elements). The crucial point for vacuum stability in GR is rather that $s$ is non-propagating, so it is not associated to creation/annihilation operators and cannot appear on the external lines (just as $A_0$ in electrodynamics).}

\vspace{1mm}\noindent
{\bf GR limit for the \Sch solution}. After having checked the recovery of the GR limit in Minkowski space, let us consider the GR limit for the \Sch solution, by studying the
static spherically symmetric solution of the RT model. A typical issue of  massive gravity theories is that they become non-linear when $r$ is smaller than  a distance, the Vainshtein radius,  which is parametrically larger than the \Sch radius $r_S$ of the source; e.g.
$r_V=(GM/m^4)^{1/5}$ in the theory defined by adding a Fierz-Pauli mass term to the Einstein-Hilbert action \cite{Vainshtein:1972sx,Deffayet:2001uk}, and  $r_V=(GM/m^2)^{1/3}$ \cite{ArkaniHamed:2002sp} in
the dRGT theory \cite{deRham:2010ik,deRham:2010kj}.  For $m={\cal O}(H_0)$ and $M=\msun$, we have   $(GM/m^2)^{1/3}\sim 100$~pc. Since linearized theory only holds for $r>r_V$, in massive gravity  in the whole range of distances probed by  solar system and laboratory experiments the linearized expansion is not valid, and one must  show that a Vainshtein mechanism is at work, i.e. that the inclusion of classical non-linearities restore the continuity with GR at $r\ll r_V$. Explicit examples of this type have indeed been found for the dRGT theory \cite{Koyama:2011xz,Koyama:2011yg}.

For the RT model, however, the situation is much simpler, and the limit $m\ra 0$ of the \Sch solution is smooth. The \Sch solution in  the RT model has been worked out in~\cite{Kehagias:2014sda}.
In the limit $r\gg r_S$, the result for the metric is
\be
ds^2=-A(r)dt^2 +B(r)dr^2 +r^2(d\theta^2+\sin^2\theta\,  d\phi^2)\, ,
\ee
where
\bees
A(r)&=&1-\frac{r_S}{r}\[1+\frac{1}{3}(1-\cos mr)\]\, ,\label{NewtA}\\
B(r)&=&1+\frac{r_S}{r}\[1-\frac{1}{3}(1-\cos mr)+\frac{1}{3}mr \sin m r\]\, ,\label{NewtB}
\ees
 In the limit $mr\ll 1$ (but still $r\gg r_S$), \eqs{NewtA}{NewtB} give
\be\label{Afinal}
A(r)\simeq 1-\frac{r_S}{r}\(1+\frac{m^2r^2}{6}\)\, ,
\ee
and (to first order in $r_S/r$) $B(r)=1/A(r)$.\footnote{The solution for the auxiliary field $U=-\iBox R$ is given simply by $U(r)=(r_S/r)\cos mr$. For the auxiliary field $S_{\mu}(x)$, in spherical coordinates  only the component $S_r(r)$ is non-vanishing, and furthermore depends only on $r$. It is convenient to define $V(r)$ from $S_r(r) =B^{1/2}(r)r V(r)$. The (not very illuminating) solution for $V(r)$ is given in eq.~(3.8) of \cite{Kehagias:2014sda}, and reduces to $V(r)\simeq -r_S/(2 r)$ for $r\gg r_S$.}
For comparison, in massive gravity   the analogous computation gives \cite{Vainshtein:1972sx,Hinterbichler:2011tt}
\be\label{AFP}
A(r)=1-\frac{4}{3}\frac{r_S}{r}\(1-\frac{r_S}{12m^4r^5}\)\, .
\ee
The factor $4/3$ in front of $r_S/r$ is due to the extra contribution coming from the exchange of the helicity-0 graviton, and  gives rise to the vDVZ discontiuity. In contrast, no vDVZ discontinuity is present in
\eq{Afinal}. Furthermore, in 
\eq{AFP} the correction blows up as  $r$ decreases, and for $r\sim r_V=(GM/m^4)^{1/5}$ it becomes of the order of the leading term, signaling the breakdown of the linearized approximation. In \eq{Afinal}, in contrast, the correction becomes smaller and smaller as $r$ decreases, and perturbation theory is valid at all scales $r\ll m^{-1}$, until we arrive at $r\simeq r_S$, where eventually also GR becomes non-linear.

In conclusion, in the RT model (as well as in the RR model, where the analysis is very similar), in static situations GR is smoothly recovered, with correction $O(m^2 r^2)$. Given that $m$ is of order $H_0$, these corrections are utterly negligible for all $r$ of order of solar system scale or smaller; e.g. $m^2 r^2\sim 10^{-30}$ for $r$ of order of the Earth-Sun distance. Even on galactic scales these corrections to GR  are totally irrelevant, with $m^2 r^2\sim 10^{-17}$ for $r=10$~kpc.

\subsubsection{Limits on time variation of $G_{\rm eff}$ from Lunar Laser Ranging}\label{sect:LLR}

The above results show that, in a static situation, the RT and RR models recover all successes of GR at short scales.  As  was pointed out in \cite{Barreira:2014kra},  this is not yet sufficient to guarantee that these models are viable at solar system scales. Another crucial test comes from the limit on the time variation of Newton's constant from Lunar Laser Ranging (LLR). The current observational result is
$\dot{G}/G=(7.1\pm 7.6)\times 10^{-14}\, {\rm yr}^{-1}$~\cite{Hofmann:2018myc}.
This measurement is so accurate that, even if performed at the Earth-Moon scale over the last few decades, it provides significant constraints on cosmological models. Indeed,  if we rewrite this limit in terms of the Hubble parameter today, using  $H_0\simeq 
h_0\times (9.777 752\, {\rm Gyr})^{-1}$,  we get
\be\label{dotGsuGH0}
\frac{\dot{G}}{G}= (0.99\pm 1.06)\times 10^{-3}\, \( \frac{0.7}{h_0}\) \, H_0\, .
\ee
Quite generally, in modified gravity models Newton's constant becomes time dependent on cosmological scale.  The scale for the time variation today is  given by $H_0$, so on cosmological scales one typically finds
$\dot{G}/G\simeq H_0$. If, in a given modified gravity model, this result holds also down to  the scale of the solar system and of the Earth-Moon system, then the bound (\ref{dotGsuGH0}) is violated and the 
model is ruled out. 

In the case of the RT model, however, we have seen in \eq{GeffGRTlargek} that $G_{\rm eff}$ reduces to $G$ at small scales. Therefore, it has no time dependence and the RT model satisfies trivially the LLR limit.
The situation is different for the RR model (and for other modified gravity models, see app.~\ref{sect:difficulties}). Indeed, in the RR model,  for sub-horizon modes, one finds~\cite{Dirian:2014ara,Barreira:2014kra}
\be\label{Geff1suk2}
\frac{G_{\rm eff}(t)}{G} =  \[1-\frac{1}{3}m^2 \bar{S}(t)\]^{-1}\, \[ 1 +{\cal O}\(\frac{1}{\hat{k}^2}\)\]\, ,
\ee
where $\bar{S}(t)$ is the background cosmological solution for the auxiliary field $S$.  This dependence on $S$  can be traced to the term $2 S G_{\mu\nu}$ in $K_{\mu\nu}$, see \eq{defKmn}. If one plugs here the solution for $\bar{S}(t)$ corresponding to the FRW background, one finds that 
$G_{\rm eff}(t)/G$ is of order $H_0$, and the bound (\ref{dotGsuGH0}) is violated. In this case one cannot appeal to  non-linear screening mechanisms,  since we have seen that the RR model (just as the RT model) has a smooth limit $m\ra 0$, so the linearized expansion can be trusted.

Of course, the FRW metric has no direct relevance for the Earth-Moon system. The latter, just as the solar system, does not expand with the Hubble flow. However, the point is that a scalar field, such as $S$, that evolves on a background that interpolates between the \Sch solution at short scales and the FRW solution at large distances, in general inherits a time dependence on small scales from the matching with the solution at large distances. As an extreme example,  in GR one can consider  the Einstein-Straus space-time, in which, inside a sphere of  radius $r_0$,   the metric is taken to be exactly the static \Sch metric generated by the mass $M$, while in the exterior  it is given by a FRW solution with energy density $\rho$ (see e.g. \cite{Carrera:2008pi,Laarakkers:2001js} for review). The two metrics are then matched by requiring that the induced metric on the boundary surface $\Sigma$ agrees on the two sides. This fixes the
matching radius $r_0$, that, with respect to the \Sch coordinates of the interior, turns out to be given by
$M=(4/3)\pi r_0^3\rho$, where $\rho$ is the energy density in FRW. 
In this case the solution for the metric is exactly static in the interior region, so it describes a limiting case in which the cosmological expansion in the inner region is perfectly screened. Nevertheless, if one studies the propagation of a scalar field obeying the equation $\Box\phi=0$ in this metric, one finds that the solution for the field in the inner region is time dependent~\cite{Laarakkers:2001js}. This is due to the fact that we must impose a matching condition for the scalar field at the surface $\Sigma$, and in this way the field inherits a time dependence even in the inner region. 

For the RR model, a detailed analysis of the solution for the scalar field $S$ in a background that interpolates between the static  solution at short distances and FRW at large distances has been performed in
\cite{Belgacem:2018wtb}.
A useful way of studying the problem  is to follow the time evolution of  the auxiliary fields $U$ and $S$ of the RR model, starting   before the epoch of structure formation. At that time the FRW metric holds everywhere, and $U$ and $S$ evolve with time according to the cosmological background solutions
$\bar{U}(t)$ and $\bar{S}(t)$. As  structures form and become non-linear, the   analysis of
\cite{Belgacem:2018wtb} shows that the 
solutions for $U$ and $S$ remain  of the form 
\be
U(t,\vx) = \bar{U}(t) + \delta U (t,\vx)\, ,\qquad S(t,\vx) = \bar{S}(t)+ \delta S(t,\vx)\, ,
\ee 
where
$\delta U (t,\vx)$ and $\delta S(t,\vx)$ remain small perturbations of $\bar{U}(t) $ and $\bar{S}(t)$, respectively. In essence, the physical reason behind this result is that, even when structures become non-linear, e.g. in the formation of galaxies, clusters, etc., 
the metric perturbation $\Phi$ never become large.  In non-linear structure formation are  rather the second spatial derivatives of $\Phi$ that become large compared to their values in the linear regime, in particular the Laplacian of $\Phi$, which is related to the density contrast and can become huge; however, the spatial derivatives of $\Phi$ never enter in the equations that govern the dynamics of the auxiliary fields $U$ and $S$. Indeed, in a perturbed FRW metric, to first order in $\Phi$,  the explicit  expression of the d'Alembertian is
\be
\Box U= -(1+2\Phi) (\ddot{U} + 3 H \dot{U})  - 4 \dot{\Phi} \dot{U} + a^{-2} (1-2\Phi)  \nabla^2 U\, ,
\ee 
so spatial derivatives of $\Phi$ do not appear.
As a result, non-linear structure formation does not stop the time evolution that  the auxiliary fields inherited from the earlier epoch described by a spatially homogeneous FRW solutions.
Near massive bodies, the perturbations 
$\delta U (t,\vx)$, $\delta S (t,\vx)$ just reduce to the static solutions $U(r)$, $S(r)$ studied in the previous subsection, and remain small as long as $r$ is larger than the \Sch radius of the massive bodies (recall for instance that $U(r)=(r_S/r)\cos mr$, which is much smaller that one for $r\gg r_S$).
So, in the end, at the Earth-Moon system scale, the solution for $S$ is, with good approximation, the sum of the cosmological and static solutions, $S(t,\vx)=\bar{S}(t)+S_{\rm static}(r)$.\footnote{This was also shown to happen exactly in models, such as galileons or k-essence, in which a field $\varphi$ has a shift symmetry $\varphi\ra \varphi+{\rm const}$. In this case, thanks to the shift symmetry, near the present epoch $t_0$ the equation of motion of the field admits an exact solution with separation of variables of the form 
$\varphi(t,r)=\varphi_{\rm static}(r)+\varphi_{\rm cosmo}(t_0)+(t-t_0)\dot{\varphi}_{\rm cosmo}(t_0)$~\cite{Babichev:2011iz}. See also \cite{Tsujikawa:2019pih} for a related example.}  A study of purely static solutions misses the term $\bar{S}(t)$, because assumes from scratch that the solution is time-independent. This time dependence induces a time-dependence of the Newton's constant, such  that the RR model violates the limit (\ref{dotGsuGH0}). This rules out the RR model. We will see in app.~\ref{sect:difficulties} that this problem affects also other nonlocal models that were proposed in the literature.

As we have seen in \eq{GeffGRTlargek}, in the RT model, in contrast, the effective Newton's constant on small scales reduces to $G$,  and looses all dependence on the auxiliary fields, so it passes without problems
also the LRR constraint.

\subsection{Tensor perturbations and modified GW propagation}

Until now we have studied the cosmological consequences of the theory at the level of background evolution and scalar perturbations. We now turn to tensor perturbations, i.e. gravitational waves (GWs) propagating in FRW. We will see, following \cite{Belgacem:2017ihm,Belgacem:2018lbp,Belgacem:2019lwx}, that the RT model has striking predictions in the tensor sector, that could be detected in the near future by GW detectors.

\subsubsection{Tensor perturbations in GR}

Let us begin by recalling that, in GR,  the evolution of tensor perturbations over FRW is governed by the equation 
\be\label{4eqtensorsect}
\tilde{h}''_A+2{\cal H}\tilde{h}'_A+k^2\tilde{h}_A=16\pi G a^2\tilde{\s}_A\, ,
\ee
where $\tilde{h}_A(\eta, \vk)$ are  the Fourier modes of the GW amplitude, and we use the index $A=+,\times$ to label the two polarizations. We are using now conformal time $\eta$, related as usual to cosmic time $t$ by $dt=a(\eta)d\eta$, and   $a(\eta)$ is the scale factor. In this section
the prime denotes the derivative with respect to cosmic time $\eta$, and  ${\cal H}=a'/a$. The source term $\tilde{\s}_A(\eta, \vk)$ is related to the helicity-2 part of the anisotropic stress tensor (see e.g.~\cite{Maggiore:2018sht}). In the following we will be interested in the free propagation between source and observer, and we will set it to zero.
It is convenient to introduce a field $\tilde{\chi}_A(\eta, \vk)$ from
\be\label{4defhchiproofs}
\tilde{h}_A(\eta, \vk)=\frac{1}{a(\eta)}  \tilde{\chi}_A(\eta, \vk)\, .
\ee
Then  \eq{4eqtensorsect} becomes
\be\label{4propchiproofs1}
\tilde{\chi}''_A+\(k^2-\frac{a''}{a}\) \tilde{\chi}_A=0\, .
\ee
For modes well inside the horizon, such as the GWs targeted by ground-based and space-born detectors, the term $a''/a\sim 1/\eta^2$ is totally negligible with respect to $k^2$;
for instance, for a GW with a frequency  $f\sim 10^2$~Hz, as typical of ground-based interferometers,
$(k\eta)^{-2}\sim (500\, {\rm km}/H_0^{-1})^2\sim 10^{-41}$. We can then neglect the term $a''/a$ in \eq{4propchiproofs1}, which then becomes a standard 
a  wave equation for $\tilde{\chi}_A$, that tells us that GWs propagate at the speed of light (that we have set here equal to unity).   

The factor $1/a$ in \eq{4defhchiproofs} tells us how the GW amplitude decreases as it propagates across cosmological distances, from the source to the observer. For inspiraling binaries this factor combines with other factors coming from the transformation of masses and frequency from the source frame to the detector frame (see e.g.  Section 4.1.4 of \cite{Maggiore:1900zz}), to produce  the well-known dependence of the GW amplitude 
$\tilde{h}_A(\eta, \vk)\propto 1/d_L(z)$,
where $d_L$ is the luminosity distance to the source. This is the origin of the fact that coalescing binaries are `standard sirens', i.e. their waveform allows a direct reconstruction of the luminosity distance
to the source~\cite{Schutz:1986gp,Dalal:2006qt,MacLeod:2007jd,Nissanke:2009kt,Cutler:2009qv,Sathyaprakash:2009xt,Zhao:2010sz,DelPozzo:2011yh,Nishizawa:2011eq,Taylor:2011fs,Taylor:2012db,Tamanini:2016zlh,Cai:2016sby}.
In GR, for a cosmological model with energy density $\rde(z)$, the relation between luminosity distance and redshift is
\be\label{dLem}
d_L(z)=\frac{1+z}{H_0}\int_0^z\, 
\frac{d\tilde{z}}{\sqrt{\oma (1+\tilde{z})^3+\ora (1+\tilde{z})^4+\rde(\tilde{z})/\rho_0 } }\, .
\ee
Therefore, a simultaneous measurement of $d_L$ and of the redshift $z$ (with an electromagnetic counterpart, or the study of the $d_L-z$ relation with statistical methods) allows us to get cosmological information. In particular, 
for sources at small redshift,  $z\ll  1$, \eq{dLem} reduces to  the Hubble law $d_{L}(z)\simeq H^{-1}_0z$, so from a measurement at such redshifts we can get a measurement of $H_0$.
This has indeed been possible with the detection of the binary neutron star (BNS) coalescence GW170817, which is at a redshift $z\simeq  0.01$, and has given the value $H_0=70.0^{+12.0}_{-8.0}\,\, {\rm km}\, {\rm s}^{-1}\, {\rm Mpc}^{-1}$~\cite{Abbott:2017xzu}. The detection of coalescences at higher redshift could  in principle allow us to access also the DE equation of state.

\subsubsection{Tensor perturbations in modified gravity}

As we will see, the free propagation of tensor perturbations in the RT model is governed by an equation of the form
\be\label{prophmodgrav}
\tilde{h}''_A  +2 {\cal H}[1-\delta(\eta)] \tilde{h}'_A+k^2\tilde{h}_A=0\, ,
\ee
for a given function $\delta(\eta)$.
It is however instructive to first work out the implications of \eq{prophmodgrav} with a generic function $\delta(\eta)$, since this equations appears  in many other modified gravity models. 
Indeed, in a generic modified gravity model both the ``friction term" $2 {\cal H}\tilde{h}'_A$ and the term $k^2\tilde{h}_A$ in \eq{4eqtensorsect} are modified. As we will recall below, the models  that modify the $k^2\tilde{h}_A$ term predict a speed of gravity different from the speed of light. 
The observation of  GW170817 and of the associated GRB has  set a limit   $|c_{\rm gw}-c|/c< O(10^{-15})$ \cite{Monitor:2017mdv}, so such models are ruled out.\footnote{Although it could still in principle happen  that  there is dependence on wavenumber that  allows for $c_{\rm gw}\neq c$ for modes $k$ well below the frequencies probed by LIGO/Virgo and restore $c_{\rm gw}=c$ to sufficient accuracy at LIGO/Virgo frequencies. This could be motivated in some models \cite{deRham:2018red}.} In particular, a large class of  Horndeski theories and other modifications of GR have been ruled out by this limit~\cite{Creminelli:2017sry,Sakstein:2017xjx,Ezquiaga:2017ekz,Baker:2017hug}. It turns out that the models that survive this constraint still modify the friction term. A propagation equation of the form (\ref{prophmodgrav}) was indeed first found in some scalar-tensor theories of Horndeski type 
\cite{Saltas:2014dha,Lombriser:2015sxa,Arai:2017hxj,Amendola:2017ovw} and in the RR nonlocal model~\cite{Belgacem:2017cqo,Belgacem:2017ihm}. In  \cite{Belgacem:2019pkk} it was shown that it also takes place in many other Horndeski-type theories that pass the test on speed of gravity (such as $f(R)$ theories, Jordan-Brans-Dicke, galileon cosmology, etc.), in Degenerate Higher Order Scalar-Tensor (DHOST) theories, and in  bigravity.
Similar effects take place in theories with extra dimensions, as originally found in~\cite{Deffayet:2007kf} (see also \cite{Pardo:2018ipy}), although in this case they are due to the loss of gravitons to the bulk and, in general, are not described by \eq{prophmodgrav} (see also \cite{Gleyzes:2014rba} for a discussion a modified GW propagation within  the  effective field theory approach to dark energy,  
\cite{Nishizawa:2017nef,Garoffolo:2019mna} for  general formalisms for testing gravity with GW propagation, and  \cite{Linder:2018jil,Dalang:2019fma,DAgostino:2019hvh} for further related work in the context of scalar-tensor theories).

Let us then study first the general consequences of \eq{prophmodgrav} (we  closely follow the discussion in \cite{Belgacem:2017ihm,Belgacem:2018lbp}).
We proceed as in the GR case, except that now,
to eliminate the friction term, we must   introduce $\tilde{\chi}_A(\eta, \vk)$ from 
\be\label{4defhchiproofsRR}
\tilde{h}_A(\eta, \vk)=\frac{1}{\tilde{a}(\eta)}  \tilde{\chi}_A(\eta, \vk)\, ,
\ee
where $\tilde{a}$ now satisfies
\be\label{deftildea}
\frac{\tilde{a}'}{\tilde{a}}={\cal H}[1-\delta(\eta)]\, .
\ee
Then we get 
\be
\tilde{\chi}''_A+\(k^2-\frac{\tilde{a}''}{\tilde{a}}\) \tilde{\chi}_A=0\, .
\ee 
Once again, inside the horizon the term $\tilde{a}''/\tilde{a}$ is totally negligible. The remaining equation,
\be\label{eqforchi}
\tilde{\chi}''_A+k^2\tilde{\chi}_A=0\, ,
\ee 
shows that  GWs still propagate at the speed of light. This is a consequence of the fact that the  term $k^2\tilde{\chi}_A$ in \eq{prophmodgrav} is the same as in GR. If the coefficient of this term had been different, we would get a speed of GWs $c_{\rm gw}\neq c$.

As we see from \eq{4defhchiproofsRR}, the effect of the modified friction term is that now
the amplitude of $\tilde{h}_A$ is proportional to $1/\tilde{a}$ rather than $1/a$. Then, in the propagation from the source to the observer, the amplitude  is multiplied by a factor 
$\tilde{a}_{\rm emis}/\tilde{a}_{\rm obs}\equiv \tilde{a}(z)/\tilde{a}(0)$,
instead of a factor $a_{\rm emis}/a_{\rm obs}=a(z)/a(0)$, where the labels refer to the emission time (at redshift $z$) and the observation time, at redshift zero, respectively. Therefore
\be\label{dLtilde}
\tilde{h}_A\propto  \frac{\tilde{a}(z)}{\tilde{a}(0) }\, \frac{a(0)}{a(z) }\, 
\frac{1}{d_L(z)}=
 \frac{\tilde{a}(z)}{a(z)} \frac{1}{d_L(z)}
\, ,
\ee
where $d_L(z)$ is the usual notion of luminosity distance (note that,  since only the ratios $\tilde{a}(z)/\tilde{a}(0)$ and $a(z)/a(0)$ enter, without loss of generality we can choose the normalizations $\tilde{a}(0)=a(0)=1$).
\Eq{dLtilde} motivates the introduction of a `GW luminosity distance' $\dgw(z)$~\cite{Belgacem:2017ihm}, related to the standard  
luminosity distance appropriate for electromagnetic signals, that we henceforth denote by $d_L^{\,\rm em}(z)$,  by $d_L^{\,\rm gw}(z)=[a(z)/\tilde{a}(z)]\, d_L^{\,\rm em}(z)$.
Rewriting \eq{deftildea} as
$ (\log a/\tilde{a})'=\delta(\eta) {\cal H}(\eta)$ and integrating, we get~\cite{Belgacem:2017ihm}
\be\label{dLgwdLem}
d_L^{\,\rm gw}(z)=d_L^{\,\rm em}(z)\exp\left\{-\int_0^z \,\frac{dz'}{1+z'}\,\delta(z')\right\}\, .
\ee
In modified gravity, the quantity extracted from a measurement of the GW amplitude of  a coalescing binary is  $d_L^{\,\rm gw}(z)$, rather than $d_L^{\,\rm em}(z)$. To avoid misunderstandings, notice that the actual distance traveled by GWs from the source to the observer is the same as the distance traveled by electromagnetic signals.  \Eq{dLgwdLem} is simply a convenient way of expressing the fact that, in modified gravity, the amplitude of the GW decreases in a different way during the propagation, so that, for a coalescing binary, the observed amplitude, rather than depending only on $\dem(z)$ and on the inclination of the orbit, as in GR, it further depends on $\delta(z)$, in such a way that the combined dependence on $\dem(z)$ and $\delta(z)$  can be reabsorbed into the quantity $\dgw(z)$.\footnote{A different effect is provided by the fact that, in brane models,  a gravitational signal can travel along geodesics in the  extra dimensions, while electromagnetic signals are  confined to the (3+1)-dimensional brane. This can  lead to delays between the arrival time of a GW  and the associated  electromagnetic signal~\cite{Chung:1999xg,Caldwell:2001ja,Visinelli:2017bny}.}

\subsubsection{Predictions of the RT model}

We  now discuss  GWs  in the RT  model, focusing  on the signal from coalescing binaries at cosmological distances.\footnote{See \cite{Belgacem:2018lbp} for a discussion of how modified GW propagation affects the ISW effect.} First of all, notice that  
this model only changes the gravitational part of the action but not the matter action, so the coupling to matter is unchanged, and at the linearized level, is still given by the usual $\hmn\TMN$ coupling. Thus, the source term in \eq{4eqtensorsect} is not affected. Furthermore we have seen that, at short scales,  such as the 
distance between the two bodies in a coalescing binary,
the RT model reduces to  GR  to huge accuracy, so there is no appreciable modification to the orbital dynamics of a  binary system, and the waveform produced by a coalescing binary in the  region far from the source (where the $1/r$ GW behavior sets in, but still the expansion of the Universe can be neglected) is the same as in GR. In the signal received by a coalescing binary, the only difference will then come from the free propagation of the GW from the source to the observer, across cosmological distances.

\begin{figure}[t]
\centering
\includegraphics[width=0.42\textwidth]{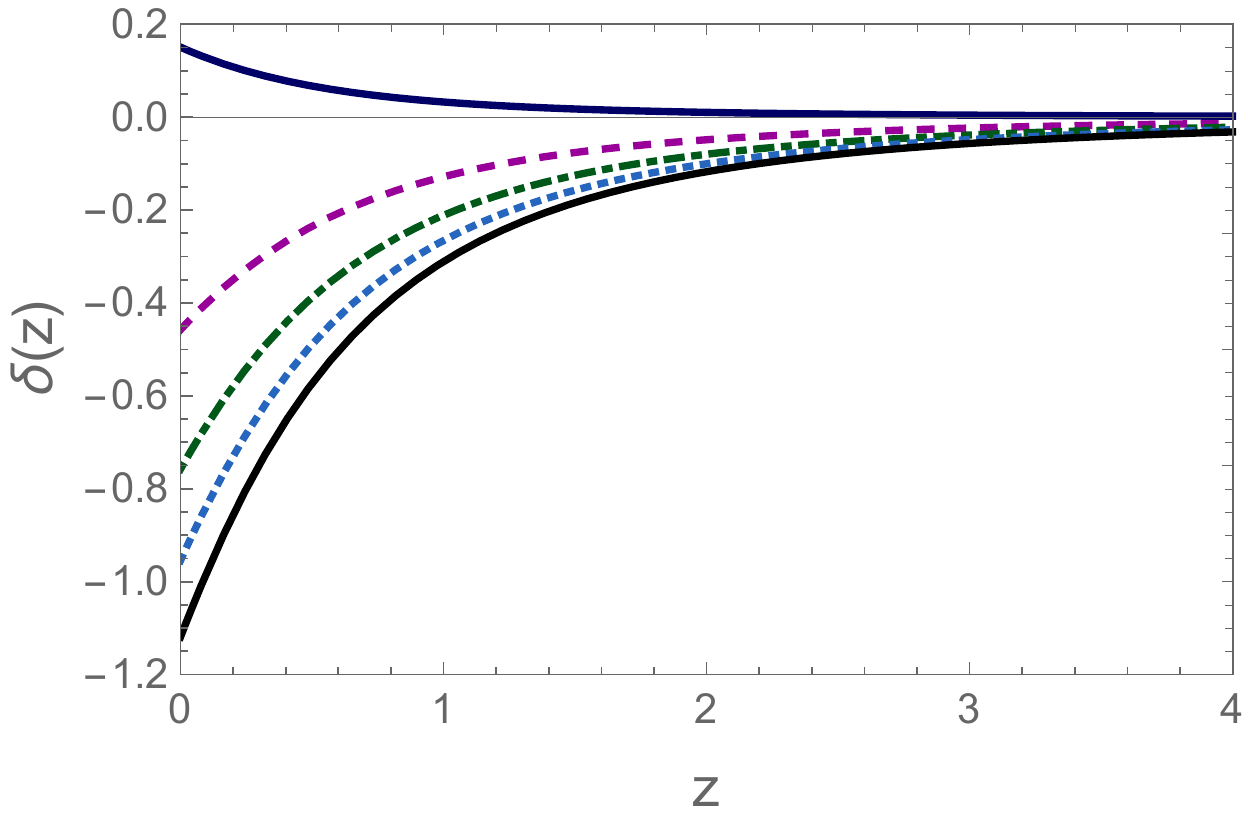}
\includegraphics[width=0.42\textwidth]{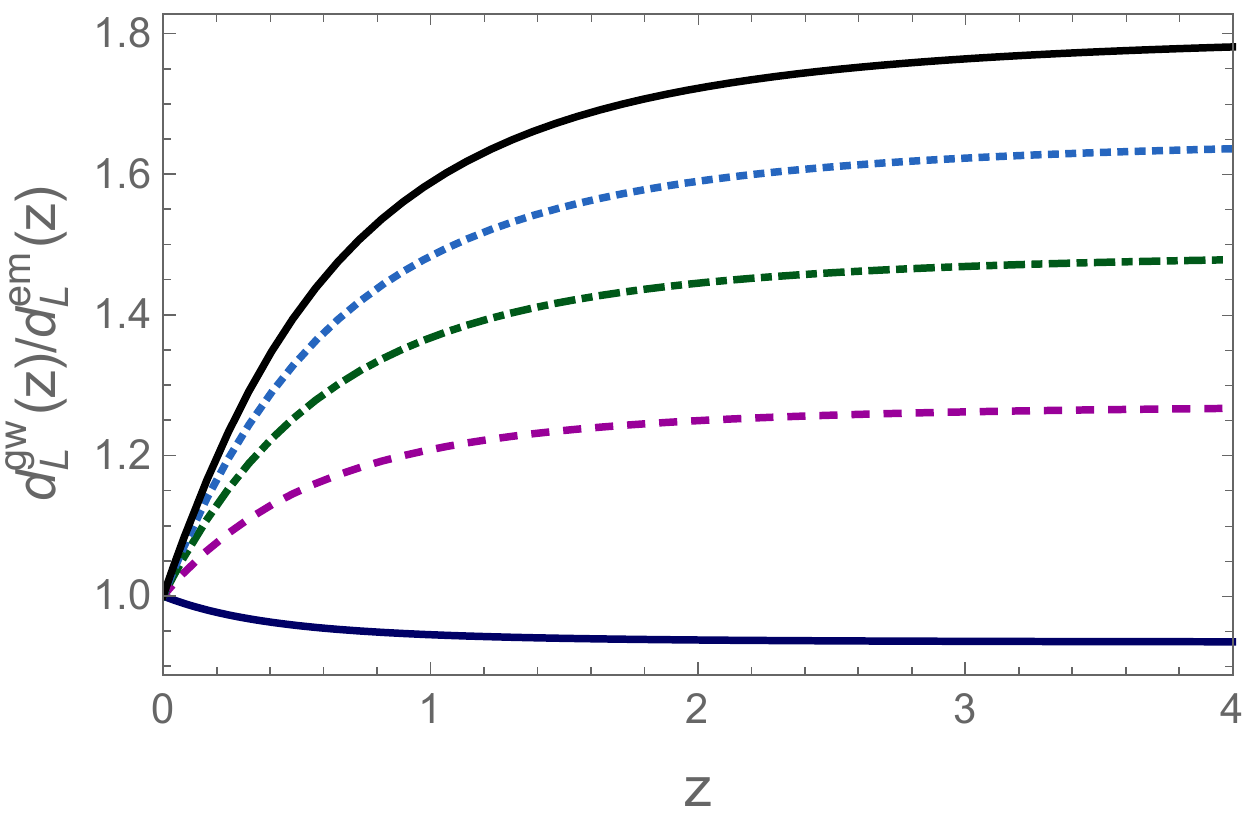}
\caption{The functions $\delta(z)$ (left panel) and $\dgw(z)/\dem(z)$ (right panel),
for the minimal RT model (blue solid line) and  for RT with $\Delta N=34$ (magenta, dashed), $\Delta N=50$ (green, dot-dashed) and $\Delta N=64$ (cyan, dotted) and  $\Delta N=100$  (black solid line).
}
\label{fig:deltadgw}
\end{figure}

The equation governing the free propagation of tensor perturbations in the RT model has been computed  in ~\cite{Dirian:2016puz}, and is\footnote{Note that eq.~(5.1) of ref.~\cite{Dirian:2016puz} was written with a different definition of the auxiliary field $V$. Denoting by $\tilde{V}$ the definition used there, by $S_{\eta}$ the $\mu=0$ component of $S_{\mu}$ in coordinates $(\eta,\vx)$ and by $S_0$ the $\mu=0$ component of $S_{\mu}$ in coordinates $(t,\vx)$, we
have $\tilde{V}\equiv S_{\eta}=a S_0 =aV/H_0$, where $V$ is the definition used here. The equation written in app.~A.1 of \cite{Dirian:2016puz}, where is described the implementation in CLASS of the perturbations of the model, are also written denoting by $V$ the quantity that we are here calling $\tilde{V}$.}
\be\label{prophmodgravRT}
\tilde{h}''_A  +[2 {\cal H}-3\gamma\bar{V}a H_0] \tilde{h}'_A+k^2\tilde{h}_A=0\, .
\ee
So  the `friction term'  $2{\cal H}\tilde{h}'_A$ is modified with respect to GR, but  the term $k^2\tilde{h}_A$ is not. Thus, first of all we see that the RT model passes the constraints from the speed of GWs. As we have mentioned, this is a non-trivial constraint that has ruled out many modified gravity theories.
\Eq{prophmodgravRT} is of the form (\ref{prophmodgrav}), with
\be\label{defdeltaRT}
\delta(\eta) =\frac{3\gamma\bar{V}(\eta)H_0}{2 H(\eta)}\, ,
\ee
where  we  have used ${\cal H}=a H$. Recall that, for the RT model in a FRW background,  we have defined the auxiliary field $V$ from 
$V=H_0S_0$, where $S_0$ is the $\mu=0$ component [in coordinates $(t,\vx)$] of   the auxiliary four-vector field $S_{\mu}$  of the RT model, see \eq{defVZ}. Recalling the definition (\ref{defgamma}) of $\gamma$, we can also write \eq{defdeltaRT} as
\be
\delta(\eta) =\frac{m^2\bar{S}_0(\eta)}{6H(\eta)} \,  .
\ee
Using the numerical solution of  the background evolution equation of the RT model studied in section~\ref{sect:bkg}, we can therefore immediately compute  $\delta$ and $\dgw/\dem$, as functions of the redshift. The results are shown in Fig.~\ref{fig:deltadgw} (see also \cite{Belgacem:2019lwx}). These results are quite spectacular, in particular at large $\Delta N$. For instance, for $\Delta N=64$, at large $z$ the ratio $\dgw/\dem$ tends asymptotically to a value $\simeq 1.65$, corresponding to a $65\%$ deviation from GR, a truly huge effect. In the limit of large $\Delta N$ [exemplified here by the case 
$(M_{\rm infl}=10^{16}\, {\rm GeV}, \Delta N=100)$; as we mentioned, for $M_{\rm infl}=10^{16}\, {\rm GeV}$ this asymptotic curve is actually reached already at $\Delta N\,\gsim\, 70$], at large $z$ the ratio  $\dgw/\dem$ reaches a value $\simeq 1.80$, i.e. a $80\%$ deviation from GR! Similarly, at $z=0$, $\delta(0)$, in the limit of  large $\Delta N$, saturates to a value $-1.11$, so, in \eq{prophmodgrav}, near $z=0$ the term $1-\delta(0)\simeq 2.11$ is more than twice the GR value.

\begin{figure}[t]
\centering
\includegraphics[width=0.46\textwidth]{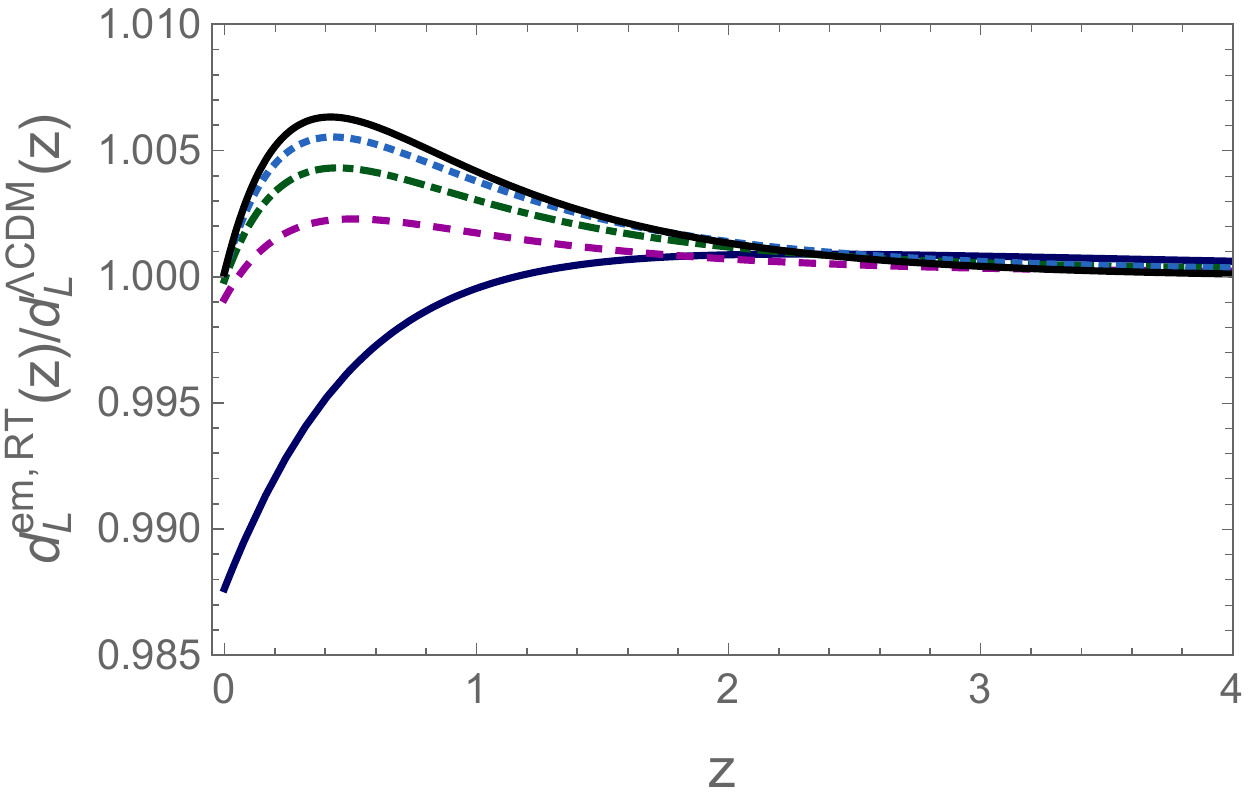}
\caption{The ratio of  $\dem(z)$ computed in the RT model to the luminosity distance of $\Lambda$CDM
for the minimal RT model (blue solid line) and  for RT with $\Delta N=34$ (magenta, dashed), $\Delta N=50$ (green, dot-dashed) and $\Delta N=64$ (cyan, dotted) and  $\Delta N=100$  (black solid line), using for each model its own mean values of $H_0$ and $\oma$.
}
\label{fig:demRToverdLCDM}
\end{figure}

This is very surprising because we have seen that, for all values of $\Delta N$, the RT model differs from $\Lambda$CDM by less that $1\%$ at the level of background evolution (see fig.~\ref{fig:Hubble}), and by a few percent to below percent level, depending on wavenumber, for the scalar perturbations, see e.g. Figs.~\ref{fig:Geff}-\ref{fig:growth}. This  is indeed what  allows the model to fit well the current cosmological observations. One would have then naturally guessed that also in the tensor perturbation sector the differences would be of the same order. Instead, for large $\Delta N$, they are much bigger, a very good news for GW experiments.

For comparing the RT model to $\Lambda$CDM the relevant quantity, rather than the ratio of $\dgw$ to $\dem$, both computed within the RT model, is actually the ratio of $\dgw$, computed in  the RT model, to the luminosity distance $d_L^{\Lambda{\rm CDM}}$ computed in $\Lambda$CDM (for which the notion of electromagnetic and GW luminosity distance coincide), and in which, in each model, the respective mean values of the parameters $H_0$ and $\oma$ are used. However, the results for
$d_L^{\rm gw, RT}/d_L^{\Lambda{\rm CDM}}$ turn out to be practically  the same as the results shown in the right panel of Fig.~\ref{fig:deltadgw}. This can be seen by writing
\be
\frac{d_L^{\rm gw, RT} (z)}{d_L^{\Lambda{\rm CDM}} (z)}=
\(\frac{d_L^{\rm gw, RT} (z)}{d_L^{\rm em, RT} (z)}\)\times 
\(\frac{d_L^{\rm em, RT} (z)}{d_L^{\Lambda{\rm CDM}} (z)}\)\, ,
\ee
where, for clarity, we have denoted by $d_L^{\rm gw, RT} (z)$
the GW luminosity distance $\dgw$ in the RT model.
The first factor on the right-hand side is the quantity that we have already shown in the right panel of Fig.~\ref{fig:deltadgw}. The second factor is shown in Fig.~\ref{fig:demRToverdLCDM}, and we see that is very close to one; in particular, for the RT model with large $\Delta N$, it reaches at most a value of order $1.006$ for $\Delta N=100$ near $z\simeq 0.3$, and then quickly goes asymptotically to values of order $1.001$. This can be understood observing that the ratio ${d_L^{\rm em, RT} (z)}/{d_L^{\Lambda{\rm CDM}} (z)}$ is determined by two factors. First, by the different mean values of $H_0$ and $\oma$ between the RT model with the given $\Delta N$ and $\Lambda$CDM; second, by the different redshift dependence of the DE density, or, equivalently, the different DE equation  of state $\wde(z)$. However, we have seen in  Table~\ref{tab:results} that Bayesian parameter estimation gives for the RT model values of $H_0$ and $\oma$ very close to those of $\Lambda$CDM, particularly at large $\Delta N$; furthermore, as discussed in 
\cite{Belgacem:2018lbp}, the change in the value of these parameters goes precisely in the direction to cancel the effect in the change of the DE equation of state. This is due to the fact that  Bayesian parameter estimation in practice requires the model to fit some fixed distance scales at large redshifts, such as the scales given by the CMB peaks or by the BAO oscillations; thus, if, compared to  $\Lambda$CDM,  one changes $\wde(z)$ in the  direction of giving, say, a larger (electromagnetic) luminosity distance at large redshift, $H_0$ and $\oma$ change in the direction such that they partially  compensate for this change. As a result the electromagnetic luminosity distance, particularly at moderate to large values $z$, changes very little. Thus, the difference in the GW luminosity distance of the RT model, compared to $\Lambda$CDM, in practice is entirely given by the effect of modified GW propagation, while the DE equation of state and the difference in $H_0$ and $\oma$ among RT and $\Lambda$CDM have a negligible effect.

\begin{figure}[t]
\centering
\includegraphics[width=0.42\textwidth]{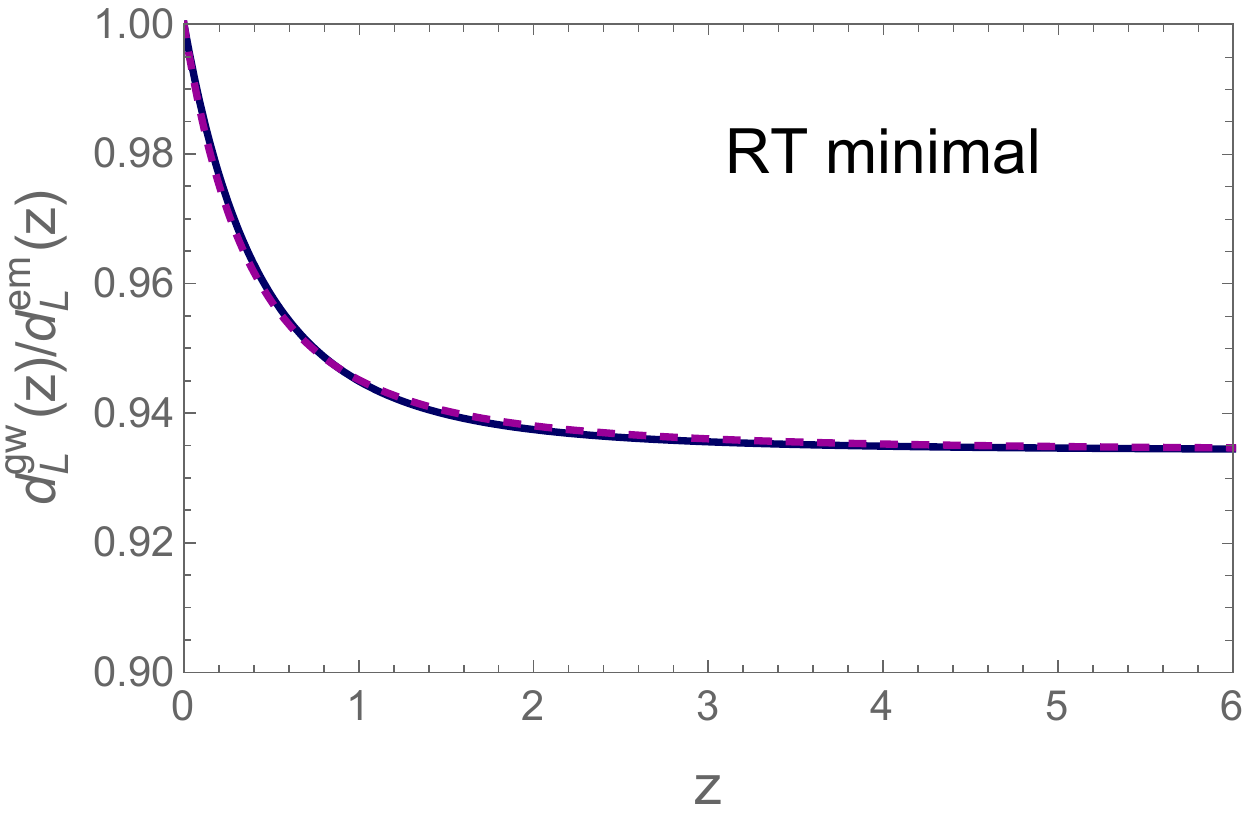}
\includegraphics[width=0.42\textwidth]{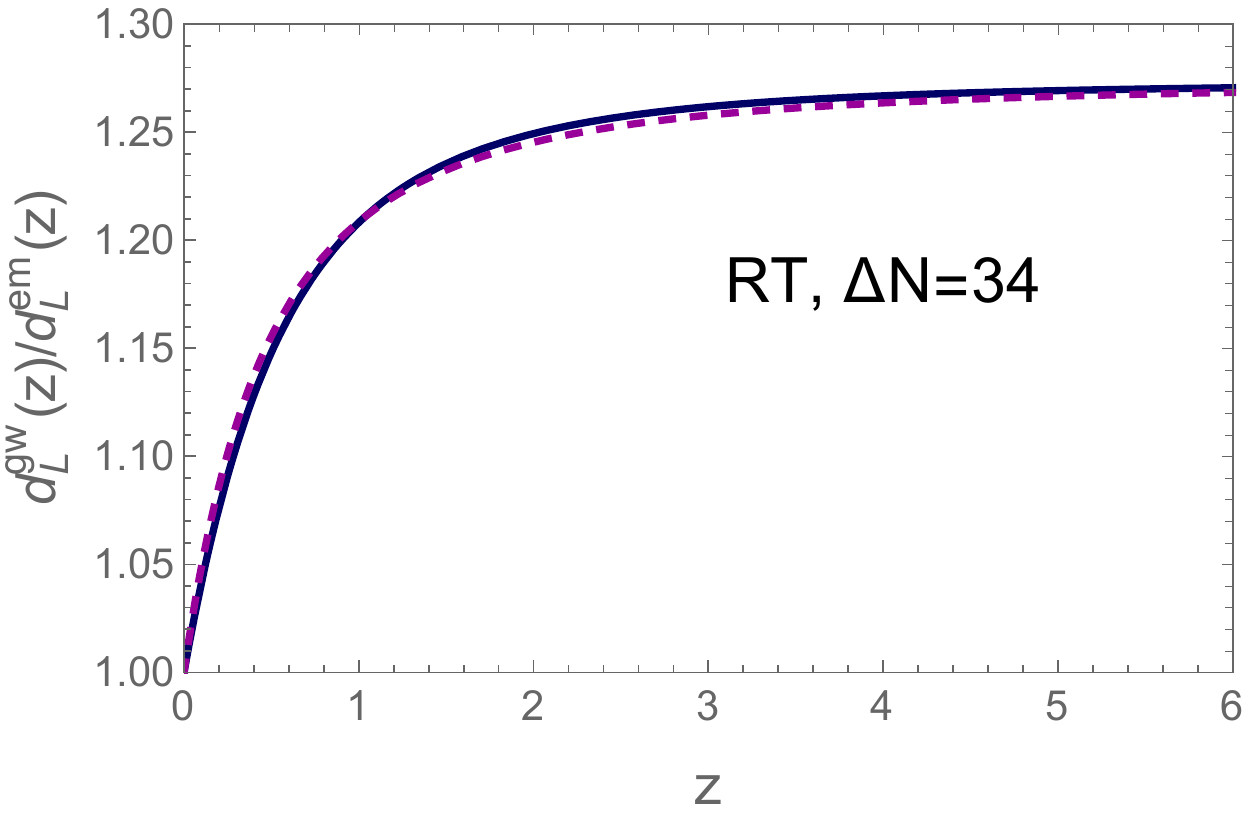}
\includegraphics[width=0.42\textwidth]{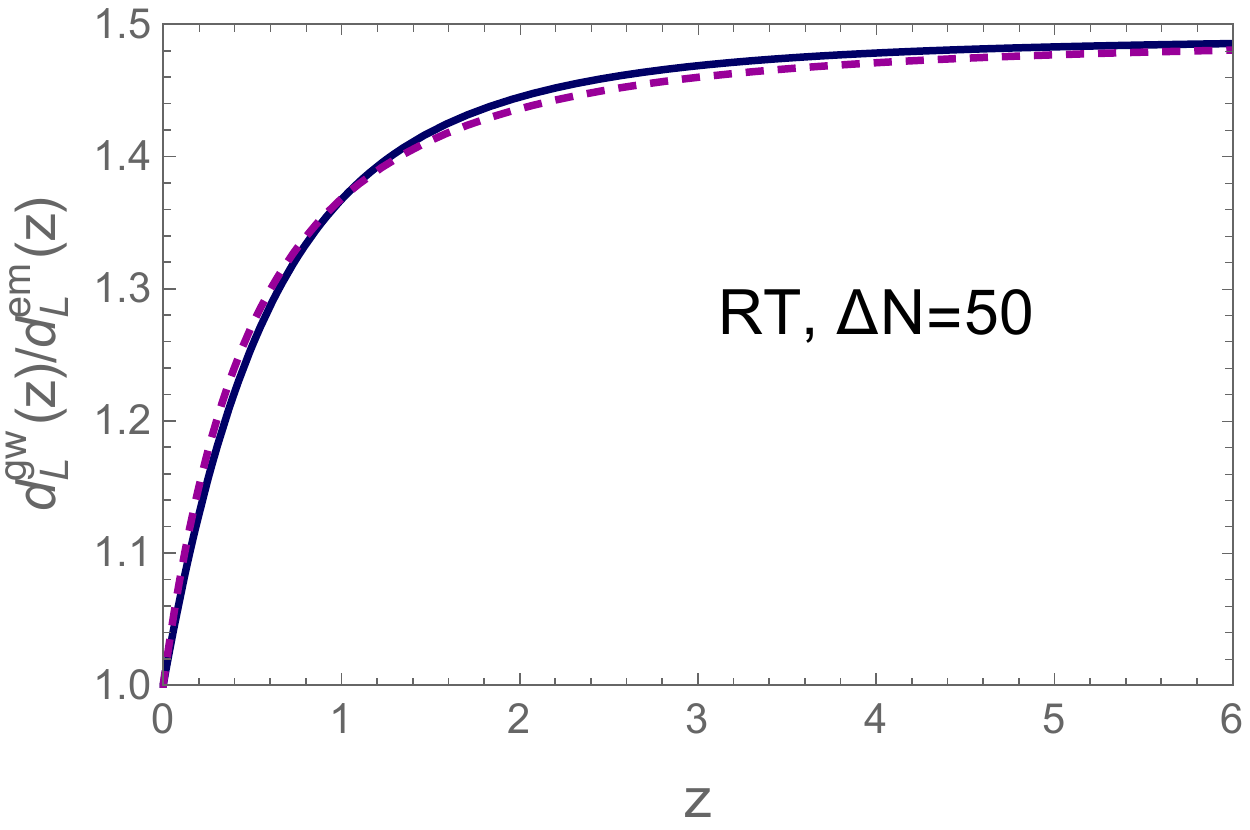}
\includegraphics[width=0.42\textwidth]{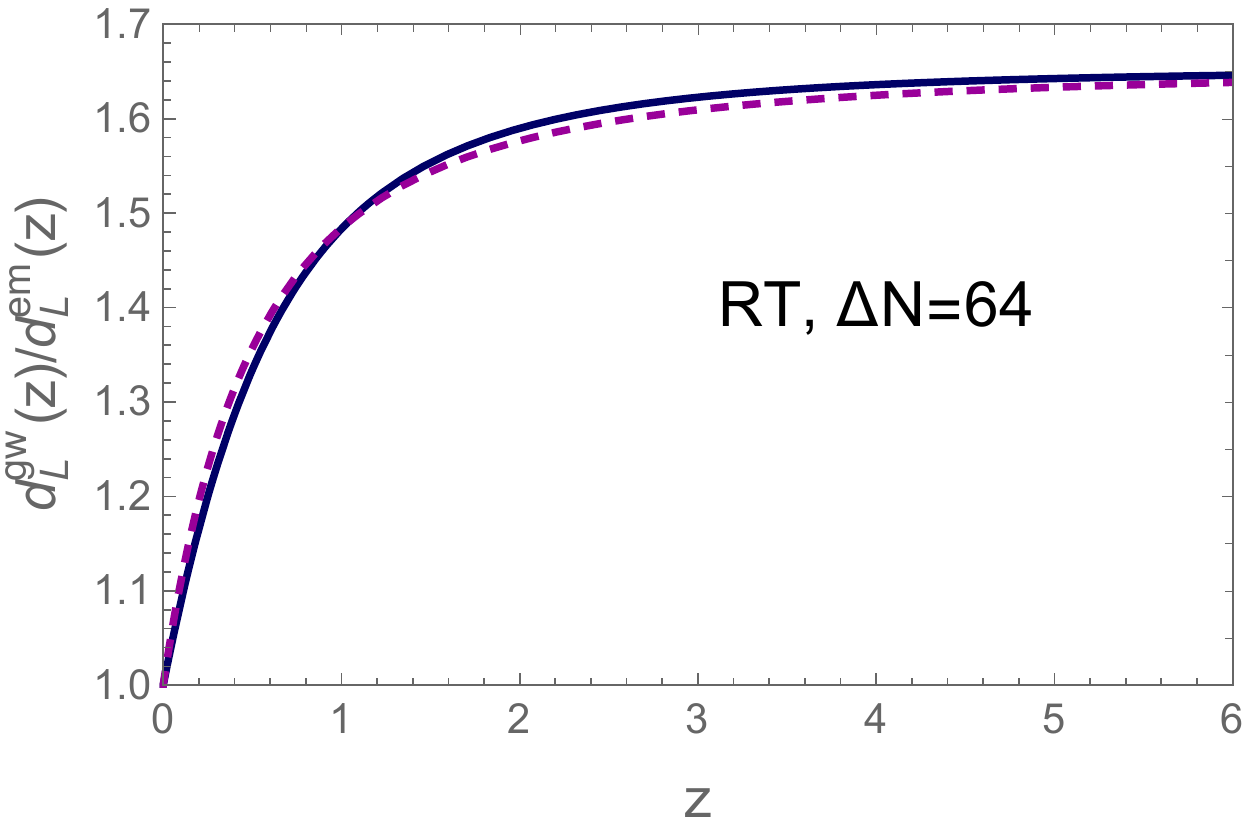}
\caption{The function $\dgw(z)/\dem(z)$ from the numerical integration (blue solid line), compared with the fit (\ref{eq:fitz}) (magenta, dashed).
Upper left panel: for  the minimal RT model; upper right:   RT with $\Delta N=34$; lower left: RT with  $\Delta N=50$; lower right: RT with  $\Delta N=64$.
}
\label{fig:fitdgw}
\end{figure}

As discussed in \cite{Belgacem:2018lbp}, the $z$ dependence of the ratio $\dgw/\dem$ is easily understood observing that, by definition, at $z\ra 0$ we must have $\dgw/\dem\ra 1$ because, if the distance to the source goes to zero, there can be no effect from modified GW propagation. At large $z$,  $\dgw/\dem$ goes to a constant because, in the RT model, as in most other modified gravity model, the emergence of dark energy is a relatively recent phenomenon, so the modifications to GR, and hence the function $\delta(z)$ in \eq{prophmodgrav}, go to zero at large redshifts. As a consequence, at large $z$ the integral in \eq{dLgwdLem} saturates to a constant value. As shown in Fig.~\ref{fig:fitdgw},
the numerical results for $\dgw(z)/\dem(z)$ are extremely well fitted by the simple parametrization~\cite{Belgacem:2018lbp}
\be\label{eq:fitz}
\frac{d_L^{\,\rm gw}(z)}{d_L^{\,\rm em}(z)}
=\Xi_0 +\frac{1-\Xi_0}{(1+z)^n}
\, ,
\ee
in terms of two parameters $\Xi_0$ and $n$.  This parametrization reproduces the fact  that, at $z=0$, $d_L^{\,\rm gw}(z)/d_L^{\,\rm em}(z)=1$,  while at large redshift $d_L^{\,\rm gw}(z)/d_L^{\,\rm em}(z)$ goes to a constant value $\Xi_0$. The index $n$ determines the rate at which this asymptotic value is reached. The best-fit values of $\Xi_0$ and $n$ are given in Table~\ref{tab:Xi0n}.

\begin{table*}[t]
\centering
\begin{tabular}{|c|c|c|c|c|c|c|}
\hline
		 &  RT, minimal & $\Delta N=34$ & $\Delta N=50$ & $\Delta N=64$ & $\Delta N=100$ \\ \hline		 
$\Xi_0$      & $0.93$         & $1.27 $       & $1.49$           & $1.65$   & $1.80$          \\ 
$n$            &  $2.59$         & $2.08$        &  $2.00$          & $1.95$   & $1.91$          \\   
$\delta(0)$ & $0.15$       &  $-0.46$      &  $-0.76$         & $-0.95$  & $-1.12$        \\
$\delta(0)/(1-\Xi_0)$ &2.29 & 1.67       & 1.54               & 1.46       & 1.39  \\
\hline
\end{tabular}
\caption{Values of $\Xi_0$,  $n$, $\delta(0)\equiv \delta (z=0)$ and $\delta(0)/(1-\Xi_0)$
for the RT model with various values of $\Delta N$.
The results have been obtained using for each model its own mean values for $\oma$ and $h_0$ from Table~\ref{tab:results}.
\label{tab:Xi0n} }
\end{table*}

Observe that the  simple parametrization (\ref{eq:fitz}) reproduces the numerical results extremely well. Indeed, 
comparing  with Fig.~\ref{fig:fitwde}, we see that it works
much better than the $(w_0,w_a)$ parametrization for the equation of state. This is due to the fact that 
\eq{eq:fitz} catches correctly both the $z\ra 0$ limit and the large $z$ limit.\footnote{Indeed, it was found in~\cite{Belgacem:2019pkk} that this parametrization fits very well the results of all other modified gravity models studied there, such as 
various Horndeski-type theories and  DHOST theories. The only exception is given by
bigravity, where it was found that, as a function of redshift,     $d_L^{\,\rm gw}(z)/d_L^{\,\rm em}(z)$
has a series of   oscillations due to the interaction between the two metrics.}
The corresponding  parametrization for the function $\delta(z)$ is obtained inverting \eq{dLgwdLem} to get 
\be
\delta(z)=-(1+z)\frac{d}{dz}\log\( \frac{d_L^{\,\rm gw}(z)}{d_L^{\,\rm em}(z)} \)
\, .
\ee
Using \eq{eq:fitz} for $d_L^{\,\rm gw}(z)/d_L^{\,\rm em}(z)$ gives

\begin{figure}[t]
\centering
\includegraphics[width=0.42\textwidth]{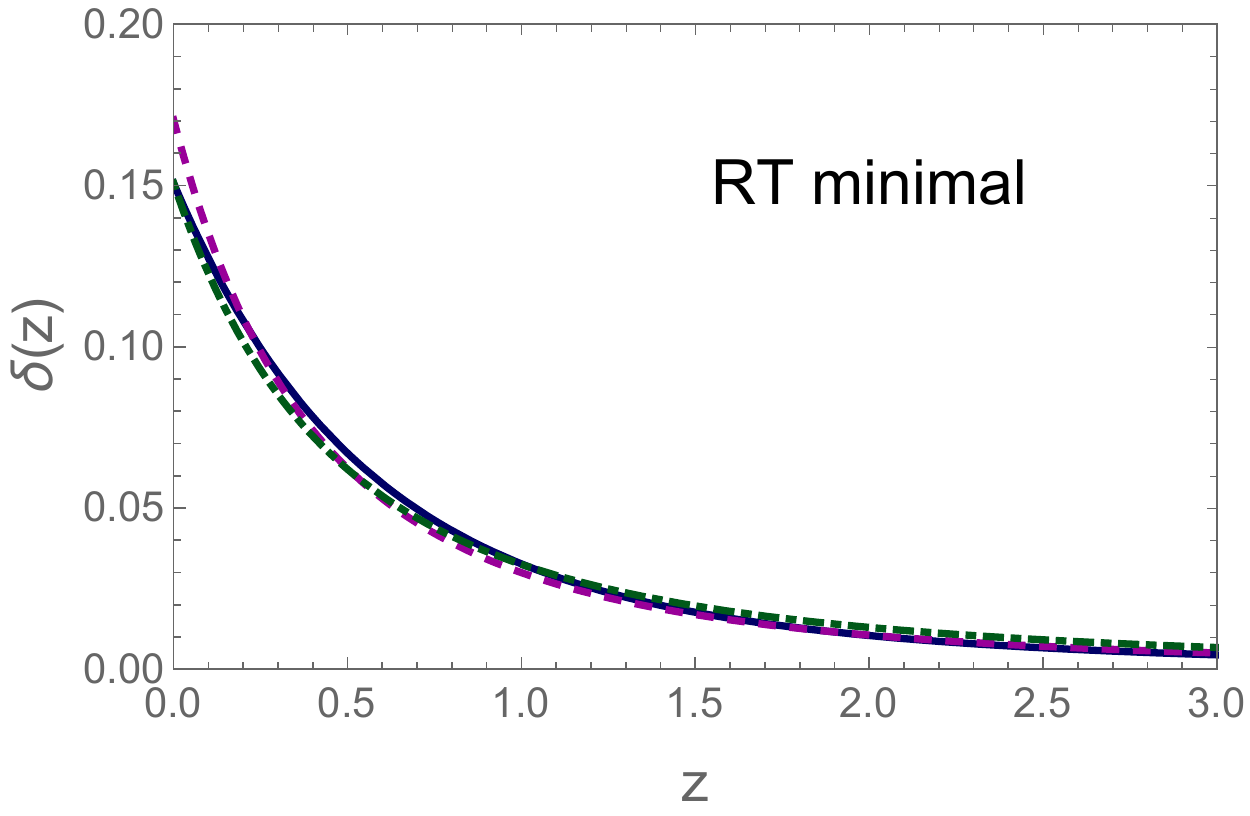}
\includegraphics[width=0.42\textwidth]{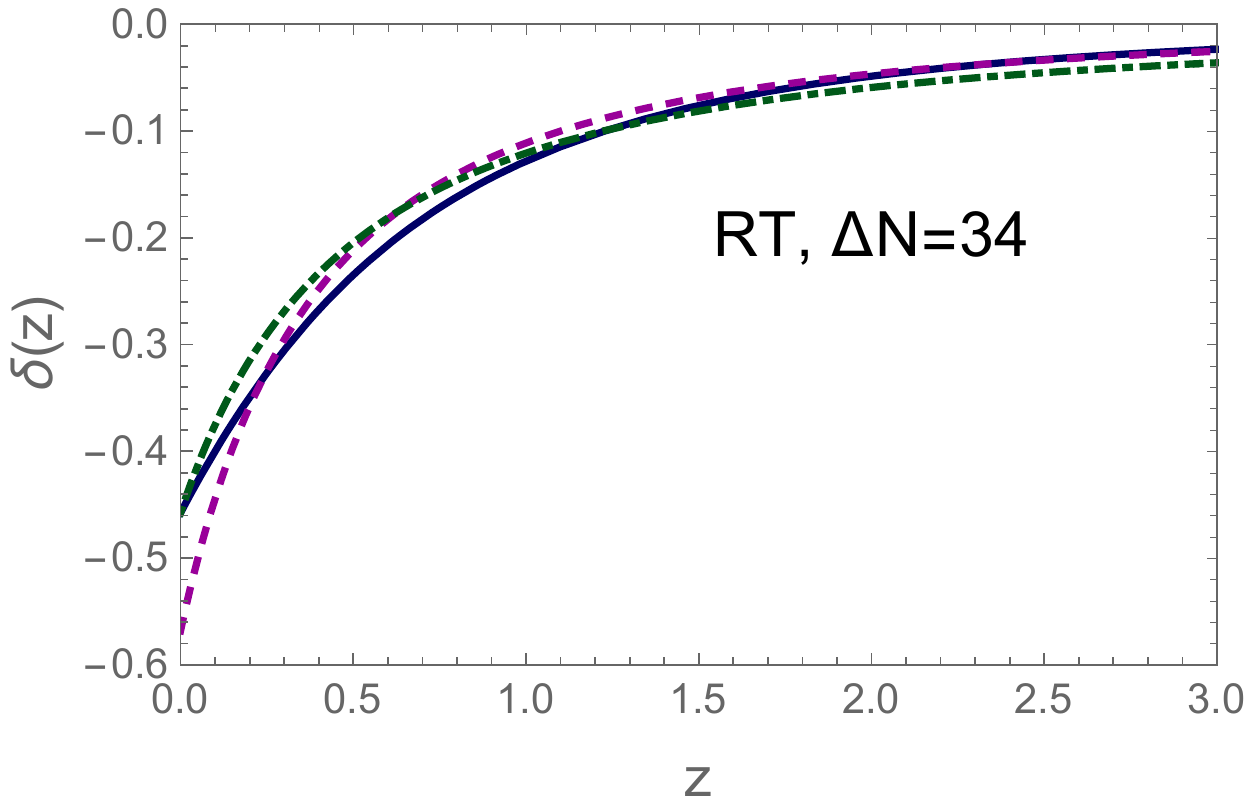}
\includegraphics[width=0.42\textwidth]{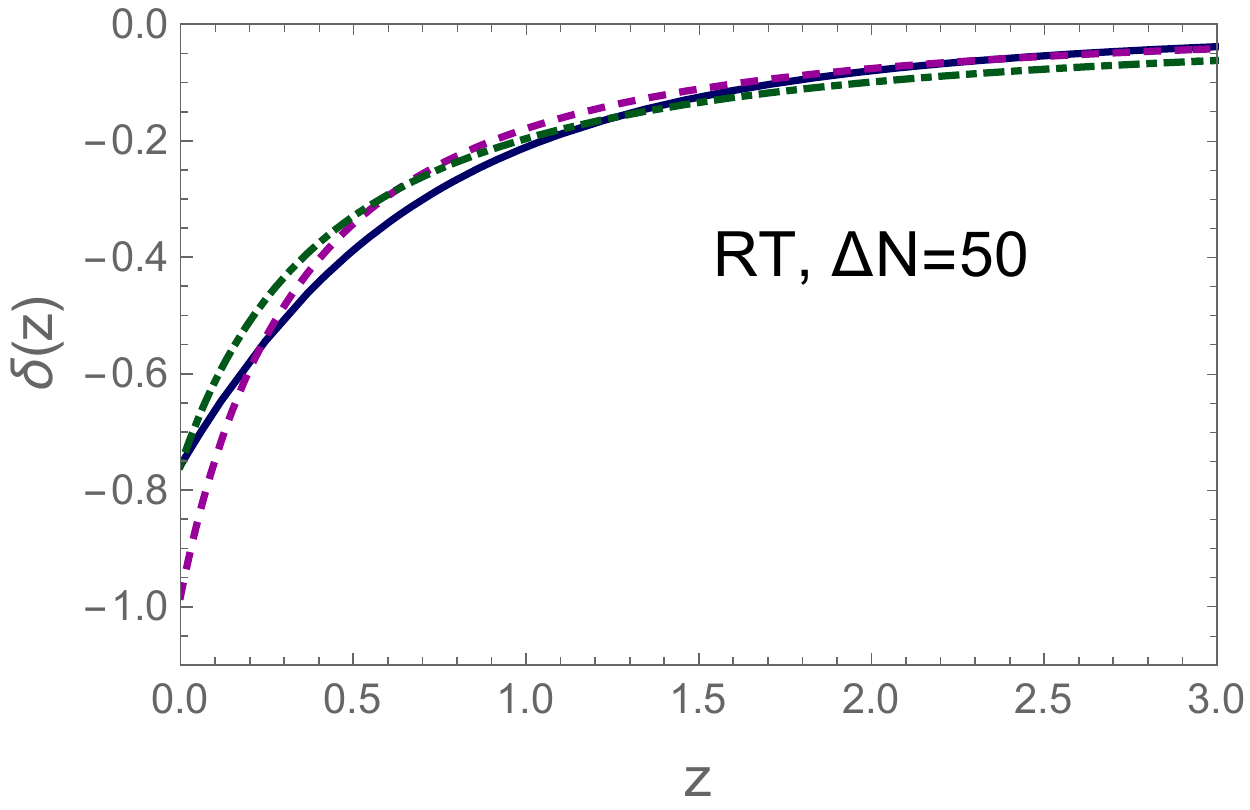}
\includegraphics[width=0.42\textwidth]{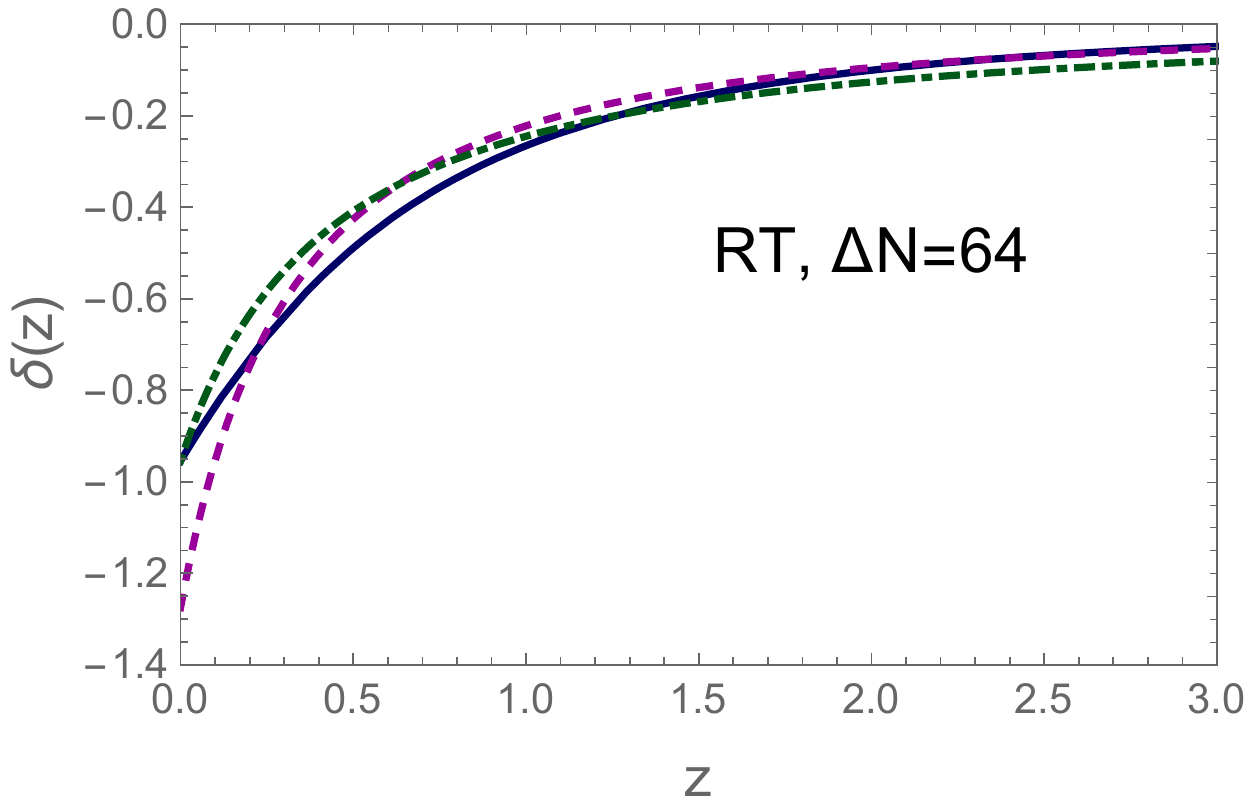}
\caption{The function $\dgw(z)/\dem(z)$ from the numerical integration (blue solid line), compared to the parametrization  (\ref{eq:fitz}) with the value of $n$ obtained from the best fit to $\dgw(z)/\dem(z)$ (magenta, dashed), and with $n=\delta(z=0)/(1-\Xi_0)$ (green, dot-dashed).
Upper left panel: for  the minimal RT model; upper right:   RT with $\Delta N=34$; lower left: RT with  $\Delta N=50$; lower right: RT with  $\Delta N=64$.
}
\label{fig:fitdelta}
\end{figure}

\be\label{paramdeltaz}
\delta(z)=\frac{n  (1-\Xi_0)}{1-\Xi_0+ \Xi_0 (1+z)^n}
\, .
\ee
Fig.~\ref{fig:fitdelta} compares the numerical result for $\delta(z)$ (blue solid line) with the fit (\ref{paramdeltaz}), using the same values of $\Xi_0$ and $n$ as in Table~\ref{tab:Xi0n} (magenta, dashed lines). We see that the $(\Xi_0,n)$ parametrization provides a fit to $\delta(z)$  less good than to $d_L^{\,\rm gw}(z)/d_L^{\,\rm em}(z)$, particularly near $z=0$. This is due to the fact that, for $d_L^{\,\rm gw}(z)/d_L^{\,\rm em}(z)$, the $(\Xi_0,n)$ parametrization catches correctly both the value in $z=0$  and the large $z$ limit; thus, as long as $d_L^{\,\rm gw}(z)/d_L^{\,\rm em}(z)$ is smooth in between, it is natural to find a value of $n$ such that the parametrization  (\ref{eq:fitz})  performs well. In contrast, the value of $\delta(z=0)$ is not automatically reproduced by the parametrization (\ref{paramdeltaz}), and indeed we see from the figures that in this region the parametrization is not accurate. For instance, the numerical integration gives the values of $\delta(0)\equiv \delta (z=0)$ shown in Table~\ref{tab:Xi0n}, while the parametrization (\ref{paramdeltaz}) would incorrectly predict
$\delta (0)\simeq \{0.17, -0.57, -0.98, -1.27,  -1.53\}$.
Note  that, with the parametrization (\ref{eq:fitz},\ref{paramdeltaz}), we have
\be\label{deltazeta0}
\delta(0)=n(1-\Xi_0)\, .
\ee
This suggests that, after having fixed $\Xi_0$ so to reproduce exactly the large-$z$ behavior of 
$\dgw(z)/\dem(z)$, rather then choosing $n$ from a best fit to $\dgw(z)/\dem(z)$, we could choose $n=\delta(0)/(1-\Xi_0)$, so that the parametrization (\ref{paramdeltaz}) reproduces exactly the  value of $\delta(0)$. The values obtained in this way are given in the last line of Table~\ref{tab:Xi0n}. If one uses these values of $n$, the  fit to $\dgw(z)/\dem(z)$ significantly degrades, but the fit to $\delta(z)$ becomes more accurate, and is shown as the green dot-dashed lines in Fig.~\ref{fig:fitdelta}. In general, since the directly observable quantity is $\dgw(z)/\dem(z)$, it is more important to have a simple and accurate  analytic representation for it, rather than for $\delta(z)$. Of course, for an accurate comparison with the data, one can also use directly the results of the numerical integration, which are obtained very quickly.

For comparison, the result for the RR model is also of the form (\ref{prophmodgrav}), except that the function $\delta$ is  given by  
\be
\delta=\frac{ 3\gamma\,  d\bar{V}/d\log a}{2(1 - 3\gamma \bar{V})}\, ,
\ee
and now $V=H_0^2S$, where $S$ is the auxiliary field of the RR model, defined by $U=-\iBox R$ and $S=-\iBox U$. The numerical integration then gives again a result very well fitted by \eq{eq:fitz}, with $\Xi_0\simeq
0.97$ and $n\simeq 2.5$~\cite{Belgacem:2018lbp}. However, contrary to the RT model, here the deviation from GR is only about $3\%$.

\subsubsection{Energy density of GWs and conservation of graviton number}

The fact that the GW amplitude in FRW does not scale as $1/a$ raises a question. As we will recall below, in GR the fact that in FRW $h\propto 1/a$ ensures that the GW energy density $\rho_{\rm GW}$ scales as $1/a^4$; in turn, this is consistent with an interpretation of a GW as a collection of massless graviton, whose comoving number density (i.e. number per unit volume in comoving coordinates) is conserved. Indeed,
the fact that the graviton number per comoving volume is conserved means that  the graviton number per physical volume scales as $1/a^3$, while the fact that the graviton is massless implies that its energy  scales as $1/a$, giving overall the $1/a^4$ behavior of $\rho_{\rm GW}$. One might then wonder whether the
scaling $h\propto 1/\tilde{a}$ is an indication that the  (comoving) graviton number is not conserved in the RT model. We will see here that, in fact, even in the RT model $\rho_{\rm GW}$ scales as $1/a^4$ and therefore the comoving number density of gravitons  is still conserved.
To this purpose, one must  realize that the expression of
$\rho_{\rm GW}$ in a generic modified gravity model is different from the GR expression. Let us first recall how things work in GR. 
We consider tensor perturbations over the FRW metric. Using conformal time, we  write
\be\label{4ds2confdeltaehij}
ds^2= a^2\[-d\eta^2
+\( \d_{ij}+\hTTij \) dx^idx^j\]\, .
\ee
It is convenient to expand the Fourier transform of 
$\hTTij$  in the basis of the polarization tensors,
\be\label{4thTTijpol}
\tilde{h}^{\rm TT}_{ij}(\eta,\vk)=\sum_{A=+,\times} e_{ij}^A(\hat{\vk}) \tilde{h}_A(\eta, \vk)\, ,
\ee
where the polarization tensors are normalized as $e^A_{ij}(\hat{\vk})e^{A'}_{ij}(\hat{\vk})=2\delta^{AA'}$. Expanding the Einstein-Hilbert action to second order in $\hTTij$
one  then finds (see e.g. sect.~21.3.4 of \cite{Maggiore:2018sht})
\bees
S_2[h]&=&\frac{1}{32\pi G}\,  \sum_A  \int d^3x\hspace{0.2mm}d\eta\,  a^2
\[ \pa_{\eta}h_A\pa_{\eta}h_A - \pa_{k}h_A\pa_{k}h_A\]\nn\\
&=&-\frac{1}{2}\,  \sum_A  \int d^4x\, \sqrt{-\bar{g}}\, \bar{g}^{\mu\nu}\pam \varphi_A\pan \varphi_A
\, ,\label{S2hvarphi}
\ees
where $\bar{g}_{\mu\nu}=a^2\emn$ is the background FRW metric in $(\eta,\vx)$ coordinates, and 
\be\label{4varphiAhA}
\varphi_A(\eta,\vx)=\frac{1}{\sqrt{16\pi G}}\, h_A(\eta,\vx)\, .
\ee
The action governing the two polarization amplitudes $h_A$ is therefore the same as the curved-space action of two canonically-normalized scalar fields $\varphi_A$. The variation of the action (\ref{S2hvarphi}) gives \eq{4eqtensorsect} (with the left-hand side equal to zero, unless we add also the matter action). At the same time, from this action we can  get the energy-momentum tensor of GWs, 
\bees
\tmn&\equiv& -\frac{2}{\sqrt{-\bar{g}}}\,\langle \frac{\d S_2[h]}{\d\gbMN}\rangle=
 \sum_A \langle \pam\varphi_A\pan\varphi_A-\gmn \frac{1}{2}\gRS\parho\varphi_A\pas\varphi_A\rangle\nn\\
 &=&\frac{1}{16\pi G}\sum_A \langle \pam h_A\pan h_A-\gmn \frac{1}{2}\gRS\parho h_A\pas h_A\rangle \, ,\label{TmnGW}
\ees
where $\langle \ldots \rangle$ denotes the spatial average over several wavelengths of the GWs, or the temporal average over several periods (see e.g. sect.~1.4 of \cite{Maggiore:1900zz}).
We denote by $t_{\eta\eta}$ the $\mu=\nu=0$ component of $\tmn$ in coordinates $(\eta,\vx)$ and by
$t_{tt}\equiv t_{00}$ the $\mu=\nu=0$ component of $\tmn$ in coordinates $(t,\vx)$. From 
$t_{\eta\eta}(d\eta)^2=t_{tt} (dt)^2$ and $dt=ad\eta$
if follows that $t_{00}=t_{\eta\eta}/a^2$, so \eq{TmnGW} gives
\be\label{t001a2}
t_{00}=\frac{1}{32\pi G}\, \frac{1}{a^2}  \sum_A  \langle (\pa_{\eta}h_A )^2+(\pa_ih_A )^2 \rangle\, .
\ee
On a plane wave the terms $\langle(\pa_{\eta}h_A )^2 \rangle $ and $ \langle (\pa_ih_A )^2 \rangle$ are equal. From \eq{eqforchi}, for wavelengths well inside the horizon, i.e. for  $k\eta\gg 1$, $\tilde{\chi}_A(\eta,\vk)\propto \sin (k\eta+\alpha)$, with $\alpha$ a phase.
Therefore $\tilde{h}_A(\eta,\vk)\propto \sin (k\eta+\alpha)/a(\eta)$ and, again for   $k\eta\gg 1$,
\be\label{paetah}
\pa_{\eta}h_A(\eta,\vk)\propto \frac{k\cos(k\eta+\alpha)}{a(\eta)} \[ 1+O\(\frac{1}{k\eta}\) \]\, .
\ee 
In $\langle(\pa_{\eta}h_A )^2 \rangle $  the term $\cos^2(k\eta+\alpha)$, averaged over several periods, simply gives a factor $1/2$, so $\langle(\pa_{\eta}h_A )^2 \rangle \propto 1/a^2$
and, from \eq{t001a2}, it 
then follows that $\rho_{\rm gw}= t_{00}$ is proportional to $1/a^{4}$, as indeed we expect for any form of radiation. 

Let us now see how the situation changes in the RT model. The propagation equation is now given by \eq{prophmodgrav}. Using \eq{deftildea} we see that it can be obtained from the GR equation with the replacement $a(\eta)\ra\tilde{a}(\eta)$. It 
can then be formally obtained from the variation of a quadratic action obtained replacing $a(\eta)\ra\tilde{a}(\eta)$ in
\eq{S2hvarphi}, i.e. from\footnote{More precisely, this is the action that reproduces the linearized equations of motions of the RT model, after having substituted the auxiliary fields with their own solutions of the equations of motion. It is therefore a `reduced' action for the $h_A$ variables only.}
\be\label{S2RTtensor}
S^{\rm RT}_2[h]=\frac{1}{32\pi G}\,  \sum_A  \int d^3x\hspace{0.2mm}d\eta\,  \tilde{a}^2
\[ \pa_{\eta}h_A\pa_{\eta}h_A - \pa_{k}h_A\pa_{k}h_A\]\, .
\ee
Introducing an effective Newton's constant from
\be
\frac{1}{\tilde{G}(\eta)}\equiv \frac{1}{ G}\, \frac{\tilde{a}^2(\eta)}{a^2(\eta)}
\ee
we can rewrite \eq{S2RTtensor} as
\be\label{S2RTtensorGeff}
S^{\rm RT}_2[h]=\,  \sum_A  \int d^3x\hspace{0.2mm}d\eta\,  \frac{1}{32\pi \tilde{G}(\eta)}\,  a^2
\[ \pa_{\eta}h_A\pa_{\eta}h_A - \pa_{k}h_A\pa_{k}h_A\]\, .
\ee
Thus, as far as tensor perturbations are concerned, at the quadratic level the RT model can be obtained from GR with the replacement $G\ra \tilde{G}(\eta)$. Note that $\tilde{G}(\eta)$ plays the role of an effective Newton's constant for tensor perturbations only. As we saw in section~\ref{sect:indicators}, scalar perturbations are governed by a different effective Newton's constant, that we denoted as $G_{\rm eff}(\eta,k)$, and which, contrary to  $\tilde{G}(\eta)$, depends also on the wavenumber $k$.

Repeating the above derivation of the energy-momentum tensor of GWs, \eq{TmnGW} becomes
\be\label{TmnGWRT}
\tmn=\frac{1}{16\pi  \tilde{G}(\eta)}\sum_A \langle \pam h_A\pan h_A-\gmn \frac{1}{2}\gRS\parho h_A\pas h_A\rangle \, ,
\ee
simply because the variation $\d S_2[h]/\d\gbMN$ is insensitive to the time dependence of $\tilde{G}(\eta)$. The energy density $\rho_{\rm gw}=t_{00}$ is then given by
\be
\rho_{\rm gw}=\frac{1}{16\pi \tilde{G}(\eta)}\, \frac{1}{a^2}  \sum_A  \langle (\pa_{\eta}h_A )^2\rangle\, .
\ee
Notice that the $1/a^2$ factor comes from the transformation from $t_{\eta\eta}$ to $t_{tt}$, i.e. from the relation $dt=ad\eta$. This is determined by the FRW background metric, so it still involves $a$ rather than $\tilde{a}$.
In contrast, $h_A\propto \sin(k\eta+\alpha)/\tilde{a}$ and therefore now, for $k\eta\gg 1$,  $\pa_{\eta}h_A\propto k\cos(k\eta+\alpha)/\tilde{a}$, which replaces \eq{paetah}.
Again, the  term $\cos^2(k\eta+\alpha)$ averages to $1/2$, so in the end the time dependence of 
$\rho_{\rm gw}$ is
\be
\rho_{\rm gw}\propto \frac{1}{16\pi \tilde{G}(\eta)}\, \frac{1}{a^2\tilde{a}^2}
=\frac{1}{16\pi G} \, \frac{1}{a^4}\, .
\ee
Therefore, once taken into account the fact that the modification of the Einstein equations implies also a modification of the formula for the GW energy-momentum tensor,  we find that, in FRW, the GW energy density of the RT model still scales as $1/a^4$, despite modified GW propagation. Therefore, the energy density  still corresponds to that of an ensemble of massless gravitons, whose number density in comoving coordinate is constant (so that the number density in physical coordinates scales as $1/a^3$) and whose energy scales as $1/a$. From the derivation, it is also clear that this result is not specific to the RT model, but holds for any modified gravity model where the equation of tensor perturbations can be written in the form (\ref{prophmodgrav}).
Notice also that the redshift of the graviton frequency  $\omega\propto 1/a$, or of the wavelength as $\lambda\propto a$, are kinematical properties that depend only on the background metric, and are the same in GR and in the RT model. As discussed in \cite{Belgacem:2018lbp}, in the RR model again $\rho_{\rm GW}\propto 1/a^4$, and in this case the effective Newton constant $\tilde{G}$ for the tensor perturbations is the same as 
the effective Newton's constant $G_{\rm eff}$ in the scalar sector.

This result also gives useful guidance for attempts at deriving the RT model from  a fundamental local theory. In particular, it rules out the possibility that the RT model could be derived from a theory with extra dimensions in which  gravitons are lost to a higher-dimensional bulk,  see the discussion in sect.~\ref{sect:extradim},
and rather points toward the dynamical mass generation mechanisms discussed in sect.~\ref{sect:dynmass}.


\subsection{Comparison with the sensitivity of current and future GW detectors}

We next compare the predictions for modified GW propagation of the RT model with the sensitivities of current and future GW detectors, elaborating on the analysis in~\cite{Belgacem:2018lbp,Belgacem:2019tbw,Belgacem:2019lwx,Belgacem:2019zzu} for ground-based detectors and in \cite{Belgacem:2019pkk} for LISA.

\subsubsection{The Advanced LIGO/Virgo/Kagra network} 

We first consider the 
network of second-generation (2G) GW detectors formed by Advanced LIGO  Hanford and Livingston,   Advanced Virgo, KAGRA and LIGO India (HLVKI), assumed to be all at target sensitivity.
In \cite{Belgacem:2019tbw}  mock catalogs of binary neutron stars (BNS) detections have been produced for this network, using state-of-the art models for the cosmic star formation rate,  for the  extra-galactic population of neutron star binaries  and for the delay  between binary formation and merger~\cite{2012PhRvD..86l2001R,2014PhRvD..89h4046R,Regimbau:2014nxa, 2015PhRvD..92f3002M,2016PhRvD..93b4018M,2017PhRvL.118o1105R,2018PhRvL.120i1101A,2015MNRAS.447.2575V,Madau:2014bja,Vitale:2018yhm}, and fixing the overall normalization  using  the local coalescence rate estimated from the O1 LIGO observation run and the O2 LIGO/Virgo observation run~\cite{LIGOScientific:2018mvr}.  Assuming  a duty cycle of 80\% and a network SNR threshold level $\rho_{\rm  threshold}=12$, it was found that the HLVKI network will detect between $O(60)$ and $O(80)$ BNS/yr, depending on the assumptions on star formation rate and distribution of neutron star masses. Of these, only about 1-2 events per year are expected to have a detected gamma ray burst (GRB) counterpart, assuming that
Fermi-GBM can make a coincident detection and that \emph{Swift} can slew to the combined GW/GRB error box and identify an X-ray counterpart. More electromagnetic counterparts could in principle be detected with just optical/IR/UV telescopes, without a GRB trigger, although their number is more difficult to estimate.

\begin{table}[t]
\centering
\begin{tabular}{|c|c|c|c|c|}
\hline
$z$ & $d_L^{\rm gw}$ (Mpc) & $\Delta d_L^{\rm gw}$ (Mpc) & $\Delta d_L^{\rm gw}/d_L^{\rm gw}$ & $\Delta\delta(0)$ \\ \hline
0.029271 & 134.815 & 4.000  &0.030 & 1.36\\
0.035195 & 157.475 & 5.636   &0.036&1.30\\
0.060585 & 283.567 & 18.706 &0.066&1.25\\
0.066283 & 316.373 & 14.509 &0.046&0.84\\
0.071053 & 327.381 & 20.085 &0.061&1.00\\
0.071730 & 342.952 & 16.957 &0.049&0.83\\
0.076180 & 341.595 & 22.360 &0.065&0.99\\
0.081819 & 418.469 & 30.238 &0.072&1.00\\
0.088698 & 396.734 & 25.757 &0.065&0.84\\
0.091869 & 402.590 & 34.170 &0.085&1.03\\
0.094237 & 406.423 & 31.472 &0.077&0.93\\
0.095288 & 432.996 & 36.423 &0.084&0.99\\
0.099956 & 491.071 & 31.721 &0.065&0.75\\
0.102531 & 461.627 & 36.858 &0.080&0.88\\
0.114869 & 626.939 & 43.010 &0.068&0.68\\
\hline
\end{tabular}
\caption{The events in a given realization of the mock catalog of joint GW-GRB detections  for the HLVKI network, over 10~yr of simulated data. The
`measured' luminosity distance is obtained from the redshift assuming $\Lambda$CDM as fiducial model, 
and scattering randomly the fiducial values of $d^{\rm gw}_L(z)$ according to a Gaussian distribution with a width equal to the error $\Delta \dgw(z)$  (from ref.~\cite{Belgacem:2019tbw}). 
In the last column we give the corresponding error on the measurement of $\delta(0)$ from each single source, assuming a $1\%$ error on the electromagnetic luminosity distance.
\label{tab:cat2G}}
\end{table}

A sample catalog of simulated GW-GRB coincidences is given in Table~\ref{tab:cat2G} (from Table~23 of  \cite{Belgacem:2019tbw}) which shows  15 joint GW-GRB coincidences detected in 10 years of simulated data.\footnote{Such a long time span is somewhat optimistic, but, given the rate of 1-2 joint GW-GRB events per year, is necessary to build a statistically significant sample.}
The first three columns of the table show the redshift of the source, which has been extracted randomly from the appropriate distribution, its luminosity distance  (which, being measured from the GW signal, is in principle a GW luminosity distance, $\dgw$, if we do not assume GR), and 
the  expected observational error on the luminosity distance $\Delta \dgw$
(which depends on the network sensitivity and on  the source orbital inclination and position in the sky with respect to the network, also extracted randomly). The `measured' value of $\dgw(z)$ is obtained from the redshift assuming $\Lambda$CDM as fiducial model, 
and scattering randomly this fiducial value according to a Gaussian distribution with a width equal to the error $\Delta \dgw(z)$. From the fourth column we see that, for the sources at the lowest redshifts, $\dgw$ can be measured to $(3-4)\%$ accuracy (depending in particular on the source inclination and position in sky with respect to the network), while, for the largest 
redshifts in the catalog, around $z\simeq 0.1$, the accuracy on $\dgw$ is about  $(7-8)\%$.\footnote{For comparison, GW170817 was at $z\simeq 0.01$ and its luminosity distance, as measured from the GW signal, was 
$\dgw=40^{+8}_{-14}$~Mpc~\cite{TheLIGOScientific:2017qsa}. The corresponding value of $\Delta \dgw/\dgw$, of order $27\%$, is much larger than those in
Table~\ref{tab:cat2G},  because it reflectes the detectors  
sensitivities during the O2 run, while in Table~\ref{tab:cat2G} the five detectors are taken at target sensitivity. Furthermore, the event was near a blind spot of Virgo, so Virgo could contribute to the source localization but not to the estimate of the other source parameters, so only the two LIGO detectors contributed to the estimate of $d_L$.}

For sources at small redshift, as appropriate for the values of $z$ in Table~\ref{tab:cat2G},
\eq{dLgwdLem} becomes
\be\label{eq:fitlowz}
\frac{d_L^{\,\rm gw}(z)}{d_L^{\,\rm em}(z)}=1-z \delta(0)+{\cal O}(z^2)\, ,
\ee
so in this limit we are actually sensitive to $\delta(0)\equiv \delta(z=0)$. The comparison between the predictions of a model and the data can therefore be performed without making use of any parametrization for $\dgw(z)/\dem(z)$, and
simply comparing directly the predictions of the model for $\delta(0)$ with the expected error on $\dgw(z)/\dem(z)$. For the RT model, with different values of $\Delta N$, the predictions for $\delta(0)$ were given in Table~\ref{tab:Xi0n}. 

Observe that, in \eq{eq:fitlowz},  the deviation of $d_L^{\,\rm gw}(z)/d_L^{\,\rm em}(z)$ from 1 is  proportional to $z$.
Limits on $\delta(0)$ from  GW170817 were obtained in \cite{Belgacem:2018lbp}. In this case, given the small redshift $z=0.01$, it is clear that one cannot obtain stringent limits, and the best result found in
\cite{Belgacem:2018lbp} was $\delta(0)=-7.8^{+9.7}_{-18.4}$.\footnote{The recently announced  detection, GW190425~\cite{Abbott:2020uma}, has a redshift  $z=0.03^{+0.01}_{-0.02}$ (assuming $\Lambda$CDM). The event is classified as a NS-NS, although the possibility that one or both binary
components of the system are BHs cannot be ruled out from the GW data.
No counterpart has been observed to date. The event has very poor angular localization because it was confidently detected only in a single detector.}
Let us now  estimate  the observational error on $\delta(0)$ that could be obtained  from individual
detections with characteristics such as those in Table~\ref{tab:cat2G}.  From \eq{eq:fitlowz},
\be\label{Deltadelta}
\Delta \delta(0)\simeq \frac{1}{z}\, \[ \frac{\Delta\dgw}{\dgw}+\frac{\Delta\dem}{\dem}\]\, .
\ee
The relative error on $\dgw$ is given in Table~\ref{tab:cat2G}. For 
the relative error on $\dem$ we observe that,
given a measurement of the redshift from an electromagnetic counterpart, $\dem$ is in principle determined by the fiducial cosmology, and in particular, at these redshifts,  by the value of $H_0$. From Table~\ref{tab:results}, the error $\Delta H_0/H_0$ is below the $1\%$ level (and one can imagine that this accuracy will further improve in the next few years).\footnote{Of course, a crucial issue here is the discrepancy  between the value of $H_0$ obtained from CMB+BAO+SNe in $\Lambda$CDM (or in the RT model, which is very close) and the value from local measurements~\cite{Riess:2019cxk,Wong:2019kwg}. Here we perform our estimates assuming the value of $H_0$ and $\Delta H_0$ from  CMB+BAO+SNe.} Note also that the redshifts in Table~\ref{tab:cat2G} are  sufficiently large that the peculiar velocity of the host galaxy, typically of  order $v\sim 200\, {\rm km}/{\rm s}$, gives a small error on the determination of the cosmological redshift, that can be neglected. So,  we assume for definiteness a relative error $\Delta \dem/\dem =1\%$ for all the 
events shown in Table~\ref{tab:cat2G} (since this is in any way subleading with respect to the error on
$\Delta\dgw/\dgw$, the precise value  assumed is not very important).
In this way we obtain the estimates for $\Delta \delta(0)$ given in the last column of Table~\ref{tab:cat2G}. We see that the accuracy obtained from the various individual  detections are quite comparable in this range of redshift, with on average slightly more accurate measurements at higher redshift, since  the average increase of the observational error  $\Delta\dgw/\dgw$  with redshift is more than compensated by the factor $1/z$ in \eq{Deltadelta}.

The error $\sigma$ obtained combining the errors $\sigma_i$ of the individual measurements is given as usual by  $1/\sigma^2=\sum_i 1/\sigma_i^2$. Comparing with  Table~\ref{tab:Xi0n}, we see that the prediction $\delta(0)\simeq -1.11$ of the RT model in the large $\Delta N$ limit could be detected at the $3\sigma$ level with about 9 BNS with counterpart, which could be collected in 6 years of data taking. Verifying the predictions for smaller values of $\Delta N$ would require more data, but in any case beyond this point one would enter in a regime where a detection of $\delta(0)$ is in principle possible already at 2G detectors. Of course, the estimate of the number of detected electromagnetic counterparts is subject to uncertainties in the modelisation of the emissions mechanism, that will hopefully be further clarified by the ongoing and future LIGO/Virgo/KAGRA observational runs. 
 Another interesting possibility is given by the detection of NS-BH binaries. These can be seen to larger  distances, because of the higher BH mass, and would therefore be very useful for testing modified GW propagation. Theoretically, it is not known whether NS-BH coalescences have a significant electromagnetic emission. Currently, a few NS-BH candidates have been reported  in the O3 LIGO/Virgo run (see \url{https://gracedb.ligo.org}), but apparently no counterpart has been observed. The other option that should be explored is the possibility of using BNS without counterparts to learn about modified GW propagation, using statistical methods, such as those based on a probabilistically assignment of the host galaxy~\cite{Schutz:1986gp}  (see \cite{DelPozzo:2011yh} for recent Bayesian approach tuned to 2G detectors), or on the fact that the NS mass function is relatively narrow~\cite{Taylor:2011fs,Taylor:2012db}. 

A  more accurate way of estimating  the sensitivity to modified GW propagation, that fully    
accounts for the partial degeneracies of $\delta(0)$ [or  of $(\Xi_0,n)$],  with the other cosmological parameters, and in particular with $H_0$, $\oma$ and the dark energy equation of state [at least  as described by the $w_0$ or $(w_0,w_a)$ parameters], is to run
a MCMC where
the above catalog of mock GW detections is used in conjunction with CMB, BAO and SNa data. This has been performed in~\cite{Belgacem:2019tbw}, where it was found that, fitting the simulated data with the parametrization (\ref{eq:fitz}) (where we 
set for definiteness $n\simeq 2.5$),  with the above 15 mock detections the HLVKI network would determine $\Xi_0$ to an accuracy $\Delta\Xi_0\simeq 0.125$. From \eq{deltazeta0}, this implies $\Delta\delta(0)\simeq 0.31$, to be compared with the value $\Delta\delta(0)\simeq 0.24$ found by combining the error of all 15 mock measurements of
$\delta(0)$ in Table~\ref{tab:cat2G} according to  $1/\sigma^2=\sum_i 1/\sigma_i^2$. Notice that, while the MCMC  takes into account more accurately the degeneracies with the other cosmological parameters, which is what eventually leads to a slightly larger estimates of $\Delta\delta(0)$, its chains converge only  with a sufficiently large set of mock GW events
(which is the reason why in \cite{Belgacem:2019tbw} was used a catalog corresponding to 10 yr of data taking). In contrast, the simpler estimate of $\Delta\delta(0)$
presented in Table~\ref{tab:cat2G} gives an idea of the contribution of individual detections, as a function of redshift.

\begin{figure}[t]
\centering
\includegraphics[width=0.42\textwidth]{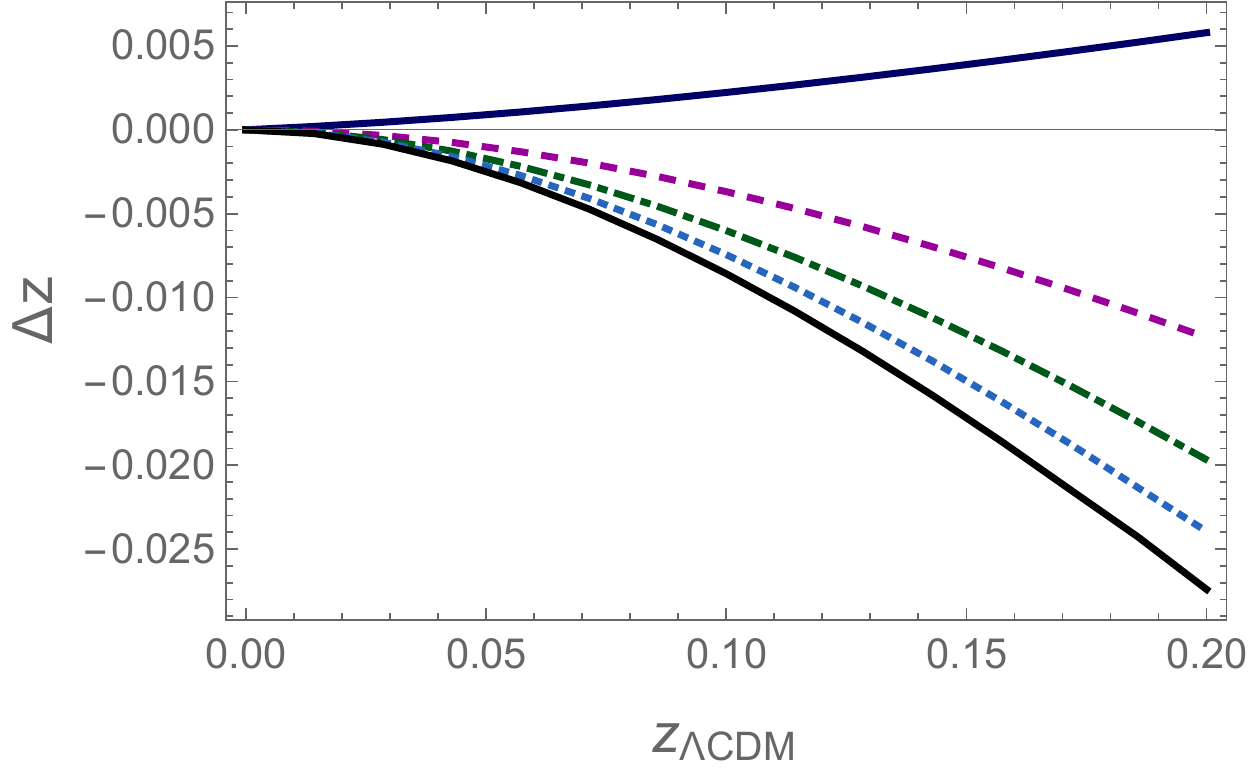}
\includegraphics[width=0.42\textwidth]{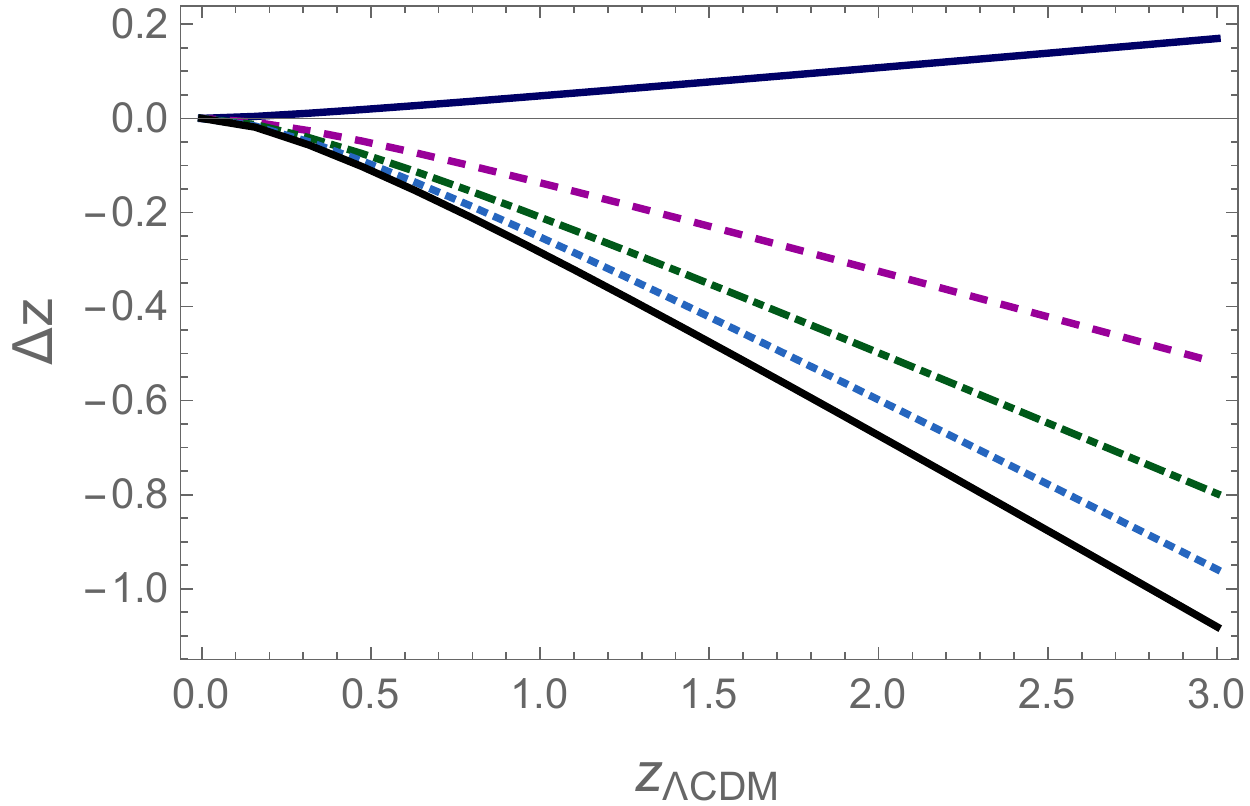}
\caption{The change in the actual redshift of the source  $\Delta z\equiv z_{\rm RT}-z_{\Lambda{\rm CDM}}$, compared to the $\Lambda$CDM prediction $z_{\Lambda{\rm CDM}}$,
 as a function of $z\equiv z_{\Lambda{\rm CDM}}$, for the minimal RT model (blue solid line) and  for RT with $\Delta N=34$ (magenta, dashed), $\Delta N=50$ (green, dot-dashed), $\Delta N=64$ (cyan, dotted) and  $\Delta N=100$  (black solid line).  Left panel: in  the range $z<0.2$, relevant for BNS at 2G detectors. Right panel: up to $z=3$.\label{fig:Deltaz_vs_z_max}}
\end{figure}

As already pointed out in~\cite{Belgacem:2019lwx}, the above results also have potentially important implications for the search of the electromagnetic counterpart to a GW detection, since they imply that the actual electromagnetic luminosity distance of the source, and hence its redshift, will be different from that 
inferred from the GW detection assuming GR. Indeed, if the correct theory is $\Lambda$CDM, given a best-fit value value $D$ of the luminosity distance to the source measured with GWs, the corresponding best-fit value of the redshift 
$z_{\Lambda{\rm CDM}}$ is predicted to be given by 
\be
\dem (z_{\Lambda{\rm CDM}})=D\, .
\ee
In contrast, if the correct description of Nature is given by the RT model, the best-fit value for redshift of the source, $z_{\rm RT}$, is given by
\be\label{zRTD}
\dgw (z_{\rm RT})=D\, .
\ee
Combining these relations we can determine the function $z_{\rm RT}(z_{\Lambda{\rm CDM}})$. In the left panel of 
Fig.~\ref{fig:Deltaz_vs_z_max} we show $\Delta z\equiv z_{\rm RT}-z_{\Lambda{\rm CDM}}$ as a function of $z\equiv z_{\Lambda{\rm CDM}}$,
for the minimal RT model and  for RT with $\Delta N=34,50,64,100$. The right panel shows the result  up to 
$z=3$, which is relevant for BNS at the Einstein Telescope (see section~\ref{sect:ET}) and, up to $z\simeq 1$,  for NS-BH binaries at the HLVKI network. The left panel provides an enlargement of the region up to  $z=0.2$, which is the range relevant for BNS at 2G detectors. At $z<0.2$, for individual detections the difference $\Delta z$ in the theoretical prediction of the redshift will be of the order of the error box induced by the observational error on $\dgw$, as it is clear from the fact that the observational errors on $\delta(0)$ in Table~\ref{tab:cat2G} are of the order of the prediction of the RT model with large $\Delta N$. For larger values of $z$, the difference can, however, become very significant.
This plot is another way to present the prediction of the RT model, complementary to Fig.~\ref{fig:deltadgw}. It's relevance is particularly clear for the search of the counterpart with telescopes. For instance, from the right panel of Fig.~\ref{fig:Deltaz_vs_z_max} we see that,
for a GW event for which $\Lambda$CDM would predict a redshift $z=3$, the RT model with very large $\Delta N$  predicts that  telescopes should rather search for the counterpart by targeting galaxies at $z\simeq 2$. 

\begin{figure}[t]
\centering
\includegraphics[width=0.45\textwidth]{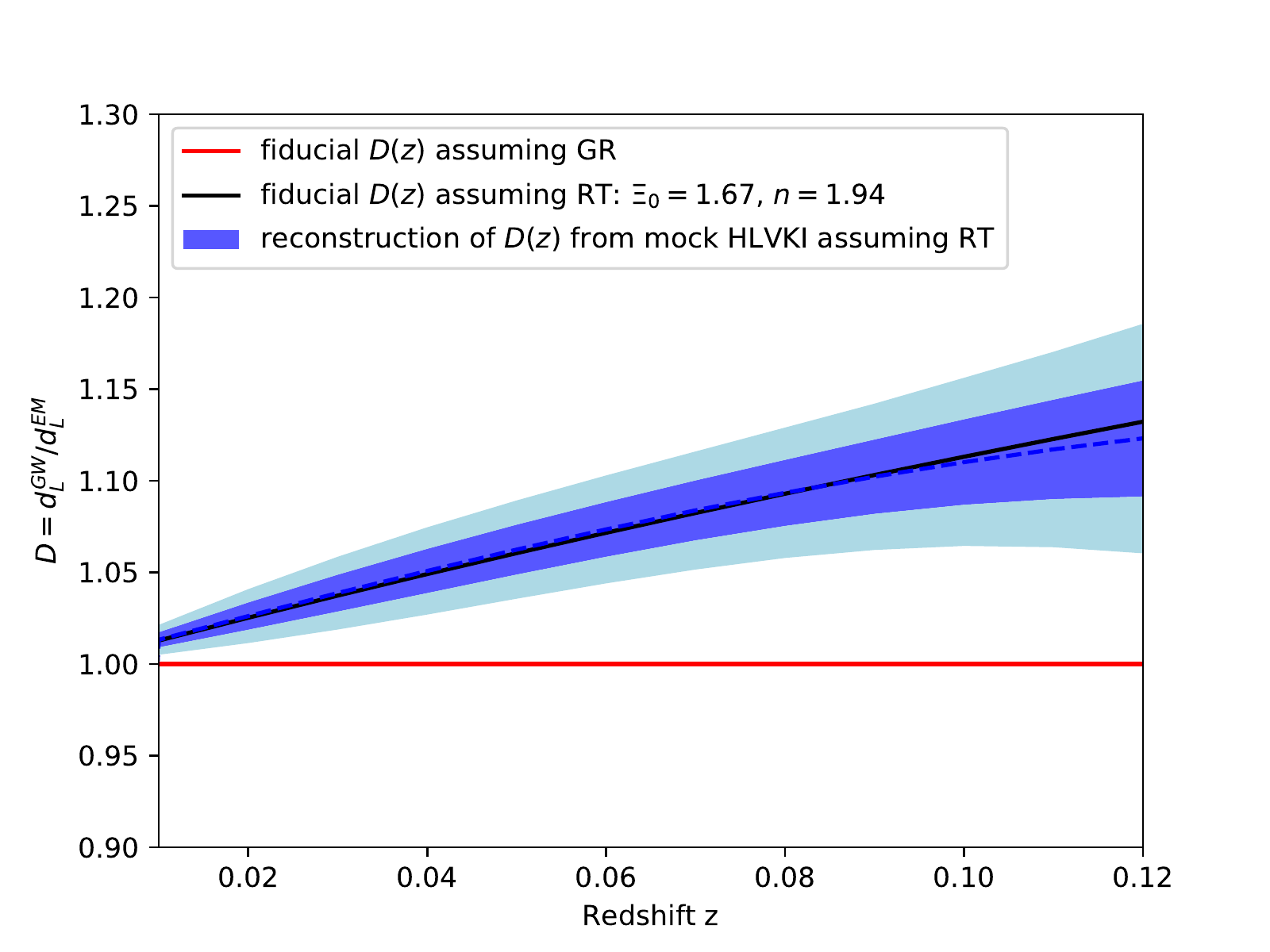}
\caption{Gaussian process reconstruction of $\dgw/\dem$, using the RT model with $\Delta N=64$ as  fiducial cosmology, using GW events at the HLVKI network with GR counterpart for $\dgw$, and DES supernovae for $\dem$. The blue and light blue regions correspond to $68\%$ and $95\%$ confidence levels, respectively. From \cite{Belgacem:2019zzu}.}
\label{fig:results_MGHLVKI}
\end{figure}

Finally, it is also interesting to see how well one can reconstruct the ratio $\dgw(z)/\dem(z)$ from the data, without assuming any parametrization for it, such as \eq{eq:fitz}. This can be done using the technique of gaussian processes, that allows the reconstruction of a function directly from the data. Several applications of gaussian processes in cosmology have been discussed   in~\cite{Seikel:2012uu,Seikel:2012cs,Yahya:2013xma,Busti:2014dua,Busti:2015aqa,Cai:2015zoa,Cai:2015pia,Cai:2016vmn,Cai:2016sby,Cai:2017yww}. In  \cite{Belgacem:2019zzu} this technique has been applied to the reconstruction of $\dgw(z)/\dem(z)$. For $\dgw$ has been used the same catalog of mock joint
GW-GRB detections  shown in Table~\ref{tab:cat2G}, where the GW events are detected
at the HLVKI network and the GRB counterparts are observed by Fermi-GBM and Swift. For $\dem(z)$ 
were considered simulated measurements  from DES supernovae, with the data generated as in  \cite{Cai:2015zoa}, with a redshift range $0.05<z <1.2$, 
and the errors on  $\dem$ estimated as in \cite{Bernstein:2011zf}. Fig.~\ref{fig:results_MGHLVKI} shows the result of the reconstruction; we see that, with these datasets, the prediction of the RT model with $\Delta N=64$ (used as fiducial in the figure) is very clearly distinguished from 
the prediction $\dgw(z)/\dem(z)=1$ of $\Lambda$CDM.

\subsubsection{3G detectors: Einstein Telescope and Cosmic Explorer}\label{sect:ET}

We next consider  third-generation (3G) ground-based interferometers currently under study, such as the Einstein Telescope (ET) in Europe~\cite{Punturo:2010zz} and  Cosmic Explorer (CE) in the US~\cite{Dwyer:2014fpa,Reitze:2019iox}, that could start to be operative in the mid 2030s. These detectors  will have the potential of exploring the Universe with GWs to truly cosmological distances, guaranteeing an extraordinary output in astrophysics, cosmology and fundamental physics~\cite{Sathyaprakash:2012jk,Maggiore:2019uih}.

For instance ET, even as a single detector, with be able to detect the coalescence of  compact binaries with total mass  $(20-100)~\msun$, as typical of BH-BH or BH-NS binaries,  up to redshift $z\sim 20$ and higher. By comparison, in the catalog of  detections from the O1 and O2 Advanced LIGO/Virgo runs, the farthest BH-BH event is at $z\simeq 0.5$ and, at final target sensitivity, 2G detectors should reach $z\simeq 1$. The corresponding rates will be of order $10^6$ events per year. For binary neutron stars, ET will detect them out to  $z\simeq 2-3$, which allows us to reach the peak of the star formation rate and therefore detect the vast majority of coalescing BNS throughout the Universe; by comparison,  at final target sensitivity, 2G detectors should reach $z\simeq 0.2$. The expected rate of BNS at ET was  computed in \cite{Belgacem:2019tbw} using state-of-the art models for the formation and evolution  of neutron star binaries, and is found to be 
between $6.2\times 10^4$ and $6.9\times 10^4$ events per year, having assumed a duty cycle of $80\%$. This corresponds to  $(0.8-0.9)\times 10^5$ events in one year of actual data.\footnote{Previous estimates for BNS 
\cite{Sathyaprakash:2009xt} were slightly higher, ${\cal O}(10^5-10^6)$~BNS/yr. This is partly due to the fact that in  \cite{Belgacem:2019tbw} has been used a threshold of 12 for the network SNR, obtained by combining the three arms of ET, while previous work typically used a  threshold of 8.}
In order to use these BNS as standard sirens one  either needs an electromagnetic counterpart, or one must use statistical methods. Here we focus on BNS with electromagnetic counterparts. We consider for definiteness mock catalog for ET, but similar estimates hold for CE.\footnote{Cosmic Explorer can reach a much greater distance for BNS, up to $z\simeq 8$. However, since the peak of the star formation rate is at $z\sim 2-3$, 
most of the coalescing  BNS will be seen already at the distances accessible to ET. Furthermore, beyond
$z\sim 1.5-2$ it will be  very difficult to detect an electromagnetic counterpart even with a GRB.
Thus, for BNS with counterpart the estimates for CE will be basically the same as for ET.
See also \cite{Belgacem:2019tbw} for the prediction  of BNS rates in a network with two CE and one ET detector.}

Refs.~\cite{Stratta:2017bwq,Belgacem:2019tbw} have estimated
estimated  the expected number and the redshift distribution of coincidences between  GW events at ET  and the electromagnetic signal observed at a GRB detector with the characteristics of the proposed  THESEUS mission~\cite{Amati:2017npy,Stratta:2018ldl}, that could be in operation at the same time as 3G  detectors.
Depending on the assumptions made, the estimated number of joint GW-GRB detections is between
$O(15)$ and $O(50)$ per year.
In Table~\ref{tab:relerr_ET_gaussian_real} (from  \cite{Belgacem:2019tbw}) we show some properties of a sample catalog, obtained assuming 10 yr of data taking. More counterparts could be obtained from future large telescopes that will be able to monitor large regions of the sky from the radio, optical to the X-ray (see \cite{Sathyaprakash:2019rom,Maggiore:2019uih} for discussion), although realistic estimates are difficult to obtain because they also depend on issues such as the prioritization  that will be given to the follow-up of GW signals.

\begin{table}[]
\centering
\begin{tabular}{|c|c|c|c|c|}
\hline
redshift  &       number of joint         & mean & mean  & \\
    bin        &  GW-GRB events &  redshift    &      $\Delta \dgw/\dgw$   &$\Delta\Xi_0$  \\ \hline
(0 , 0.1) & 4 &0.07108 & 0.00868  &0.11\\
(0.1 , 0.2) & 24 &0.15001 &0.01784 & 0.09\\
(0.2 , 0.3) & 24 &0.24043 &0.02558 &0.09\\
(0.3 , 0.4) & 27 &0.35355 & 0.03529 &0.09 \\
(0.4 , 0.5) & 28 &0.44966 & 0.04843 &0.10\\
(0.5 , 0.6) & 9 & 0.53785 & 0.05646  &0.10\\
(0.6 , 0.7) & 14 & 0.64540 &0.05329 &0.09\\
(0.7 , 0.8) & 13 &0.73793 &0.05493 &0.09\\
(0.8 , 0.9) & 8 &0.85497 &0.06413  &0.10\\
(0.9 , 1.0) & 4 & 0.93702 & 0.06257 &0.09\\
(1.0 , 1.1) & 6 & 1.05334 & 0.06494 &0.09\\
(1.1 , 1.2) & 3 &1.15162 &0.06749 &0.09\\
(1.2 , 1.3) & 1 &1.25943 & 0.07373 &0.10 \\
(1.3 , 1.4)& -- &-- & -- &--\\
(1.4 , 1.5) &2 & 1.45375 &0.07851&0.10\\
(1.5 , 1.6) & 1 & 1.58407 &0.07577&0.09 \\
(1.6 , 1.7) & 1 & 1.62843 & 0.07947 &0.10\\
\hline
\end{tabular}
\caption{Number of event and  mean value  of the observational error   $\Delta\dgw/\dgw$ in different redshift bins,  for  a specific realization of the catalog of joint GW-GRB detections, assuming 10 yr of data (from  \cite{Belgacem:2019tbw}). In the last column we give an estimate of the error on $\Xi_0$ from an individual source in the given frequency bin.
\label{tab:relerr_ET_gaussian_real}}
\end{table}

We can now estimate the accuracy that can be obtained on modified GW propagation from individual events such as those in Table~\ref{tab:relerr_ET_gaussian_real}. In the lowest redshift bin, say $z\, \lsim \, (0.1-0.2)$, we are in a situation similar to that studied above for 2G detectors, and we can use $\delta(0)$ as observable. However, now
$\Delta \dgw/\dgw$ is below $1\%$, and one can also easily imagine that, by the time ET will be operative, the accuracy on $H_0$ will have further improved, so we assume $\Delta \dem/\dem =0.5\%$. We then find  that each single event with counterpart at, say, $z=0.2$, will allow us to measure $\delta(0)$ to an accuracy of about $0.15-0.20$. Comparing with the predictions in Table~\ref{tab:Xi0n}, we see that just one event will be sufficient to detect at $5\sigma$ the predictions of the RT model with large $\Delta N$!

For sources at not too small redshift, we must rather use the full expression (\ref{eq:fitz}). According to 
Table~\ref{tab:Xi0n}, we fix $n=2$ (the precise value has limited importance for the analysis)
and keep only  $\Xi_0$ as the parameter labeling the predictions. From \eq{eq:fitz},
\be
\frac{\Delta\dgw}{\dgw}+\frac{\Delta\dem}{\dem}=\Delta\Xi_0 \(1-\frac{1}{(1+z)^n}\)\, .
\ee
Setting for definiteness $\Delta \dem/\dem=0.5\%$, for a single source with a  redshift given by the third column in  Table~\ref{tab:relerr_ET_gaussian_real} and a value of $\Delta \dgw/\dgw$ as in the fourth column, we get the accuracy $\Delta\Xi_0$ given in the last column. We see that, independently of redshift, each individual detection would provide a measurement of $\Xi_0$ at the $(9-10)\%$ level. Thus, 
the predictions for $\Xi_0$ given in Table~\ref{tab:Xi0n}, that  for large $\Delta N$ differ by the GR result by as much as  $80\%$, could be tested at more than $5\sigma$ with just a single joint GW-GRB detection. Combining the errors on $\Xi_0$ from each of the 169 events in Table~\ref{tab:Xi0n} (taking into account the number of events per bin) we get an overall error $\Delta\Xi_0\simeq 0.7\%$.
This agrees with the result obtained in \cite{Belgacem:2019tbw},
using the full catalogs corresponding to 10 yr of data, and performing a MCMC to take into account more precisely  the degeneracies
with $H_0$, $\oma$ and $w_0$ (and using, conservatively, the current datasets on  CMB, BAO and SNe), where it was found   that $\Xi_0$ can be determined to an accuracy of  $1\%$. Clearly, with respect to the size of the deviation of $\Xi_0$ from the GR value, which can be as large as $80\%$, this is a remarkable precision.

\begin{figure}[t]
\centering
\includegraphics[width=0.45\textwidth]{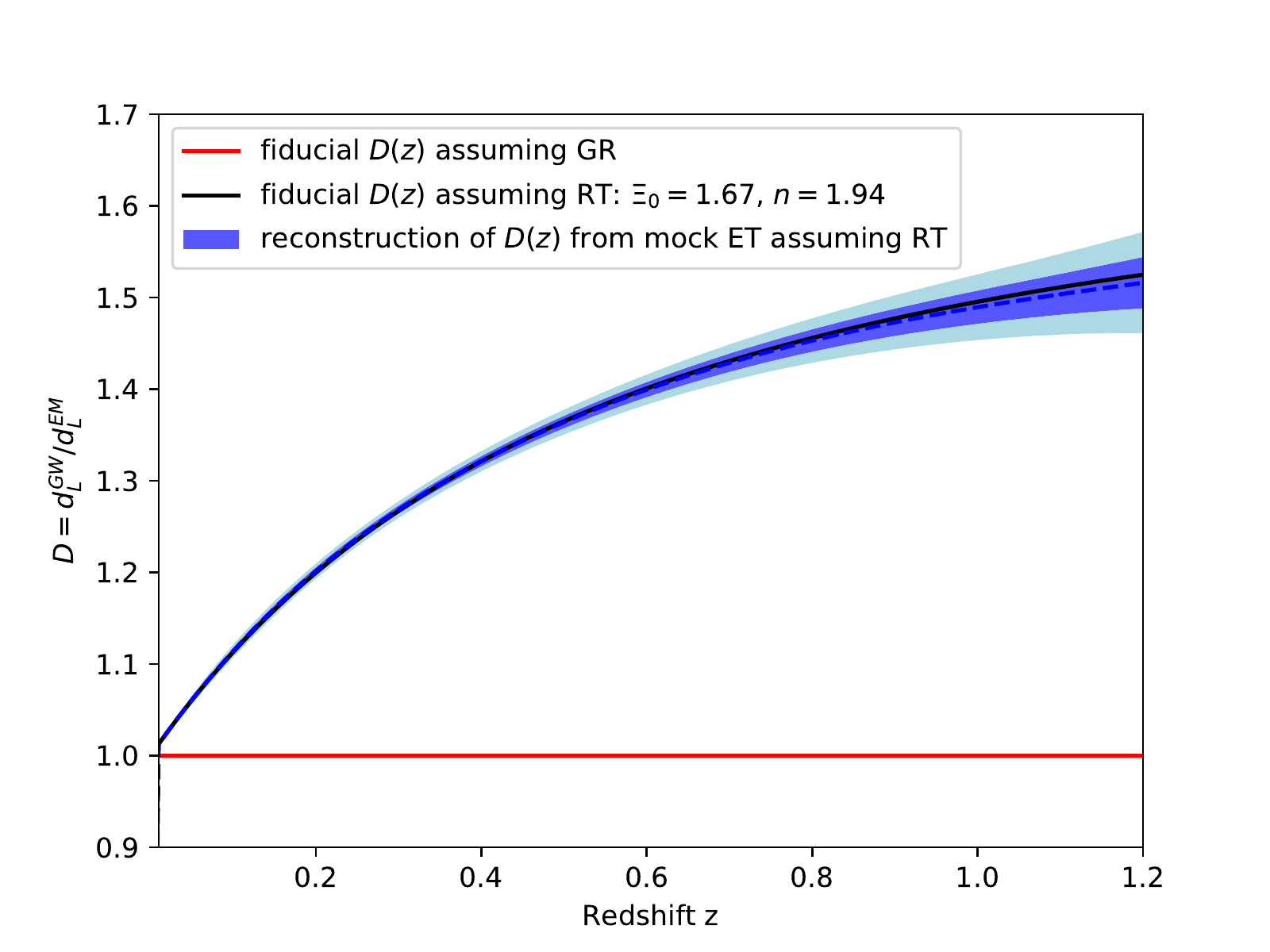}
\caption{As in Fig.~\ref{fig:results_MGHLVKI}, using the mock ET and DES catalogs for the RT fiducial cosmology. From \cite{Belgacem:2019zzu}.}
\label{fig:results_MGET}
\end{figure}

Fig.~\ref{fig:results_MGET} shows the result of a gaussian process reconstruction of $\dgw(z)/\dem(z)$
using mock ET and DES catalogs, using as  fiducial the RT model with $\Delta N=64$. We see that, for a deviation from GR of this size,
$\dgw(z)/\dem(z)$ would be reconstructed with exquisite precision, allowing not only to prove GR wrong on cosmological scales, but also to pinpoint with accuracy the properties of the alternative model, in this case the parameter $\Delta N$ that characterizes the RT model.

\subsubsection{LISA}

A   study of the accuracy to $\Xi_0$ that could be obtained with the space interferometer LISA has been presented in \cite{Belgacem:2019pkk}. In this case  the coalescences of  supermassive BH binaries plays the role of standard sirens, since they are believed to  merge in a gas rich environment that is expected to  power electromagnetic emission, resulting in a detectable electromagnetic counterpart. The corresponding mock catalogs were generated by using advanced models for galaxy formation and merger, and  different scenarios for the seeds of the
massive black holes  and for the delays between galaxy merger and massive black hole merger, resulting in three main scenarios (heavy seeds and no delay, heavy seeds with a specific prescription for the delay, and light seeds due to pop III stars). For each scenario were simulated 22  catalogs corresponding to the nominal 4~yr of the LISA mission. For each catalog was then performed a quick Fisher matrix cosmological analysis assuming $\Lambda$CDM, and was then  selected, for each scenario, the median catalog among all ranked 22 catalogs as the representative catalog for the corresponding astrophysical model. 
The corresponding catalogs had
 32, 12 and 9  sources  for heavy seeds and no delay, heavy seeds with delay, and light seeds, respectively.
On these catalogs, were then run full MCMC to constrain $\Xi_0$, including again CMB, BAO and SNe to reduce the degeneracies with $\oma$, $H_0$ and $w_0$. Furthermore, two different assumptions on the accuracy of the estimate of the source redshift were considered, and denoted as `optimistic' and `realistic', respectively.
The resulting estimate was $\Delta\Xi_0=(1-2)\%$ with the optimistic assumption on the estimate of the source redshift, and
$\Delta\Xi_0=(2-4)\%$ with the realistic assumption. In all cases, we are again in a situation where the predictions of the RT model  given in Table~\ref{tab:Xi0n}, that for
large $\Delta N$ differ from GR between $30\%$ and $80\%$, are very clearly detected. 

While the contribution of a single source cannot be obtained from a MCMC (where a large number of sources is obtained to get the convergence of the chains), it can be estimated by comparing the results from catalogs with a different number of sources. For the realistic scenario for the estimate of the source redshift, 
from Table~2 of \cite{Belgacem:2019pkk} we see that LISA could measure $\Xi_0$ with an error
$\Delta\Xi_0\simeq\{0.023,0.036,0.044\}$ in the three catalogs containing, respectively, $N=\{32,12,9\}$ events. These numbers are well reproduced by~\cite{Belgacem:2019lwx}   
\be
\Delta\Xi_0\simeq 0.13/\sqrt{N}\, ,
\ee
so  each SMBH event gives a measure of $\Xi_0$ with an average accuracy of about $13\%$.
Using the optimistic scenario for redshift determination we rather get
$\Delta\Xi_0\simeq 0.06/\sqrt{N}$.
Thus, even a single SMBH event at LISA could be sufficient to detect the effect predicted by the RT model with large $\Delta N$.

\begin{figure}[t]
\centering
\includegraphics[width=0.45\textwidth]{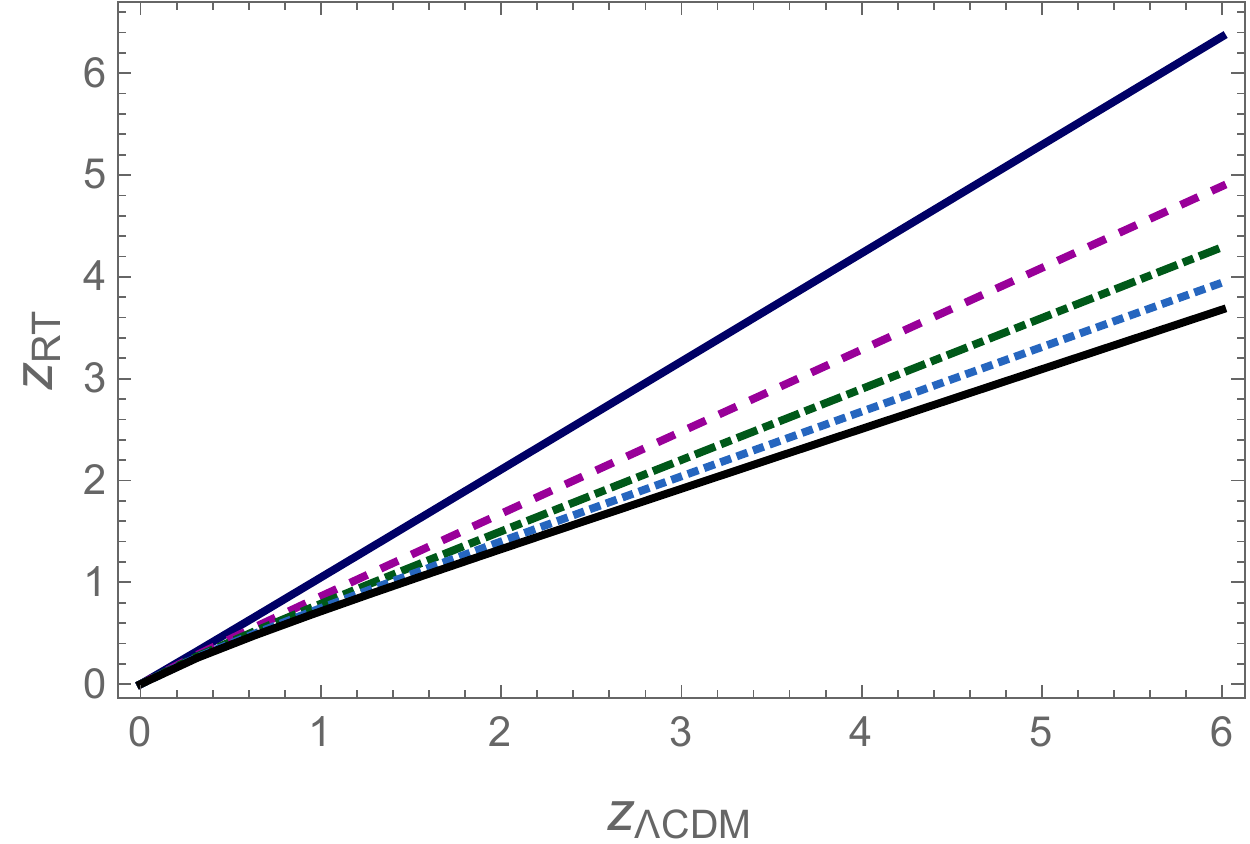}
\includegraphics[width=0.45\textwidth]{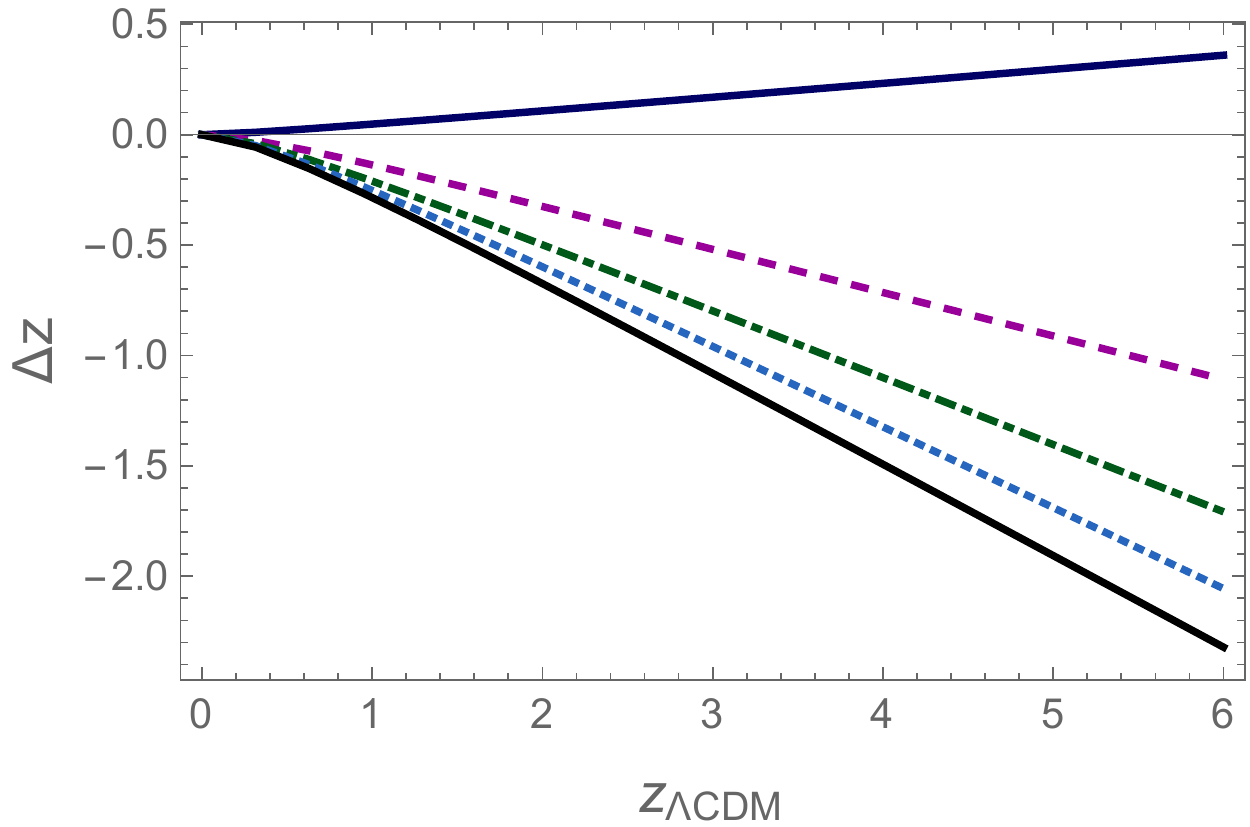}
\caption{$z_{RT}$ as a function of $z_{\Lambda{\rm CDM}}$ (left panel) and the difference
$\Delta z=z_{RT}-z_{\Lambda{\rm CDM}}$, as a function of $z_{\Lambda{\rm CDM}}$ (right panel)
for the minimal RT model (blue solid line) and  for RT with $\Delta N=34$ (magenta, dashed), $\Delta N=50$ (green, dot-dashed), $\Delta N=64$ (cyan, dotted) and  $\Delta N=100$  (black solid line).
}
\label{fig:zRTvszLCDM}
\end{figure}

As already  pointed out in the discussion after \eq{zRTD},
another aspect of the effect is that, if the RT model provides the correct description of Nature, the actual redshift of the source, as determined through electromagnetic observation, would turn out to be very different from that inferred from a measurement of $\dgw$ and interpreted 
using $\Lambda$CDM. For LISA this effect is particularly remarkable since, as we see from Fig.~12 of  \cite{Belgacem:2019pkk}, the mock catalogs of supermassive BH binary coalescences observed at LISA include events up to   $z\simeq 6$ and higher. This effect is shown in Fig.~\ref{fig:zRTvszLCDM}, where the left panel shows that redshift of the source predicted by the RT model as a function of the redshift predicted by  $\Lambda$CDM, and the right panel shows the difference $\Delta z=z_{RT}-z_{\Lambda{\rm CDM}}$, as a function of $z_{\Lambda{\rm CDM}}$ (i.e. the same as Fig.~\ref{fig:Deltaz_vs_z_max}, but on a range of redshifts appropriate to supermassive BH binaries at LISA).    For instance, for a given measurement of the luminosity distance through GWs for which $\Lambda$CDM would predict, say, a redshift $z=6$, the RT model with very large $\Delta N$ rather predicts that the source will be found, by electromagnetic observations, at $z\simeq 3.7$, a rather striking difference.

\section{Conclusions}

We have discussed in detail a modification of gravity on cosmological scales, summarizing and extending 
previous work by our group. The model is based on a clear and well-defined theoretical framework. Rather than introducing extra degrees of freedom, such as extra scalar, vector or tensor  fields, or extra polarization for the graviton, as in typical modified gravity models, the basic idea is that  long-distance modifications to the dynamics of gravity are induced by infrared quantum effects in GR itself. This means that the proper tool is no longer the action of the theory but the corresponding quantum effective action. In quantum field theory, at a fundamental level,  actions are local functionals of the fields; however,
whenever the theory contains massless particles (such as the graviton in GR), or particles that are light with respect to the relevant energy scale, the corresponding quantum effective action also contains non-local terms. We have seen how, with non-local terms, we can construct a mass  term for a gauge field without violating gauge invariance, and  mass terms for the gravitational field that do not violate diffeomorphism invariance.  Linearizing  the theory  over flat space, we have seen that two independent mass terms can be constructed: one for the conformal mode, and one for the transverse-traceless mode $\hTTij$, see \eq{Gamma2hhssm}. Our basic assumption is that  the conformal mode indeed becomes massive, while  $\hTTij$ remains massless. 

The idea that the conformal mode becomes massive (while  $\hTTij$ stays massless) currently has the status of a conjecture, which  is  difficult to verify from first principles since it involves non-perturbative physics. Nevertheless, we have seen that numerical results from Euclidean quantum gravity on the lattice and from causal dynamical triangulations, as well as  analytic computations using functional  renormalization group equations, give some support for the hypothesis of a dynamical mass generation. The fact that the resulting mass scale is indeed associated to the conformal mode is also suggested again by numerical results from causal dynamical triangulations, and by several arguments that show that the conformal mode is the most problematic one in the infrared.
Once one assumes the validity of this conjecture, leading to the linearized action (\ref{SEH2nlmass}), 
the covariantization of this linearized nonlocal theory leads quite naturally to two different possibilities, that we have called the RT and RR models. We have seen that eventually the RR model is ruled out phenomenologically, while the RT model (\ref{RT}) has been the main focus of our paper.

We have then explored in details the observational consequences of the RT model. Constructing a model that  fulfills all observational constraints and gives predictions testable in the near future is in general very difficult, as has been learned from the explicit study of several modified gravity models. Here, one should also appreciate that, once accepted the underlying assumptions spelled out above, the theory is basically fixed (apart from the two options given by the RR and RT model), 
and has the same number of parameters as $\Lambda$CDM, with a new mass scale $m$ replacing the cosmological constant (plus, as we have seen for the RT model, a single constant $\Delta N$ which reflects all our ignorance on initial conditions). Thus, the theory either is consistent with observations  or it doesn't. We do not have the freedom of playing with arbitrary functions, as for instance in scalar-tensor theories of the Horndeski type. 

At the phenomenological level, the RT model turns out to have  a number of remarkable properties:

\begin{itemize}

\item Its cosmological solutions, at the background level, show an accelerated expansion at the present cosmological epoch, without the need for a cosmological constant. In other words, giving a mass to the conformal mode provides a possible explanation for the observed accelerated expansion of the Universe and for the origin of dark energy. 

\item The fact that dark energy starts to dominate just at the present cosmological epoch is obtained by choosing  a value for the mass scale $m$, or, more precisely, for the scale 
$\Lambda_{\rm\scriptscriptstyle RT} \sim  (\mplr m)^{1/2}$, which is the fundamental scale that is  generated dynamically. In this sense, the model does not solve the coincidence problem. However, 
even if $\Lambda_{\rm\scriptscriptstyle RT}$ cannot be predicted (just as we cannot predict the value of
$\Lambda_{\rm\scriptscriptstyle QCD}$ in strong interactions), the required numerical value, of the order  of the meV, is not particularly surprising from the point of view of quantum field theory (contrary, e.g., to theories that introduces a fundamental mass scale of order $H_0\sim 10^{-33}$~eV). Furthermore, 
such a dynamically generated mass scale  is  a renormalization group invariant, so there is no problem of technical naturalness.

\item Scalar perturbations over the FRW background are stable and remain small during the whole cosmological evolution. This is already a non-trivial property, that has ruled out many modified gravity models. Furthermore, the scalar perturbations of the RT model are very close to those of  $\Lambda$CDM, which in the end allows the model to fit current cosmological data well, while still being potentially distinguishable with future missions.

\item A full MCMC analysis shows that the model (for all values of $\Delta N$) fits CMB, BAO, SNe, measurements of $H(z)$  and structure formation data at the same level as $\Lambda$CDM.

\item The model reduces to GR at small scales, without the need of invoking non-linear screening mechanisms, and therefore passes all the constraints from solar system and laboratory experiments. It furthermore complies with the limit on the time variation of the Newton's constant from Lunar Laser Ranging. As we have seen,  this is  in general non-trivial even when the static solution has the correct GR limit (and, indeed, it is this bound that rules out the RR model).

\item The sector of cosmological tensor perturbations (i.e. GWs propagating over a FRW background) provides a great surprise. In the RT  model GW propagation across cosmological distances is different from GR, so that the relation between the luminosity distance extracted from a coalescing binary and the redshift is modified, giving rise to the notion of  `GW luminosity distance' $\dgw(z)$. This has been found to be common to all modified gravity models. What is remarkable for the RT model (in particular for large $\Delta N$) is the size of the effect, that, at the redshifts accessible to future GW detectors such as third-generation ground based detectors such as Einstein Telescope and Cosmic Explorer, or the space interferometer LISA,  
could lead to deviations from GR as large as $80\%$. We have seen that, with these detectors, even the detection of a  single standard siren with electromagnetic counterpart would be sufficient to detect the effect at more than $5\sigma$. The effect is smaller at the redshifts accessible to the second-generation network made by advanced LIGO/Virgo and KAGRA, but still could be potentially within reach even at these detectors, over several years of data taking.

\end{itemize}

\vspace{5mm}\noindent
{\bf Acknowledgments.} 
The work  of E.B., A.F., S.F. and M.M. is supported by the  Swiss National Science Foundation and  by the SwissMap National Center for Competence in Research. The work of Y.D. is supported by  Swiss National Science Foundation and by  a Consolidator Grant of the European Research Council (ERC-2015-CoG grant 680886). We thank Jes\'us Torrado for his excellent guidance with using Cobaya.

\appendix

\section{Difficulties of alternative nonlocal models}\label{sect:difficulties}

A natural question is whether it is possible to construct other nonlocal models that share the good phenomenological properties of the RT model. We will see that, in fact, this is very difficult, and this will allow us to better appreciate the results presented above. To organize the discussion, it can be useful to follow the path that actually lead to the formulation of the RT model and of other variants, and see what conditions eliminated the various alternatives (see also \cite{Maggiore:2016gpx}).

A first nonlocal model associated to a mass scale was proposed, on purely phenomenological grounds, by Arkani-Hamed,  Dimopoulos,  Dvali, and Gabadadze~\cite{ArkaniHamed:2002fu}  and consisted in modifying the Einstein equations  into
\be\label{degrav}
\(1-\frac{m^2}{\Box}\)\Gmn=8\pi G\,\Tmn\, ,
\ee
where $m$ is the new mass scale.\footnote{Actually, in \cite{ArkaniHamed:2002fu} the model was  presented as a modification of GR that is acausal on cosmological scales. As we have discussed
in sect.~\ref{sect:causality}, causality is, however, preserved once this is  understood as the equation of motion derived from a quantum effective action for the in-in vacuum expectation value of the metric, which automatically ensures that the Green's function in the $\iBox$ operator is the retarded one.}
This model was  proposed  to introduce the 
degravitation idea, namely the idea the vacuum energy density, even if present, does not gravitate. In fact, at least performing naively the inversion of the nonlocal operator, \eq{degrav} can be rewritten as $\Gmn =8\pi G\, (\Box-m^2)^{-1}\Box\Tmn$. Therefore the low-momentum modes of $\Tmn$ are filtered out and in particular a term in $\Tmn$ due to a cosmological constant does not contribute.

However, a  drawback of \eq{degrav} is that the energy-momentum tensor is not automatically conserved, since in curved space  $\n^{\mu}$ does not commute with $\Box$ and therefore with $\iBox$. As a consequence,  the Bianchi identity $\n^{\mu}\Gmn=0$ no longer ensures $\n^{\mu}\Tmn=0$. In \cite{Jaccard:2013gla} it was then observed that it is possible to cure this problem by making use of the decomposition (\ref{transv}) to extract the transverse part of the tensor
$\iBox\Gmn$.
 One can then modify \eq{degrav}  into
\be\label{GmnT}
\Gmn -m^2\(\iBox\Gmn\)^{\rm T}=8\pi G\,\Tmn\, ,
\ee
and energy--momentum conservation  becomes automatic. However, the  cosmological evolution of this model  turned out to be unstable, already at the background level~\cite{Maggiore:2013mea,Foffa:2013vma}, so this model  is not phenomenologically viable. This instability is due to the fact that, once written the model in local form introducing some auxiliary fields, the latter have unstable modes already during RD and MD (see in particular app.~A of \cite{Foffa:2013vma}). Then, any small deviation from the standard FRW solution will quickly be amplified and lead to a completely different evolution, inconsistent with the observations.
It was then realized in \cite{Maggiore:2013mea} that this instability is related to the action of the $\iBox$ operator on a tensor such as $\Gmn$ or $\Rmn$, and is absent if it acts on a scalar such as $R$. This led to the RT model (\ref{RT}). Trying to work out a similar model at the level of the quantum effective  action, rather than of the equations of motion, led to the RR model ({\ref{RR})~\cite{Maggiore:2014sia}, which, at least at the level of cosmology, shared all good properties of the RT model. Its  cosmological solutions were studied in detail in
\cite{Maggiore:2014sia,Maggiore:2016gpx,Belgacem:2017cqo} (see also \cite{Nersisyan:2016hjh} for a different branch of solutions).

A natural generalization of the RR model is  given by the quantum effective action
\be\label{actionTotal}
\Gamma=\frac{\mplr^2}{2}\int d^4 x \sqrt{-g}\,
\left[R-\mu_1 R\frac{1}{\Box^2}R-\mu_2 C^{\mu\nu\rho\sigma}\frac{1}{\Box^2}C_{\mu\nu\rho\sigma}-\mu_3\RMN\frac{1}{\Box^2}\Rmn
\right]\,,
\ee
where $\mu_1$, $\mu_2$ and $\mu_3$ are  parameters with dimension of  $({\rm mass})^2$, and $C_{\mu\nu\rho\sigma}$ is the Weyl tensor. This extension was studied in \cite{Cusin:2015rex}, where it was found that
the term $\RMN\Box^{-2}\Rmn$ is ruled out since it gives  instabilities in the cosmological evolution at the background level, again due to the behavior of the auxiliary fields. The Weyl-square term instead does not contribute to the background evolution, since the Weyl tensor vanishes in FRW, and it also has well-behaved scalar perturbations. However, quite interestingly, it was ruled out by the fact that its tensor perturbations are unstable, showing the the stability of perturbations is in general a non-trivial requirement both in the scalar and in the tensor sector.

The realization that both models that survived, RT and RR, had the physical meaning of a mass for the conformal mode~\cite{Maggiore:2015rma} then suggested to focus the attention on models with such a meaning. One interesting variant of the RR model  is given by
\be\label{actDelta4}
\Gamma_{\Delta_4}=\frac{\mplr^2}{2}\int d^4x\, \sqrt{-g}\, \[ R-\frac{m^2}{6} R\frac{1}{\DP} R\]\, .
\ee
where $\DP$ is the Paneitz operator,
\be\label{defDP}
\Delta_4\equiv\Box^2+2\RMN\n_{\mu}\n_{\nu}-\frac{2}{3}R\Box+\frac{1}{3}\gMN\n_{\mu} R\n_{\nu}\, ,
\ee
and whose linearization over Minkowsli space is the same as the RT or RR models.
This is the operator that enters in the conformal anomaly in four dimensions and, from the point of view of conformal invariance, is the natural generalization of the d'Alembertian from two to four dimensions.  This model had a viable cosmological evolution \cite{Cusin:2016nzi}, although its prediction for the equation of state of DE, $w_0\simeq -1.31$, already seemed off with respect to the observations, as was indeed confirmed from a MCMC analysis in \cite{Belgacem:2017cqo}. In any case, what definitely ruled out the model was the realization that, in the tensor sector,  it predicts a speed of GWs different from the speed of light~\cite{Belgacem:2017cqo}. These examples show that the requirements of having a  viable background evolution, stable  scalar perturbations, good fit to the cosmological observations, stable  tensor perturbations, and $c_{\rm gw}=c$,  all provide  non-trivial tests, potentially able to rule out a model.

Finally, as we mentioned in section~\ref{sect:LLR}, limits on the time variation of Newton's constant ruled out also the RR model \cite{Barreira:2014kra,Belgacem:2018wtb}. The detailed analysis in \cite{Belgacem:2018wtb} is very general, and makes it clear that the same situation will happen in any nonlocal model in which the effective Newton's constant at short scales depends on the auxiliary fields of the theory. This therefore  applies also to  models of the form
\be\label{Gm2n}
\Gamma=\frac{\mplr^2}{2}\int d^{4}x \sqrt{-g}\,\[ R- \(\frac{m^2}{\Box}\)^n R\]\, ,
\ee
corresponding to a running of Newton's constant, that 
have been studied, at the level of  background evolution,  in \cite{Vardanyan:2017kal} for $n=1$ and in \cite{Amendola:2017qge} for $n=2$, where it was found  that their cosmological evolution  appears to be in principle viable, at least at the background level (in particular the model with $n=1$ has an evolution very close to that of the RR model, up to the present epoch). In contrast, we have seen that the RT model passes the LLR limit because at short scales $G_{\rm eff}$ loses any dependence on the auxiliary fields and  reduces to $G$, see \eq{GeffGRTlargek}. 

Last but not least, a different but related line of research is given by non-local models that are not associated to a mass scale, and whose development predated that of the nonlocal models associated to a mass scale on which we have focused. The underlying physical motivation is again that IR divergences could  generate, through non-perturbative effects, the relevant nonlocal terms in the quantum effective action. The first nonlocal gravity model of this type was proposed by Wetterich~\cite{Wetterich:1997bz}, and was based on the quantum effective action, 
\be\label{GammaWett}
\Gamma=\frac{\mplr^2}{2}\int d^4x \sqrt{-g}\,\[ R-\lambda R\iBox R\]\, .
\ee
Since $\iBox R$ is dimensionless, the associated  constant $\lambda$ is also dimensionless. The model, however, did not produce a viable cosmological evolution~\cite{Wetterich:1997bz}. Deser and Woodard \cite{Deser:2007jk,Deser:2013uya}  (see \cite{Woodard:2014iga} for review) considered a more general nonlocal model of the form 
\be\label{GammaDW}
\Gamma_{\rm DW}=\frac{\mplr^2}{2}\int d^4x \sqrt{-g}\,\[ R- Rf(\iBox R)\]\, ,
\ee
with $f(X)$ a dimensionless function, which  was tuned so to obtain the desired background evolution. Requiring that the cosmological evolution closely mimics that of $\Lambda$CDM leads to a function well-fitted by~\cite{Deffayet:2009ca}
\be\label{f(X)}
f(X)=0.245\[ \tanh\(0.0350(X+16.5)+0.032 (X+16.5)^2+0.003 (X+16.5)^3\)-1\]\, ,
\ee
for $X\equiv \iBox R <0$. To comply with solar system constraints, the proponents of the model set $f(X)=0$ for $X>0$.  The argument suggested in  ref.~\cite{Deser:2013uya,Woodard:2014iga} for this choice was that   in a cosmological setting  (where the time derivative dominates) $X\simeq (-\pa_t^2)^{-1}R$ is negative   because of the minus sign in  $-\pa_t^2$, while it is positive in the regime dominated by structure formation, where the spatial derivatives dominate and $X\simeq (\n^2)^{-1}R$.
It was however shown in \cite{Belgacem:2018wtb} that this is not correct, and $X$ is always negative, even in a static situation. Thus, the Deser-Woodard model lacks a screening mechanism and is ruled out by the comparison with observations at the solar system scale. In \cite{Deser:2019lmm} then  Deser and Woodard proposed a variant of their model constructed with a function $f(Y)$ of the variable
$Y=\iBox\gMN\pam X\pan X$, where again $X= \iBox R$, which indeed changes sign between static and time-dependent solutions, and again postulated that $f(Y)=0$ for $Y<0$. The consequences of the model have not been throughly investigated, in particular stability of the solutions, etc.; however, apart from a certain convolutedness of the model, one can anticipate potential problems with LLR similar to those that ruled out the RR model.


\bibliographystyle{utphys}
\bibliography{myrefs}

\end{document}